\documentclass{article}
\pdfoutput=1

\usepackage{arxiv}
\usepackage[utf8]{inputenc}
\usepackage[T1]{fontenc}    % use 8-bit T1 fonts

\usepackage[sort&compress,numbers]{natbib}

\usepackage{graphicx}
\usepackage{subfig}       % subfloat
\graphicspath{{figures/}} % list folders that latex will search for images

\usepackage{mathtools}  % loads »amsmath«
\usepackage{psfrag}       % psfrag for eps editing
\usepackage{amsmath,bm,amsfonts,mathrsfs}   % math related packages
\usepackage{shortcuts}    %user defined shortcuts
\usepackage{color}

\usepackage{hyperref}       % hyperlinks
\usepackage{url}            % simple URL typesetting
\usepackage{enumitem}

\usepackage{algorithm,algorithmic}

\newcommand*\lbar[1]{\overline{#1}}

\usepackage[color=green!20]{todonotes}

\title{Bayesian neural networks for weak solution of PDEs with uncertainty quantification}

\author{
  Xiaoxuan Zhang$^1$, ~Krishna Garikipati$^{1,2,3}$ \thanks{Corresponding author. E-mail address: krishna@umich.edu \hfill \today} \\[3mm]
  $^1$Department of Mechanical Engineering, University of Michigan, United States \\
  $^2$Department of Mathematics, University of Michigan, United States \\
  $^3$Michigan Institute for Computational Discovery \& Engineering, University of Michigan, United States \\
}

\begin{document}

\maketitle

\begin{abstract}
  Solving partial differential equations (PDEs) is the canonical approach for understanding the behavior of physical systems.
However, large scale solutions of PDEs using state of the art discretization techniques remains an expensive proposition.
In this work, a new physics-constrained neural network (NN) approach is proposed to solve PDEs without labels, with a view to enabling high-throughput solutions in support of design and decision-making.
Distinct from existing physics-informed NN approaches, where the strong form or weak form of PDEs are used to construct the loss function, we write the loss function of NNs based on the discretized residual of PDEs through an efficient, convolutional operator-based, and vectorized implementation.
We explore an encoder-decoder NN structure for both deterministic and probabilistic models, with Bayesian NNs (BNNs) for the latter, which allow us to quantify both epistemic uncertainty from model parameters and aleatoric uncertainty from noise in the data.
For BNNs, the discretized residual is used to construct the likelihood function.
In our approach, both deterministic and probabilistic convolutional layers are used to learn the applied boundary conditions (BCs) and to detect the problem domain.
As both Dirichlet and Neumann BCs are specified as inputs to NNs, a single NN can solve for similar physics, but with different BCs and on a number of problem domains. 
The trained surrogate PDE solvers can also make interpolating and extrapolating (to a certain extent) predictions for BCs that they were not exposed to during training.
Such surrogate models are of particular importance for problems, where similar types of PDEs need to be repeatedly solved for many times with slight variations. 
We demonstrate the capability and performance of the proposed framework by applying it to different steady-state and equilibrium boundary value problems with physics that spans diffusion, linear elasticity, and nonlinear elasticity.

\end{abstract}


\section{Introduction}

Solving partial differential equations (PDEs) is crucial for many scientists and engineers to understand the behavior of different physical systems.
Popular numerical methods to solve PDEs include, but are not limited to, the finite element method (FEM), finite difference method, finite volume method, Fourier method, and other methods, where each has its own advantages and limitations.
Among those methods, FEM is arguably the most widely used method due to its flexibility for solving problems with complex geometrical and irregular shapes. 
However, when a large number of simulations need to be carried out, such as homogenization, optimization, or inverse problems, these numerical methods can sometimes be very expensive, thus efficient surrogate models are needed.

With the drastically increasing computational power of graphics processing units (GPUs), machine learning has gained great popularity in a wide range of applications in fields, such as computer vision, speech recognition, and anomaly detection, etc. 
It has also emerged as a powerful approach among data-driven methods for surrogate modeling in many material, physics, and engineering applications \cite{Ramprasad2017-material-informatics-review,Bock2019Kalidindi-ML-CM-review}, such as material screening \cite{Meredig2014-screen-materials,Wolverton2016Ward-screen-material-property}, constitutive modeling \cite{Hashash2004-NN-constitutive,Chinesta2019Ibanez-hybrid-constitutive-modeling,Sun+Wang2019-game-constitutive,Huang2020Learning-constitutive-DNN-indirect-observation}, scale bridging \cite{Brockherde2017DFT-MD,Teichert2019Garikipati-ML-bridge,Teichert2020Scale-bridge-active-learning}, and system identification \cite{Brunton2016Kutz-system-id,Wang2019Garikipati-CMAME-Variational-System-Identification,Wang2020System-inference-Covid19,Wang2020Variational-System-ID-sparse-data,Wang2020Garikipati-Xun-perspective-system-id-bayesian}, effective material properties prediction \cite{Kalidindi+Cecen2018-CNN-Structure-property,Li2019Zhuang-effective-ME-DNN,Agrawal2018Yang-Composites-S-P-deep-learning,Kondo2017CNN-ionic-conductivity}, non-linear material response characterization \cite{Hambli2011multiscale-bone-with-NN,Bessa2017-data-driven-framework-elasticity-inelasticity,Jones2019Frankel-Oligocrytal-behavior-CNN,Sun2018Wang-homogenization,Yvonnet2015Le-RVE-elasticity,Yvonnet2018Lu-NN-RVE-graphene,Zhang2020Garikipati-CMAME-ML-RVE}, etc. 
Attempts to use machine learning techniques to learn the full field solution of physical systems are explored in \cite{Zhu2018Zabaras-UQ-Bayesian-ED-CNN,Winovich2019Lin-ConvPDE-UQ,Bhatnagar2019CNN-encoder-decoder,Li2020Reaction-diffusion-prediction-CNN}, and many others.
For example, \cite{Zhu2018Zabaras-UQ-Bayesian-ED-CNN} proposed a Bayesian approach to convolutional neural networks (CNNs) to study the flow problem in heterogeneous media with uncertainty quantification. 
CNNs are also used to predict the velocity and pressure fields for aerodynamic problems in \cite{Bhatnagar2019CNN-encoder-decoder} and predict concentration distribution for single-species reaction-diffusion systems in \cite{Li2020Reaction-diffusion-prediction-CNN}.
Those approaches generally require a large amount of full-field solutions as the training data, either from experiments or direct numerical simulations (DNSs), which might not be easily available or expensive to obtain. 

Another thrust in using machine learning techniques in the computational mechanics, materials, and physics community is to solve PDEs with little or no pre-labeled data \cite{Lagaris1998NN-PDEs,Han2018Solve-high-dimension-PDE-deep-learning,Sirignano2018DGM-solve-PDEs,Raissi2019physics-informed-forward-inverse-jcp,Zhu2019Perdikaris-Physics-PDE-CNN,Geneva2020Zabaras-JCP-auto-regressive-NN-PDE,Yang2021BPINNs-PDE,Berg2018unified-deep-ann-PDE-complex-geometries,Sun2020Surrogate-modeling-flow-physics-constrained,Samaniego2020energy-approach-PDE-ML}.
For example, high-dimensional free-boundary PDEs are solved with fully connected NNs by reformulating PDEs as backward stochastic differential equations in \cite{Han2018Solve-high-dimension-PDE-deep-learning}. 
The. Deep Galerkin Method proposed in \cite{Sirignano2018DGM-solve-PDEs} is used to solve high-dimensional free-boundary PDEs with neural networks (NNs), which satisfy the differential operators, initial condition (IC), and boundary conditions (BCs).
The Physics-informed Neural Networks (PINNs) approach has been proposed to solve different transient systems \cite{Raissi2019physics-informed-forward-inverse-jcp}.
In this approach, the strong form of PDEs, either using the continuous or discrete time derivative, is constructed and serves as part of the loss function, which further consists of contributions from the IC and BCs \cite{Raissi2019physics-informed-forward-inverse-jcp}.
Various extensions of PINNs have been made to solve different systems \cite{Jin2020NSFnets-PINN,Pang2019fPINNs,Meng2020PPINN,Geneva2020Zabaras-JCP-auto-regressive-NN-PDE,Yang2021BPINNs-PDE,Wang2020DL-free-boundary,Jagtap2020cPINN-discrete-domain-conservation-law}.
Surrogate PDE solvers based on the weak/variational formulation are also studied in \cite{Zang2020Weak-adversarial-network-PDE,Chen2020Friedrichs-learning-weak-PDE,Khodayi-mehr2019VarNet,Li2020D3M-domain-decomposition,Kharazmi2021hp-VPINNs}.

When solving PDE systems, it is important to quantify uncertainties from different sources, such as geometry representation, BCs, material parameters, and others, to better understand the systems and make reliable predictions. 
The sources of uncertainties can generally be categorized as either epistemic and aleatory, where the former can be reduced by gathering more data or using a more sophisticated model, and the latter is less prone to reduction \cite{Kiureghian2009Ditlevsen-aleatory-or-epistemic}.
With machine learning techniques, probabilistic models can be constructed to easily quantify the uncertainty.
Uncertainty quantification (UQ) with surrogate PDE solvers has been investigated in \cite{Karniadakis2019Zhang-Quantify-UQ-dropout,Yang2019Perdikaris-Adversarial-UQ,Geneva2020Zabaras-JCP-auto-regressive-NN-PDE,Yang2021BPINNs-PDE}.
For example, \cite{Karniadakis2019Zhang-Quantify-UQ-dropout} employs the dropout techniques proposed in \cite{Gal2016Dropout} to quantify uncertainties of a surrogate solver that combines  arbitrary polynomial chaos with PINNs.
In \cite{Yang2019Perdikaris-Adversarial-UQ}, probabilistic PINNs are constructed based on latent variable models and trained with an adversarial inference process for UQ.
A Bayesian framework is used in \cite{Geneva2020Zabaras-JCP-auto-regressive-NN-PDE} for UQ, where the posterior distribution of the surrogate model parameters is constructed based on the Stochastic Weight Averaging Gaussian technique proposed in \cite{Maddox2019SWAG}.
A Bayesian PINN is proposed in \cite{Yang2021BPINNs-PDE} for UQ, where PINNs are used to construct the likelihood function and either the Hamiltonian Monte Carlo or  variational inference (VI) techniques are used to estimate the posterior distribution.

With properly trained surrogate PDE solvers, there is interest in  using them in problems, such as homogenization or optimization, to rapidly predict the response of the same PDE systems, but with different IC or BCs, and potentially even on different problem domains.
However, such goals are in general difficult to achieve with existing surrogate approaches, which typically enforce only one specific set of BCs via the loss function. 
It is very challenging to make predictions for new sets of BCs with such surrogate solvers without re-training them.
In this work, we aim to address such challenges by proposing a new physics-constrained NN to solve PDEs, where the BCs are specified as inputs to the NNs.
Motivated by the FEM, which uses the weak formulation to completely define a physical system described by the governing PDEs and the associated BCs, and solves the problem based on the discretized residual, we construct the discretized residual of PDEs from NN predicted solutions to form the loss function to train NNs, through an efficient, convolutional operator-based, and vectorized residual calculation implementation.
As shown in Fig. \ref{fig:NN}, the weak PDE loss layers are independent from the NN that serves as the surrogate PDE solver and do not introduce any new trainable parameters.
Such features offer us great flexibility to choose the NN architecture. 
We focus on an encoder-decoder NN structure, which has been investigated for other physical systems \cite{Zhu2018Zabaras-UQ-Bayesian-ED-CNN,Winovich2019Lin-ConvPDE-UQ,Bhatnagar2019CNN-encoder-decoder}.
We studied both deterministic and probabilistic models, with Bayesian NNs (BNNs) for the latter.
The encoder-decoder structure can be easily adopted to the BNNs with the modularized probabilistic layers provided in the TensorFlow Probability (TFP) library.
In our approach, deterministic/probabilistic convolutional NN layers are used to learn the applied BCs (both Dirichlet and Neumann) and to detect the problem domains through carefully designed input data structure. 
Thus, with our approach, a single NN can be used to simultaneously solve different BVPs that are governed by the same PDEs but on different domains with different BCs.
In addition, similar to other surrogate PDE solvers, our approach is also label free.
Furthermore, the trained surrogate solvers can make predictions for interpolated and extrapolated (to a certain extent) BCs that they were not exposed to during training.

In our BNNs, each model parameter is sampled from a posterior distribution. 
We solve for the posterior distribution of model parameters with the VI method instead of the Markov Chain Monte Carlo (MCMC) sampling, as the latter involves expensive iterative inference steps and is not suitable for systems with a large number of parameters \cite{Blei2017Variational-inference-review,Kingma2014Welling+Variational-bayes}.
In our work, the likelihood function is constructed based on the discretized PDE residuals.
The BNNs allow us to quantify both epistemic uncertainty from model parameters and aleatoric uncertainty from noise in the data.
In our study, an additive noise is applied to the NN predicted solution as in \cite{Luo2020Bayesian-deep-learning,Wang2020Garikipati-Xun-perspective-system-id-bayesian,Geneva2020Zabaras-JCP-auto-regressive-NN-PDE,Zhu2018Zabaras-UQ-Bayesian-ED-CNN,Zhu2019Perdikaris-Physics-PDE-CNN} to represent the aleatoric uncertainty.
Such an additive noise represents potential errors from various sources, such as discretization error \cite{Geneva2020Zabaras-JCP-auto-regressive-NN-PDE}, geometry representation, boundary conditions, and material parameter measurement, among others.

The proposed framework is a generalized approach that is applicable to both steady-state and transient problems.
In this work, we present the details of this new framework, and its application for the steady-state diffusion, linear elasticity, and nonlinear elasticity.
We defer the investigation of transient problems to a subsequent work.
To the authors' best knowledge, this is the first attempt to simultaneously solve PDEs on different domains with different BCs with a single surrogate solver, with the further feature of UQ. 
In this study, the problem domains are represented via pixels on a square background grid for simplicity. 
Thus, the boundary of a domain is not a smooth curve, but has a pixel-level resolution.
One can refer to \cite{Bhatnagar2019CNN-encoder-decoder,Li2020Reaction-diffusion-prediction-CNN,Gao2020PhyGeoNet-PDE-on-Irregular-domain} and many others for strategies to map complex and irregular domain onto a regular grid mesh. 
Such geometry transformation can be taken into account in the proposed PDE loss layers with the isoparametric mapping concept of the FEM via the shape functions, though it is not the focus of this work.

The rest of the paper is organized as follows.
In Section~\ref{sec:bvp-definition}, we briefly summarize the general mathematical description of the type of physical systems that is studied in this work.
The structures of discretized residual constrained NNs used in this work are presented in Section~\ref{sec:NN}.
Section~\ref{sec:NN-FEM} provides the details of an efficient implementation of the discretized residual calculation.
Section~\ref{sec:data} covers the data structure of NN inputs, domain/boundary detection, setup of BVPs, and NN training procedures.
Detailed simulation results are presented in Section~\ref{sec:results}, where steady-state diffusion, linear elasticity, and non-linear elasticity are studied.  
Concluding remarks and perspectives are offered in Section~\ref{sec:conclusion}.

\section{Problem definition} \label{sec:bvp-definition}

In this work, we are interested in solving the steady-state diffusion, linear elasticity, and nonlinear elasticity problems with discretized residual constrained NNs.
These three physical systems are described by a general elliptic PDE on a domain $\Omega$ with the Dirichlet BC on $\Gamma^{\Bvarphi}$ and the Neumann BC on $\Gamma^{\Bk}$ as
\begin{equation}
  \begin{aligned}
    \nabla \cdot \BA(\Bvarphi) = \Bzero \quad & \text{on} \quad \Omega, \\
    \Bvarphi (\BX) = \bar{\Bvarphi} (\BX)  \quad & \text{on} \quad  \Gamma^{\Bvarphi}, \\
    \Bk (\BX) = \bar{\Bk} (\BX)  \quad & \text{on} \quad  \Gamma^\Bk,
  \end{aligned}
  \label{eq:general-pde}
\end{equation}
where $\Bvarphi(\BX)$ represents the location-dependent unknown variable and $\BX$ is the coordinate.
The overall boundary of the continuum body satisfies $\Gamma = \Gamma^{\Bvarphi} \bigcup \Gamma^{\Bk}$ and $\Gamma^{\Bvarphi} \bigcap \Gamma^{\Bk}=\emptyset$.
It is worth mentioning that even though bold typeface $\Bvarphi$, $\BA$, and $\Bk$ are used in \eref{eq:general-pde}, depending on the degree of freedoms (DOFs) of each physical system, they can represent either scalar, vector, or tensor fields.
For example, in the diffusion problem, $\Bvarphi$, $\BA$, and $\Bk$ represent the compositional order parameter (scalar), the diffusive flux (vector), and the outward flux (scalar), respectively. 
Whereas in elasticity problems, $\Bvarphi$, $\BA$, and $\Bk$ represent the deformation field (vector), the stress field (second-order tensor), and the surface traction (vector), respectively. 
The details of each system are provided in the numerical simulation section.

The weak form of \eref{eq:general-pde} states: For variations $\Bomega$ satisfying $\forall \Bomega \in \scrV$ with $\scrV = \left\{ \Bomega | \Bomega=\Bzero~\text{on}~\Gamma^{\Bvarphi} \right\}$, seek trial solutions $\Bvarphi \in \scrS $ with $ \scrS = \left\{ \Bvarphi|\Bvarphi=\bar{\Bvarphi}~\text{on}~\Gamma^{\Bvarphi} \right\}$ such that
\begin{equation}
  \int_\Omega  \nabla \Bomega \cdot \BA(\Bvarphi) ~ dV - \int_{\Gamma^\Bk} \bar{\Bk} \cdot \Bomega ~dS =0.
  \label{eq:general-pde-weak}
\end{equation}
Eq. \eref{eq:general-pde-weak} is obtained by multiplying (\ref{eq:general-pde}$_1$) with $\Bomega$, integrating by parts, and then incorporating the Neumann BC in (\ref{eq:general-pde}$_3$).
For the diffusion problem, $\Bomega$ is a scalar field. For elasticity problems, $\Bomega$ is a vector field.

To obtain the approximate solutions of \eref{eq:general-pde-weak}, finite-dimensional approximations of $\Bomega$ and $\Bvarphi$, denoted by $\Bomega^h$ and $\Bvarphi^h$, are constructed with $\forall \Bomega^h \in \scrV^h = \left\{ \Bomega^h | \Bomega^h=\Bzero~\text{on}~\Gamma^{\Bvarphi} \right\}$ and $\Bvarphi^h \in \scrS^h = \left\{ \Bvarphi^h|\Bvarphi^h=\bar{\Bvarphi}~\text{on}~\Gamma^{\Bvarphi} \right\}$. 
The discretized terms $\Bomega^h$, $\nabla\Bomega^h$, and $\Bvarphi^h$ are computed as
\begin{equation}
  \Bomega^h = \BN \Bd_{\Bomega},
  \quad
  \nabla\Bomega^h = \BB \Bd_{\Bomega},
  \quad
  \text{and}
  \quad
  \Bvarphi^h = \BN \Bd_{\Bvarphi} 
  \label{eq:discretized-form-u}
\end{equation}
in terms of the nodal solutions $\Bd_{\Bomega}$ and $\Bd_{\Bvarphi}$, the basis functions $\BN$, and the gradient matrix $\BB = \nabla\BN$.
Inserting \eref{eq:discretized-form-u} into \eref{eq:general-pde-weak} we obtain the discretized residual by a sum over subdomains $\Omega^e$ and their associated boundary $\Gamma^{e}$ as
\begin{equation}
  \BR = \sum_{e=1}^{n_\text{elem}} \left\{ \int_{\Omega^e} \BB^T \BA(\Bvarphi^h) dV - \int_{\Gamma^{e,\Bk}} \BN^T \bar{\Bk}~dS \right\}
  \label{eq:discretized-residual}
\end{equation}
where $n_\text{elem}$ represents the total number of subdomains.
The volume and surface integrations in \eref{eq:discretized-residual} are evaluated numerically via Gaussian quadrature rules.
In this work, the problem domain $\Omega$ is represented by images. 
The adjacent pixels in images are used to form the subdomain $\Omega^e$, whose connectivity information is preserved automatically by the image data.
The values at each pixel of the image are treated as nodal values. 
A more detailed discussion on constructing the subdomains based on image pixels is provided in Section \ref{sec:NN-FEM}.

\section{Discretized residual constrained neural networks}\label{sec:NN}
In this section, we present the formulation of discretized residual constrained deterministic/probabilistic NNs for solving PDEs for given BCs without labels. 

\subsection{Deterministic neural networks}
\begin{figure}[t!]
  \centering
  \includegraphics[width=1.0\linewidth]{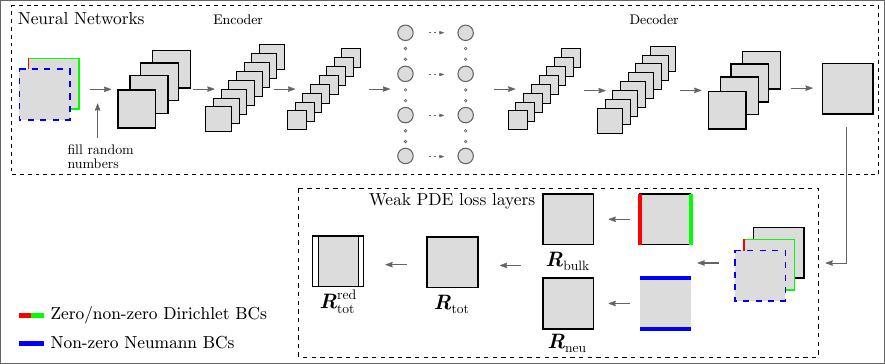}
  \caption{Illustration of the discretized residual constrained NNs, which consist of an encoder-decoder NN structure and a group of layers to compute the PDE-related loss. The NNs can be either deterministic or probabilistic. The weak PDE loss layers take both NN inputs and outputs as their inputs. The Dirichlet BCs from the NN inputs are applied to the NN predicted solution to compute the bulk residual. The Neumann residual is directly computed based on the Neumann BCs from the inputs\protect\footnotemark.}
  \label{fig:NN}
\end{figure}

\footnotetext{Material parameters can also be specified as inputs to the NNs to further enrich the flexibility of the proposed approach.
For simplicity, only fixed material parameters are considered in the numerical sections in this work.}

We first consider deterministic NNs, whose parameters $\BTheta$ are represented by single values instead of distributions as in probabilistic NNs.
In our approach, the NNs take image data that contains information of both Dirichlet and Neumann BCs as inputs and output the full field weak solutions of the PDEs that are associated to the input BCs.
As shown in \eref{eq:discretized-residual}, the discretized residual of PDEs consists of two contributions, one bulk term and one surface term.
We propose the weak PDE loss layers, which are discussed in detail in Section \ref{sec:NN-FEM}, to compute the residual in the bulk and on the Neumann boundary.
As illustrated in Fig. \ref{fig:NN}, the weak PDE loss layers are constructed  based on NN predicted solutions.
Those weak PDE loss layers only contain forward calculations based on the FEM without introducing any new parameters to be optimized.
Since the weak PDE loss layers are independent of the NN that serves as the surrogate PDE solver, and they also do not introduce any new trainable parameters, there is flexibility in choosing the NN architecture. 
As shown in Fig. \ref{fig:NN}, we focus on an encoder-decoder NN structure, which has been investigated for other physical systems \cite{Zhu2018Zabaras-UQ-Bayesian-ED-CNN,Winovich2019Lin-ConvPDE-UQ,Bhatnagar2019CNN-encoder-decoder}.

When using mini-batch optimization to train the discretized residual constrained deterministic NNs over a dataset $\calD$, the batch loss $\calL_i$ is written in terms of the reduced total residual $\BR_\text{tot}^\text{red}$, as illustrated in Fig. \ref{fig:NN}, as
\begin{equation}
  \calL_i = \frac{1}{N} \sum_{n=1}^{N} \left(\BR_\text{tot}^\text{red}( \calD_i, \BTheta) \right)^2,
  \label{eq:loss-deterministic}
\end{equation}
for each mini-batch $i=1,2,\cdots,M$ with $N$ indicating the size of data in each mini-batch.
The detailed training process of the discretized residual constrained NNs is discussed in Section \ref{sec:NN-training}.

\subsection{Probabilistic neural networks}
\subsubsection{Background}
For probabilistic NNs, we consider  BNNs, whose model parameters $\BTheta$ are stochastic and  sampled from a posterior distribution $P(\BTheta | \calD)$ instead of being represented by single values as in deterministic NNs. 
The posterior distribution $P(\BTheta | \calD)$ is computed based on the Bayes' theorem
\begin{equation}
  P(\BTheta | \calD) = \frac{P(\calD|\BTheta) P(\BTheta)}{P(\calD)},
  \label{eq:bayes}
\end{equation}
where $\calD$ denote the i.i.d.~observations (training data) and $P$ represents the probability density function.
In \eref{eq:bayes}, $P(\calD|\BTheta)$ is the likelihood, $P(\BTheta)$ is the prior probability, and $P(\calD)$ is the evidence function, respectively.
The likelihood is the probability of $\calD$ given $\BTheta$, which describes the probability of the observed data for given parameters $\BTheta$.
A larger value of $P(\calD | \BTheta )$ means the probability of the observed data is larger, implying that $\BTheta$ is more reasonable.
The prior needs to be specified before the Bayesian inference process \cite{gelman2013bayesian}.

To compute the posterior distributions of $\BTheta$, one can use  popular sampling-based methods, such as MCMC.
However, the sampling method involves expensive iterative inference steps and would be difficult to use when datasets are large or models are very complex \cite{Kingma2014Welling+Variational-bayes,Liu2016Wang-Stein-variational-gradient-descent,Blei2017Variational-inference-review}.
Alternatively, we can use the VI, which approximates the exact posterior distribution $P(\BTheta | \calD)$ with a more tractable surrogate distribution $Q(\BTheta)$ by minimizing the Kullback-Leibler (KL) divergence \cite{Liu2016Wang-Stein-variational-gradient-descent,Blei2017Variational-inference-review,Graves2011Varational-inference}
\begin{equation}
  Q^* = \text{arg~min}~D_\text{KL}(Q(\BTheta)||P(\BTheta | \calD)).
  \label{eq:kl-divergence}
\end{equation}
Compared with MCMC, the VI is faster and easier to scale to large datasets. 
We therefore explore it in this work, even though it is less rigorously studied than MCMC \cite{Blei2017Variational-inference-review}. 
The KL divergence is computed as 
\begin{equation}
  D_\text{KL}(Q(\BTheta)||P(\BTheta | \calD)) = \mathbb{E}_Q[\log Q(\BTheta)] - \mathbb{E}_Q[\log P(\BTheta, \calD)]  + \log P(\calD),
  \label{eq:kl-divergence-calculation}
\end{equation}
which requires computing the logarithm of the evidence, $\text{log}P(\calD)$ in \eref{eq:bayes} \cite{Blei2017Variational-inference-review}.
Since $P(\calD)$ is hard to compute, it is challenging to direct evaluate  the objective function in \eref{eq:kl-divergence}.
Alternatively, we can optimize the so-called evidence lower bound (ELBO)
\begin{equation}
  \begin{aligned}
  \text{ELBO}(Q) &= \mathbb{E}_Q[\log P(\BTheta, \calD)] - \mathbb{E}_Q[\log Q(\BTheta)] \\
   & = \mathbb{E}_{Q}[\log P(\calD | \BTheta )] - \left( \mathbb{E}_{Q}[\log Q(\BTheta)] - \mathbb{E}_{Q}[\log P(\BTheta)]  \right) \\
   & = \mathbb{E}_{Q}[\log P(\calD | \BTheta )] - D_\text{KL}\left( Q(\BTheta) ||P(\BTheta)  \right).
  \end{aligned}
  \label{eq:elbo}
\end{equation}
which is equivalent to the KL-divergence up to an added constant. 
So, the loss function for the BNN is written as
\begin{equation}
  \calL = D_\text{KL}\left( Q(\BTheta) ||P(\BTheta)  \right) - \mathbb{E}_{Q}[\log P(\calD | \BTheta )],
  \label{eq:bnn-loss-exact}
\end{equation}
which consists of a prior-dependent part and a data-dependent part, 
with the former being the KL-divergence of the surrogate posterior distribution $Q(\BTheta)$ and the prior $P(\BTheta)$, and the latter being the negative log-likelihood cost.

\subsubsection{Flipout}

Different methods are available for training NNs with stochastic weights, such as weight perturbation \cite{Graves2011Varational-inference,Blundell2015Weight-Uncertainty-NN,Wen2018GrosseFlipout}, activation perturbation \cite{Ioffe2015Szegedy-batch-normalization}, reparameterization \cite{Kingma2014Welling+Variational-bayes}, and many others.
In this work, we follow a specific weight perturbation method, the so-called Flipout, proposed in \cite{Wen2018GrosseFlipout}. 
Compared with other weight perturbation algorithms that suffer from high variance of the gradient estimates because the same perturbation is shared in a mini-batch for all training examples, Flipout is an efficient method, which decorrelates the gradients in a mini-batch by implicitly sampling pseudo-independent weight perturbation for each example, and thus reduces the variance of NNs with stochastic weights \cite{Wen2018GrosseFlipout}.
This method can be efficiently implemented in a vectorized manner with unbiased stochastic gradients.

A brief description of Flipout is summarized here. Readers are directed to Ref. \cite{Wen2018GrosseFlipout} for details.
Flipout assumes that the perturbations of different weights are independent, and the perturbation distribution is symmetric around zero.
Under such assumptions, the perturbation distribution is invariant to element-wise multiplication by a random sign matrix.
To minimize the loss $\calL$, the distribution of $Q(\BTheta)$ can be described in terms of perturbations with $W=\lbar{W}+\Delta W$, where $\lbar{W}$ and $\Delta W$ are the mean and a stochastic perturbation for $\BTheta$, respectively.
Flipout uses a base perturbation $\widehat{\Delta W}$ shared by all examples (training data points) in a mini-batch, and arrives at the perturbation for individual example by multiplying $\widehat{\Delta W}$ with a different rank-one sign matrix
\begin{equation}
  \Delta W_n =  \widehat{\Delta W} \circ r_n s_n^t,
\end{equation}
where the subscript $n$ indicates an individual example in a mini-batch, and $r_n$ and $s_n$ are entries of random vectors uniformly sampled from $\pm 1$.
Using different perturbations for each example in a mini-batch rather than an identical perturbation for all the example in a mini-batch ensures the reduction of the variance of the stochastic gradients in Flipout during training.
For BNNs, the $\lbar{W}$ and $\widehat{\Delta W}$ are the mean and standard deviation of the posterior distribution $Q(\BTheta)$, which are obtained via backpropagation with stochastic optimization algorithms.
For mini-batch optimization, the batch loss is written as 
\begin{equation}
  \calL_i = \frac{1}{M} D_\text{KL}\left( Q(\BTheta) ||P(\BTheta)  \right) - \mathbb{E}_{Q}[\log P(\calD_i | \BTheta^{(i)} )],
  \label{eq:bnn-loss-batch}
\end{equation}
for each mini-batch $i=1,2,\cdots,M$ \cite{Blundell2015Weight-Uncertainty-NN}.
With \eref{eq:bnn-loss-batch}, we have $\calL = \sum_i \calL_i$.
Following \cite{Blundell2015Weight-Uncertainty-NN}, Monte Carlo (MC) sampling is used to approximate the expectation in \eref{eq:bnn-loss-batch} as 
\begin{equation}
  \calL_i \approx \frac{1}{M} D_\text{KL}\left( Q(\BTheta) ||P(\BTheta)  \right) - \frac{1}{N} \sum_{n=1}^{N} \log P(\calD_i^n | \BTheta^{(i)} ),
  \label{eq:bnn-loss-approx-batch}
\end{equation}
where $N$ is the size of each mini-batch dataset, and $\BTheta^{(i)}$ denotes the $i$th batch sample drawn from the posterior distribution $Q(\BTheta)$. 
Even though only one set of parameters $\BTheta^{(i)}$ is drawn from $Q(\BTheta)$ for each mini-batch, the perturbation approach proposed by Flipout ensures that parameters are different for the individual example $\calD_i^n$ to calculate the log-likelihood cost.
Probabilistic dense layers and convolutional layers with the. Flipout weight perturbation technique have been implemented in the TFP Library \footnote{\href{https://www.tensorflow.org/probability/api_docs/python/tfp/layers}{www.tensorflow.org/probability/api\_docs/python/tfp/layers}} and are used to construct the BNNs in this work.

\subsubsection{Neural network structure and loss function}
As the probabilistic layers are implemented in the TFP library in a modularized form, we can easily construct the discretized residual constrained BNNs to have a similar encoder-decoder architecture, as shown in Fig. \ref{fig:NN},  as the deterministic model but with all weights being drawn from probability distributions. 
The loss of the BNNs is given in \eref{eq:bnn-loss-exact}. 
The probabilistic layers in the TFP library automatically calculate the prior-dependent KL-divergence and add it to the total loss.

The data-dependent loss is accounted for by the likelihood function. 
In general, data contains noise that leads to aleatoric uncertainty, which cannot be reduced by training the surrogate model with more observations.
Additive noise, which is independent of the data and is commonly treated as Gaussian,  is often added to the output of the surrogate model to construct the likelihood function \cite{Luo2020Bayesian-deep-learning,Wang2020Garikipati-Xun-perspective-system-id-bayesian,Geneva2020Zabaras-JCP-auto-regressive-NN-PDE,Zhu2018Zabaras-UQ-Bayesian-ED-CNN,Zhu2019Perdikaris-Physics-PDE-CNN}.
Such an additive noise represents potential errors from various sources, such as discretization error \cite{Geneva2020Zabaras-JCP-auto-regressive-NN-PDE}, geometry representation, boundary conditions, and material parameter measurement.
Assuming a Gaussian noise $\Bepsilon \sim \calN(\Bzero, \Sigma_1\BI)$ with a zero-mean and a pre-specified constant covariance $\Sigma_1$\footnote{Aleatoric uncertainty can further be categorized into homoscedastic uncertainty and heteroscedastic uncertainty \cite{Kendall2017Gal-both-uncertainty}. We assume $\Sigma_1$ being the former case for simplicity, which stays constant for different inputs. The latter is useful in cases where output noise depends on the model inputs.}, the NN output $\By$ is written as
\begin{equation}
    \By = \Bf(\Bx, \BTheta) + \Bepsilon,
    \label{eq:f-surrogate}
\end{equation}
where $\Bf(\Bx, \BTheta)$ represents the surrogate NNs.
For discretized residual constrained NNs, the likelihood function is constructed based on the residual value, rather than NN predicted solutions. 
The expected value of point-wise residual is zero, which states that the governing PDEs are weakly satisfied at each location.
This ensures that the proposed surrogate PDE solvers are label free.
With the noise $\Bepsilon$ in \eref{eq:f-surrogate} propagating through the residual calculation, the likelihood function\footnote{For systems where nonlinear operations are involved in the residual calculation, the residual noise distribution is in general non Gaussian even if the noise in the NN outputs is assumed to be Gaussian. Under the conditions that $\Sigma_1$ is small and the nonlinear operations are smooth and approximately linear locally, we assume that the noise distribution of the residual is approximately Gaussian.} is written as
\begin{equation}
  P(\BR_\text{tot}^\text{red}|\Bzero, \Sigma\BI) = \prod_{k=1}^{K} \calN \left( R_\text{tot}^\text{red,$k$}|0, \Sigma_2 \right)
  \label{eq:likelihood-function}
\end{equation}
where $k$ indicates the pixel location with $K$ total pixels.
As it is challenging to directly calculate $\Sigma_2$ via error propagation based on $\Sigma_1$, we treat $\Sigma_2$ as a learnable parameter to be optimized base on the NN loss.
In \eref{eq:likelihood-function}, $\Sigma_2$ essentially serves as a threshold for the residual to converge to. 
The batch-wise loss of the residual constrained BNNs has the following format
\begin{equation}
  \calL_i \approx \frac{1}{M} D_\text{KL}\left( Q(\BTheta) ||P(\BTheta)  \right) - \frac{1}{N} \sum_{n=1}^{N} \sum_{k=1}^{K} \log \left( \calN \left( R_\text{tot}^\text{red,$k$}(\calD_i^n,\BTheta^{(i)}) | 0, \Sigma_2 \right) \right).
  \label{eq:loss-probabilistic}
\end{equation}
The detailed training process of the residual constrained BNNs is discussed in Section \ref{sec:NN-training}.

\subsubsection{Uncertainty quantification}

The BNNs allow us to quantify both epistemic uncertainty from model parameters and aleatoric uncertainty from noise in the data.
With the discretized residual constrained BNNs, the posterior predictive distribution $P(\By^*|\Bx^*, \calD)$ for a specific testing data point $\{\Bx^*, \By^*\}$ is expressed as \cite{Zhu2018Zabaras-UQ-Bayesian-ED-CNN, Luo2020Bayesian-deep-learning}
\begin{equation}
  \begin{aligned}
  P(\By^*|\Bx^*, \calD) & = \int P(\By^*|\Bx^*, \BTheta) P(\BTheta | \calD) d \BTheta \\
  & \approx \int P(\By^*|\Bx^*, \BTheta) Q(\BTheta) d \BTheta, \\
  \end{aligned}
  %\label{}
\end{equation}
which can be numerically evaluated via MC sampling as
\begin{equation}
  \begin{aligned}
    P(\By^*|\Bx^*, \calD) & \approx \frac{1}{S} \sum_{s=1}^{S}  P(\By^*|\Bx^*, \BTheta^s) 
    \quad \text{where} \quad \BTheta^s \sim Q(\BTheta), \\
  \end{aligned}
  %\label{}
\end{equation}
with $s$ indicating each sampling.
To represent the uncertainty, we compute the statistical moments of $\By^*$ via the predictive expectation 
\begin{equation}
  \mathbb{E}[\By^*|\Bx^*, \calD] \approx \frac{1}{S} \sum_{s=1}^{S} \Bf(\Bx^*, \BTheta^s)
\end{equation}
and the predictive variance 
\begin{equation}
  \begin{aligned}
    \text{Var}[\By^*|\Bx^*, \calD] 
    & = \mathbb{E}[(\By^* + \Bepsilon)^2] - (\mathbb{E}[\By^* + \Bepsilon])^2 \\
  & \approx \frac{1}{S} \sum_{s=1}^{S} \left(  \Bf(\Bx^*, \BTheta^s)\Bf^T(\Bx^*, \BTheta^s)  + \Sigma_1\BI \right)  
  - \left( \frac{1}{S} \sum_{s=1}^{S} \Bf(\Bx^*, \BTheta^s)  \right)\left(\frac{1}{S} \sum_{s=1}^{S} \Bf(\Bx^*, \BTheta^s)  \right)^T. \\
  \end{aligned}
\end{equation}

\section{Efficient implementation of the residual calculation} \label{sec:NN-FEM} 

\begin{figure}[h!]
    \centering
    \includegraphics[width=1.0\linewidth]{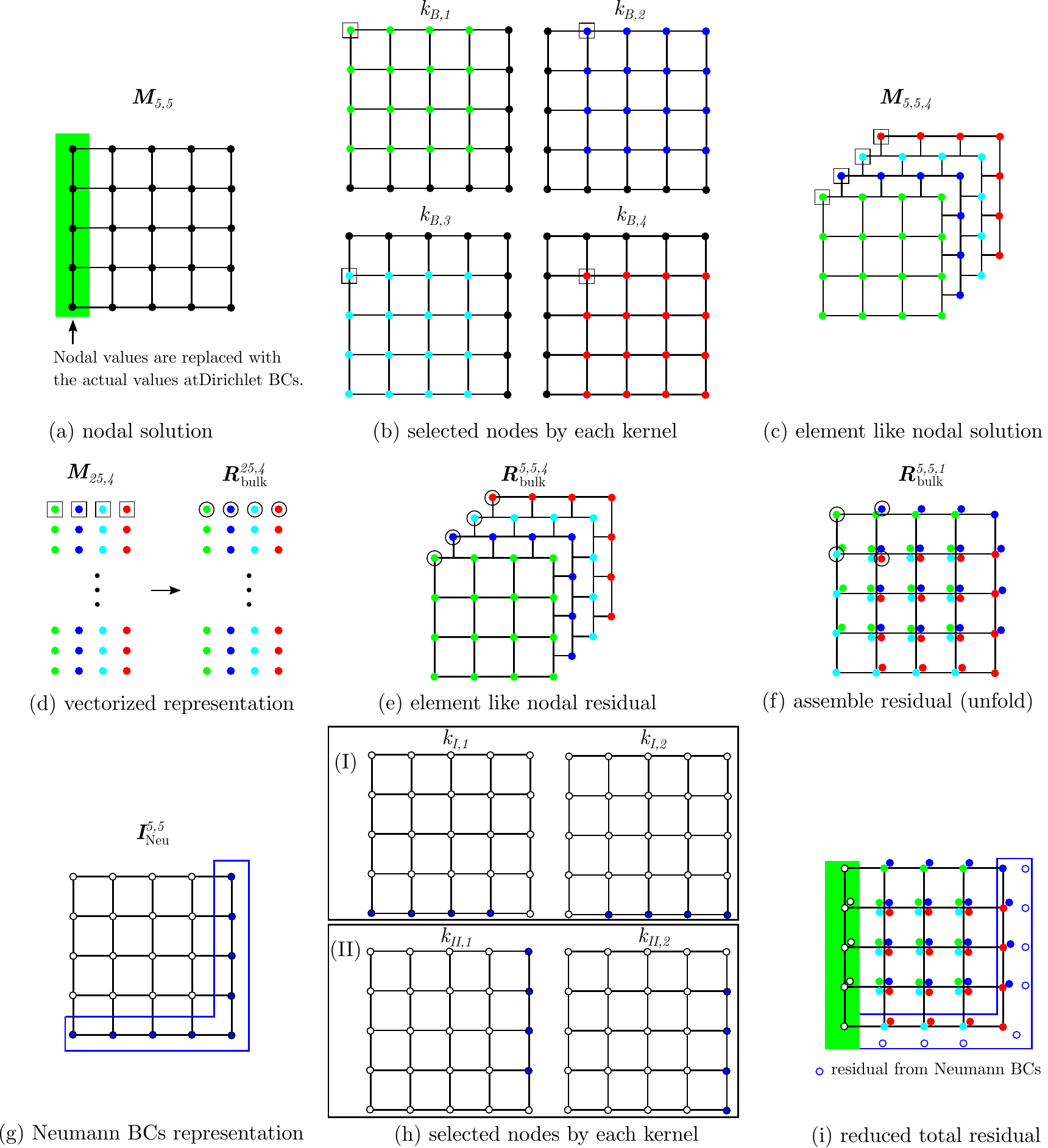}
    \caption{Illustration of the implementation steps of computing the residual in the weak PDE loss layers. Paddings of zeros are not shown in (c,d,e).}
    \label{fig:schematic-weak}
\end{figure}

In this section, we describe the implementation details of the weak PDE loss layers.
We heavily utilize the convolutional operation, and the vector/matrix/tensor operations to achieve numerical efficiency.
Readers are directed to our source code for additional details\footnote{\href{https://github.com/mechanoChem/mechanoChemML}{github.com/mechanoChem/mechanoChemML}}. 
As shown in Fig. \ref{fig:NN}, the weak PDE loss layers take both NN inputs (BCs information) and outputs (NN predicted solution) as their inputs.
The data structure to represent the BCs is discussed in detail in Section \ref{sec:data-representation}.
A schematic  of the major implementation steps of the weak PDE loss layers is shown in Fig. \ref{fig:schematic-weak}. 
We choose a steady state diffusion problem with a scalar unknown at each node for illustration purpose, with Dirichlet BCs being applied on the left side, non-zero Neumann BCs being applied on the bottom and right sides, and zero Neumann BCs on the top.
Assuming that the output of the NN shown in Fig. \ref{fig:NN} is a $5\times5$ matrix (an image with $5\times5$ pixels), denoted as $\BM_{5,5}^\text{NN}$\footnote{In the source code, $\BM_{5,5}^\text{NN}$ is stored as $\BM_{5,5,1}^\text{NN}$ with a third dimension of 1, which indicates the DOF per node. For elasticity problems, the third dimension has a size of 2. Here, we drop the ``1'' to simplify the notations.} with the value of each entry being the actual concentration, $\BM_{5,5}^\text{NN}$ is equivalent to the nodal solution on a domain, which is discretized with 4x4 elements, as shown in Fig. \ref{fig:schematic-weak}(a).
The implementation procedure is summarized in the Algorithm Box \ref{algo:compute-residual}

\subsection{Dirichlet BCs}
The channel of NN inputs with Dirichlet BCs information is denoted as $\BI_\text{D}^{5,5}$. 
To enforce the Dirichlet BCs, we replace the nodal values of $\BM_{5,5}^\text{NN}$ at the location of Dirichlet boundary with the actual values of $\BI_\text{D}^{5,5}$ to obtain a new matrix, denoted as $\BM_{5,5}$, as indicated by the green color in Fig. \ref{fig:schematic-weak}(a).
The Dirichlet BCs are then automatically incorporated into the residual vector during the bulk residual calculation discussed in the next Section.

\subsection{Bulk residual}

The matrix representation of the nodal solution automatically contains the element connectivity information of the mesh.
To compute the bulk residual, we first apply convolutional operations to $\BM_{5,5}$ with the following kernels
\begin{equation}
  k_{B,1} =
\begin{bmatrix}
1 & 0\\
0 & 0\\
\end{bmatrix}, 
\quad
k_{B,2} =
\begin{bmatrix}
0 & 1\\
0 & 0\\
\end{bmatrix}, 
\quad
k_{B,3} =
\begin{bmatrix}
0 & 0\\
1 & 0\\
\end{bmatrix}, 
\quad
k_{B,4} =
\begin{bmatrix}
0 & 0\\
0 & 1\\
\end{bmatrix}.
  \label{eq:kernel-node-to-bulk-elem}
\end{equation}
Each convolutional operation results in a matrix with a size of $5\times5$\footnote{The resulting matrix size is $4\times4$. Zero paddings are used to ensure the resulted matrix with a dimension of $5\times5$. Keeping the matrix size unchanged during the convolutional operations is not necessary and might require a small amount of extra floating-point operations, but it is less prone to errors if we handle matrices with a fixed size.}, which corresponds to the selected nodes, as highlighted with colors in Fig. \ref{fig:schematic-weak}(b).
With these four convolutional operations, we now have a matrix with a size of $5\times5\times4$ ($\BM_{5,5,4}$), as shown in Fig. \ref{fig:schematic-weak}(c).
We then reshape the matrix to an array $25\times4$ ($\BM_{25,4}$), as shown in Fig. \ref{fig:schematic-weak}(d).
Each row of $\BM_{25,4}$ corresponds to the local nodal solution vector inside one finite element, the subdomain $\Omega^e$ in \eref{eq:discretized-residual}, which can then be used to efficiently evaluate the residual via the matrix-vector operation.

To evaluate the residual of the steady-state diffusion problem with $2\times2$ Gauss points, the B-operator matrix in \eref{eq:discretized-residual-diffusion} has a size of $4\times2\times4$ (\# of Gauss points $\times$ dimensions $\times$ \# of nodes), denoted as $\BB_{4,2,4}$, with its transpose denoted as $\BB_{4,4,2}^T$.
The bulk residual at each Gauss point $i$ is evaluated as
\begin{equation}
(\BR_\text{bulk}^{25,4})^i = \omega_i D \BM_{25,4} \BB_{i,4,2} \BB_{i,2,4} 
  \label{eq:bulk-residual-matrix-vector-evaluation}
\end{equation}
with $\omega_i$ denoting the weights. 
The total bulk residual is computed as
\begin{equation}
\BR_\text{bulk}^{25,4} = \sum_{i=1}^{4} \BR_\text{bulk}^i,
\end{equation}
as shown in Fig. \ref{fig:schematic-weak}(d). $\BR_\text{bulk}^{25,4}$ is then reshaped to $\BR_\text{bulk}^{5,5,4}$, and stored in the element-like form, as shown in Fig. \ref{fig:schematic-weak}(e).

Next, we use the \verb tf.roll ~function to unfold the element-like residual to the correct nodal position, as shown Fig. \ref{fig:schematic-weak}(f), with
\begin{equation}
  \begin{aligned}
    \BR_\text{bulk}^{5,5,0:1} & = \BR_\text{bulk}^{5,5,0:1} \\
    \BR_\text{bulk}^{5,5,1:2} & = \text{tf.roll} (\BR_\text{bulk}^{5,5,1:2}, [1], [2])\\
    \BR_\text{bulk}^{5,5,2:3} & = \text{tf.roll} (\BR_\text{bulk}^{5,5,2:3}, [1], [1])\\
    \BR_\text{bulk}^{5,5,3:4} & = \text{tf.roll} (\BR_\text{bulk}^{5,5,3:4}, [1,1], [1,2])\\
  \end{aligned}
\end{equation}
The assemble operation in \eref{eq:discretized-residual-diffusion} for the bulk integration is now achieved by the \verb tf.reduce_sum ($\BR_\text{bulk}^{5,5,4}$) function without looping over all the elements to get $\BR_\text{bulk}^{5,5,1}$ as done traditionally in the FEM.
Readers are directed to our source code for the implementation of the linear/non-linear elasticity problems.

\begin{algorithm}[t!]
  \caption{Residual calculation for the steady-state diffusion example. \label{algo:compute-residual}}
  \textbf{Bulk residual with applied Dirichlet BCs:} $\BR_\text{tot}$
  \begin{algorithmic}[1]
    \STATE Apply Dirichlet BCs to NN predicted solutions $\BM^\text{NN}_{5,5}$ by replacing the nodal values at the corresponding locations to obtain $\BM_{5,5}$ (Fig. \ref{fig:schematic-weak}a).
    \STATE Convert $\BM_{5,5}$ from nodal value representation to a four-node element representation $\BM_{5,5,4}$ by convolutional operations with kernels $k_{B,1}, k_{B,2}, k_{B,3}$, and $k_{B,4}$ (Fig. \ref{fig:schematic-weak}b, \ref{fig:schematic-weak}c). For elasticity problems, NN predicted solutions have two channels to represent both $u_x$ and $u_y$. The same four kernels are applied to both channels, resulting in the element representation $\BM$ with a third dimension of 8 instead of 4.
    \STATE Get the vectorized representation $\BM_{25,4}$ with each row being the local nodal solutions for one element (Fig. \ref{fig:schematic-weak}d).
    \STATE Compute bulk residual $\BR_{25,4}$ for each element (Fig. \ref{fig:schematic-weak}d). Readers are directed to our source code for details of the bulk residual calculation of linear/nonlinear elasticity systems.  
    \STATE Switch back to matrix representation of element-like nodal residual $\BR_\text{bulk}^{5,5,4}$ (Fig. \ref{fig:schematic-weak}e).
    \STATE Assemble bulk residual $\BR_\text{bulk}^{5,5,1}$ (Fig. \ref{fig:schematic-weak}f).
  \end{algorithmic}
  \textbf{Residual at Neumann BCs:} $\BR_\text{Neu}$
  \begin{algorithmic}[1]
    \STATE Use kernels $k_{I,1}, k_{I,2}$ and $k_{II,1}, k_{II,2}$ to construct two groups of two-node surface elements $\BI^{5,5,2}_\text{Neu,I}$ and $\BI^{5,5,2}_\text{Neu,II}$. 
    For elasticity problems, we have four groups of surface elements with two for $T_x$ and two for $T_y$.
    \STATE Get the vectorized representation of surface elements $\BI^{25,2}_\text{Neu,I}$ and $\BI^{25,2}_\text{Neu,II}$.
    \STATE Compute residuals $\BR_\text{Neu,I}^{25,2}$ and $\BR_\text{Neu,II}^{25,2}$ at Neumann BCs.
    \STATE Switch back to matrix representation of element-like nodal residual $\BR_\text{Neu,I}^{5,5,2}$ and $\BR_\text{Neu,II}^{5,5,2}$.
    \STATE Assemble residual at Neumann BCs $\BR_\text{Neu}^{5,5,1}$.
  \end{algorithmic}
  \textbf{Reduced total residual:} $\BR_\text{tot}^\text{red}$
  \begin{algorithmic}[1]
    \STATE Create a mask matrix $\BM_\text{bulk}^{5,5}$ based on $\BI_\text{D}^{5,5}$ to represent the pixel locations with valid bulk residual values. The entries of $\BM_\text{bulk}^{5,5}$ are zero for the components of $\BI_\text{D}^{5,5}$ with a value of $-1$, which indicates the margins between actual problem domain and the background grid (see more details in Section \ref{sec:data-representation}). For the steady-state diffusion examples, all entries of  $\BM_\text{bulk}^{5,5}$ are one.
    \STATE Create a reverse mask matrix $\BM_\text{D,rev}^{5,5}$ based on $\BI_\text{D}^{5,5}$ to represent the pixel locations that are not at the Dirichlet boundary. The entries of $\BM_\text{D,rev}^{5,5}$ is zero at the entry locations of $\BI_\text{D}^{5,5}$ with a value larger than zero.
    \STATE Compute total residual $\BR_\text{tot}$ based on \eref{eq:compute-total-residual-example}.
    \STATE Multiply (element-wise) $\BR_\text{tot}$ with $\BM_\text{D,rev}^{5,5}$ and $\BM_\text{bulk}^{5,5}$ to get $\BR_\text{tot}^\text{red}$.
  \end{algorithmic}
\end{algorithm}

\subsection{Neumann BCs}
One channel of the inputs that contains purely Neumann BCs, denoted as $\BI^{5,5}_\text{Neu}$, is shown in Fig. \ref{fig:schematic-weak}(g), where the matrix contains only non-zero items at the non-zero Neumann boundary locations. 
The Neumann residual needs to be evaluated within surface elements. 
Similar to computing the bulk residual, we apply convolutional operations to $\BI^{5,5}_\text{Neu}$ to construct surface elements. 
Two sets of kernels are used to construct two groups of surface elements, with group I for computing residual on the top and bottom edges, and group II for the left and right edges.
We use the following two kernels 
\begin{equation}
  k_{I,1} =
\begin{bmatrix}
1 & 0\\
0 & 0\\
\end{bmatrix}, 
\quad
k_{I,2} =
\begin{bmatrix}
0 & 1\\
0 & 0\\
\end{bmatrix}, 
  \label{eq:kernel-node-to-x-surf-elem}
\end{equation}
to construct surface elements $\BI^{5,5,2}_\text{Neu,I}$ for the first group, with the selected nodal information being shown Fig. \ref{fig:schematic-weak}(h-I), and the following kernels 
\begin{equation}
  k_{II,1} =
\begin{bmatrix}
1 & 0\\
0 & 0\\
\end{bmatrix}, 
\quad
k_{II,2} =
\begin{bmatrix}
0 & 0\\
1 & 0\\
\end{bmatrix}.
  \label{eq:kernel-node-to-y-surf-elem}
\end{equation}
to construct surface elements $\BI^{5,5,2}_\text{Neu,II}$ for the second group, with the selected nodal information being shown Fig. \ref{fig:schematic-weak}(h-II). 

Similar to the bulk residual calculation, we form two matrices, $\BI^{25,2}_\text{Neu,I}$ and $\BI^{25,2}_\text{Neu,II}$, to compute the Neumann residual.
We use 2 Gauss points for surface integration.
The shape function $\BN$ in \eref{eq:discretized-residual-diffusion} has a size of $2\times2$ (\# of Gauss points $\times$ \# of nodes), denoted as $\BN_{2,2}$.
We evaluate the Neumann residual at each Gauss point $i$ via
\begin{equation}
  (\BR_\text{Neu,I}^{25,2})^i = \omega_i \BI^{25,2}_\text{Neu,I} \BN_{i,2} \BN_{i,2}
  \quad \text{and} \quad
  (\BR_\text{Neu,II}^{25,2})^i = \omega_i \BI^{25,2}_\text{Neu,II} \BN_{i,2} \BN_{i,2} 
\end{equation}
with $\omega_i$ denoting the weights. The total Neumann residual is computed as
\begin{equation}
  \BR_\text{Neu,I}^{25,2} = \sum_{i=1}^{2} \BR_\text{Neu,I}^i
  \quad \text{and} \quad
  \BR_\text{Neu,II}^{25,2} = \sum_{i=1}^{2} \BR_\text{Neu,II}^i.
\end{equation}
Again, we use the \verb tf.roll ~function to unfold the element-like residual to the correct nodal position, similar to those shown Fig. \ref{fig:schematic-weak}(f), for group I
\begin{equation}
  \begin{aligned}
    \BR_\text{Neu,I}^{5,5,0:1} & = \BR_\text{Neu,I}^{5,5,0:1} \\
    \BR_\text{Neu,I}^{5,5,1:2} & = \text{tf.roll} (\BR_\text{Neu,I}^{5,5,1:2}, [1], [1])\\
  \end{aligned}
\end{equation}
and for group II
\begin{equation}
  \begin{aligned}
    \BR_\text{Neu,II}^{5,5,0:1} & = \BR_\text{Neu,II}^{5,5,0:1} \\
    \BR_\text{Neu,II}^{5,5,1:2} & = \text{tf.roll} (\BR_\text{Neu,II}^{5,5,1:2}, [1], [2]).\\
  \end{aligned}
\end{equation}
The assemble operation in \eref{eq:discretized-residual-diffusion} for the surface integration is now achieved by the \verb tf.reduce_sum ($\BR_\text{Neu,I}^{5,5,2}$) and \verb tf.reduce_sum ($\BR_\text{Neu,II}^{5,5,2}$) without looping over elements.
We obtain the final Neumann residual
\begin{equation}
  \BR_\text{Neu}^{5,5,1} = \BR_\text{Neu,I} + \BR_\text{Neu,II}.
\end{equation}
The total residual $\BR_\text{tot}$ in \eref{eq:discretized-residual-diffusion}, as shown in Fig. \ref{fig:NN}, is computed as
\begin{equation}
  \BR_\text{tot}^{5,5,1} = \BR_\text{bulk}^{5,5,1} - \BR_\text{Neu}^{5,5,1} 
  \label{eq:compute-total-residual-example}
\end{equation}
by applying the Neumann residual to the bulk residual.
To construct the deterministic loss in \eref{eq:loss-deterministic} and the likelihood function in \eref{eq:likelihood-function}, the reduced residual $\BR_\text{tot}^\text{red}$ obtained by excluding the residual at the Dirichlet boundary location from $\BR_\text{tot}$ is used, as shown Fig. \ref{fig:schematic-weak}(i).
It is worth mentioning that additional auxiliary matrix/vector/tensor operations have been introduced, which are not included in the description, to complete this efficient residual evaluation. Readers are invited to refer to our code for the detailed implementation.

\section{Data representation and numerical aspects} \label{sec:data}
In this section, we present  details on the data structure of NN inputs, domain/boundary detection, the setup of BVPs, and the NN training procedure.

\subsection{Data structure of NN inputs} \label{sec:data-representation}
\begin{figure}[t!]
  \centering
  %\psfrag{a}[c][c]{step 1}
  \includegraphics[width=0.45\linewidth]{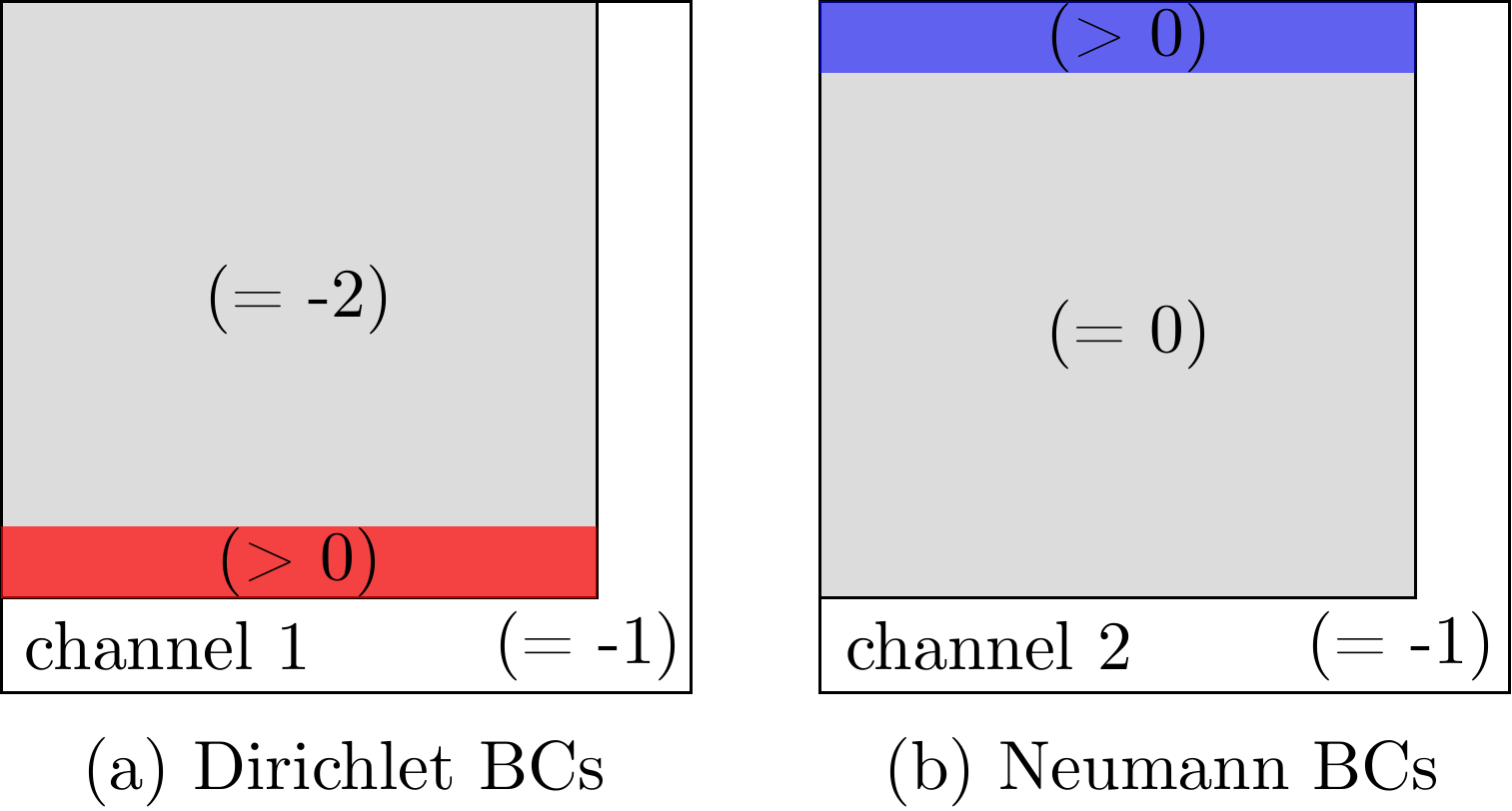}
  \caption{Illustration of the data structure of NN inputs for steady-state diffusion problem. The NN inputs contain two channels with the first one containing the Dirichlet BCs (red) and the second one containing the Neumann BCs (blue)\protect\footnotemark.
    Only the boundary locations have values that are greater than 0.
    In the first channel, the problem domain (gray color) is filled with a value of $-2$, which serves an indicator to fill with random numbers during the training process. 
    The margin part (white region) is filled with a value of $-1$. 
    The residual contribution from this region is excluded when computing $\BR_\text{tot}^\text{red}$.
    In the second channel, the problem domain is filled with a value of $0$. 
    Similarly, the margin is filled with a value of $-1$. 
  }
  \label{fig:data-diffusion-illustration}
\end{figure}
\footnotetext{For elasticity problems, the inputs contain four channels, with the first two representing Dirichlet BCs for $\bar{u}_x$ and $\bar{u}_y$ and the last two presenting Neumann BCs for $\bar{T}_x$ and $\bar{T}_y$.}

Since the discretized residual constrained NNs do not require labels for training, the NN inputs are synthetically generated  with only information on problem domains and the applied BCs.
We consider a fixed square background grid of $[0,~1]\times[0,~1]$, with \verb nx ~ and \verb ny ~ total pixels in each dimension.
For both diffusion and elasticity problems, each input data point is a three-dimensional matrix $\BI_\text{nx,ny,2$\times$DOF}$ to represent a set of BCs. The first two indices of $\BI$ indicate the pixels locations in X- and Y- directions. 
For steady-state diffusion problem with one scalar DOF per node, there are two channels in the third dimension, which contain information of Dirichlet and Neumann BCs, respectively.
For elasticity problems, there are four channels in the third dimension with the first two channels containing Dirichlet BCs in X- and Y- directions and the last two channels containing Neumann BCs in X- and Y- directions, respectively.
Data normalization between $[-1,~1]$ is used to ensure that all the physically meaningful data in our study has a value greater than 0.

The structure of the input data is illustrated in Fig. \ref{fig:data-diffusion-illustration} with the diffusion problem as an example.
In our study, the problem domain does not necessarily occupy the whole background grid, which results in the margin region as shown in Figs \ref{fig:data-diffusion-illustration} and \ref{fig:bvp-problem-setup}.
For the channel(s) containing Dirichlet BCs, the problem domain is filled with $-2$ except the Dirichlet boundary values, which is greater than 0.
The auxiliary number $-2$ serves as an indicator to be filled with random numbers during the training process. 
For the margin region, which represents the space between the background grid and the problem domain, if there is any, is filled with $-1$. 
The auxiliary number\footnote{The auxiliary numbers $-1$ and $-2$ are arbitrary choices with no physical meaning. Users can choose different values to assign to the margin and the problem domain for the inputs.} $-1$ serves as an indicator to evaluate $\BR_\text{tot}^\text{red}$ with the residual in this region being excluded.
For the channel(s) containing Neumann BCs, the problem domain is filled with a value of $0$ except the Neumann boundary values. 
When convolutional kernels operate on the problem domain, only the boundary makes a non-zero contribution. 
Similarly, the margin  is filled with a value of $-1$ for assisting the calculation of $\BR_\text{tot}^\text{red}$.
Examples of the actual inputs for steady state diffusion are shown in Fig. \ref{fig:diffusion-20bvp-results}(a,b).

\subsection{Domain/Boundary detection}
As discussed in Section \ref{sec:data-representation}, a fixed value of $-1$ is assigned to the margins.
When calculating the residual, a mask matrix is created for domain detection.
This mask matrix is created based on the information of Dirichlet BCs from the inputs and ensures that only the residual inside the actual problem domain is evaluated.
One can refer to \cite{Bhatnagar2019CNN-encoder-decoder,Li2020Reaction-diffusion-prediction-CNN,Gao2020PhyGeoNet-PDE-on-Irregular-domain} and many others for strategies to map complex and irregular domain into a regular grid mesh. 
Such geometry transformation can be easily taken into account in the proposed PDE loss layers with the isoparametric mapping concept of the FEM via the shape functions. 
The proposed approach, using a mask matrix for domain detection, should still be applicable to other parametric domain representations, though it is not the focus of this work.

In our study, each input data point represents a unique BVP for a specific problem domain with a specific set of applied BCs.
To detect the Dirichlet BCs, during the NN training, the input data is first passed to a customized Keras layer, called \verb LayerFillRandomNumber, which fills the pixel locations with values of  $-2$ in the Dirichlet BCs channel with uniformly generated random numbers in the range of $[0,~1]$.
As the problem domain is filled with random numbers, the convolutional kernels eventually only pick up and learn the actual Dirichlet boundary values.
The data structure in the Neumann BCs channel automatically ensures that the kernels learn the information of BCs, as the problem domain is filled with zeros.

\subsection{Setup of BVPs}
\begin{figure}[t!]
  \centering
  %\psfrag{a}[c][c]{step 1}
  \includegraphics[width=1.0\linewidth]{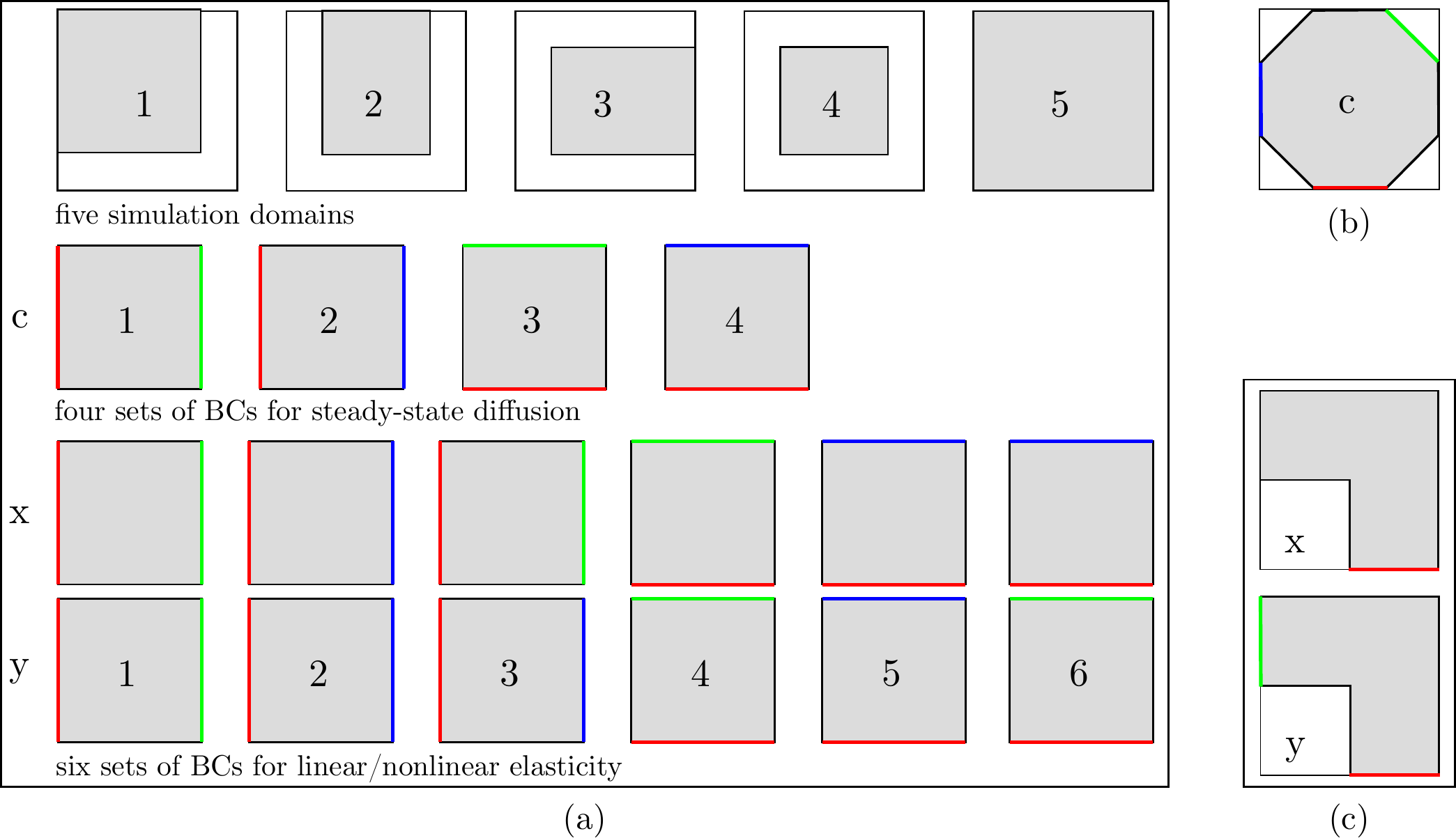}
  \caption{Illustration of the setup of BVPs on different domains for different physics. In these drawings, red represents a zero Dirichlet BC. Green represents a non-zero Dirichlet BC. Blue represents a non-zero Neumann BC. No color is assigned to Zero Neumann BCs. (a) Setup of five rectangle simulation domains of different sizes and locations on a fixed background grid with different applied BCs. For steady-state diffusion, 4 sets of BCs are assigned to each simulation domain, leading to 20 diffusion BVPs. For linear/nonlinear elasticity, 6 sets of BCs are assigned to each simulation domain, leading to 30 linear/nonlinear elasticity BVPs. (b) Setup of one diffusion BVP with an octagon simulation domain with mixed BCs. (c) Setup of one linear elasticity BVP with a L-shape simulation domain with the bottom edge fixed and the left edge loaded vertically.}
  \label{fig:bvp-problem-setup}
\end{figure}

To demonstrate the performance of our proposed method, we investigate different setups of BVPs for the three considered physical systems. 
Specifically, we consider rectangle, octagon, and L-shape simulation domains, as shown in Fig. \ref{fig:bvp-problem-setup}.
For the rectangular domain, we allow its size and location to change with respect to a fixed background grid\footnote{The fixed background grid is necessary to ensure that the same NN structure can be used to solve PDEs on different simulation domains}.
Five rectangular domains are considered, as shown in Fig. \ref{fig:bvp-problem-setup}(a).
For steady-state diffusion, four unique sets of BCs are assigned to each domain, resulting in 20 different diffusion BVPs.
For linear/nonlinear elasticity, six unique sets of BCs are assigned to each domain, resulting in 30 different linear/nonlinear elasticity BVPs.
On the octagonal domain, we solve for the steady-state diffusion problem with one specific set of BCs, as shown in Fig. \ref{fig:bvp-problem-setup}(b).
On the L-shape domain, we solve for linear elasticity with one specific set of BCs, as shown in Fig. \ref{fig:bvp-problem-setup}(c).
The NN inputs corresponding to these BVP setups are synthetically generated to train the discretized residual-constrained NNs. 
To compare the solution accuracy between NNs and DNSs, we also solve these BVPs with \verb mechanoChemFEM \footnote{Code available at \href{https://github.com/mechanoChem/mechanoChemFEM}{github.com/mechanoChem/mechanoChemFEM}.}, which is a publicly available multiphysics code developed by us based on the deal.II library \cite{dealII90}.

\subsection{NNs training} \label{sec:NN-training}

\begin{figure}[t!]
  \centering
  %\psfrag{a}[c][c]{step 1}
  \includegraphics[width=1.0\linewidth]{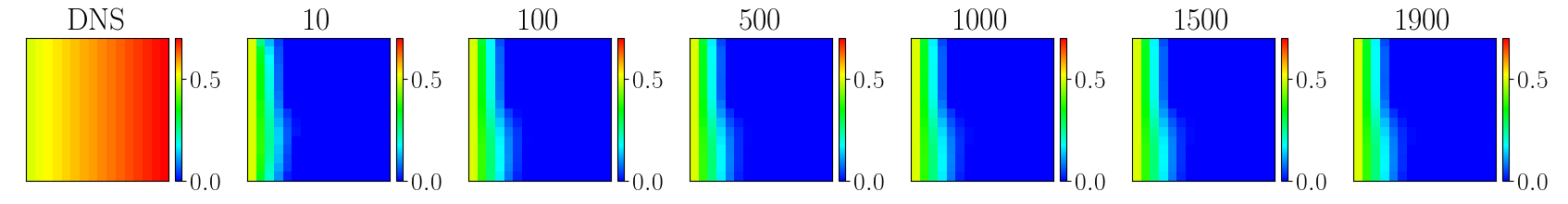}
  \includegraphics[width=1.0\linewidth]{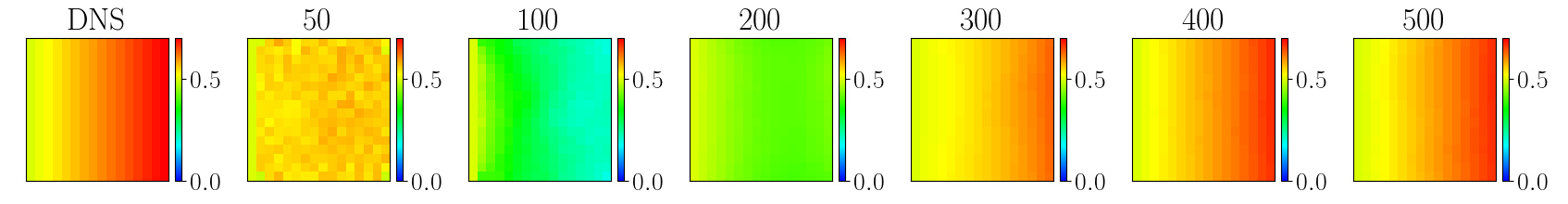}
  \caption{Illustration of the deterministic NNs predicted solution at different epochs for a diffusion BVP setup with domain id 5 and BCs id 2 (flux loading from the right edge), as shown in Fig. \ref{fig:bvp-problem-setup}(a). Top: without zero initialization. Bottom: with zero initialization for the first 100 epochs.}
  \label{fig:zero-initialization}
\end{figure}

\begin{algorithm}[t!]
  \caption{Training procedure for deterministic NNs. \label{algo:cnn-training}}
  \begin{algorithmic}[1]
    \STATE Load NN inputs with each data point being a unique set of BCs.
    \STATE Augment the inputs by duplicating them for multiple times to generate $\calD$.
    \STATE Split $\calD$ into training, validation, and testing datasets. 
    \STATE Setup the encoder-decoder deterministic NN structure, with the first layer being a customized layer to fill the locations that have values of $-2$ in $\calD$ with uniform random numbers between $[0,1]$ to ensure $\calD$ is i.i.d.
    \FOR{epoch < total epochs}
    \STATE Batch train the NNs
    \IF{use zero initialization \AND epoch < total zero initialization epochs}
    \STATE Use dummy labels with values of 0.5, which is equivalent to an actual zero before data normalization, to form the MSE loss to train the NN.
    \ELSE
    \STATE Use $\BR_\text{tot}^\text{red}$ to form the deterministic loss to train the NN.
    \ENDIF
    \ENDFOR
    \STATE Make prediction.
  \end{algorithmic}
\end{algorithm}

For deterministic NNs, a fixed learning rate is used to batch optimize the loss function \eref{eq:loss-deterministic} to solve the PDE systems.
In our study, we found that problems loaded with Dirichlet BCs converge faster than cases loaded with Neumann BCs. 
The proposed approach sometimes fails to find the correct solution for the latter case. 
This observation holds for all three considered systems. 
This is mainly because, for the latter case, it is essentially the gradient of the unknown(s) that drives the loss down instead of the unknown(s) itself as for the former case.
We demonstrate this by showing the NNs predicted solution at different epochs in Fig. \ref{fig:zero-initialization} for a diffusion BVP setup with domain id 5 and BCs id 2 (see Fig. \ref{fig:bvp-problem-setup}a) with zero concentration on the left edge and non-zero flux on the right edge.
The top row of Fig. \ref{fig:zero-initialization} shows that the NN predicted concentration very slowly changes by a front progressing from the left edge (zero Dirichlet BCs) to the right edge (flux BCs), and the solution is not yet close to the DNS results.

Such challenge arises mainly because the parameters of NNs are randomly initialized. 
NN predicted solutions at the earlier training stage are random numbers close to zero. Since data normalization is used, the NNs output scaled results with zero being equivalent to an actual value of $-1$. 
Such random outputs could easily violate the governing equations, e.g. resulting in a deformation gradient with negative determinant in nonlinear elasticity. 
Recalling that the solution vector in the FEM is initialized to zero in general, we adopt the same approach for the NNs. 
For the first few epochs, we train NNs with dummy labels with values of 0.5 (equivalent to an actual value of 0) without enforcing the PDE constraint. 
We call this as the zero initialization process.
This process helps to improve the initialization of NN parameters. 
After the zero initialization procedure is completed, the PDE constraints are enabled to train the NNs to solve the PDE systems. 
We found that this remedy drastically improves the training results. 
In addition, it also speeds up the overall training process, as shown in the bottom row of Fig. \ref{fig:zero-initialization}, where the NN predicted solutions approach the DNS results at 500 epochs, much faster than the case without the zero initialization process.
The training process\footnote{Usually, the number of unique sets of BCs is small, compared to $\calD$, which is augmented  multiple times. Thus, even though the dataset is split into training, validation, and testing groups, each group could potentially contain all the unique BCs. The difference among these dataset groups lies in the interior domain, which is filled with random numbers.} for deterministic NNs is summarized in the Algorithm Box \ref{algo:cnn-training}. 

For probabilistic NNs, we can use the proposed approach successfully solve a single BVP. 
However, when we try to solve multiple BVPs, we notice that the BNNs converge faster to purely Dirichlet problems (boundary id 1, 3 for the diffusion problem and boundary id 1, 4 for elasticity problems) than those with non-zero Neumann BCs.
Once the BNNs converges to a sub-optimal state, it is very challenging to optimize BNNs further for other BVPs with Neumann BCs.
To overcome this challenge, we first train deterministic NNs with identical architecture as the BNNs. 
Once the deterministic NNs are converged to a desired state, we then initialize the mean of the posterior distribution of parameters in the BNNs with the optimized parameters from the deterministic model.
We call this as the optimal parameter initialization process.
During the subsequent training of the BNNs, similar as in \cite{Geneva2020Zabaras-JCP-auto-regressive-NN-PDE}, we use a small learning rate to explore the local parameter space around these optimized parameters. 
The training process for BNNs is summarized in the Algorithm Box \ref{algo:bnn-training}. 

\begin{algorithm}[t!]
  \caption{Training procedure for BNNs. \label{algo:bnn-training}}
  \begin{algorithmic}[1]
    \STATE Load NN inputs with each data point being a unique set of BCs.
    \STATE Augment the inputs by duplicating them multiple times to generate $\calD$.
    \STATE Split $\calD$ into training, validation, and testing datasets. 
    \STATE Setup the encoder-decoder probabilistic NN structure, with the first layer being a customized layer to fill the locations that have values of $-2$ in $\calD$ with uniform random numbers between $[0,1]$.
    \IF{use optimal parameter initialization}
      \STATE Load the optimized parameters from the deterministic NNs to initialize the mean of the posterior distribution of BNN parameters.
    \ELSE
      \STATE Use random initialization for the posterior distribution of BNN parameters.
    \ENDIF
    \FOR{epoch < total epochs}
    \STATE Batch train the NNs
    \IF{use zero initialization \AND epoch < total zero initialization epochs \AND (\NOT use optimal parameter initialization)}
    \STATE Use dummy labels with values of 0.5, which is equivalent to an actual zero before data normalization, to form the MSE loss to train the NN.
    \ELSE
    \STATE Use $\BR_\text{tot}^\text{red}$ to form the likelihood loss to train the NN.
    \ENDIF
    \ENDFOR
    \STATE MC sampling for UQ.
  \end{algorithmic}
\end{algorithm}

\section{Numerical results}\label{sec:results}
In this section, the discretized residual constrained NNs are used to solve for the setup of BVPs presented in Section \ref{sec:bvp-definition} for steady-state diffusion, linear elasticity, and non-linear elasticity, to demonstrate the capability and performance of the proposed framework.

\subsection{Steady-state diffusion}
In this section, we use the proposed method to solve different steady-state diffusion problems.  
\subsubsection{Background}
The general description of an elliptic PDE system given \eref{eq:general-pde} is rewritten as 
\begin{equation}
  \begin{aligned}
    \nabla \cdot \BH = \Bzero \quad & \text{on} \quad \Omega, \\
    C (\BX) = \bar{C} (\BX)  \quad & \text{on} \quad  \Gamma^{C}, \\
    H = \bar{H} (\BX)  \quad & \text{on} \quad  \Gamma^H,
  \end{aligned}
  \label{eq:general-pde-diffusion}
\end{equation}
for the one species steady-state diffusion problem. 
In \eref{eq:general-pde-diffusion}, $C$ represents the compositional order parameter, $\BH$ is the diffusive flux term defined as
\begin{equation}
  \BH = - D \nabla C,
  \label{eq:diffusion-flux}
\end{equation}
with $D$ as the diffusivity, and $H$ is the outward surface flux in the normal direction\footnote{In section \ref{sec:results-diffusion-rectangle} and \ref{sec:results-diffusion-octagon}, the inward flux has a positive sign.}. 
The discretized residual function \eref{eq:discretized-residual} for steady-state diffusion is written as
\begin{equation}
    \begin{aligned}
  \BR  & = \sum_{e=1}^{n_\text{elem}} \left\{ \int_{\Omega^e} \BB^T \BH dV - \int_{\Gamma^{e,H}} \BN^T \bar{H}~dS \right\}. \\
    \end{aligned}
  \label{eq:discretized-residual-diffusion}
\end{equation}
A diffusivity $D=1.0$ is used in both DNSs and the surrogate PDE solver.

\subsubsection{Multiple rectangular domains with different BCs}\label{sec:results-diffusion-rectangle}
\begin{table}
  \centering
  \begin{tabular}{l | l | l | l}
    \hline
    Deterministic         & Probabilistic         & Size         & Layer arguments \\ \hline
    Input                 & Input                 & -            & - \\
    LayerFillRandomNumber & LayerFillRandomNumber & -            & - \\
    Conv2D                & Convolution2DFlipout  & filters = 8  & kernel (5,5), padding: same, ReLU \\
    MaxPooling2D          & MaxPooling2D          & -            & kernel (2,2), padding: same\\
    Conv2D                & Convolution2DFlipout  & filters = 16 & kernel (5,5), padding: same, ReLU \\
    MaxPooling2D          & MaxPooling2D          & -            & kernel (2,2), padding: same\\
    Conv2D                & Convolution2DFlipout  & filters = 16 & kernel (5,5), padding: same, ReLU \\
    MaxPooling2D          & MaxPooling2D          & -            & kernel (2,2), padding: same\\
    Flatten               & Flatten               & -            & - \\
    Dense                 & DenseFlipout          & units = 64   & ReLU \\
    Dense                 & DenseFlipout          & units = 64   & ReLU \\
    Reshape               & Reshape               & -            & $[4,4,4]$ \\
    Conv2D                & Convolution2DFlipout  & filters = 16 & kernel (5,5), padding: same, ReLU \\
    UpSampling2D          & UpSampling2D          & -            & size (2,2) \\
    Conv2D                & Convolution2DFlipout  & filters = 16 & kernel (5,5), padding: same, ReLU \\
    UpSampling2D          & UpSampling2D          & -            & size (2,2) \\
    Conv2D                & Convolution2DFlipout  & filters = 16 & kernel (5,5), padding: same, ReLU \\
    Conv2D                & Convolution2DFlipout  & filters = 1  & kernel (5,5), padding: same, ReLU \\
    \hline
  \end{tabular}
  \caption{Details of both deterministic and probabilistic NNs for solving 20 steady-state diffusion BVPs.}
  \label{tab:diffusion-20bvp-NNs}
\end{table}

\begin{table}
  \centering
  \begin{tabular}{l | l | l }
    \hline
    Description                     & Deterministic                 & Probabilistic         \\ \hline
    Total parameters                & 33,209                        & 66,202                 \\
    Size of $\calD$                 & 20 $\times$ Aug: $2^{10}$     & 20 $\times$ Aug: $2^{9}$      \\
    Epochs                          & 20,000                        & 5,000                 \\
    Zero initialization epochs      & 100                           & -                     \\
    Optimizer                       & Nadam                         & Nadam                 \\
    Learning Rate                   & 2.5e-4                        & 1e-8                  \\
    Batch Size                      & 256                           & 64                    \\
    $\Sigma_1$                      & -                             & 1e-8                  \\
    Initial value of $\Sigma_2$     & -                             & 1e-8                  \\
    \hline
  \end{tabular}
  \caption{Training related parameters for solving 20 steady-state diffusion BVPs. Aug: data augmentation. }
  \label{tab:diffusion-20bvp-NNs-others}
\end{table}

In this section, we use the proposed PDE constrained NNs to simultaneously solve 20 steady-state diffusion BVPs, as shown in Fig. \ref{fig:bvp-problem-setup}(a),  with a resolution of $16\times16$.
The architectures of both deterministic and probabilistic NNs and other training related NN parameters are summarized in Table \ref{tab:diffusion-20bvp-NNs} and \ref{tab:diffusion-20bvp-NNs-others}, respectively.
The NN hyperparameters are manually tuned to achieve a desired performance.
We follow the training procedures in Algorithm Boxes \ref{algo:cnn-training} and \ref{algo:bnn-training} to first train the deterministic NN with zero initialization, followed by training the BNNs with the optimal parameter initialization process.
The results of two selected BVPs are shown in Fig. \ref{fig:diffusion-20bvp-results}, with remaining results from other setups being given in Appendix \ref{appendix:diffusion-20bvps}.
The statistical moments of the BNN predictions are evaluated based on 50 MC samplings.
In Fig. \ref{fig:diffusion-20bvp-results}, BVP (i) and (ii) correspond to bc id 1 (non-zero Dirichlet loading) and bc id 2 (non-zero Neumann loading) applied to domain id 1. 
The NN inputs for both BVPs are shown in Fig. \ref{fig:diffusion-20bvp-results}(a,b), in which only the red colored regions are physically meaningful with values $>0$. 
The comparison of solutions among DNSs, the deterministic NN, and the BNN for these two BVPs is shown qualitatively in Fig. \ref{fig:diffusion-20bvp-results}(c,e), with quantitative comparison of the solution distribution along the dashed line between DNSs and the BNN given in Fig. \ref{fig:diffusion-20bvp-results}(d,f).
Such a comparison shows that the proposed method has successfully solved the BVPs with desired accuracy.
We further observe from Fig. \ref{fig:diffusion-20bvp-results}(f) that the uncertainty at the locations with the Neumann BCs is higher than other places, which is expected.

\begin{figure}[t!]
  \centering
  \subfloat[NN inputs for BVP (i)]{\includegraphics[height=0.17\linewidth]{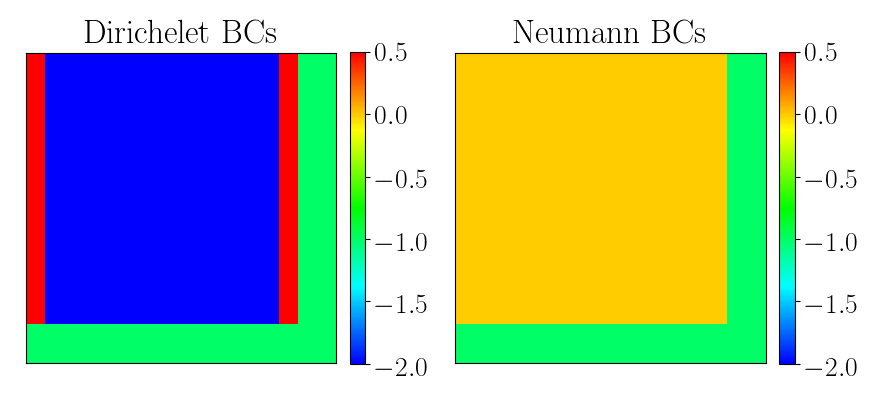}} \hspace{2cm}
  \subfloat[NN inputs for BVP (ii)]{\includegraphics[height=0.17\linewidth]{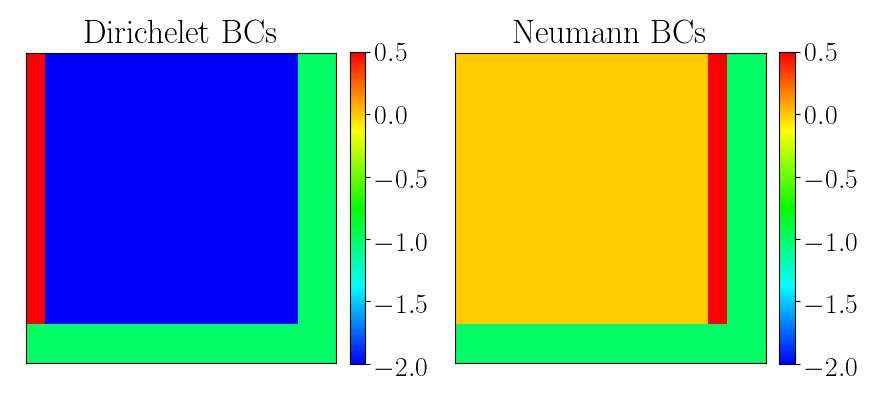}}  \\
  \subfloat[results for BVP (i)]{\includegraphics[height=0.18\linewidth]{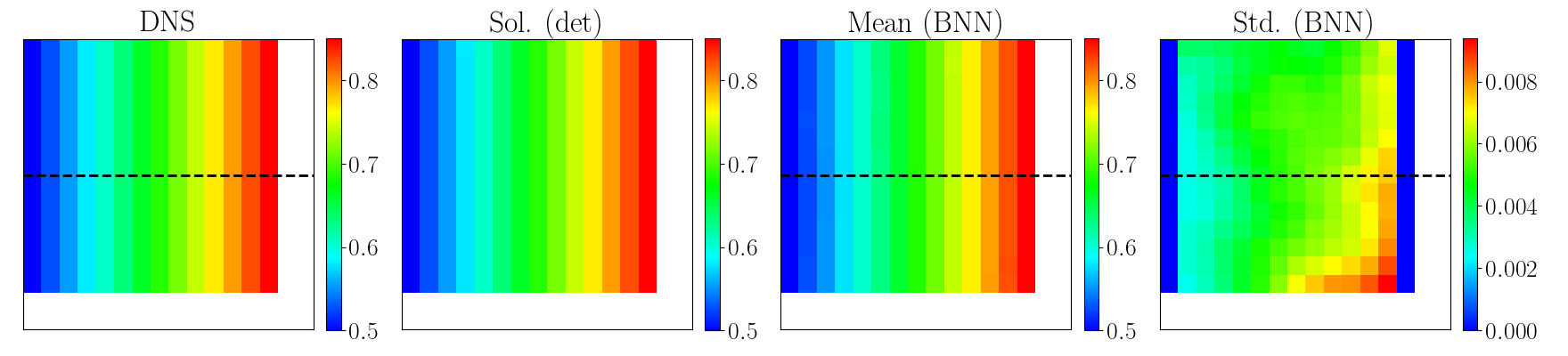}} 
  \subfloat[UQ for BVP (i)]{\includegraphics[height=0.18\linewidth]{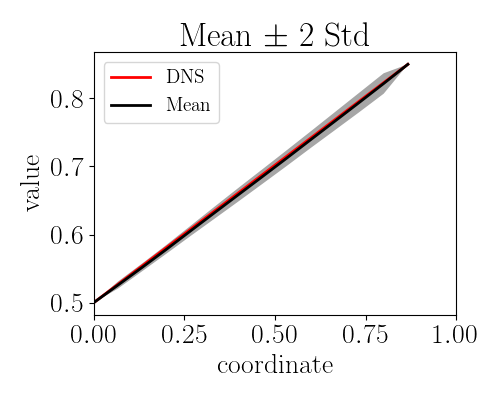}} \\
  \subfloat[results for BVP (ii)]{\includegraphics[height=0.18\linewidth]{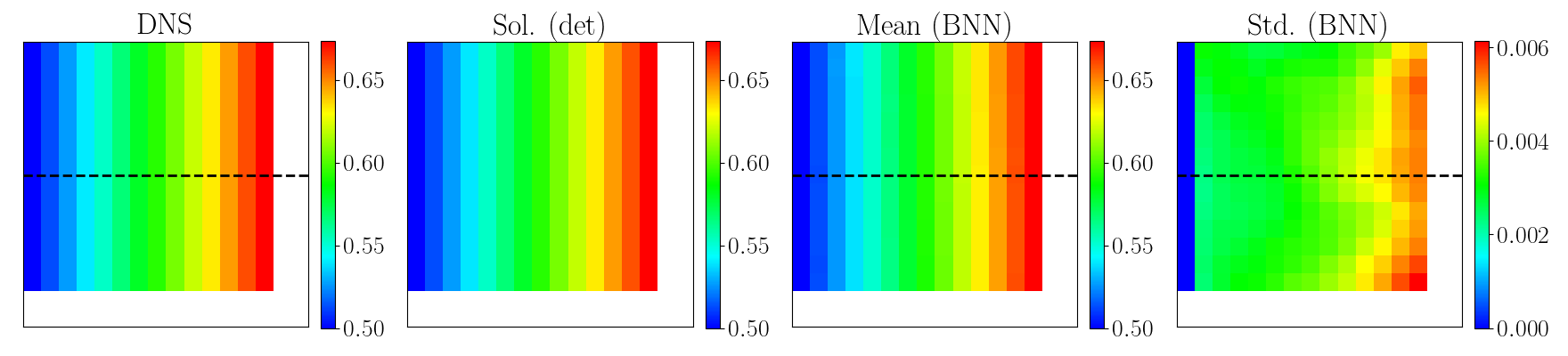}} 
  \subfloat[UQ for BVP (ii)]{\includegraphics[height=0.18\linewidth]{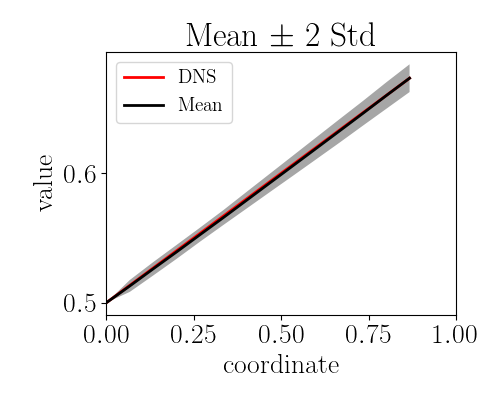}} \\ 
  \caption{Results of two selected BVPs out of the 20 steady-state diffusion BVPs with varying domains and different applied BCs simultaneously solved by a single NN with the proposed method. BVP (i) and (ii) correspond to bc id 1 and 2 for domain id 1, as shown in Fig. \ref{fig:bvp-problem-setup}(a). 
  (a, b) NN inputs for different BVPs.
  (c, e) Solutions from DNS, deterministic (det) NNs, and BNNs (Mean, Std.) for different BVPs\protect\footnotemark.
(d, f) Quantitative comparison of the solution distribution between DNS and BNNs along the dashed line.}
  \label{fig:diffusion-20bvp-results}
\end{figure}
\footnotetext{As Dirichlet BCs are enforced to NN predicted solutions, the uncertainty at the Dirichlet boundary locations is not evaluated. A zero standard deviation of the solution at these locations is shown in Fig. \ref{fig:diffusion-20bvp-results}(c,e).}

\begin{figure}[t!]
  \centering
  %\psfrag{a}[c][c]{step 1}
  \subfloat[deterministic loss]{\includegraphics[width=0.33\linewidth]{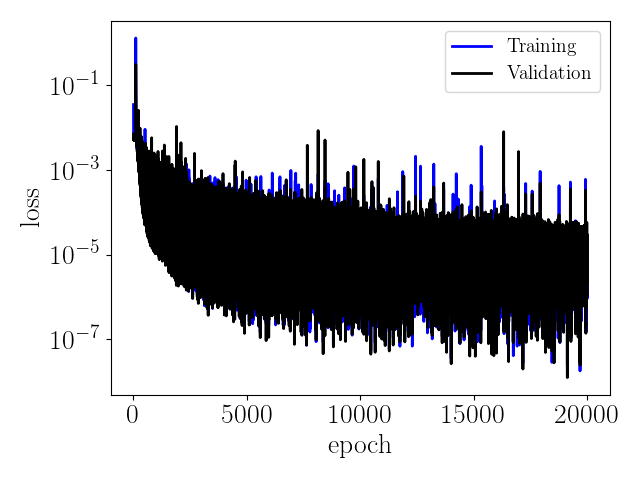}}
  \subfloat[probabilistic loss]{\includegraphics[width=0.33\linewidth]{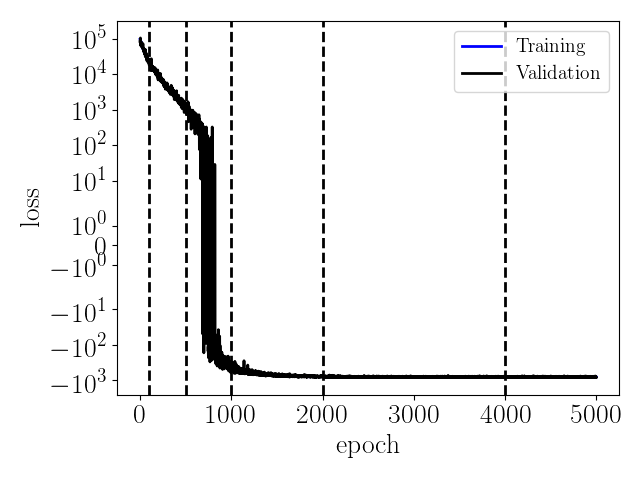}}
  \subfloat[$\Sigma_2$]{\includegraphics[width=0.33\linewidth]{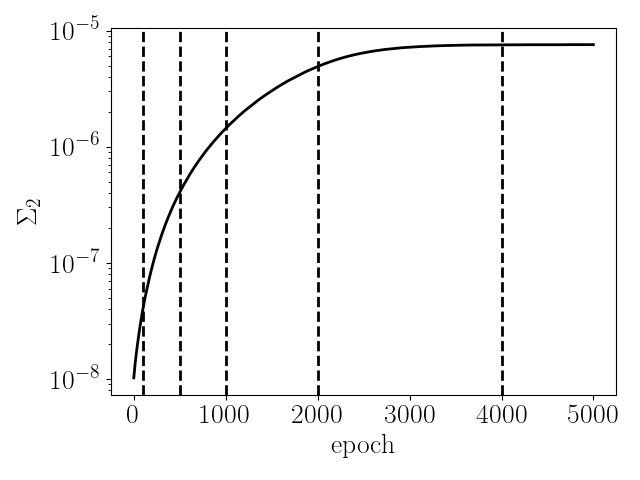}} \\
  \caption{NN training information for results shown in Fig. \ref{fig:diffusion-20bvp-results}. (a) Loss from the deterministic NN. (b) Loss from the BNN. (c) Evolution of $\Sigma_2$ from the BNN.}
  \label{fig:diffusion-20bvp-loss}
\end{figure}

\begin{figure}[t!]
  \centering
  %\psfrag{a}[c][c]{step 1}
  \subfloat[epoch 100]{\includegraphics[width=0.20\linewidth]{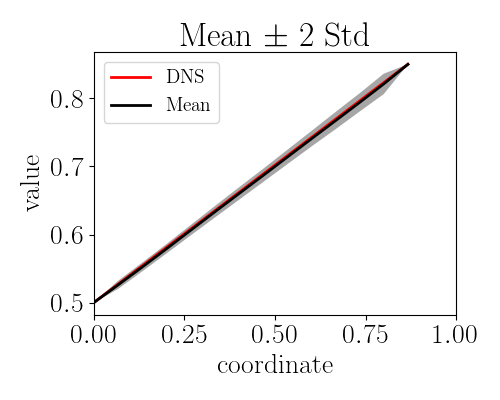}}
  \subfloat[epoch 500]{\includegraphics[width=0.20\linewidth]{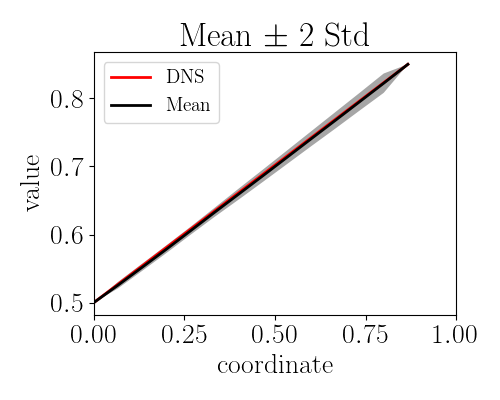}}
  \subfloat[epoch 1000]{\includegraphics[width=0.20\linewidth]{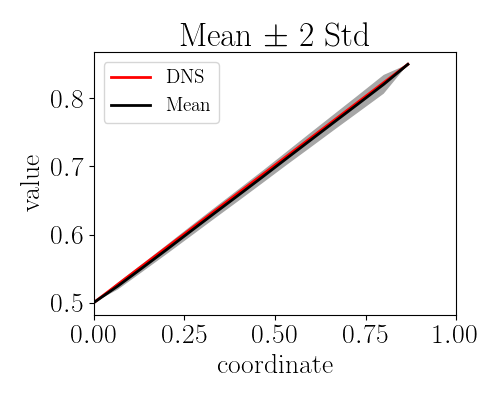}}
  \subfloat[epoch 2000]{\includegraphics[width=0.20\linewidth]{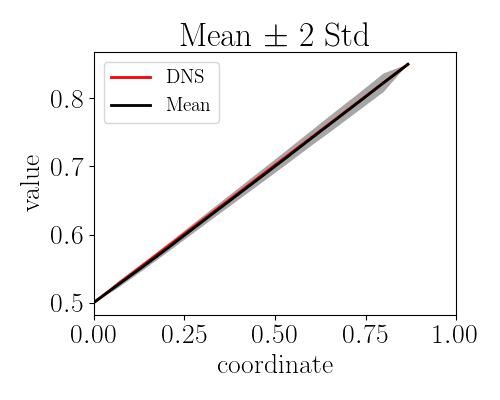}}
  \subfloat[epoch 4000]{\includegraphics[width=0.20\linewidth]{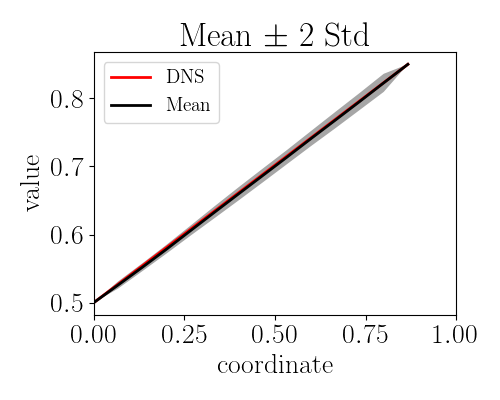}} \\
  \subfloat[epoch 100]{\includegraphics[width=0.20\linewidth]{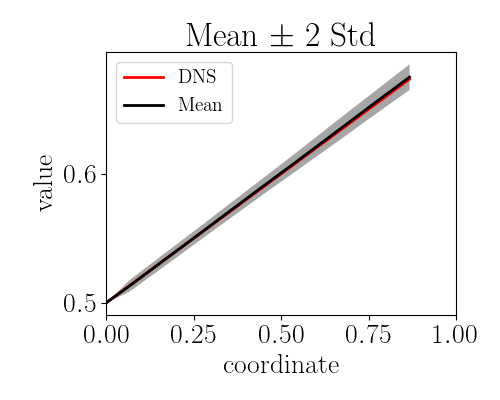}}
  \subfloat[epoch 500]{\includegraphics[width=0.20\linewidth]{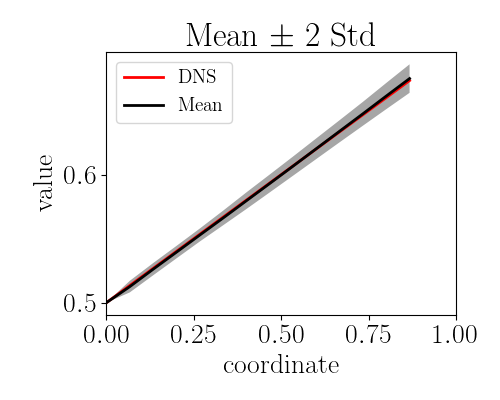}}
  \subfloat[epoch 1000]{\includegraphics[width=0.20\linewidth]{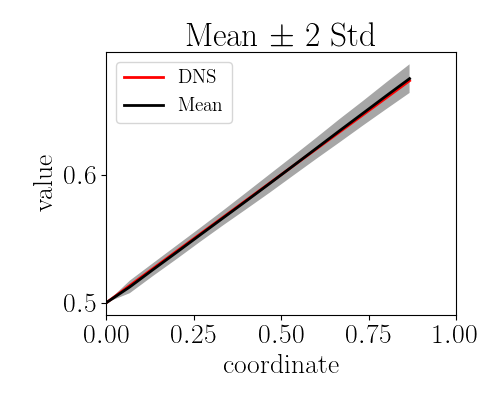}}
  \subfloat[epoch 2000]{\includegraphics[width=0.20\linewidth]{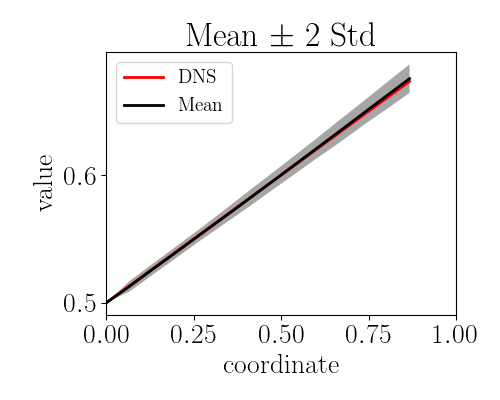}}
  \subfloat[epoch 4000]{\includegraphics[width=0.20\linewidth]{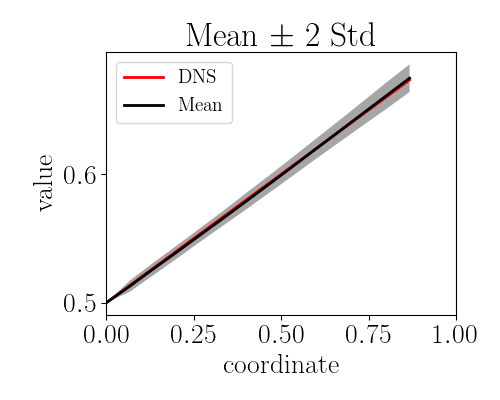}}
  \caption{Solution distribution from BNNs along the dashed line at different epochs for BVP (i) (top) and BVP (ii) (bottom).}
  \label{fig:diffusion-20bvp-bnn-vs-sigma2}
\end{figure}

The training losses for both deterministic and probabilistic NNs are given in Fig. \ref{fig:diffusion-20bvp-loss}(a, b).
The negative loss in Fig. \ref{fig:diffusion-20bvp-loss}(b) is reasonable, because the total loss of BNNs in \eref{eq:loss-probabilistic} consists two terms.
The first term in \eref{eq:loss-probabilistic} is non-negative, whereas the second term could be either positive or negative depending on values of both $\BR_\text{tot}^\text{red}$ and $\Sigma_2$.
The evolution of $\Sigma_2$ from the BNN is shown in Fig. \ref{fig:diffusion-20bvp-loss}(c), which converges to a specific value during training. 
The evolution of $\Sigma_2$ is correlated to the sign change of the BNN loss.
To further evaluate the relation between BNN predicted results and the value of $\Sigma_2$, we report the solution distribution along the dashed line for epochs 100, 500, 1000, 2000, 4000, indicated by the vertical lines in Fig. \ref{fig:diffusion-20bvp-loss}(c). 
The results for both BVP (i) and (ii) are presented in Fig. \ref{fig:diffusion-20bvp-bnn-vs-sigma2}, which shows that solutions from the BNN are stable during training regardless of the evolution of $\Sigma_2$.
Such behavior is expected as the BNN is initialized with optimal parameters from the deterministic NNs and is trained with a very small learning rate to only explore the local parameter space around the optimized parameters.
The change of the probabilistic loss in Fig. \ref{fig:diffusion-20bvp-loss}(b) is attributed to the initial value of $\Sigma_2$, which differs from its actual value. 
Based on the observation in Fig. \ref{fig:diffusion-20bvp-bnn-vs-sigma2}, it is therefore reasonable to train the BNNs for a small number of epochs to evaluate the statistical moments of BNN predicted solutions.
For the remaining simulations in Section \ref{sec:results}, the BNNs are trained for 100 epochs before evaluating the statistical moments of related quantities.

\subsubsection{Single octagon domain with mixed BCs}\label{sec:results-diffusion-octagon}
\begin{figure}[t!]
  \centering
  %\psfrag{a}[c][c]{step 1}
  \subfloat[results ($32\times32$)]{\includegraphics[width=0.90\linewidth]{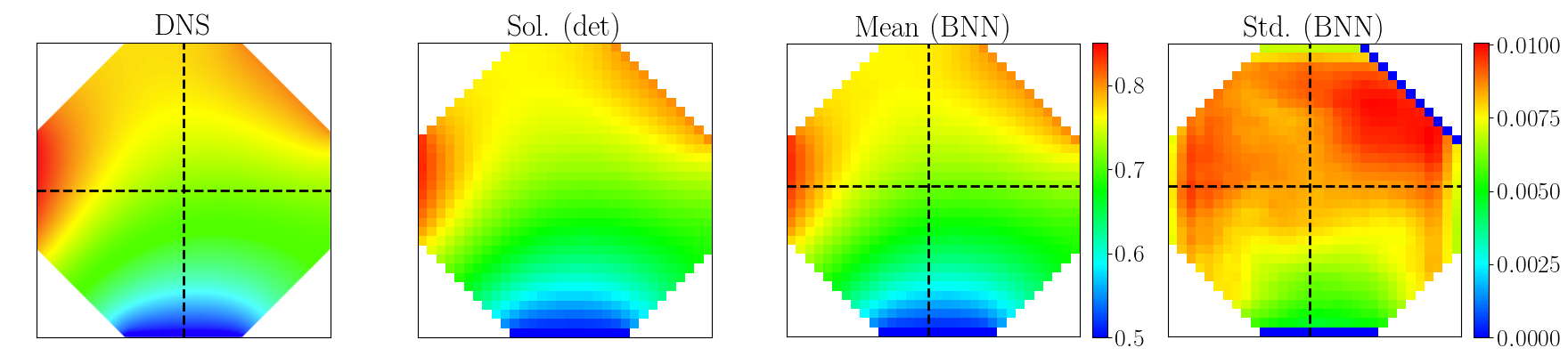}} \\
  \subfloat[results ($64\times64$)]{\includegraphics[width=0.90\linewidth]{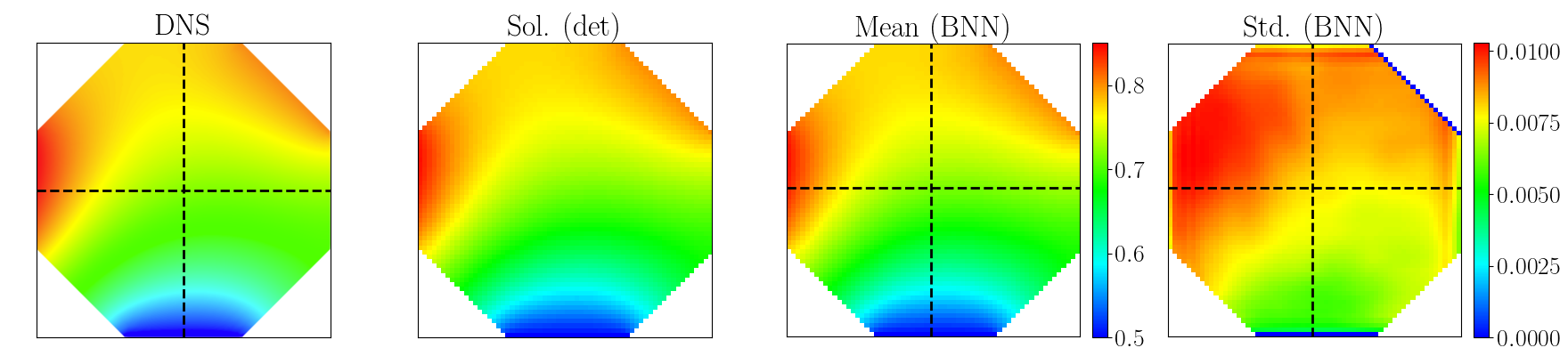}} \\
  \subfloat[UQ horizontal ($32\times 32$)]{\includegraphics[trim=0.5cm 0.0cm 0.5cm 0.0cm, clip, width=0.245\linewidth]{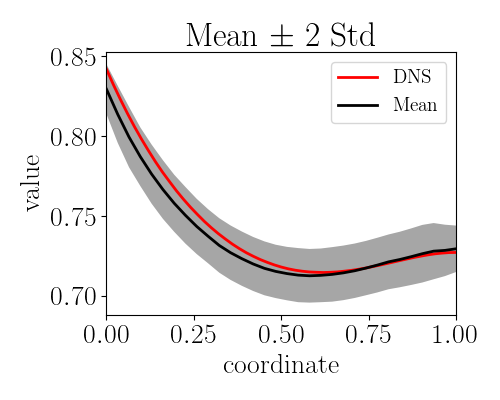}}
\subfloat[UQ horizontal ($64\times 64$)]{\includegraphics[trim=0.5cm 0.0cm 0.5cm 0.0cm, clip, width=0.245\linewidth]{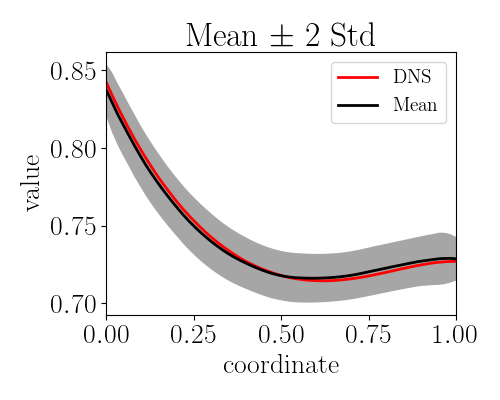}}
\subfloat[UQ vertical ($32\times 32$)]{\includegraphics[trim=0.5cm 0.0cm 0.5cm 0.0cm, clip, width=0.245\linewidth]{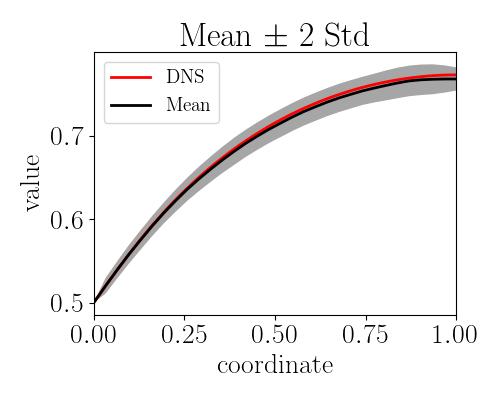}}
  \subfloat[UQ vertical ($64\times 64$)]{\includegraphics[trim=0.5cm 0.0cm 0.5cm 0.0cm, clip, width=0.245\linewidth]{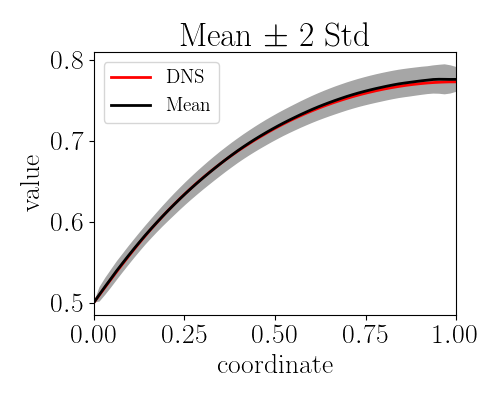}}
  \caption{Steady-state diffusion BVP on an octagonal domain with mixed BCs for different output resolutions. 
    (a, b) Solutions from DNS, deterministic (det) NNs, and BNNs (Mean, Std.) for output resolutions of $32\times 32$ and $64\times 64$, respectively.
  (c-f) Quantitative comparison of the solution distribution between DNS and BNNs along the horizontal and vertical dashed lines, which shows that the NN solutions are more accurate with a finer output resolution.}
  \label{fig:diffusion-irregular-1bvp}
\end{figure}

The example in the previous section is fairly simple and is essentially a one-dimensional problem. 
In this section, we use the proposed PDE constrained NNs to solve steady-state diffusion on an octagonal domain with mixed BCs, as shown in Fig. \ref{fig:bvp-problem-setup}(b), whose solution is nonlinear in both X- and Y- directions.
To keep the discussion concise for easy reading, the architectures of both deterministic and probabilistic NNs along with other training related information are provided in the Appendix \ref{appendix:diffusion-octagon}.
We follow the procedures described in Section \ref{sec:NN-training} to train both types of NNs with two output resolutions, $32\times32$ and $64\times64$. 

Similar to the development in Section \ref{sec:results-diffusion-rectangle}, we compare solutions among DNSs, the deterministic NN, and the BNN for these two output resolutions and show them qualitatively in Fig. \ref{fig:diffusion-irregular-1bvp}(a,b), with quantitative comparison of the solution distribution along both the horizontal and vertical dashed lines between DNSs and the BNN given in Fig. \ref{fig:diffusion-irregular-1bvp}(c-f).
Standard deviations along the bottom edge and the top right corner edge in Fig. \ref{fig:diffusion-irregular-1bvp}(a,b) are not evaluated and are assigned a value of zero, because Dirichlet BCs are enforced to NN predicted solutions along these edges. 
Fig. \ref{fig:diffusion-irregular-1bvp} shows that NN results from both output resolutions are comparable with the DNS solution. 
The comparison between Figs. \ref{fig:diffusion-irregular-1bvp}(c,e) and \ref{fig:diffusion-irregular-1bvp}(d,f) further shows that, as expected, the NN solutions are improved with a finer output resolution.
The uncertainties of the NN predicted solution along the horizontal line is larger than it along the vertical line. 
This is reasonable as the non-linearity in the solutions along the horizontal line is higher than it along the vertical line.
This example and the example in the previous section demonstrate that the proposed framework can properly recognize both regular and irregular problem domains, learn different BCs, and simultaneously solve multiple BVPs.

\subsection{Linear elasticity}
In this section, we use the proposed method to solve different linear elasticity problems.  

\subsubsection{Background}
The general description of an elliptic PDE system given \eref{eq:general-pde} is rewritten as 
\begin{equation}
  \begin{aligned}
    \nabla \cdot \Bsigma = \Bzero \quad & \text{on} \quad \Omega, \\
    \Bu (\BX) = \bar{\Bu} (\BX)  \quad & \text{on} \quad  \Gamma^{\Bu}, \\
    \BT = \bar{\BT} (\BX)  \quad & \text{on} \quad  \Gamma^\BT,
  \end{aligned}
  \label{eq:general-pde-linear-elasticity}
\end{equation}
for the linear elasticity problem. 
In \eref{eq:general-pde-linear-elasticity}, $\Bu$ represents the displacement field, $\Bsigma$ is the stress tensor, and $\BT$ is the surface traction.
Here, $\Bsigma$ is related to the strain $\Bvarepsilon=\frac{1}{2}\left( \nabla \Bu + (\nabla \Bu)^T\right)$ via the following constitutive relationship
\begin{equation}
  \Bsigma = \lambda\tr (\Bvarepsilon) \Bone + 2\mu\Bvarepsilon
  \label{eq:constitutive-linear}
\end{equation}
where $\lambda$ and $\mu$ are the Lam\'e constants, and $\Bone$ is the second-order identity tensor.
The discretized residual function \eref{eq:discretized-residual} for the linear elasticity problem is written as
\begin{equation}
  \BR = \sum_{e=1}^{n_\text{elem}} \left\{ \int_{\Omega^e} \BB^T \Bsigma dV - \int_{\Gamma^{e,T}} \BN^T \bar{\BT}~dS \right\}.
  \label{eq:discretized-residual-linear}
\end{equation}
A set of material parameters with $\lambda=14.4231$ and $\mu=9.61538$ is used in both DNSs and the surrogate PDE solver.

\subsubsection{Multiple rectangular domains with different BCs}\label{sec:linear-rectangle}

\begin{figure}[t!]
    \centering
    \subfloat[three selected BVPs on rectangle domains]{\includegraphics[height=0.20\linewidth]{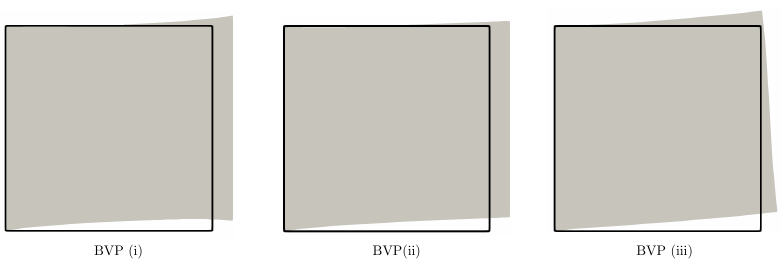}} \hspace{2cm}
    \subfloat[l-shape]{\includegraphics[height=0.20\linewidth]{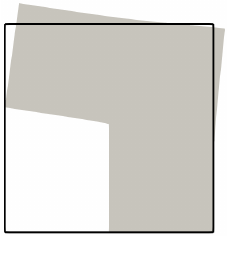}}
    \caption{Illustration of the deformed shape of selected linear elasticity BVPs. The wireframe and the gray region indicate the undeformed and deformed problem domain, respectively. (a) Three selected BVPs of out the 30 BVPs solved in section \ref{sec:linear-rectangle}. (b) The L-shape BVP solved in section \ref{sec:linear-lshape}.}
    \label{fig:linear-bvp-deformed}
\end{figure}

\begin{figure}[p!]
  \centering
  \subfloat[$u_x$ results for BVP (i)]  {\includegraphics[trim=0.1cm 0.5cm 0.1cm 0.25cm, clip,height=0.16\linewidth]{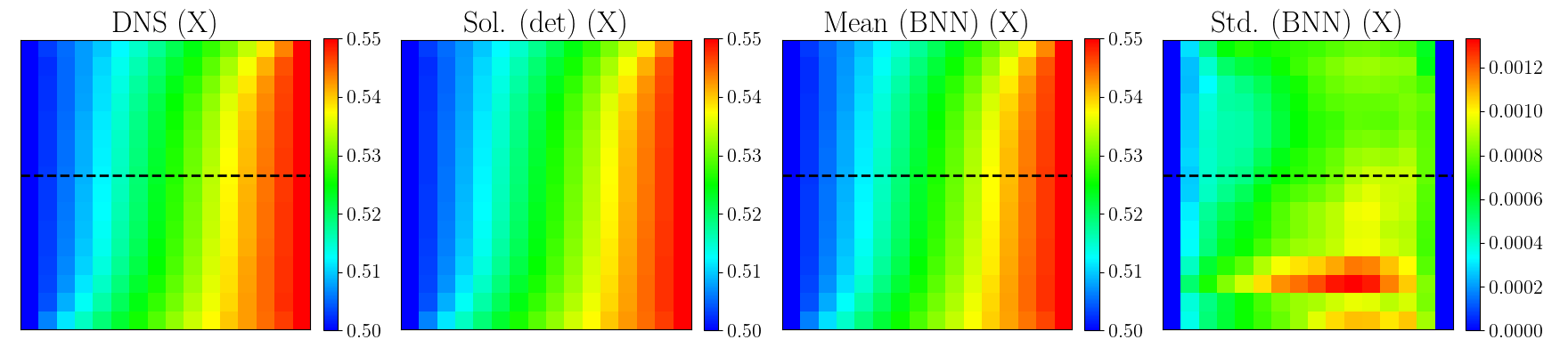}} 
  \subfloat[$u_x$ UQ for BVP (i)]       {\includegraphics[trim=0.1cm 0.5cm 0.1cm 0.25cm, clip,height=0.16\linewidth]{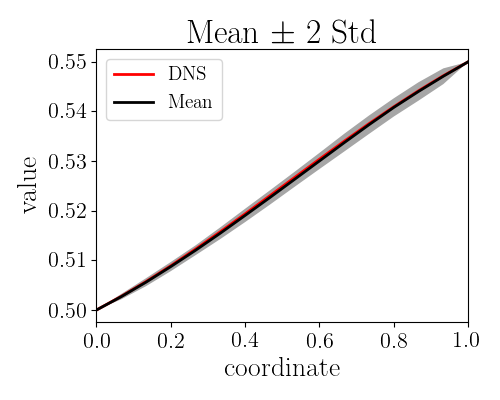}} \\[-2mm]
  \subfloat[$u_y$ results for BVP (i)]  {\includegraphics[trim=0.1cm 0.5cm 0.1cm 0.25cm, clip,height=0.16\linewidth]{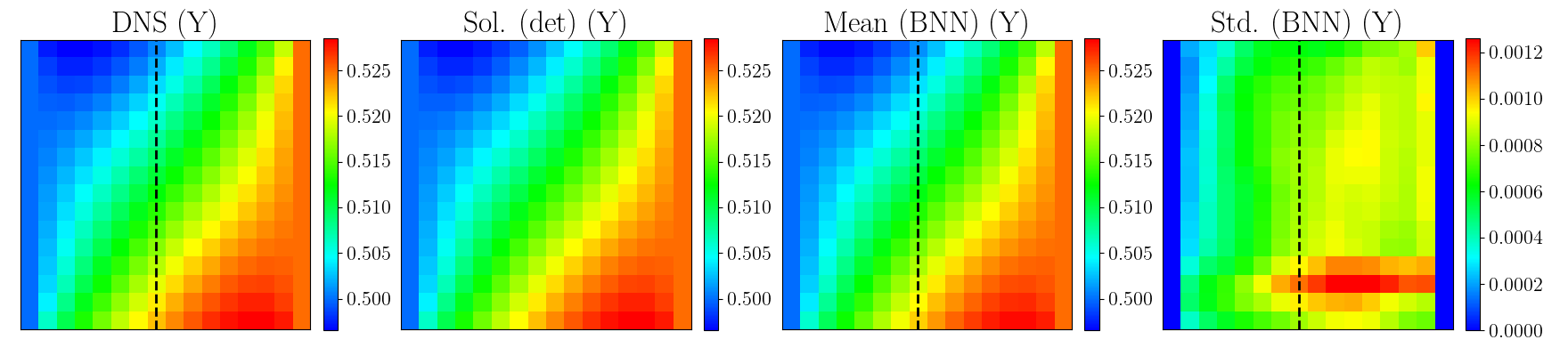}} 
  \subfloat[$u_y$ UQ for BVP (i)]       {\includegraphics[trim=0.1cm 0.5cm 0.1cm 0.25cm, clip,height=0.16\linewidth]{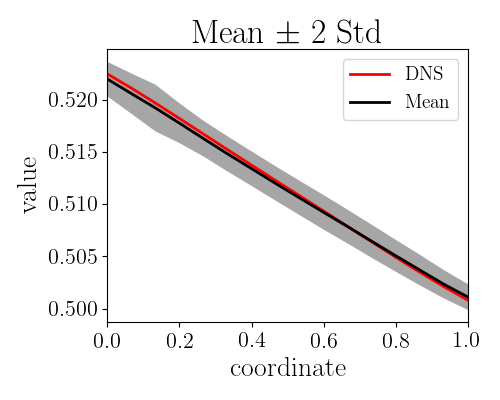}} \\[-2mm]
  \subfloat[$u_x$ results for BVP (ii)] {\includegraphics[trim=0.1cm 0.5cm 0.1cm 0.25cm, clip,height=0.16\linewidth]{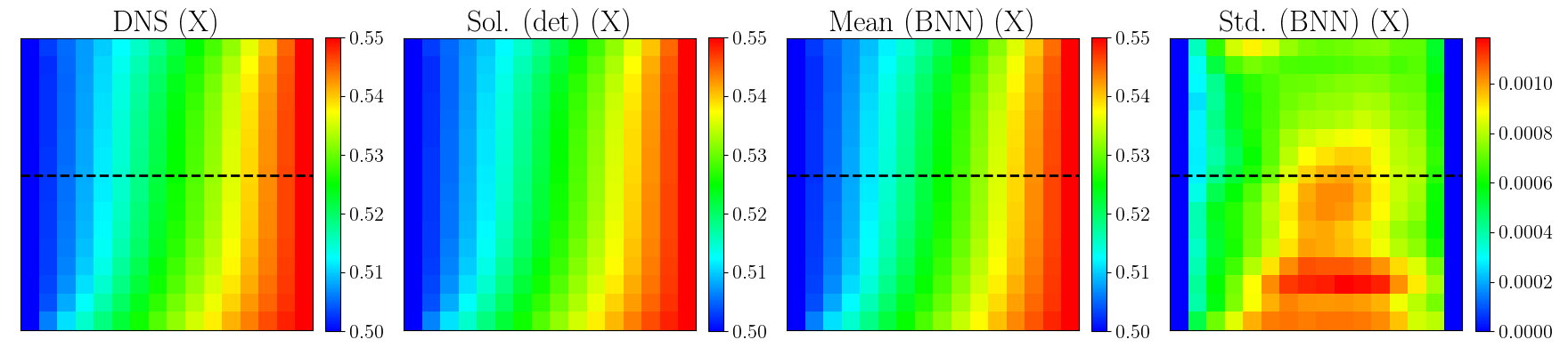}} 
  \subfloat[$u_x$ UQ for BVP (ii)]      {\includegraphics[trim=0.1cm 0.5cm 0.1cm 0.25cm, clip,height=0.16\linewidth]{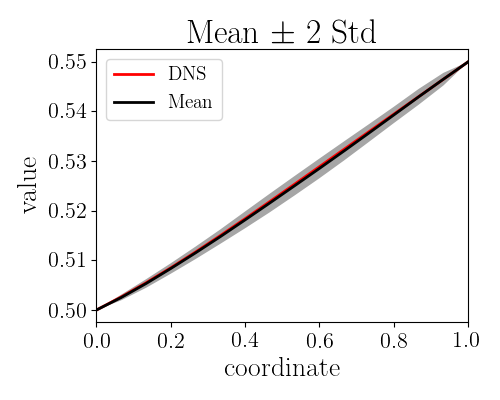}} \\[-2mm]
  \subfloat[$u_y$ results for BVP (ii)] {\includegraphics[trim=0.1cm 0.5cm 0.1cm 0.25cm, clip,height=0.16\linewidth]{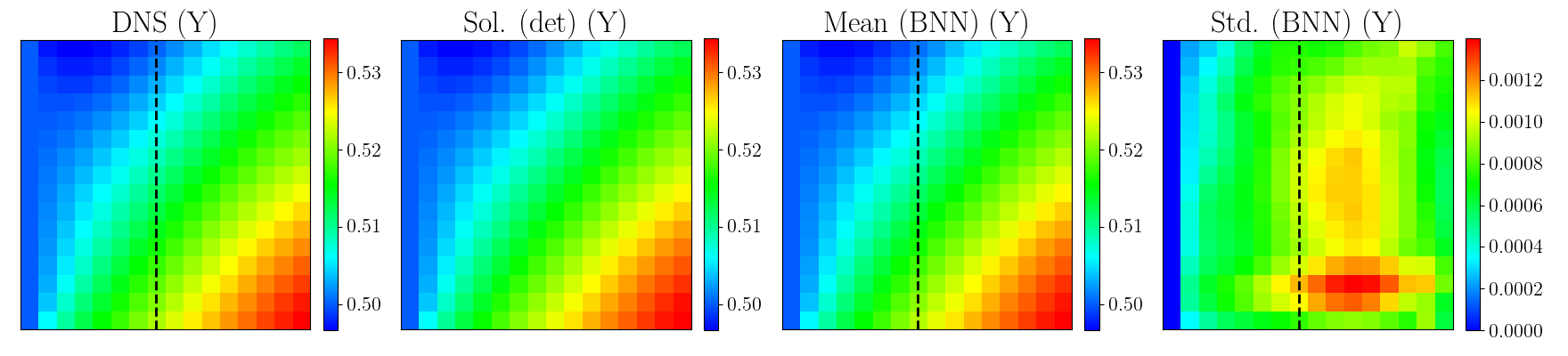}} 
  \subfloat[$u_y$ UQ for BVP (ii)]      {\includegraphics[trim=0.1cm 0.5cm 0.1cm 0.25cm, clip,height=0.16\linewidth]{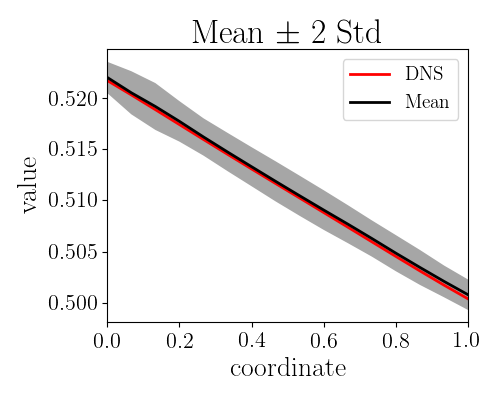}} \\[-2mm]
  \subfloat[$u_x$ results for BVP (iii)]{\includegraphics[trim=0.1cm 0.5cm 0.1cm 0.25cm, clip,height=0.16\linewidth]{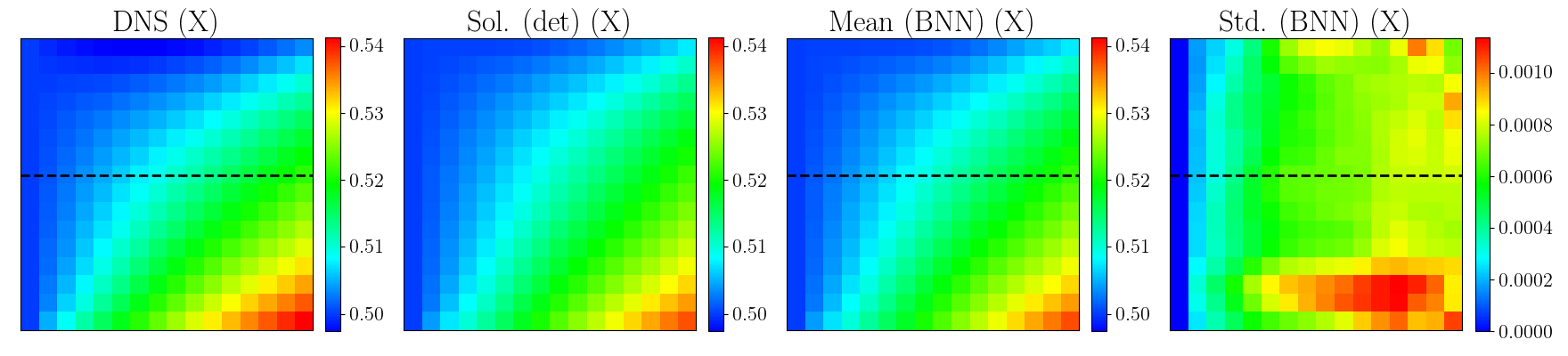}} 
  \subfloat[$u_x$ UQ for BVP (iii)]     {\includegraphics[trim=0.1cm 0.5cm 0.1cm 0.25cm, clip,height=0.16\linewidth]{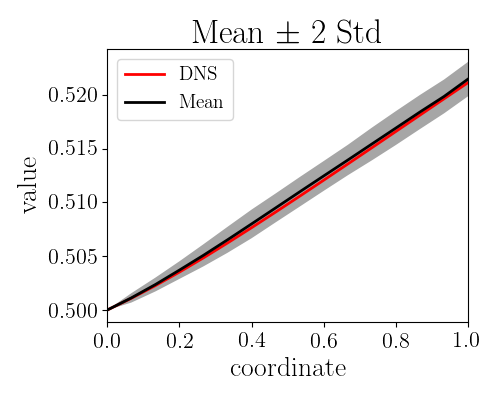}} \\[-2mm]
  \subfloat[$u_y$ results for BVP (iii)]{\includegraphics[trim=0.1cm 0.5cm 0.1cm 0.25cm, clip,height=0.16\linewidth]{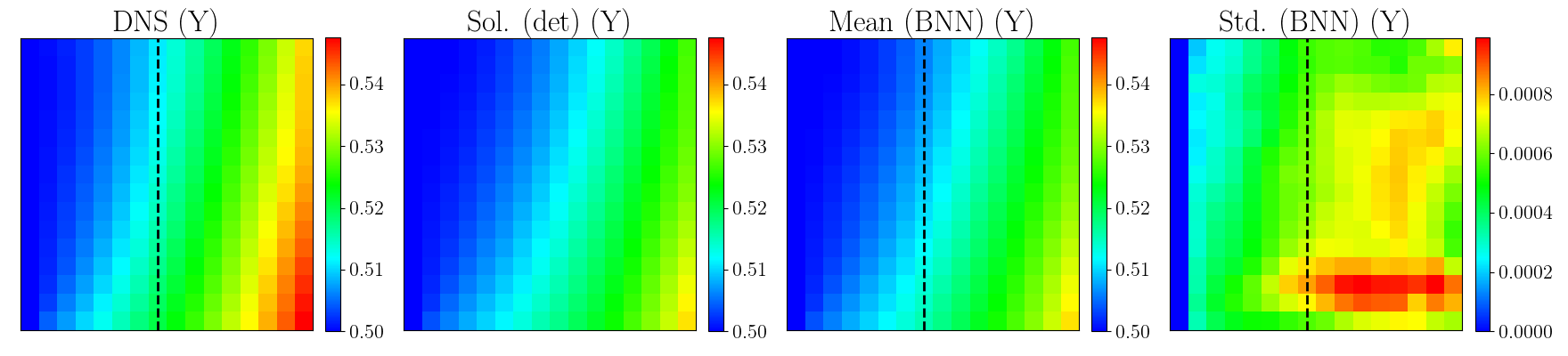}} 
  \subfloat[$u_y$ UQ for BVP (iii)]     {\includegraphics[trim=0.1cm 0.5cm 0.1cm 0.25cm, clip,height=0.16\linewidth]{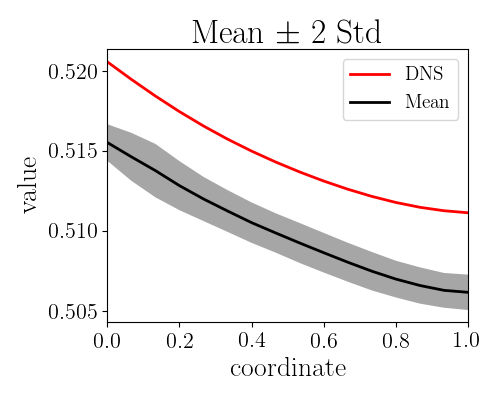}} \\
  \caption{Results of three selected BVPs out of the 30 linear elasticity BVPs with varying domains and different applied BCs simultaneously solved by a single deterministic or probabilistic NN with the proposed method. BVP (i), (ii), (iii) correspond to bc id 1, 2, and 3 for domain id 5, as shown in Fig. \ref{fig:bvp-problem-setup}(a). 
      (a, c, e, g, i, k) Solutions from DNS, deterministic (det) NNs, and BNNs (Mean, Std.) for different BVPs.
    (b, d, f, h, j, l) Quantitative comparison of the solution distribution between DNS and BNNs along the dashed lines.}
  \label{fig:linear-30bvp-results}
\end{figure}

\begin{table}
  \centering
  \begin{tabular}{l | l | l }
    \hline
    Description                     & Deterministic                 & Probabilistic         \\ \hline
    Total parameters                & 34,010                        & 67,803                 \\
    Size of $\calD$                 & 30 $\times$ Aug: $2^{9}$      & 30 $\times$ Aug: $2^{9}$      \\
    Epochs                          & 20,000                        & 100                   \\
    Zero initialization epochs      & 100                           & -                     \\
    Optimizer                       & Nadam                         & Nadam                 \\
    Learning Rate                   & 2.5e-4                        & 1e-8                  \\
    Batch Size                      & 128                           & 64                    \\
    $\Sigma_1$                      & -                             & 1e-8                  \\
    Initial value of $\Sigma_2$     & -                             & 1e-8                  \\
    \hline
  \end{tabular}
  \caption{Training related parameters for solving 30 linear elasticity BVPs. Aug: data augmentation. }
  \label{tab:linear-30bvp-NNs-others}
\end{table}

In this section, we use the proposed PDE constrained NNs to simultaneously solve 30 linear elasticity BVPs, as shown in Fig. \ref{fig:bvp-problem-setup}(a), with a resolution of $16\times16$. 
The deformed problem domains from DNSs for three representative setups are shown in Fig. \ref{fig:linear-bvp-deformed}(a).
Similar architectures of both deterministic and probabilistic NNs as summarized in Table \ref{tab:diffusion-20bvp-NNs} are used, except the last layer has two filters, representing $u_x$ and $u_y$, instead of one for the steady-state diffusion problem.
The other training related NN parameters are summarized in Table \ref{tab:linear-30bvp-NNs-others}. 
We follow the procedures described in Section \ref{sec:NN-training} to train both types of NNs. 
The NN results of three selected BVPs, as shown in Fig. \ref{fig:linear-bvp-deformed}(a),  are presented in Fig. \ref{fig:linear-30bvp-results}, with remaining results from other setups being given in Appendix \ref{appendix:linear-30bvps}.
The statistical moments of the BNN predictions are evaluated based on 50 MC samplings.
In Fig. \ref{fig:linear-30bvp-results}, BVP (i), (ii), and (iii) correspond to bc id 1 (non-zero Dirichlet loading), bc id 2 (non-zero Neumann loading), bc id 3 (mixed loading) applied to domain id 5, respectively. 
The comparison of solutions among DNSs, the deterministic NN, and the BNN for these threeBVPs is shown qualitatively in Fig. \ref{fig:linear-30bvp-results}(a,c,e,g,i,k), with quantitative comparison of the solution distribution along the dashed lines between DNSs and the BNN given in Fig. \ref{fig:linear-30bvp-results}(b,d,f,h,j,l).
Such a comparison shows that the proposed method has successfully solved most of the BVPs with desired accuracy.
The results from NNs in Fig. \ref{fig:linear-30bvp-results}(i,k,l) are slightly worse than the DNSs. 
This happens mainly because the deformation for linear elasticity is small. 
The scaled results have a narrow range of $[0.5, 0.55]$, which is challenging for NNs to learn to distinguish, particularly for purely non-zero traction loadings.
For the more challenging nonlinear elasticity case, where the deformation is large, the NNs can successfully solve such BVPs, as shown in Fig. \ref{fig:nonlinear-30bvp-results}. 
The performance difference between linear and nonlinear elasticity problems suggests that carefully choosing data normalization is important for improving the performance of the PDE constrained surrogate solvers.

\subsubsection{L-shape domain with solution interpolation}\label{sec:linear-lshape}
\begin{figure}[t!]
    \centering
    %\psfrag{a}[c][c]{step 1}
    \subfloat[$u_x$ results]{\includegraphics[height=0.17\linewidth]{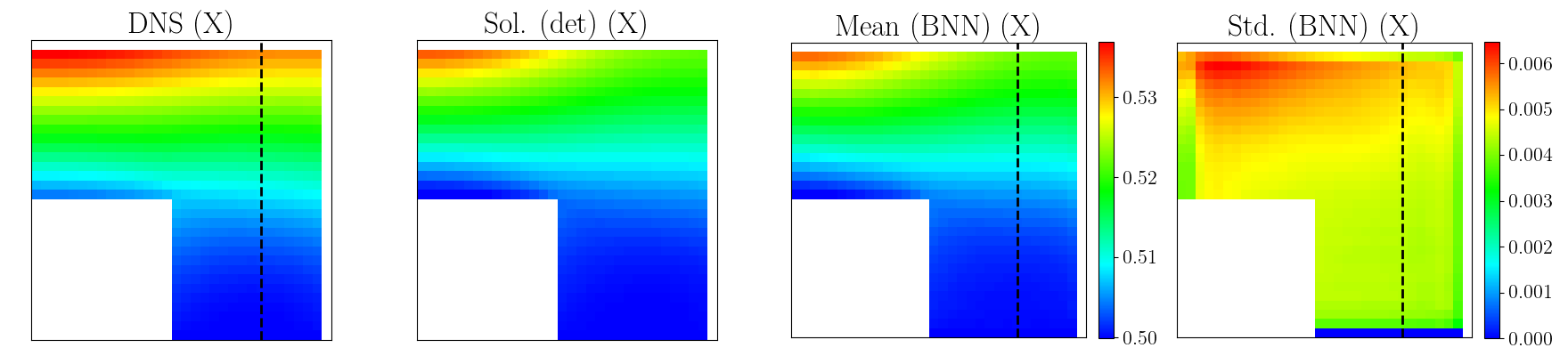}}
    \subfloat[$u_x$ UQ]{\includegraphics[height=0.17\linewidth]{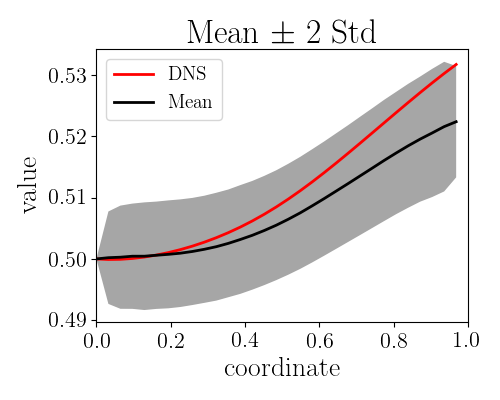}} \\
    \subfloat[$u_y$ results]{\includegraphics[height=0.17\linewidth]{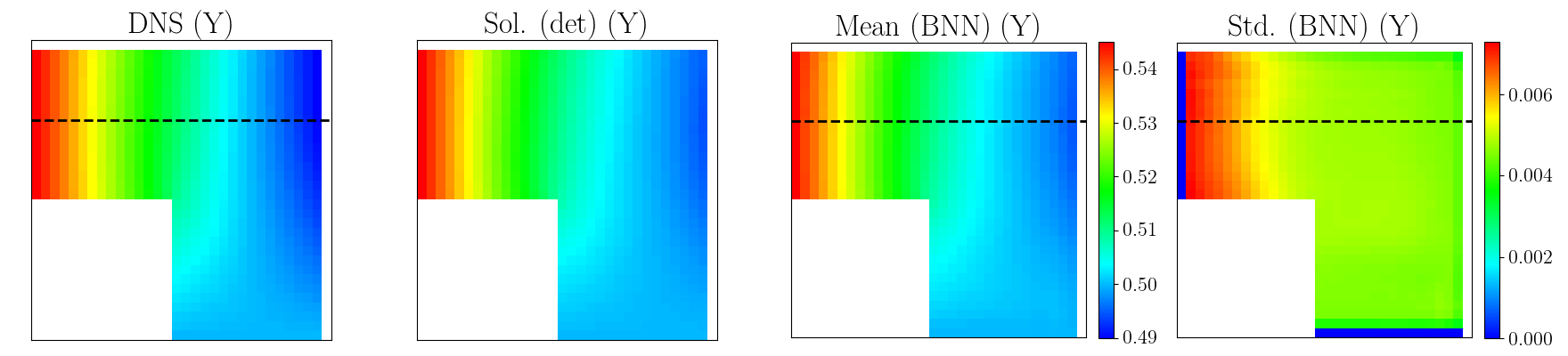}}
    \subfloat[$u_y$ UQ]{\includegraphics[height=0.17\linewidth]{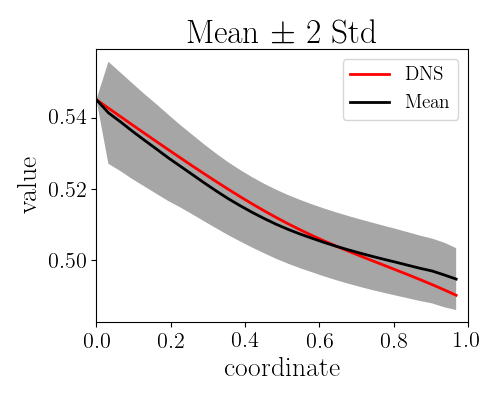}} \\
    \subfloat[$F_x$ results]{\includegraphics[width=0.245\linewidth]{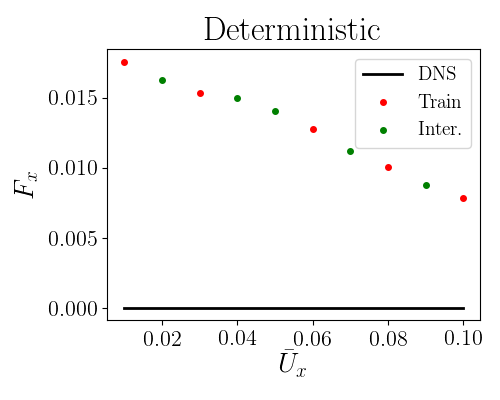}}
    \subfloat[$F_x$ results]{\includegraphics[width=0.245\linewidth]{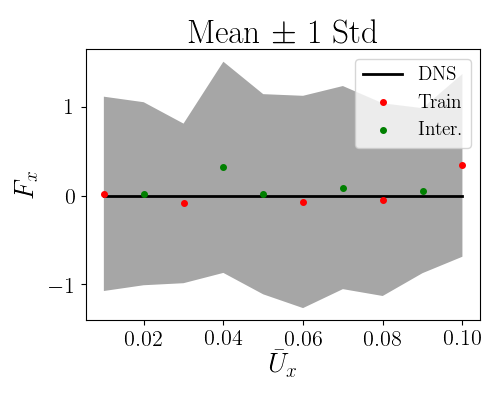}}
    \subfloat[$F_y$ results]{\includegraphics[width=0.245\linewidth]{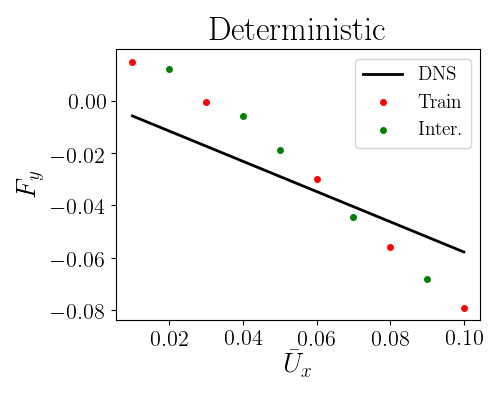}}
    \subfloat[$F_y$ results]{\includegraphics[width=0.245\linewidth]{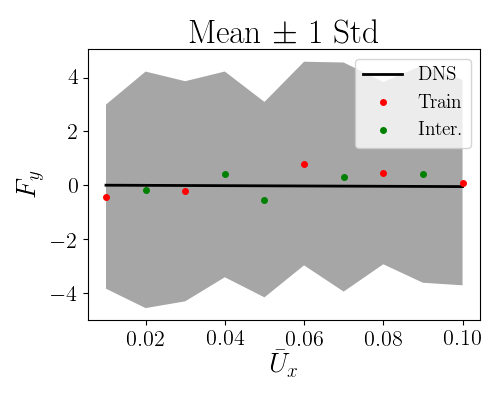}}
    \caption{Results for the L-shape BVP.
      (a, c) Solutions from DNS, deterministic (det) NNs, and BNNs (Mean, Std.) for different BVPs.
      (b, d) Quantitative comparison of the solution distribution between the DNS and BNNs along the dashed lines.
      (e, g) Comparison of reaction force in X- and Y-direction  between the DNS and deterministic NNs.
      (f, h) Comparison of reaction force in X- and Y-direction  between the DNS and BNNs. ``Inter'' indicates the interpolated prediction (see the text for an explanation.)
    }
    \label{fig:linear-lshape-results}
\end{figure}

So far, we have demonstrated the capability of the proposed method to solve PDEs on regular/irregular, and fixed/varying domains with different applied BCs. 
In this section, we use the proposed method to solve linear elasticity on a L-shape domain with fully constrained bottom edge and non-zero applied Dirichlet vertical loading on the left edge, as shown in Fig. \ref{fig:bvp-problem-setup}(c).
We focus on the linear loading regime, with the final deformed shape shown in Fig. \ref{fig:linear-bvp-deformed}(b), without accounting for any nonlinear behavior such as crack propagation \cite{winkler2001traglastuntersuchungen,Zhang2013Linder-IJNME-march-cube}.
In previous examples, for each set of BCs, only one specific set of values for non-zero Dirichlet/Neumann loadings is exposed to the NNs.
Even though the NNs can successfully learn the BCs and solve the corresponding BVPs, it remains challenging for the NNs to make interpolating/extrapolating predictions, as the NNs do not learn the physical meaning of a set of BCs based on a single loading data point.
To enforce  learning of the  physical meaning of the BCs upon the NNs, for each set of BCs, we have to expose NNs to multiple incremental loadings.
For the L-shape BVP, we create 10 input data points (see the reaction force plots in \ref{fig:linear-lshape-results}(e-h)), with each differing from each other only by the actual values at the non-zero Dirichlet BCs.
Since strains are small, we only focus on studying the interpolated predictions of the NNs, and leave the study of extrapolated predictions for the non-linear elasticity BVPs investigated in section \ref{sec:nonlinear-rectangle-interpolation}.
During training, the NNs are exposed to 5 loading cases, as indicated by the red dots in Fig. \ref{fig:linear-lshape-results}(e-h). 
The 5 loading cases for interpolating prediction are marked with green dots in Fig. \ref{fig:linear-lshape-results}(e-h). 
Again, to keep the discussion concise for easy reading, the architectures of both deterministic and probabilistic NNs along with other training related information are provided in the Appendix \ref{appendix:linear-lshape}.
We follow the procedures described in Section \ref{sec:NN-training} to train both types of NNs with an output resolution of $32\times32$.

We compare solutions among DNSs, the deterministic NN, and the BNN for the last interpolating loading step and show them qualitatively in Fig. \ref{fig:linear-lshape-results}(a,c), with quantitative comparison of the solution distribution along both the horizontal and vertical dashed lines between DNSs and the BNN given in Fig. \ref{fig:linear-lshape-results}(b,d).
Additional results from other interpolating loading steps are given in Appendix \ref{appendix:linear-lshape}.
The reaction forces in both directions at the bottom edge are shown in Fig. \ref{fig:linear-lshape-results}(e-h).
For BNNs, the statistical moments of reaction forces are evaluated based on averaging reaction forces computed from 50 MC samplings.
Fig. \ref{fig:linear-lshape-results} shows that NN results are comparable with the DNS solution. 
For this BVP setup, the reaction force in the X-direction is zero and is also very small in the Y-direction, as indicated by the DNS results in  Fig. \ref{fig:linear-lshape-results}(e,g).
The deterministic NNs predict $F_y$ in the correct range, but fail to predict $F_x$.
For the BNNs, the predicted reaction forces is quite different from the DNS, which is expected, as both $u_x$ and $u_y$ at the bottom of the geometry is very small, thus leading to high uncertainties in the reaction forces.
Still the accurate interpolating prediction shown Fig. \ref{fig:linear-lshape-results}(b,d) is very appealing for problems, such as homogenization and inverse modelling studies, where many similar BVPs with small variations need to be simulated repeatedly.

\subsection{Nonlinear elasticity}
In this section, we use the proposed method to solve different nonlinear elasticity problems.  

\subsubsection{Background}

The general description of an elliptic PDE system given \eref{eq:general-pde} is rewritten as 
\begin{equation}
  \begin{aligned}
    \nabla \cdot \BP = \Bzero \quad & \text{on} \quad \Omega_0, \\
    \Bu (\BX) = \bar{\Bu} (\BX)  \quad & \text{on} \quad  \Gamma^{\Bu}_0, \\
    \BT = \bar{\BT} (\BX)  \quad & \text{on} \quad  \Gamma^\BT_0,
  \end{aligned}
  \label{eq:general-pde-nonlinear-elasticity}
\end{equation}
for the non-linear elasticity problem with the subscript $0$ indicating the reference configuration. 
In \eref{eq:general-pde-nonlinear-elasticity}, $\Bu$ represents the displacement field, $\BP$ is the first Piola-Kirchhoff stress tensor, and $\BT$ is the surface traction.
In the non-linear elasticity problem, the deformation gradient is defined as $\BF=\Bone + \partial\Bu/\partial\BX$ with $\Bone$ being the second-order identity tensor.
The right Cauchy-Green deformation tensor is written as $\BC=\BF^{T}\BF$.
The following compressible Neo-hookean hyperelastic free energy function is considered 
\begin{equation}
  W=\frac{1}{2} \mu(\text{tr}(\BC)3-3-2 \ln(J))+\lambda \frac{1}{2}(J-1)^2,
\end{equation}
with $\mu$ and $\lambda$ as the Lam\'e constants and $J=\det(F)$.
The Piola stress tensor $\BP$ is computed as
\begin{equation}
  \BP = \frac{\partial W}{\partial \BF} 
  =  \lambda(J^2-J)\BF^{-T} +\mu(\BF-\BF^{-T}).
\end{equation}
The discretized residual function \eref{eq:discretized-residual} for the non-linear elasticity problem\footnote{
Even with the zero-initialization process, the NN outputs at early stages of training  could violate the physics, e.g. with a negative or zero determinant of the deformation gradient $J$. To ensure that the residual can be evaluated and to prevent residuals from these ``bad'' pixels values from contributing to the final loss, we regularize the loss by omitting the residual contribution with $J<0.1$ and $J>5.0$. As the training continues towards a later stage, the NN predicted solutions gradually fulfill the governing PDEs, and the regularization on $J$ will cease to function.}
is written as
\begin{equation}
  \BR = \sum_{e=1}^{n_\text{elem}} \left\{ \int_{\Omega^e} \BB^T \BP dV - \int_{\Gamma^{e,T}} \BN^T \bar{\BT}~dS \right\}.
  \label{eq:discretized-residual-nonlinear}
\end{equation}
A set of material parameters with $\lambda=14.4231$ and $\mu=9.61538$ is used in both DNSs and the surrogate PDE solver.

\subsubsection{Multiple rectangular domains with different BCs}

\begin{figure}[t!]
    \centering
    \includegraphics[width=0.75\linewidth]{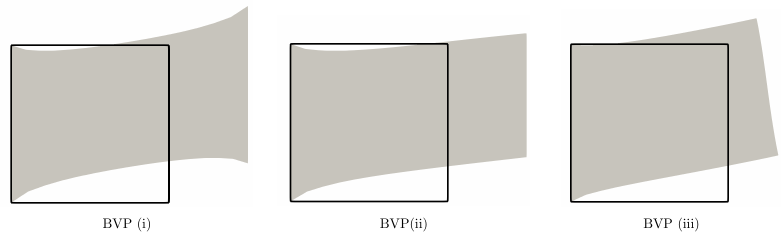}
    \caption{Illustration of the deformed shape of the three selected nonlinear elasticity BVPs. The wireframe and the gray region indicate the undeformed and deformed problem domain, respectively.}
    \label{fig:nonlinear-30bvp-deformed}
\end{figure}

\begin{figure}[p!]
  \centering
  \subfloat[$u_x$ results for BVP (i)]  {\includegraphics[trim=0.1cm 0.5cm 0.1cm 0.25cm, clip,height=0.16\linewidth]{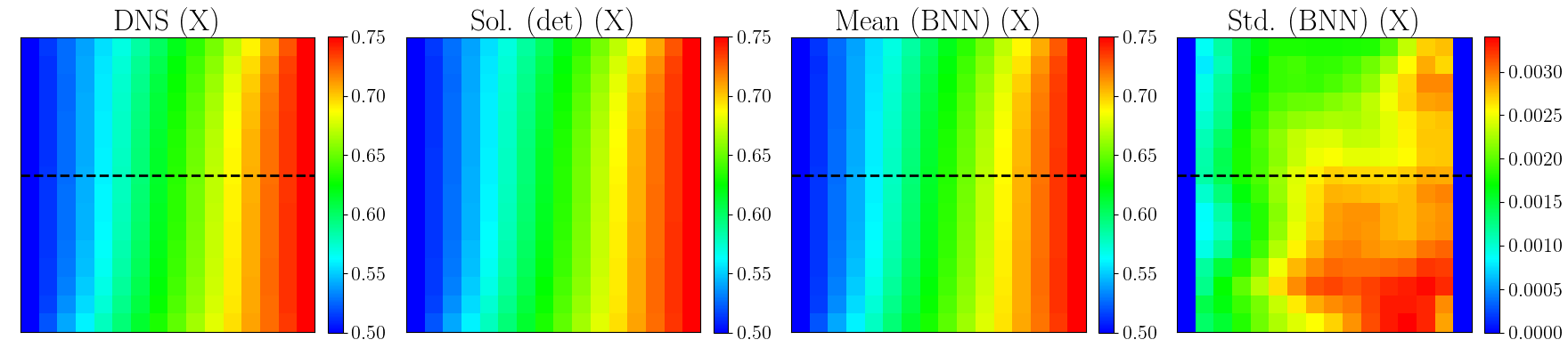}} 
  \subfloat[$u_x$ UQ for BVP (i)]       {\includegraphics[trim=0.1cm 0.5cm 0.1cm 0.25cm, clip,height=0.16\linewidth]{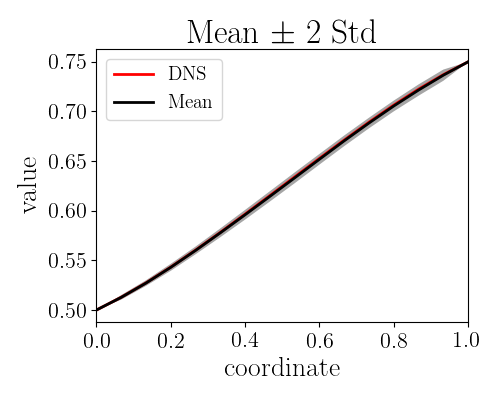}} \\[-2mm]
  \subfloat[$u_y$ results for BVP (i)]  {\includegraphics[trim=0.1cm 0.5cm 0.1cm 0.25cm, clip,height=0.16\linewidth]{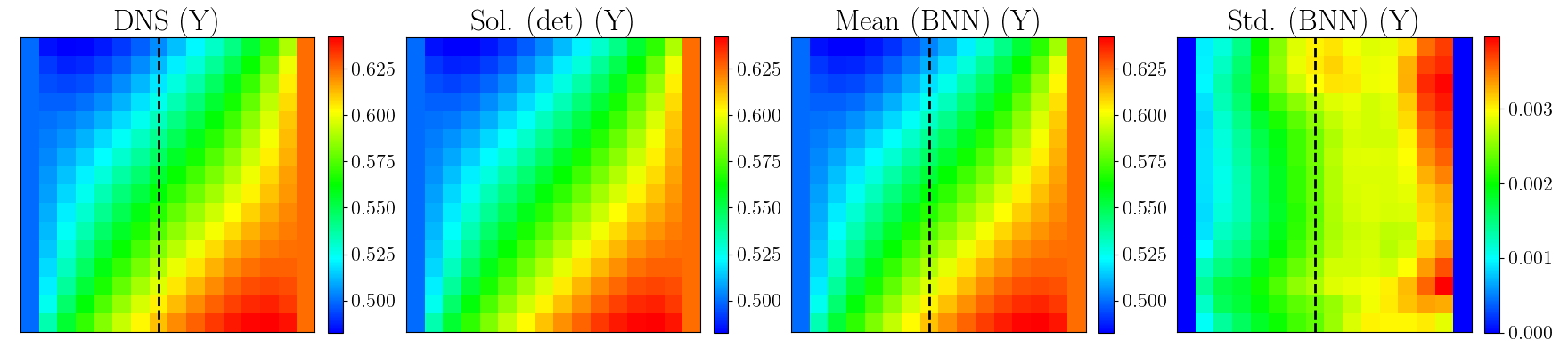}} 
  \subfloat[$u_y$ UQ for BVP (i)]       {\includegraphics[trim=0.1cm 0.5cm 0.1cm 0.25cm, clip,height=0.16\linewidth]{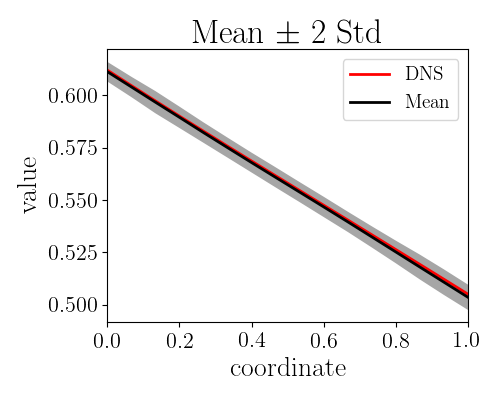}} \\[-2mm]
  \subfloat[$u_x$ results for BVP (ii)] {\includegraphics[trim=0.1cm 0.5cm 0.1cm 0.25cm, clip,height=0.16\linewidth]{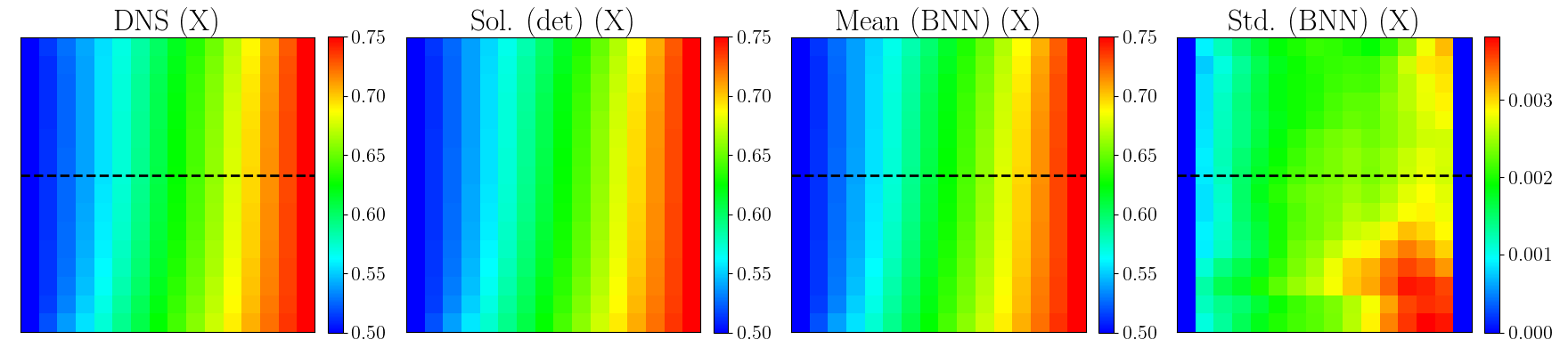}} 
  \subfloat[$u_x$ UQ for BVP (ii)]      {\includegraphics[trim=0.1cm 0.5cm 0.1cm 0.25cm, clip,height=0.16\linewidth]{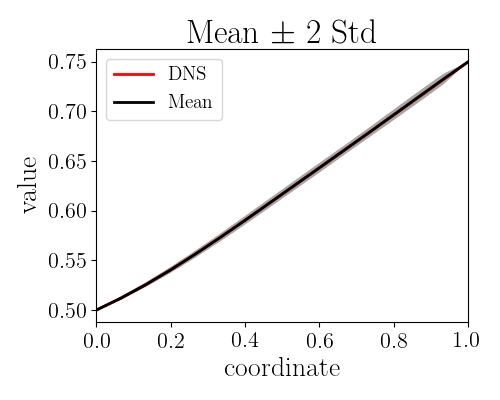}} \\[-2mm]
  \subfloat[$u_y$ results for BVP (ii)] {\includegraphics[trim=0.1cm 0.5cm 0.1cm 0.25cm, clip,height=0.16\linewidth]{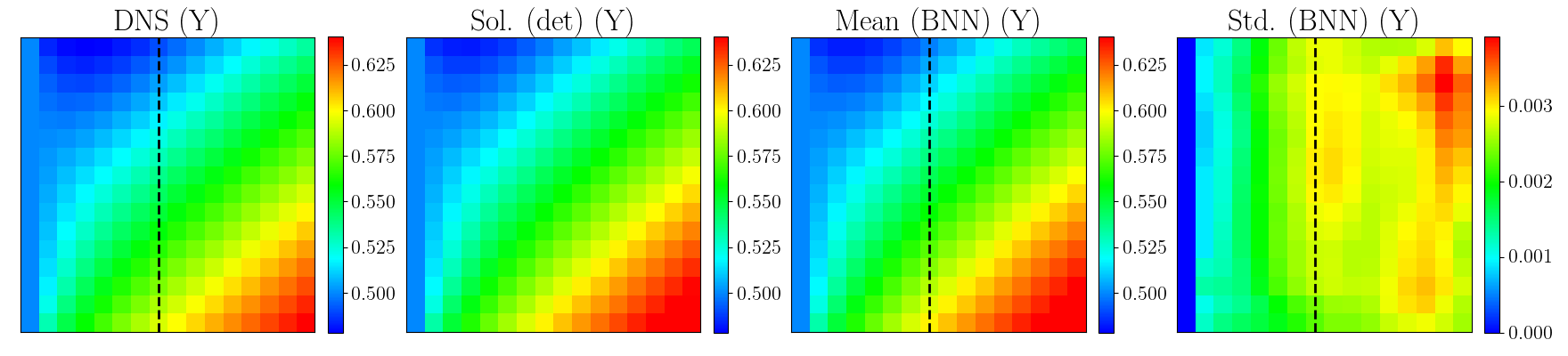}} 
  \subfloat[$u_y$ UQ for BVP (ii)]      {\includegraphics[trim=0.1cm 0.5cm 0.1cm 0.25cm, clip,height=0.16\linewidth]{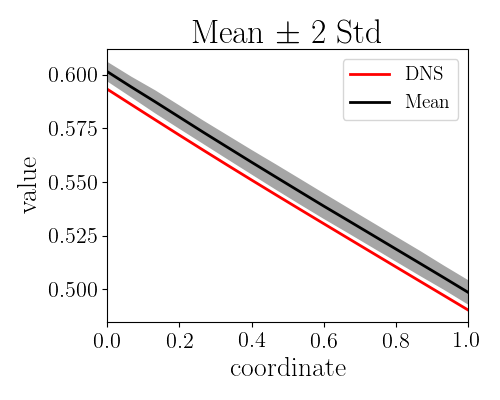}} \\[-2mm]
  \subfloat[$u_x$ results for BVP (iii)]{\includegraphics[trim=0.1cm 0.5cm 0.1cm 0.25cm, clip,height=0.16\linewidth]{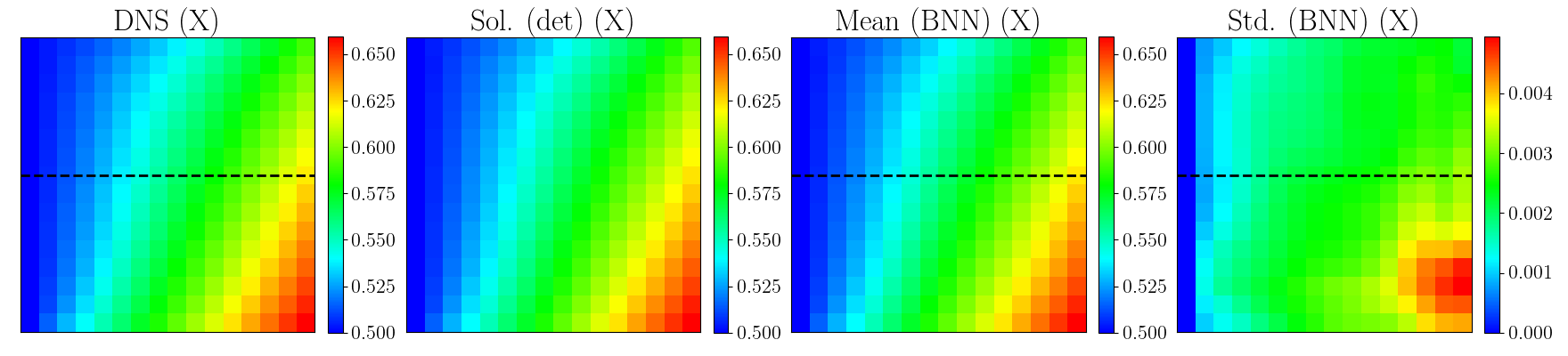}} 
  \subfloat[$u_x$ UQ for BVP (iii)]     {\includegraphics[trim=0.1cm 0.5cm 0.1cm 0.25cm, clip,height=0.16\linewidth]{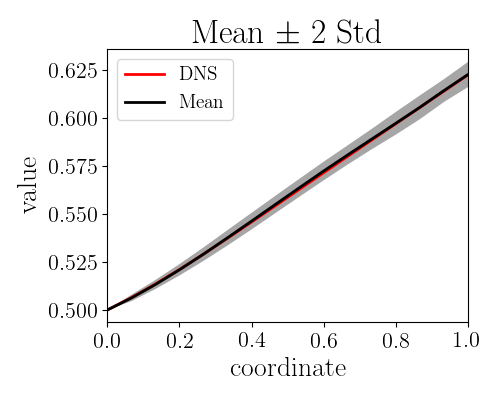}} \\[-2mm]
  \subfloat[$u_y$ results for BVP (iii)]{\includegraphics[trim=0.1cm 0.5cm 0.1cm 0.25cm, clip,height=0.16\linewidth]{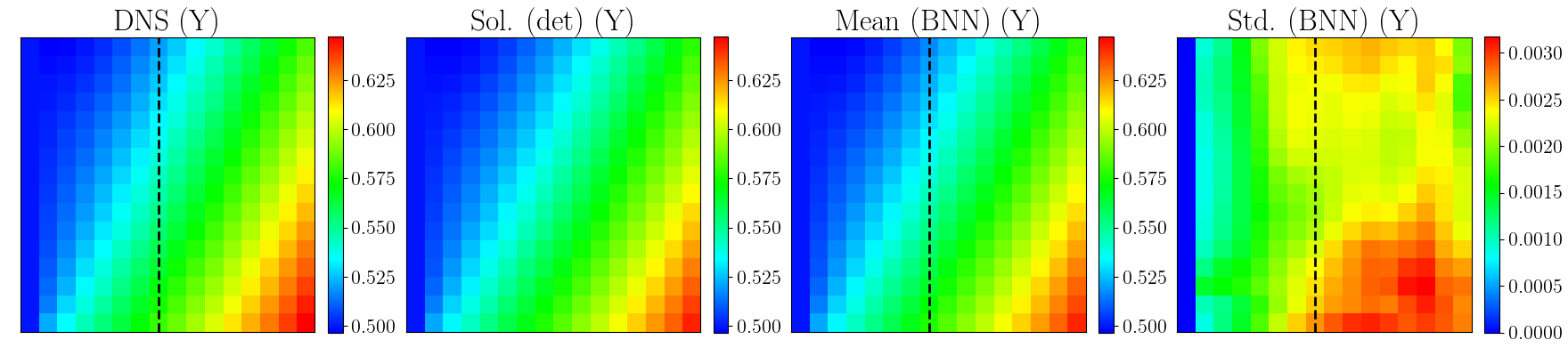}} 
  \subfloat[$u_y$ UQ for BVP (iii)]     {\includegraphics[trim=0.1cm 0.5cm 0.1cm 0.25cm, clip,height=0.16\linewidth]{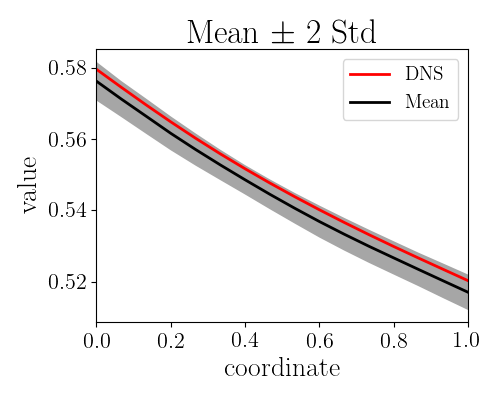}} \\
  \caption{Results of three selected BVPs out of the 30 nonlinear elasticity BVPs with varying domains and different applied BCs simultaneously solved by a single deterministic or probabilistic NN with the proposed method. BVP (i), (ii), (iii) correspond to bc id 1, 2, and 3 for domain id 5, as shown in Fig. \ref{fig:bvp-problem-setup}(a). 
      (a, c, e, g, i, k) Solutions from DNS, deterministic (det) NNs, and BNNs (Mean, Std.) for different BVPs.
  (b, d, f, h, j, l) Quantitative comparison of the solution distribution between DNS and BNNs along the dashed lines.}
  \label{fig:nonlinear-30bvp-results}
\end{figure}

In this section, we use the proposed PDE constrained NNs to simultaneously solve 30 nonlinear elasticity BVPs, as show in Fig. \ref{fig:bvp-problem-setup}(a), with a resolution of $16\times16$. 
The deformed problem domains from DNSs for three representative setups are shown in Fig. \ref{fig:nonlinear-30bvp-deformed}.
The architectures of both deterministic and probabilistic NNs and the training related NN parameters used in this section are identical to those used in section \ref{sec:linear-rectangle} for solving linear elasticity BVPs. 
We follow the procedures described in Section \ref{sec:NN-training} to train both types of NNs. 
The NN results of three selected BVPs, as shown in Fig. \ref{fig:nonlinear-30bvp-deformed},  are presented in Fig. \ref{fig:nonlinear-30bvp-results}, with the remaining results from other setups given in Appendix \ref{appendix:nonlinear-30bvps}.
The statistical moments of the BNN predictions are evaluated based on 50 MC samplings.
In Fig. \ref{fig:nonlinear-30bvp-results}, BVP (i), (ii), and (iii) correspond to bc id 1 (non-zero Dirichlet loading), bc id 2 (non-zero Neumann loading), bc id 3 (mixed loading) applied to domain id 5. 
The comparison of solutions between DNSs, the deterministic NN, and the BNN for these threeBVPs is shown qualitatively in Fig. \ref{fig:nonlinear-30bvp-results}(a,c,e,g,i,k), with quantitative comparison of the solution distribution along the dashed lines between DNSs and the BNN given in Fig. \ref{fig:nonlinear-30bvp-results}(b,d,f,h,j,l).
Such comparison shows that the proposed method has successfully solved multiple BVPs with desired accuracy.

\subsubsection{Rectangular domain with solution interpolation and extrapolation}\label{sec:nonlinear-rectangle-interpolation}

\begin{figure}[t!]
    \centering
    %\psfrag{a}[c][c]{step 1}
    \includegraphics[width=0.3\linewidth]{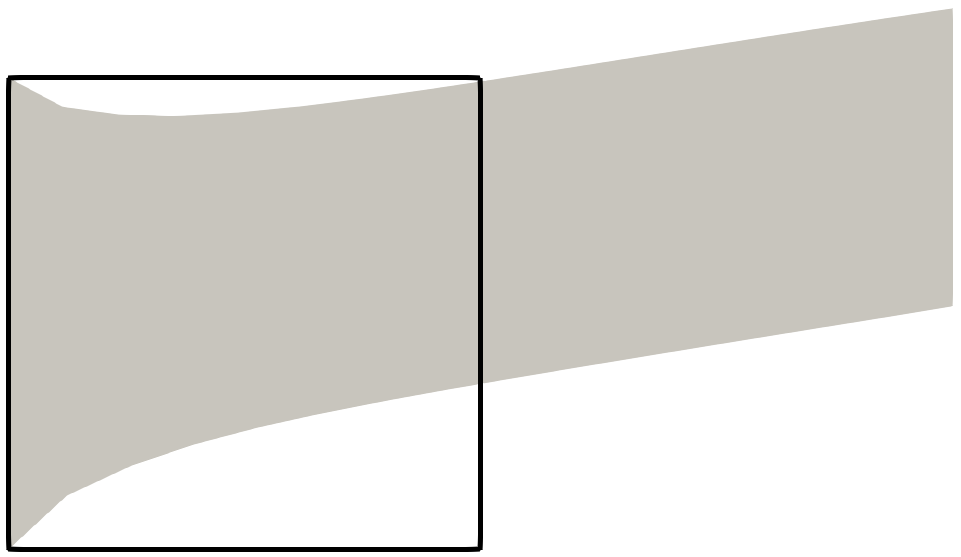}
    \caption{Illustration of the deformed shape of the nonlinear elasticity BVP for NN interpolating and extrapolating prediction with non-zero Dirichlet loading in the X-direction and non-zero Neumann loading in the Y-direction. The wireframe and the gray region indicate the undeformed and deformed problem domain, respectively.}
    \label{fig:nonlinear-inter-extra-deformed}
\end{figure}

\begin{table}[t!]
    \centering
    \begin{tabular}{c|c|c|c|c}
        \hline
        & Hardware & Software & Wall-time & Averaged $L_2$ error\\ \hline
        FEM & Intel i7-8750, 2.2GHz (use single core) & mechanoChemFEM & 110ms & -\\
        deterministic NN & GeForce GTX 1050 Ti, 4GB memory & Tensorflow & 0.22ms & 3.06e-4 \\ 
        BNN & GeForce GTX 1050 Ti, 4GB memory & Tensorflow & 0.29ms & 4.04e-4 \\ \hline
    \end{tabular}
    \caption{Comparison of the wall-clock time of finite element simulation and the NN prediction for solving the BVP in Fig. \ref{fig:nonlinear-inter-extra-deformed}. The wall-time is averaged over multiple simulations/predictions\protect\footnotemark. The averaged $L_2$ error in \eref{eq:l2-error} between DNSs and the NN prediction confirm the accuracy of the surrogate PDE solver.}
    \label{tab:wall-time-nonlinear}
\end{table}
\footnotetext{The wall time might be reduced if the used software is further optimized. However, it is generally the case that NN prediction could be orders faster than the DNS.}

\begin{figure}[t!]
    \centering
    %\psfrag{a}[c][c]{step 1}
    \includegraphics[trim=0.1cm 0.25cm 0.1cm 0.25cm, clip,width=0.245\linewidth]{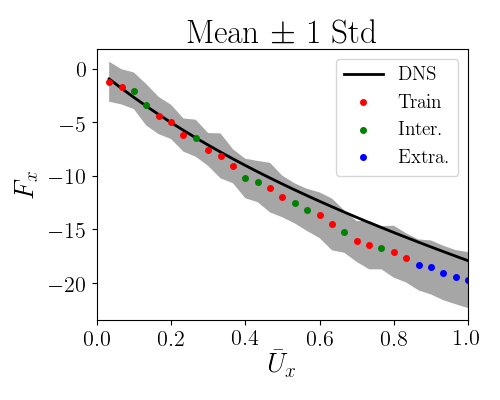}
    \includegraphics[trim=0.1cm 0.25cm 0.1cm 0.25cm, clip,width=0.245\linewidth]{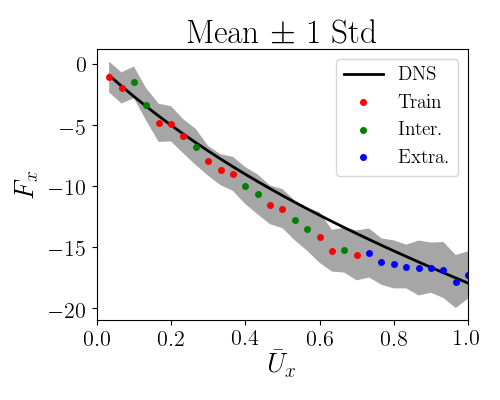}
    \includegraphics[trim=0.1cm 0.25cm 0.1cm 0.25cm, clip,width=0.245\linewidth]{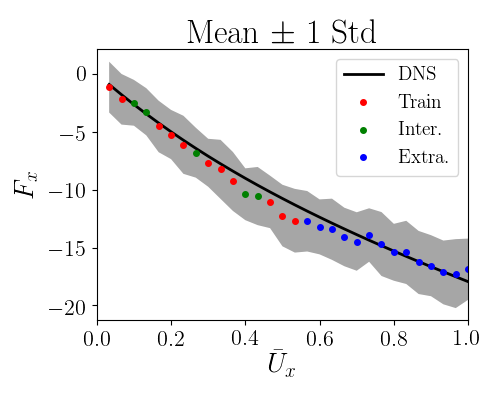}
    \includegraphics[trim=0.1cm 0.25cm 0.1cm 0.25cm, clip,width=0.245\linewidth]{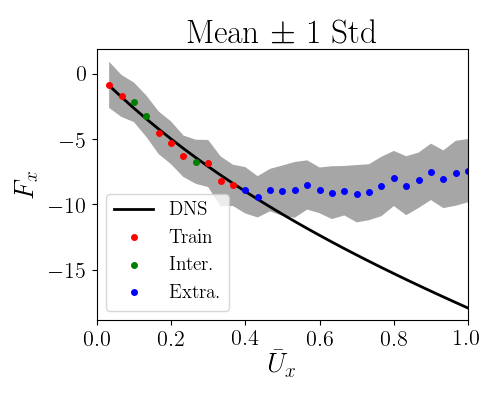} \\
    \subfloat[case (i)]{\includegraphics[trim=0.1cm 0.25cm 0.1cm 0.25cm, clip,width=0.245\linewidth]{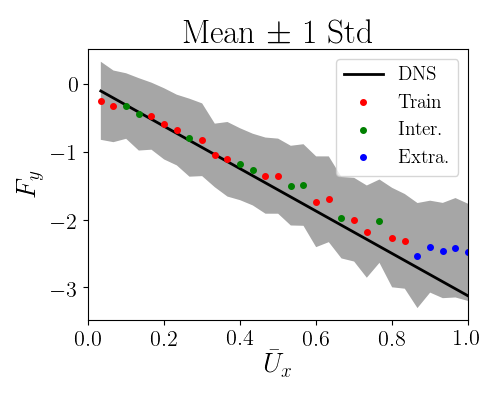}}
    \subfloat[case (ii)]{\includegraphics[trim=0.1cm 0.25cm 0.1cm 0.25cm, clip,width=0.245\linewidth]{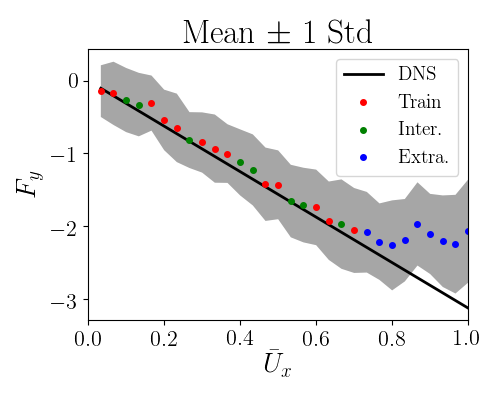}}
    \subfloat[case (iii)]{\includegraphics[trim=0.1cm 0.25cm 0.1cm 0.25cm, clip,width=0.245\linewidth]{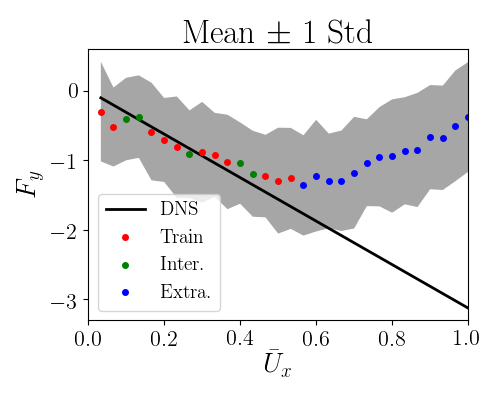}}
    \subfloat[case (iv)]{\includegraphics[trim=0.1cm 0.25cm 0.1cm 0.25cm, clip,width=0.245\linewidth]{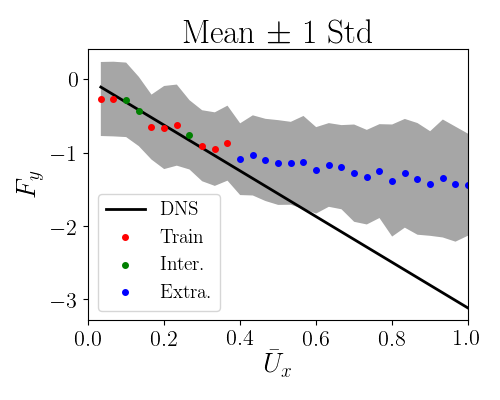}}
    \caption{Comparison of the reaction forces in both X- and Y-directions between DNSs and the BNN predicted solution. For the four different cases, the loading steps exposed to NNs during training decrease, with increased NN predicted extrapolating loading steps. The interpolating predicted reaction forces in general are good for all the four cases. The extrapolating predicted reaction forces are reasonable for a short range beyond the training range, especially for $F_x$.}
    \label{fig:nonlinear-inter-extra-F}
\end{figure}

\begin{figure}[t!]
    \centering
    %\psfrag{a}[c][c]{step 1}
    \subfloat[$u_x$ results]{\includegraphics[trim=0.1cm 0.25cm 0.1cm 0.25cm, clip,height=0.16\linewidth]{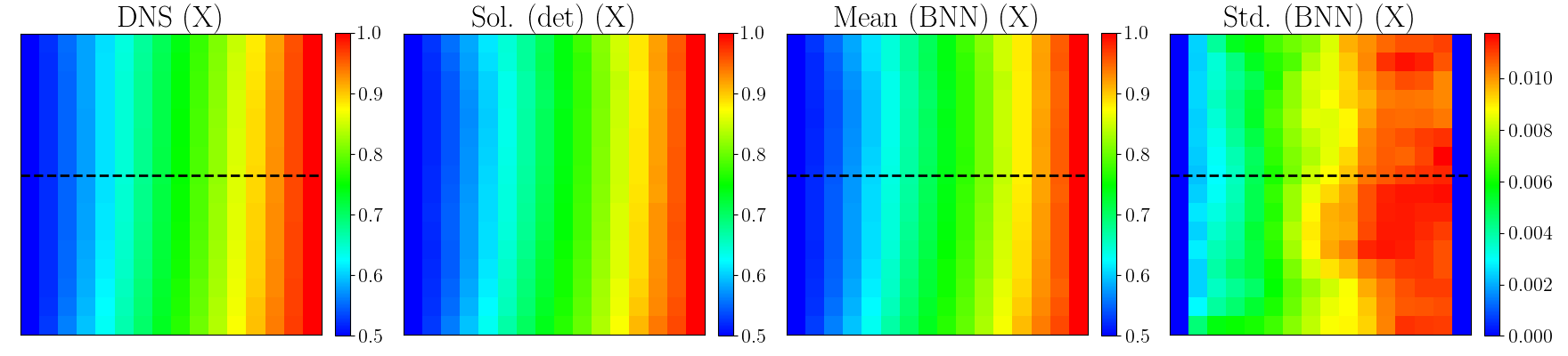}}
    \subfloat[$u_x$ UQ]{\includegraphics[trim=0.1cm 0.25cm 0.1cm 0.25cm, clip,height=0.16\linewidth]{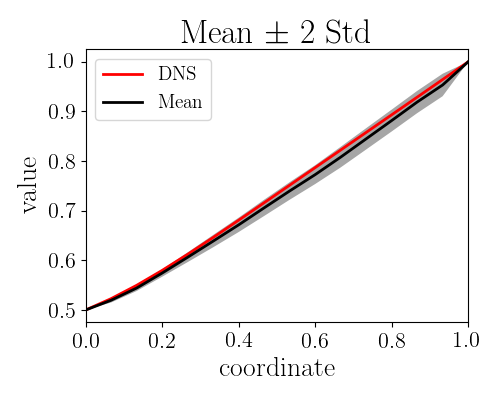}} \\
    \subfloat[$u_y$ results]{\includegraphics[trim=0.1cm 0.25cm 0.1cm 0.25cm, clip,height=0.16\linewidth]{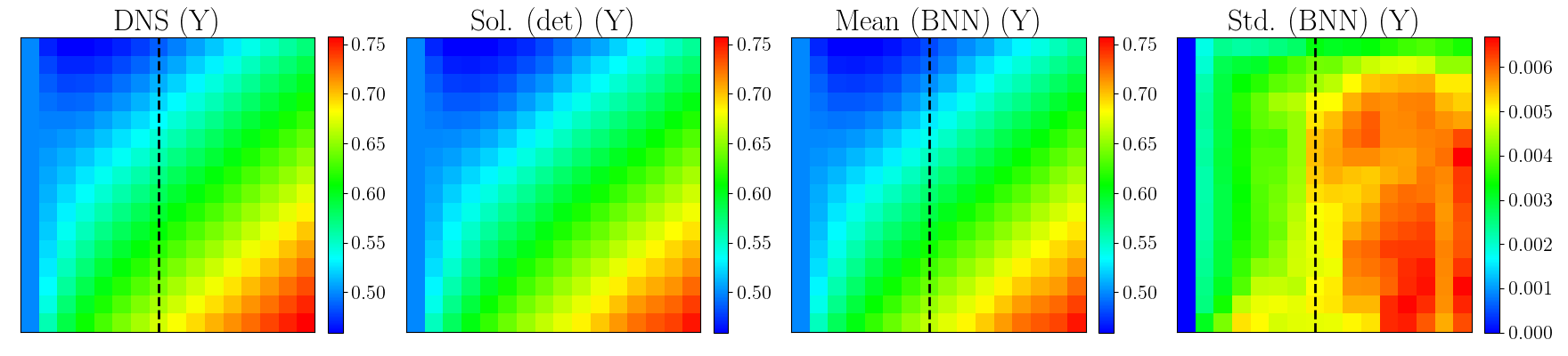}}
    \subfloat[$u_y$ UQ]{\includegraphics[trim=0.1cm 0.25cm 0.1cm 0.25cm, clip,height=0.16\linewidth]{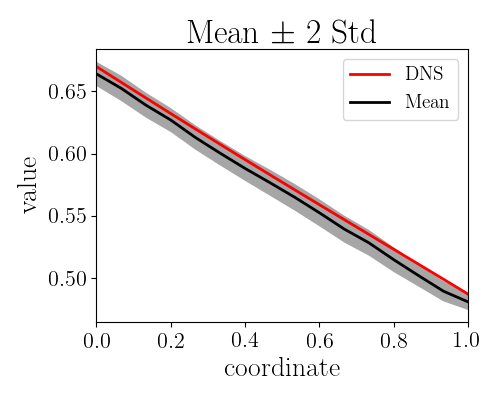}}
    \caption{Results of the last loading step with NN extrapolating prediction for case (i) in Fig. \ref{fig:nonlinear-inter-extra-F}. 
      (a, c) Solutions from DNS, deterministic (det) NNs, and BNNs (Mean, Std.) for different BVPs.
      (b, d) Quantitative comparison of the solution distribution between DNS and BNNs along the dashed lines.}
    \label{fig:nonlinear-inter-extra-results}
\end{figure}

In this section, we explore the interpolating and extrapolating capability of the proposed framework for the BVP setup shown in Fig. \ref{fig:nonlinear-inter-extra-deformed}. Both the DNS and NN solution have resolutions of $16\times16$.
The architectures of both deterministic and probabilistic NNs and the training related NN parameters used in this section are identical to those used in section \ref{sec:linear-rectangle} for solving linear elasticity BVPs. 
The domain is fixed in both directions on the left edge and is loaded with non-zero Dirichlet loading in the X-direction and non-zero Neumann loading in the Y-direction. 
As discussed in section \ref{sec:linear-lshape}, in order to enforce the learning of the boundary conditions and make interpolating prediction, the NNs need to be exposed to step-wise boundary loading.
We train the NNs in four different cases and show the reaction forces in both directions for trained, interpolated and extrapolated BCs, as shown in Fig. \ref{fig:nonlinear-inter-extra-F}.
In each case, the training dataset contains different loading steps, as indicated by the red dots in Fig. \ref{fig:nonlinear-inter-extra-F}. 
The total number of training loading steps decreases with increased case number.
Thus, the total number of loading steps with extrapolating NN prediction increases in these four cases, as indicated by the blue dots Fig. \ref{fig:nonlinear-inter-extra-F}.
The interpolating NN prediction is marked with green dots in Fig. \ref{fig:nonlinear-inter-extra-F}.
Unlike the reaction forces in the L-shape BVP, the NNs results matches the DNSs well.
We further observe that the interpolating predicted reaction forces in general are accurate for all the four cases. 
And the extrapolating predicted reaction forces are reasonable for a short range beyond the training range, especially for $F_x$.
The NN results from the last loading step in case (i) with extrapolating prediction are shown in Fig. \ref{fig:nonlinear-inter-extra-results}, from which we can observe that the extrapolating NN prediction is still very accurate.
Additional interpolating and extrapolating NN prediction results for case (i) are given in Appendix \ref{appendix:nonlinear-inter-extra-additional}.
With properly trained NNs, the surrogate model could make predictions for new BCs orders faster than the traditional numerical methods, as shown in Table \ref{tab:wall-time-nonlinear}.
The time and volume averaged $L_2$ error between DNSs and the NN predictions for case (i) is computed as
\begin{equation}
    L_2 = \frac{1}{L} \sum_{l=1}^{L} \left(  \frac{1}{K} \sqrt{\sum_{k=1}^K \left(  y_{l,k}^\text{DNS} - y_{l,k}^\text{NN} \right)^2}\right)
    \label{eq:l2-error}
\end{equation}
with $L$ ($=30$) indicating the total number of incremental loading steps (time) and $K$ ($=16\times16$) indicating the total pixels in the problem domain (volume).
As shown in Table \ref{tab:wall-time-nonlinear}, the averaged $L_2$ error is about 3.06e-4 for deterministic NNs and 4.04e-4 for BNNs. 
Compared to the unscaled DNS solution, which is in the range of $[0,~1]$, the $L_2$ error is small, which further confirms the accuracy of the surrogate PDE solver.
The prediction capability for interpolated and extrapolated BCs with good accuracy is very useful for rapidly estimating the solution of BVPs with similar physics but different BCs, particularly for homogenization and inverse problems where many similar BVPs with small variations need to be simulated repeatedly.

\section{Conclusion} \label{sec:conclusion}

In this work, an approach to solve PDEs with discretized residual constrained NNs is proposed.
Both deterministic and probabilistic NNs with an encoder-decoder structure are explored in this work, with the latter to quantify the uncertainties from model parameters and noise in the data.
An efficient NN-based implementation to calculate the discretized PDE residual is proposed. 
The NNs take a specially designed data structure, which contains information of the problem domain and the applied BCs, to solve BVPs.
The proposed approach is applied to different physical problems, including steady-state diffusion, linear and nonlinear elasticity.
Different examples for each system are considered to demonstrate the capability and performance of the proposed approach, which can simultaneously solve BVPs with varying domains and different applied BCs.
We also show the interpolation and extrapolation capability of the proposed NN solvers.
The ability to make accurate interpolated and extrapolated predictions on BCs that the NNs have not been exposed to during training is particularly useful for rapidly estimating the solution of BVPs with similar physics but different BCs, particularly for homogenization and inverse problems where many similar BVPs with small variations need to be simulated repeatedly.
A well-trained NN-based PDE solver can be easily shared and reused by users to investigate BVPs, and attains solutions much faster compared to the traditional numerical methods with acceptable error for most engineering design and  decision-making applications.
The trained NN PDE solver can be further trained to improve its accuracy and capability by using different transfer learning techniques as studied in \cite{Teichert2019Garikipati-ML-SurrogateOpt,Zhang2020Garikipati-CMAME-ML-RVE}.
The proposed approach is generalizable and can be easily applied to other physical systems.
Extending the proposed approach to transient systems is currently being investigated in a subsequent work by the authors.

\section*{Acknowledgments}
We gratefully acknowledge the support of Toyota Research Institute, Award \#849910: ``Computational framework for data-driven, predictive, multi-scale and multi-physics modeling of battery materials''.  
Computing resources were provided in part by the National Science Foundation, United States via grant 1531752 MRI: Acquisition of Conflux, A Novel Platform for Data-Driven Computational Physics (Tech. Monitor: Ed Walker). 
This work also used the Extreme Science and Engineering Discovery Environment (XSEDE) Comet at the San Diego Supercomputer Center and Stampede2 at The University of Texas at Austin's Texas Advanced Computing Center through allocation TG-MSS160003 and TG-DMR180072.

%\newpage

\appendix
\section{Supporting materials}
\subsection{Steady-state diffusion}\label{appendix:diffusion-results-extra}
\subsubsection{Multiple rectangle domains with different BCs} \label{appendix:diffusion-20bvps}
\begin{figure}[p!]
  \centering
  {\includegraphics[height=0.08\linewidth]{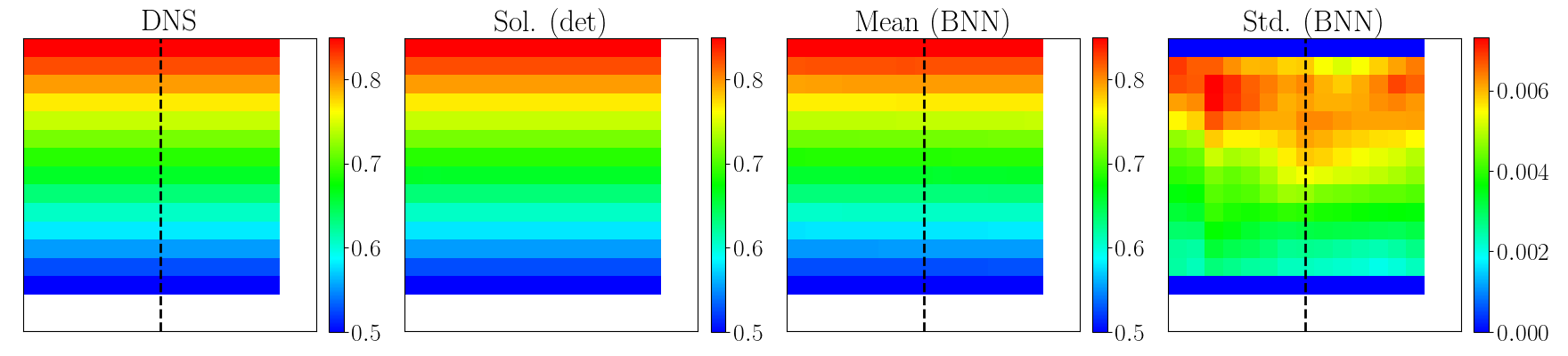}} 
  {\includegraphics[height=0.08\linewidth]{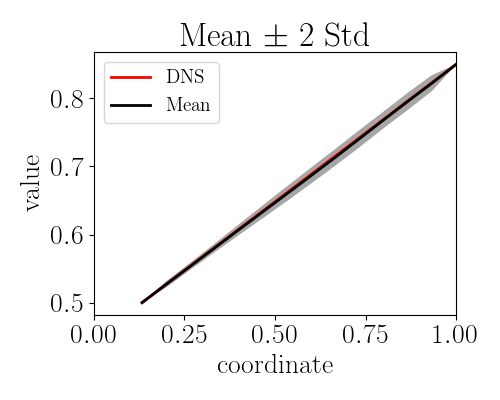}} 
  {\includegraphics[height=0.08\linewidth]{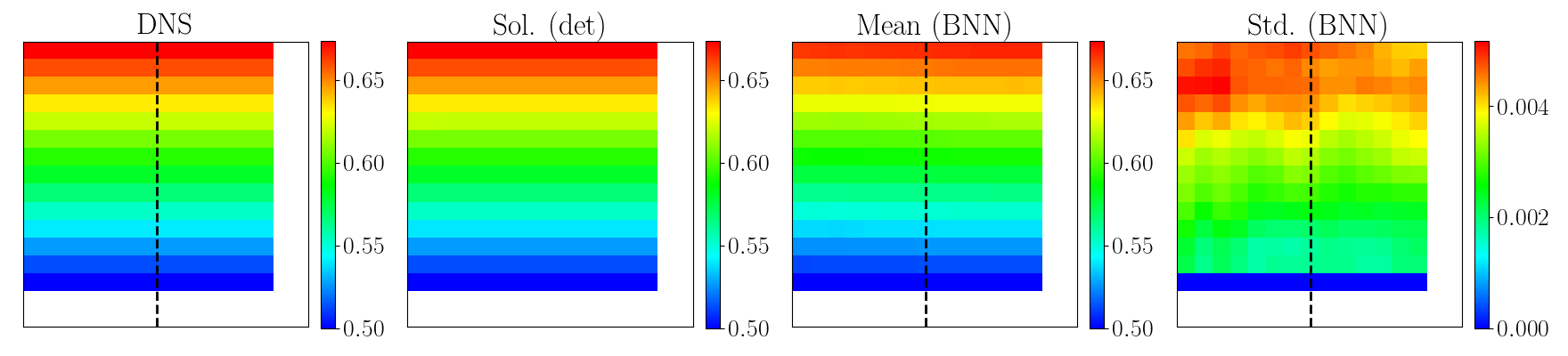}} 
  {\includegraphics[height=0.08\linewidth]{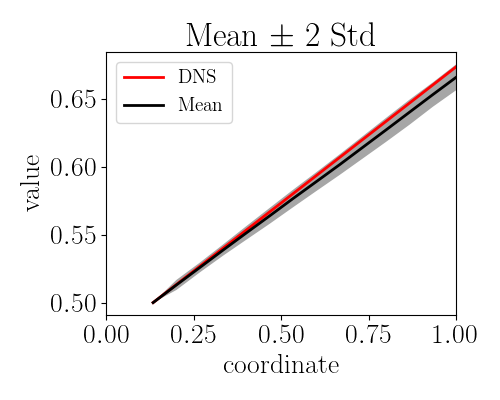}} \\ 
  {\includegraphics[height=0.08\linewidth]{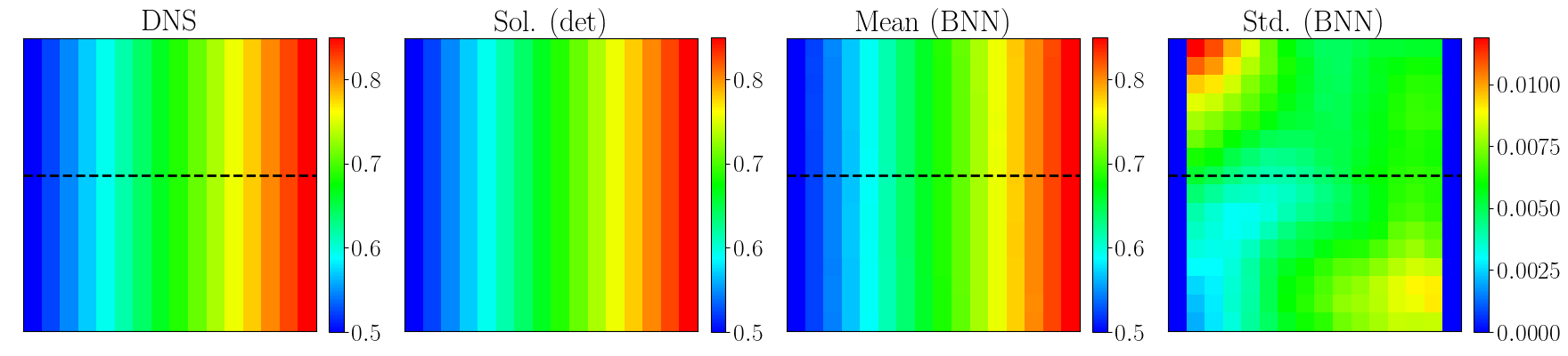}} 
  {\includegraphics[height=0.08\linewidth]{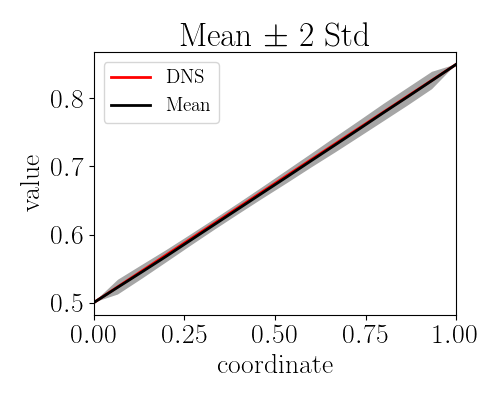}} 
  {\includegraphics[height=0.08\linewidth]{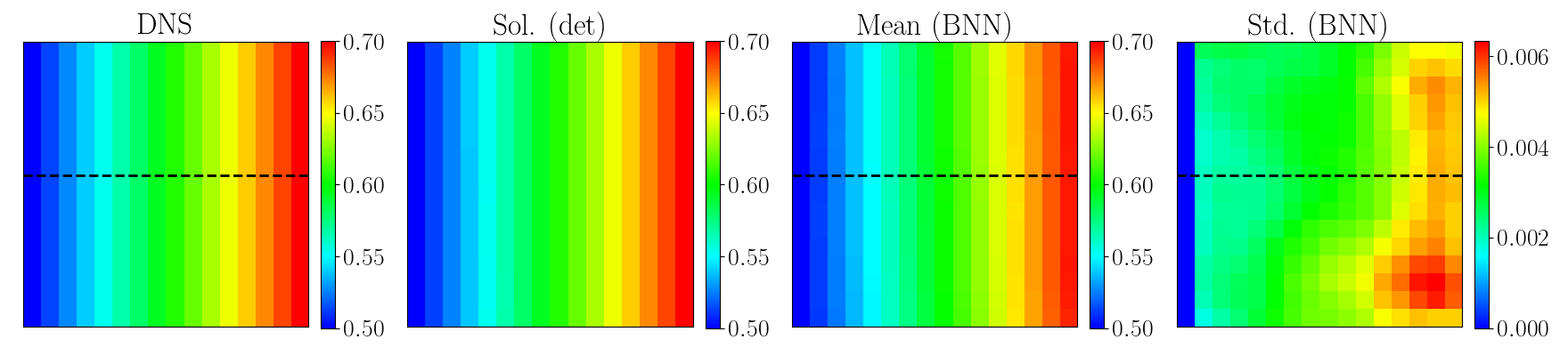}} 
  {\includegraphics[height=0.08\linewidth]{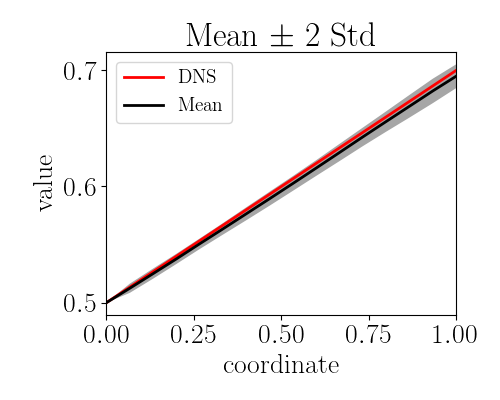}} \\
  {\includegraphics[height=0.08\linewidth]{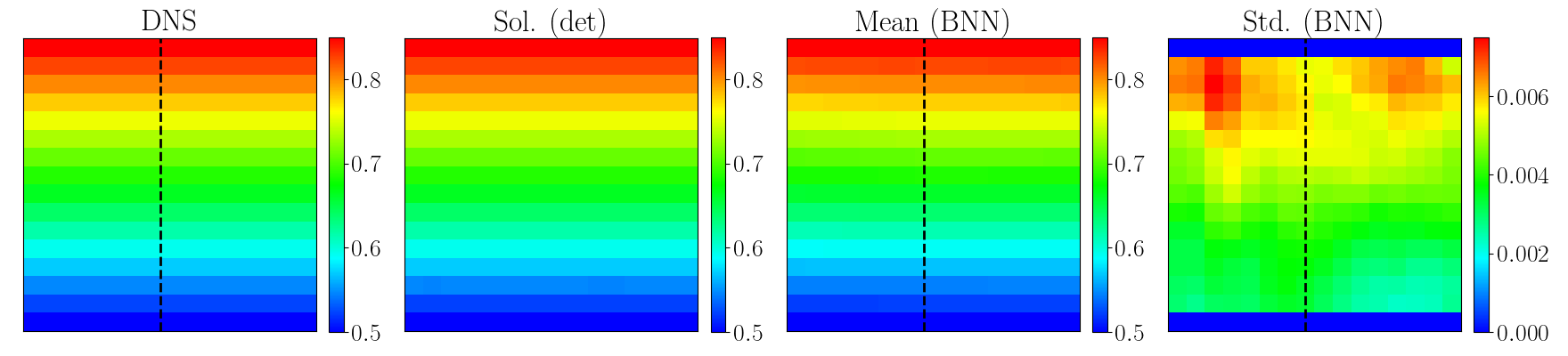}} 
  {\includegraphics[height=0.08\linewidth]{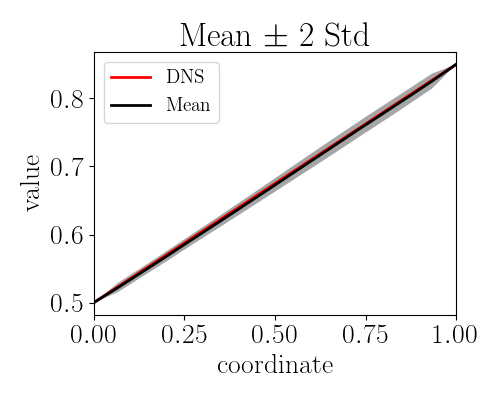}} 
  {\includegraphics[height=0.08\linewidth]{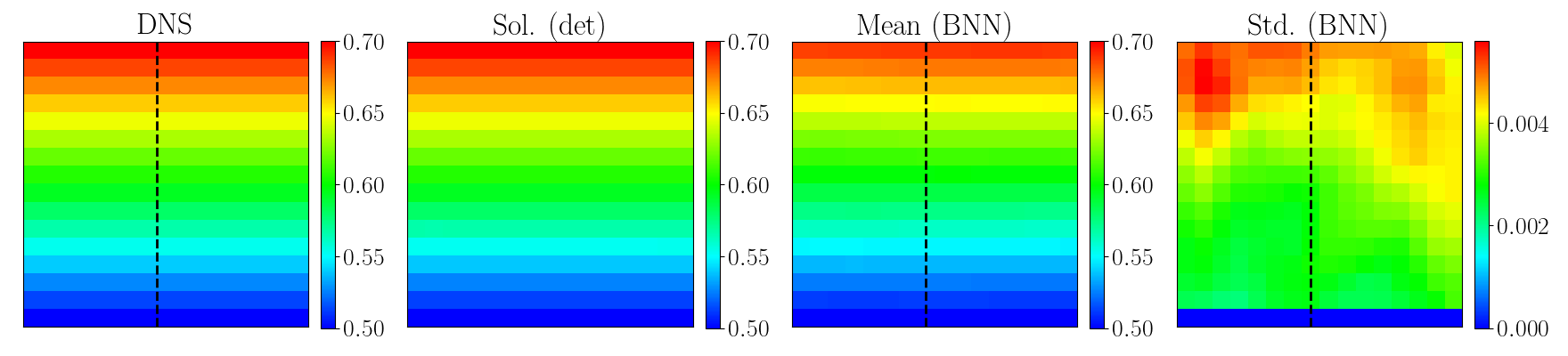}} 
  {\includegraphics[height=0.08\linewidth]{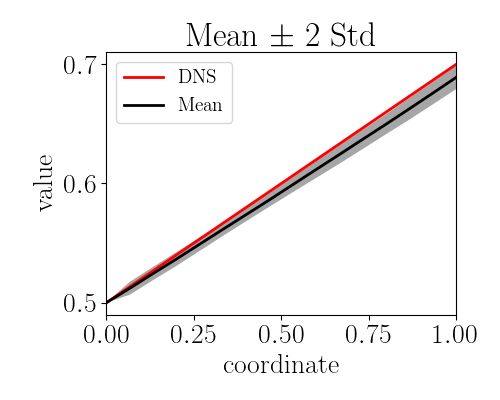}} \\
  {\includegraphics[height=0.08\linewidth]{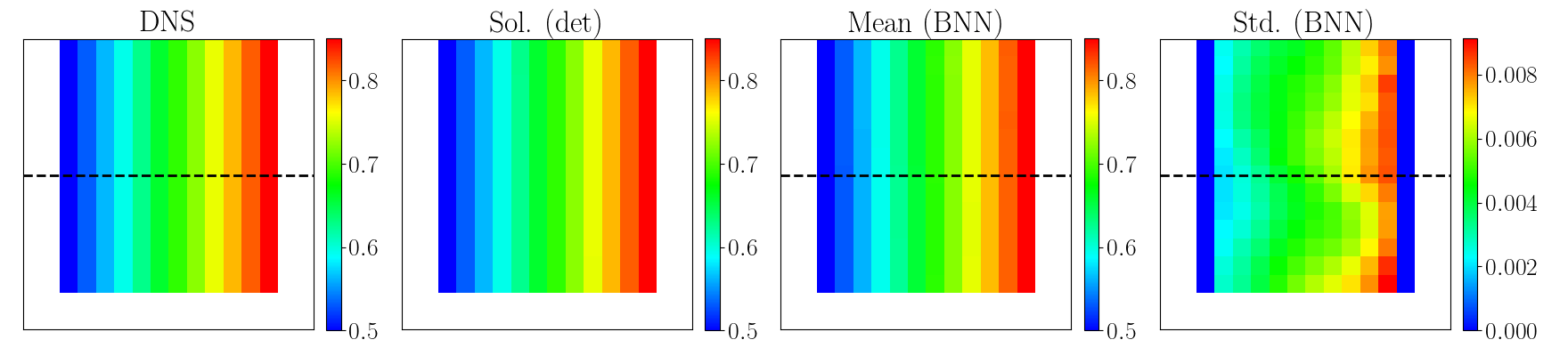}} 
  {\includegraphics[height=0.08\linewidth]{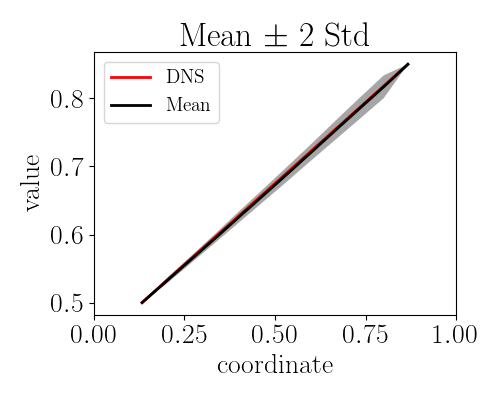}} 
  {\includegraphics[height=0.08\linewidth]{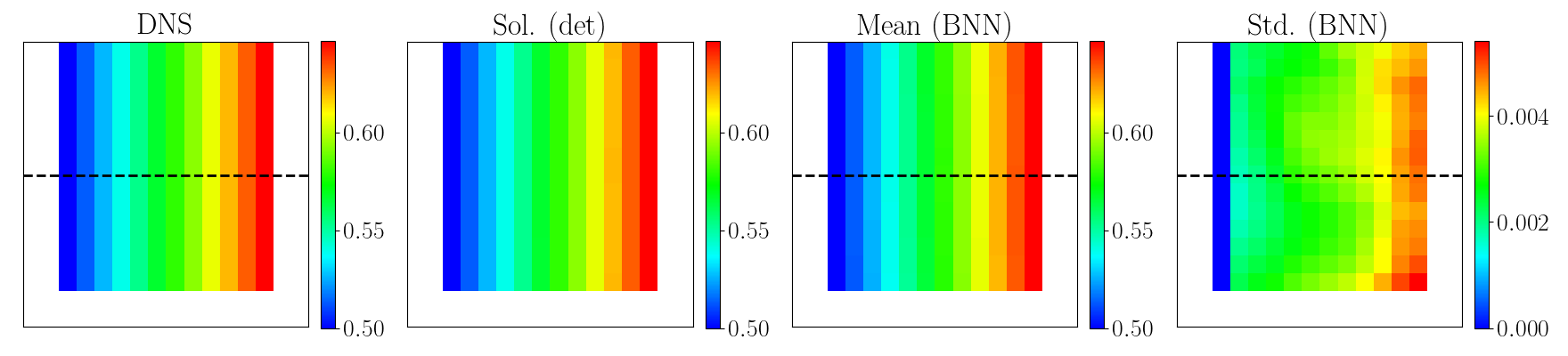}} 
  {\includegraphics[height=0.08\linewidth]{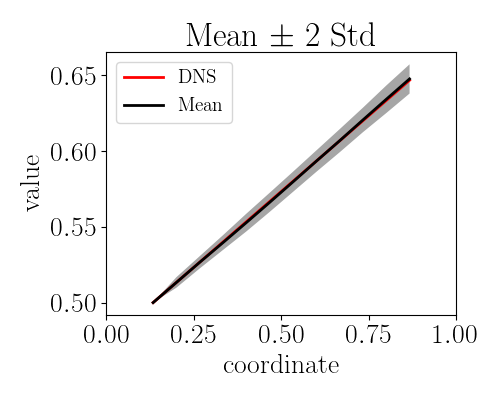}} \\
  {\includegraphics[height=0.08\linewidth]{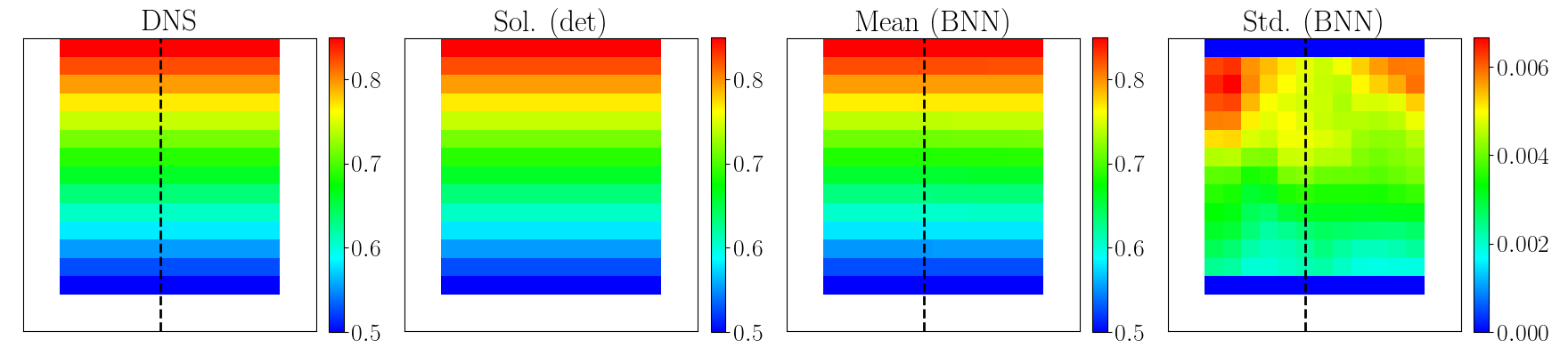}} 
  {\includegraphics[height=0.08\linewidth]{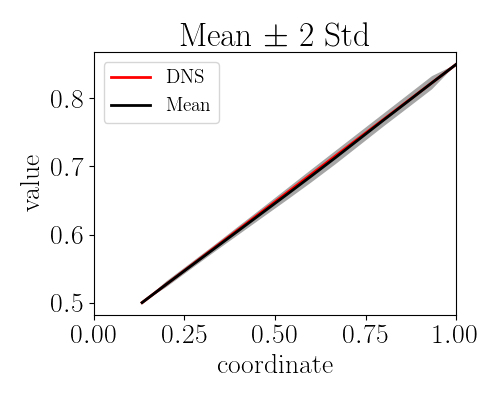}} 
  {\includegraphics[height=0.08\linewidth]{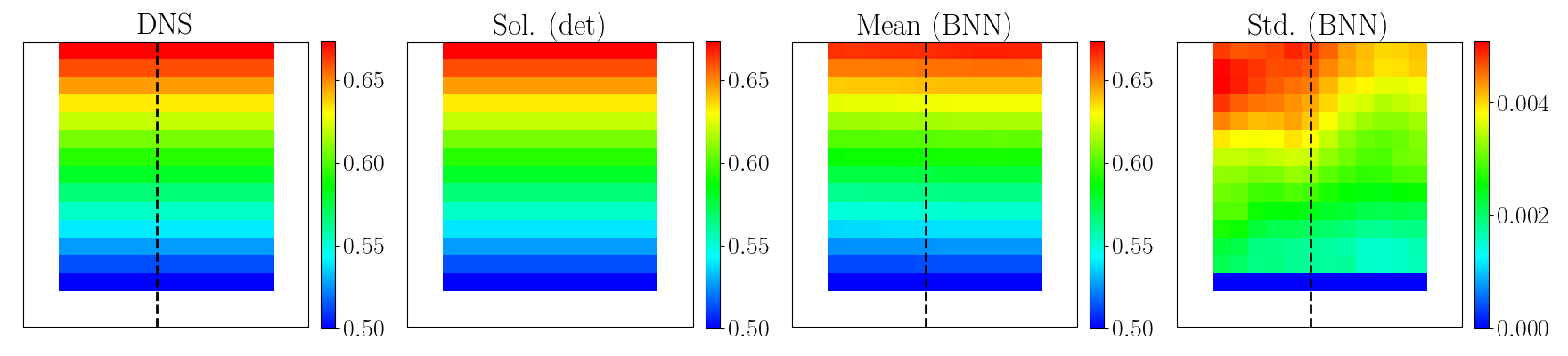}} 
  {\includegraphics[height=0.08\linewidth]{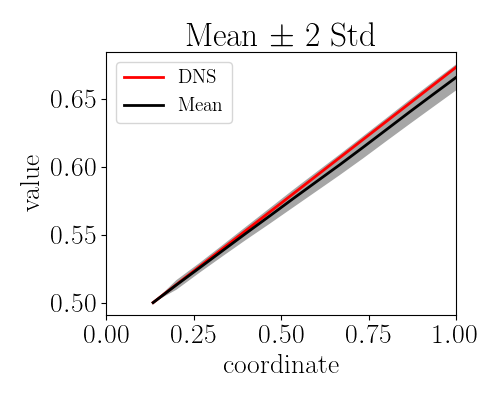}} \\
  {\includegraphics[height=0.08\linewidth]{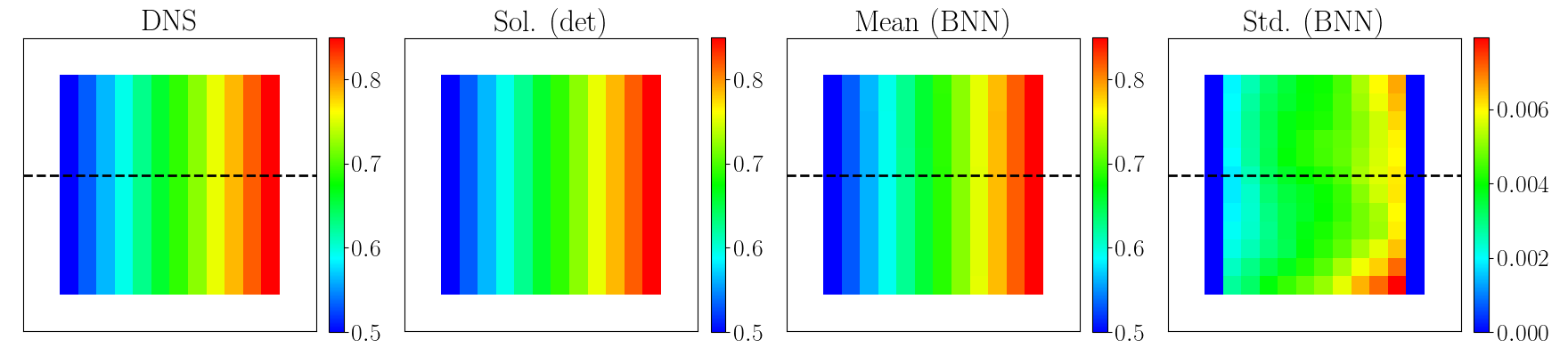}} 
  {\includegraphics[height=0.08\linewidth]{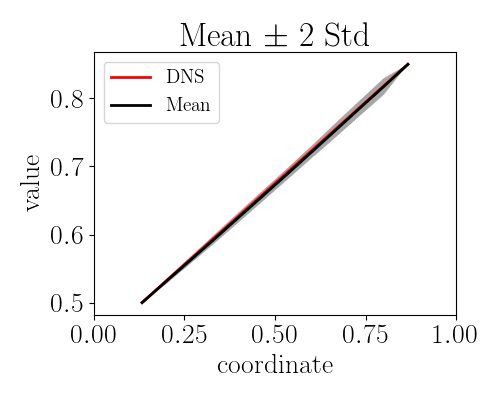}} 
  {\includegraphics[height=0.08\linewidth]{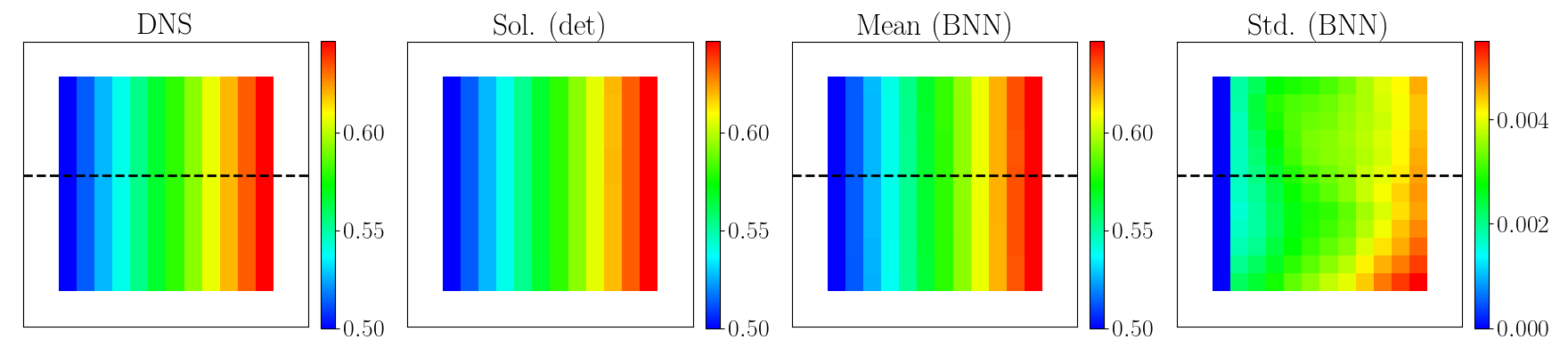}} 
  {\includegraphics[height=0.08\linewidth]{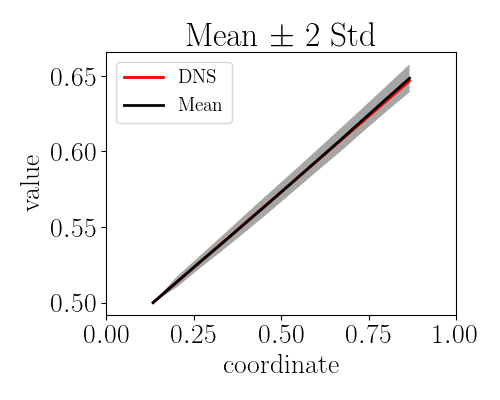}} \\ 
  {\includegraphics[height=0.08\linewidth]{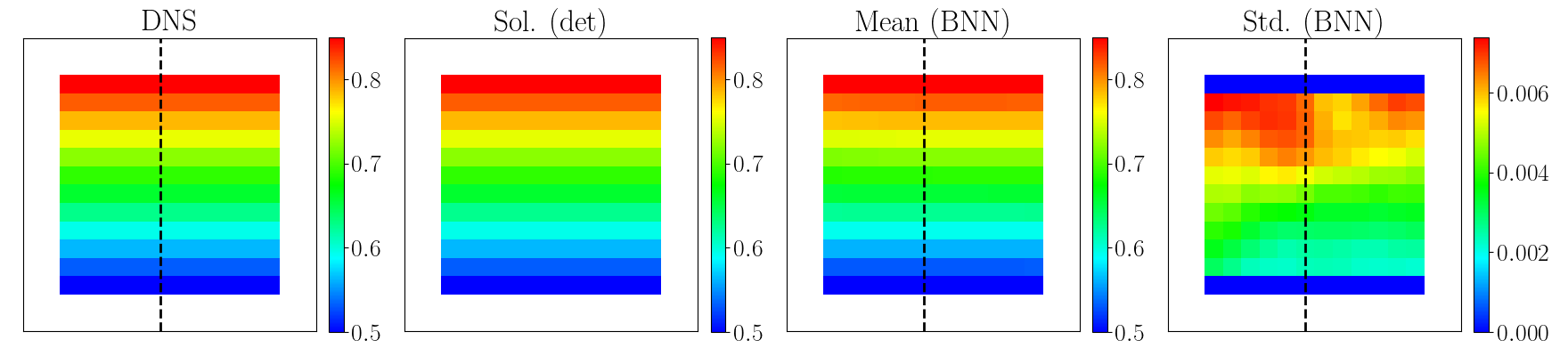}} 
  {\includegraphics[height=0.08\linewidth]{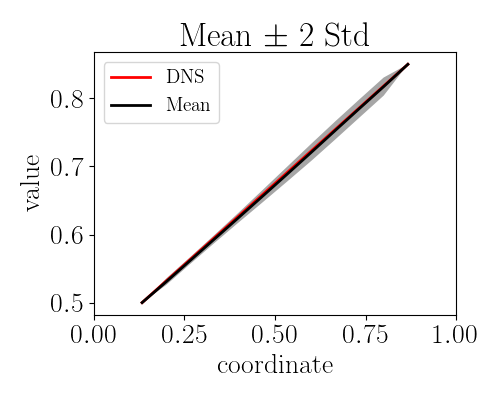}} 
  {\includegraphics[height=0.08\linewidth]{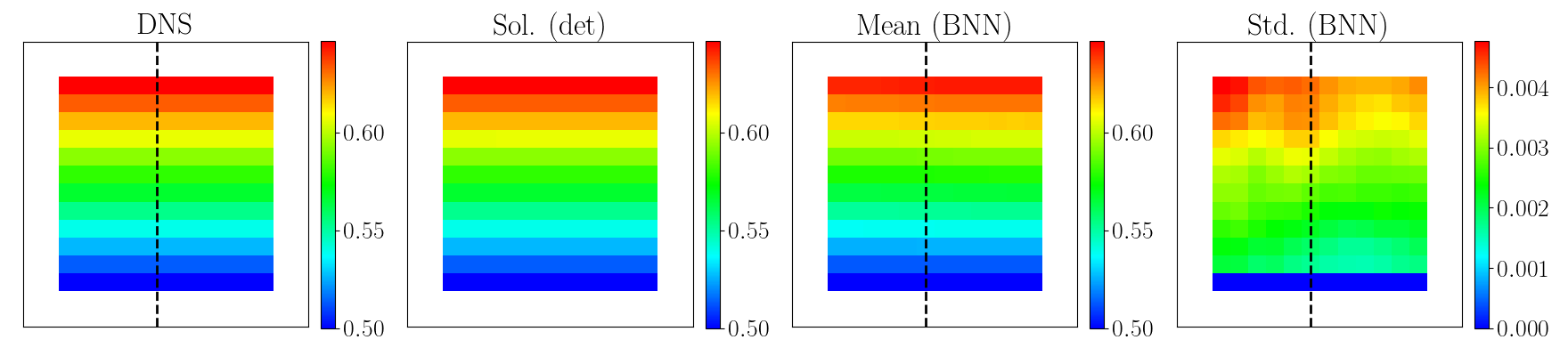}} 
  {\includegraphics[height=0.08\linewidth]{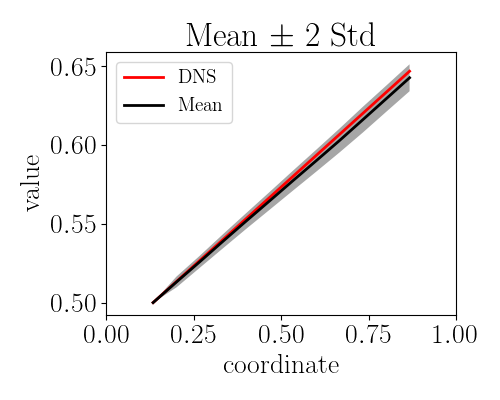}} \\
  {\includegraphics[height=0.08\linewidth]{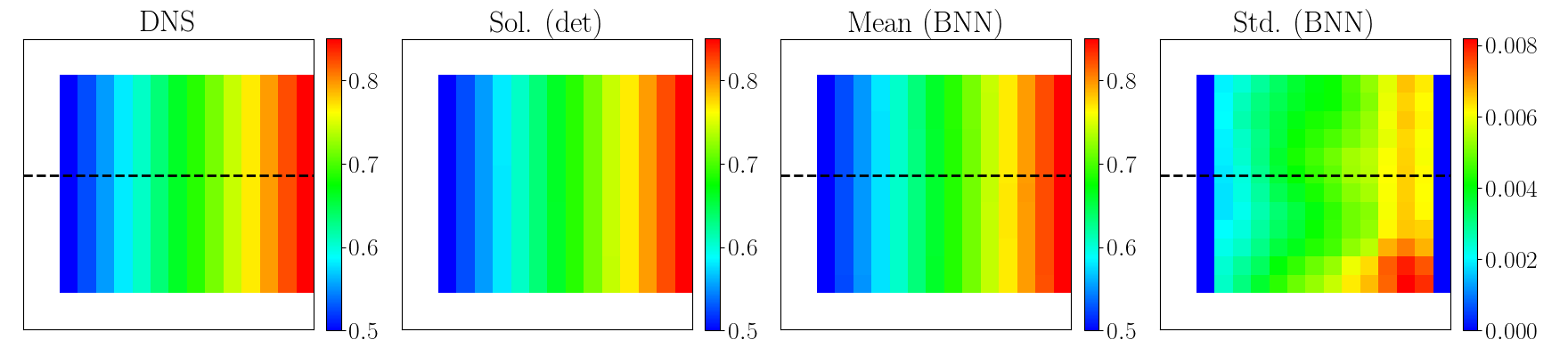}} 
  {\includegraphics[height=0.08\linewidth]{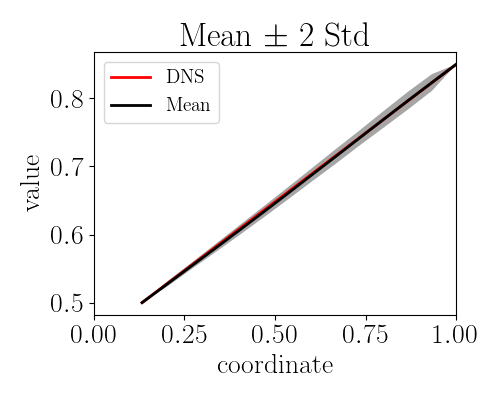}} 
  {\includegraphics[height=0.08\linewidth]{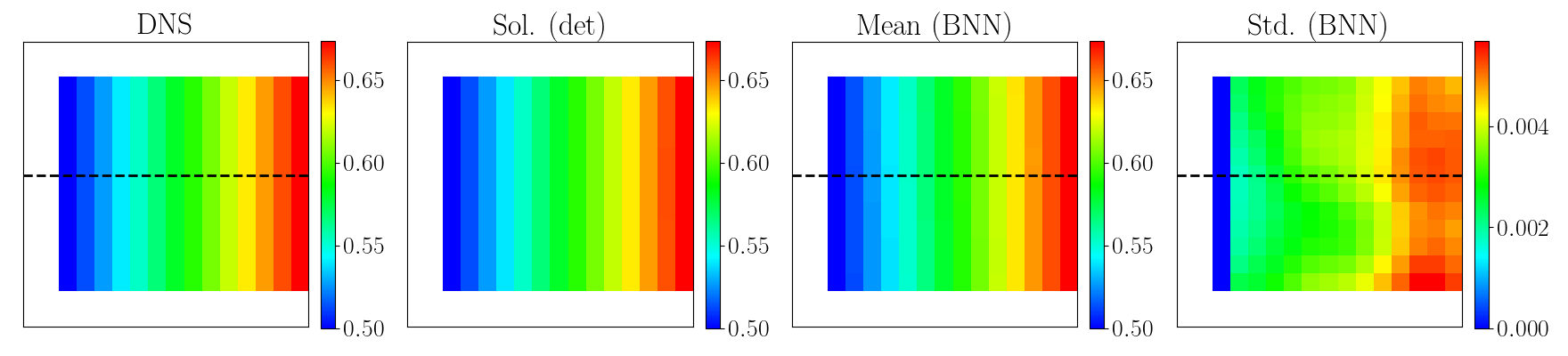}} 
  {\includegraphics[height=0.08\linewidth]{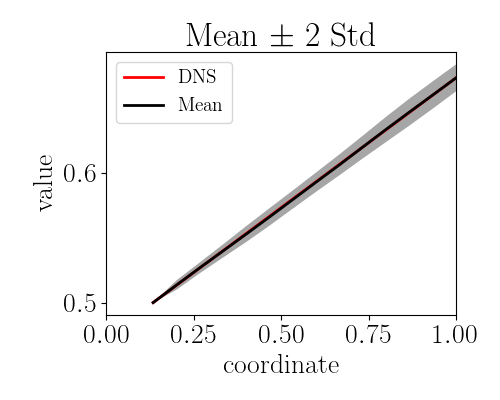}} \\
  {\includegraphics[height=0.08\linewidth]{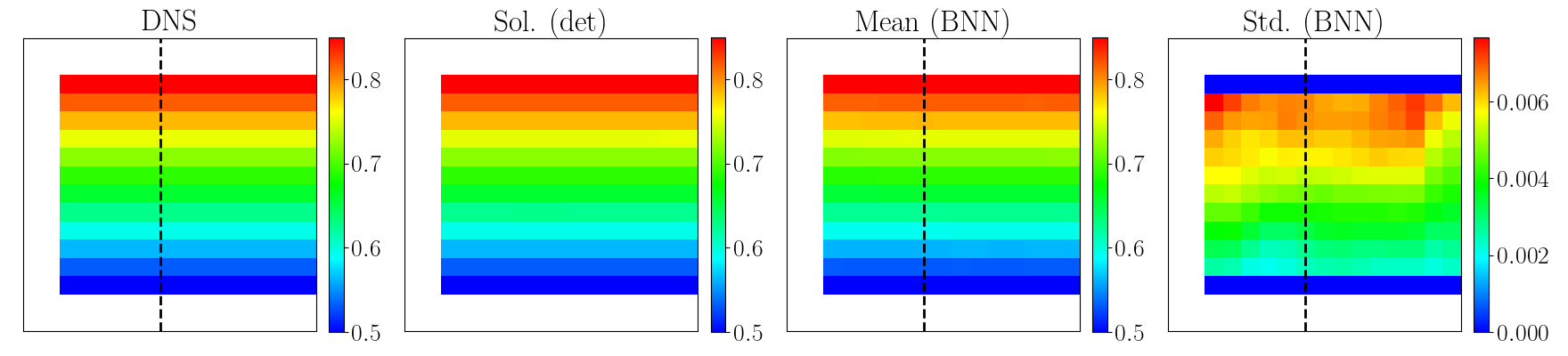}} 
  {\includegraphics[height=0.08\linewidth]{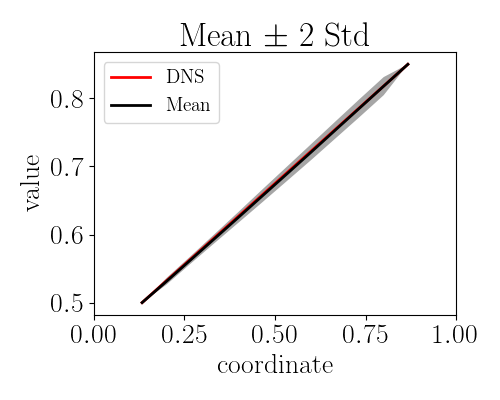}} 
  {\includegraphics[height=0.08\linewidth]{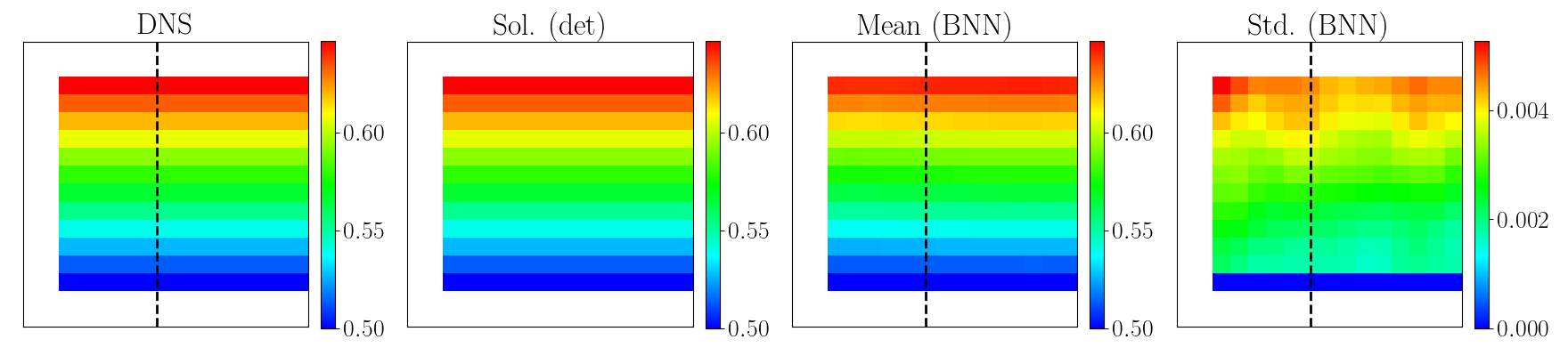}} 
  {\includegraphics[height=0.08\linewidth]{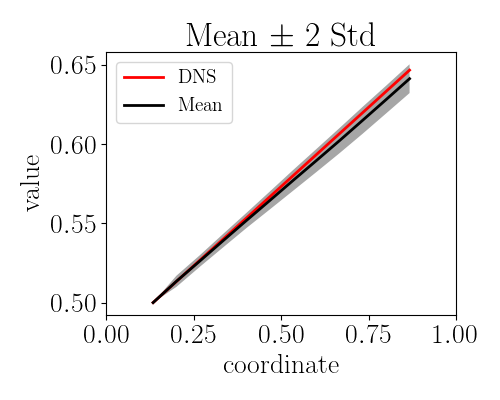}} \\
  \caption{Additional NN results for steady-state diffusion BVPs on rectangle domains.}
  \label{fig:diffusion-20bvp-results-additional}
\end{figure}

Additional results from the 20 BVPs with rectangle domains are summarized in Fig. \ref{fig:diffusion-20bvp-results-additional}.

\subsubsection{Single octagon domain with mixed BCs} \label{appendix:diffusion-octagon}
NN structure information for the octagon domain simulation are summarized in Table \ref{tab:diffusion-octagon-NNs} and \ref{tab:diffusion-octagon-NNs-others}.

\begin{table}
  \centering
  \begin{tabular}{l | l | l | l}
    \hline
    Deterministic         & Probabilistic         & Size         & Layer arguments \\ \hline
    Input                 & Input                 & -            & - \\
    LayerFillRandomNumber & LayerFillRandomNumber & -            & - \\
    Conv2D                & Convolution2DFlipout  & filters = 8  & kernel (5,5), padding: same, ReLU \\
    MaxPooling2D          & MaxPooling2D          & -            & kernel (2,2), padding: same\\
    Conv2D                & Convolution2DFlipout  & filters = 8  & kernel (5,5), padding: same, ReLU \\
    MaxPooling2D          & MaxPooling2D          & -            & kernel (2,2), padding: same\\
    Conv2D                & Convolution2DFlipout  & filters = 8  & kernel (5,5), padding: same, ReLU \\
    MaxPooling2D          & MaxPooling2D          & -            & kernel (2,2), padding: same\\
    Flatten               & Flatten               & -            & - \\
    Dense                 & DenseFlipout          & units = 32   & ReLU \\
    Dense                 & DenseFlipout          & units = 32   & ReLU \\
    Reshape               & Reshape               & -            & $[4,4,4]$ \\
    Conv2D                & Convolution2DFlipout  & filters = 8  & kernel (5,5), padding: same, ReLU \\
    UpSampling2D          & UpSampling2D          & -            & size (2,2) \\
    Conv2D                & Convolution2DFlipout  & filters = 8  & kernel (5,5), padding: same, ReLU \\
    UpSampling2D          & UpSampling2D          & -            & size (2,2) \\
    Conv2D                & Convolution2DFlipout  & filters = 8  & kernel (5,5), padding: same, ReLU \\
    UpSampling2D          & UpSampling2D          & -            & size (2,2) \\
    Conv2D                & Convolution2DFlipout  & filters = 16 & kernel (5,5), padding: same, ReLU \\
    Conv2D                & Convolution2DFlipout  & filters = 1  & kernel (5,5), padding: same, ReLU \\
    \hline
  \end{tabular}
  \caption{Details of both deterministic and probabilistic NNs for solving diffusion BVPs on the octagon domain with an output resolution of $32\times 32$.}
  \label{tab:diffusion-octagon-NNs}
\end{table}

\begin{table}
  \centering
  \begin{tabular}{l | l | l }
    \hline
    Description                     & Deterministic                 & Probabilistic       \\ \hline
    Total parameters                & 16,049                        & 31,970                 \\
    Size of $\calD$                 & 1 $\times$ Aug: $2^{12}$      & 1 $\times$ Aug: $2^{11}$      \\
    Epochs                          & 20,000                        & 100                  \\
    Zero initialization epochs      & 100                           & -                     \\
    Optimizer                       & Nadam                         & Nadam                 \\
    Learning Rate                   & 2.5e-4                        & 1e-8                  \\
    Batch Size                      & 256                           & 64                    \\
    $\Sigma_1$                      & -                             & 1e-8                  \\
    Initial value of $\Sigma_2$     & -                             & 1e-8                  \\
    \hline
  \end{tabular}
  \caption{Training related parameters for solving steady-state diffusion on the octagon domain. Aug: data augmentation. }
  \label{tab:diffusion-octagon-NNs-others}
\end{table}

\subsection{Linear elasticity}\label{appendix:linear-results-extra}
\subsubsection{Multiple rectangle domains with different BCs} \label{appendix:linear-30bvps}
\begin{figure}[p!]
  \centering
  {\includegraphics[height=0.08\linewidth]{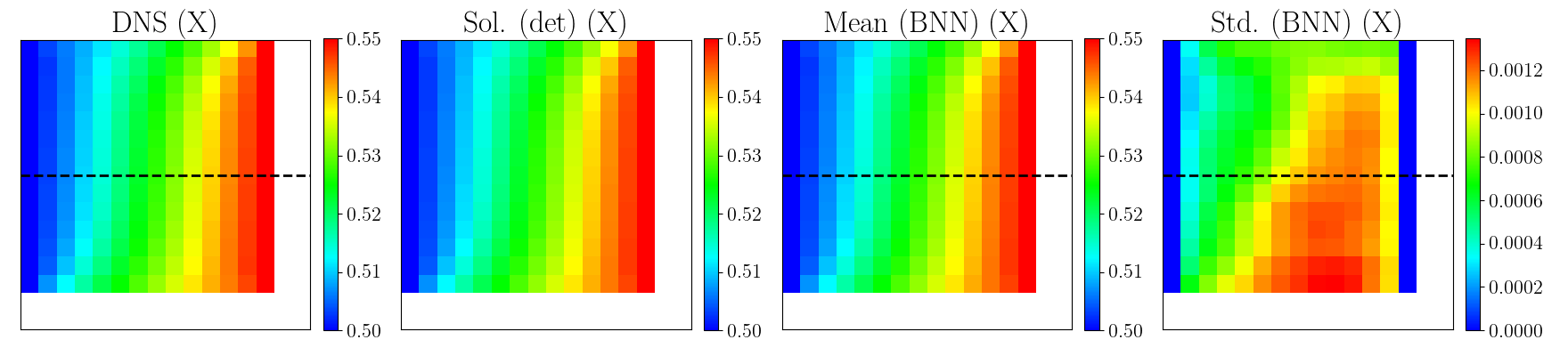}} 
  {\includegraphics[height=0.08\linewidth]{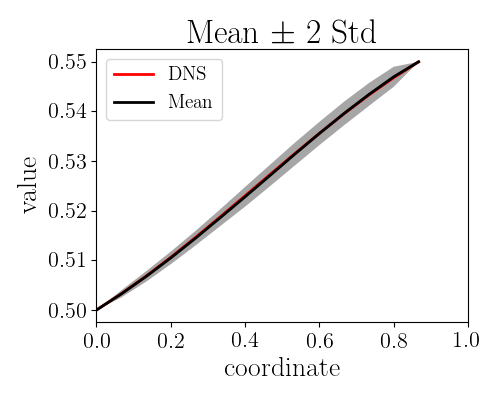}}
  {\includegraphics[height=0.08\linewidth]{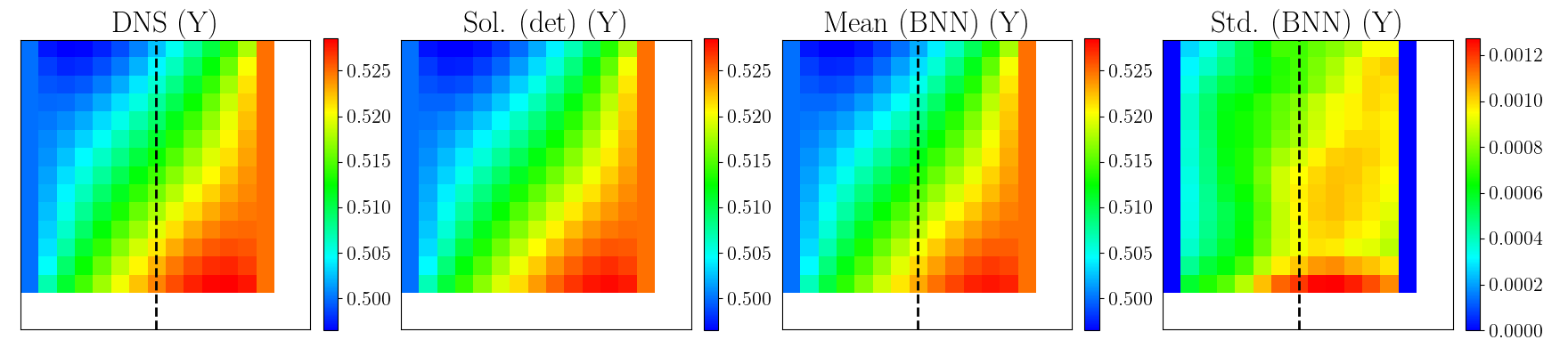}} 
  {\includegraphics[height=0.08\linewidth]{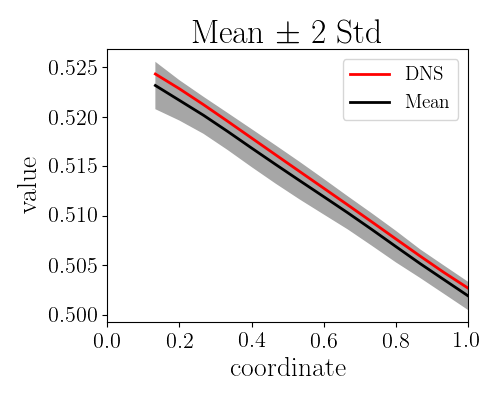}} \\
  {\includegraphics[height=0.08\linewidth]{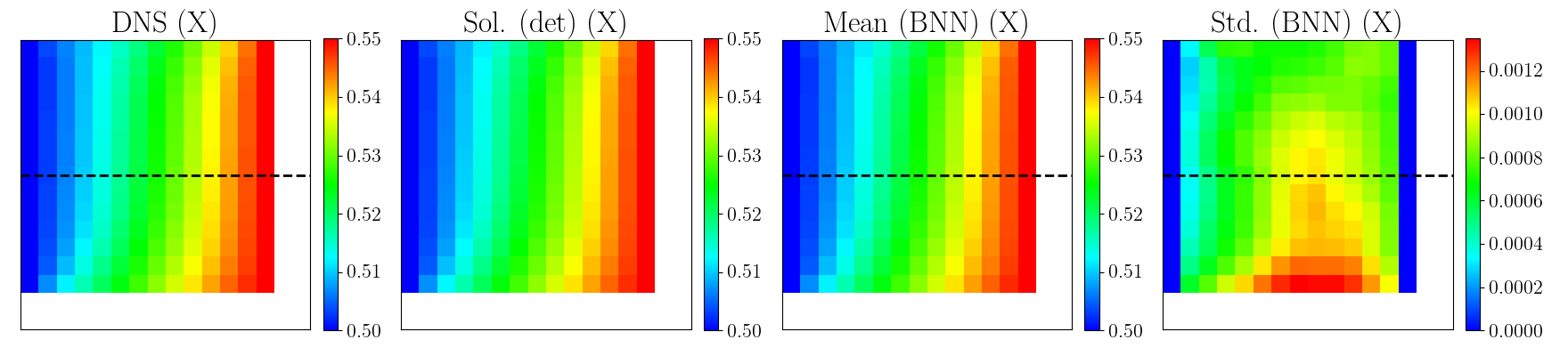}} 
  {\includegraphics[height=0.08\linewidth]{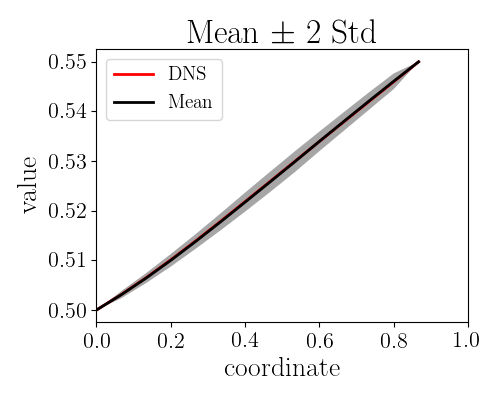}} 
  {\includegraphics[height=0.08\linewidth]{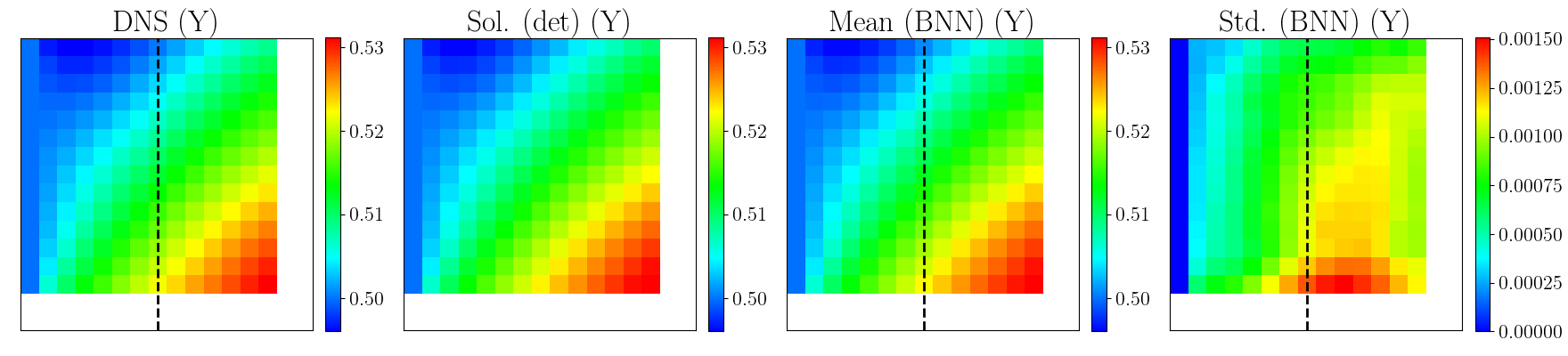}} 
  {\includegraphics[height=0.08\linewidth]{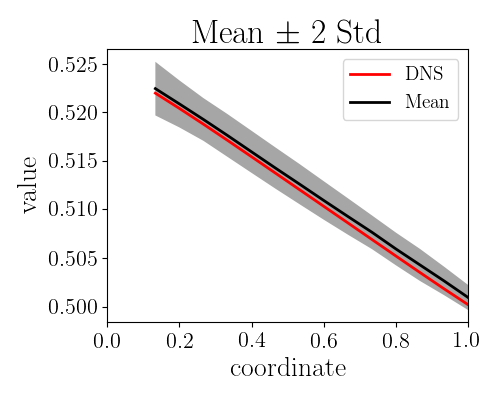}} \\
  {\includegraphics[height=0.08\linewidth]{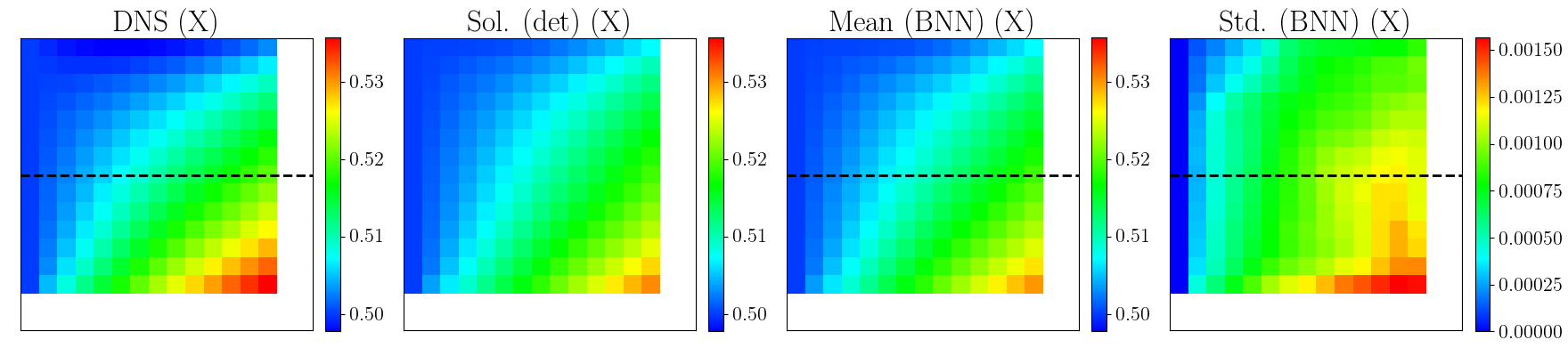}} 
  {\includegraphics[height=0.08\linewidth]{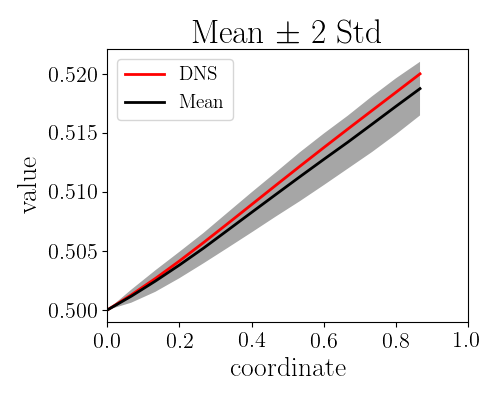}} 
  {\includegraphics[height=0.08\linewidth]{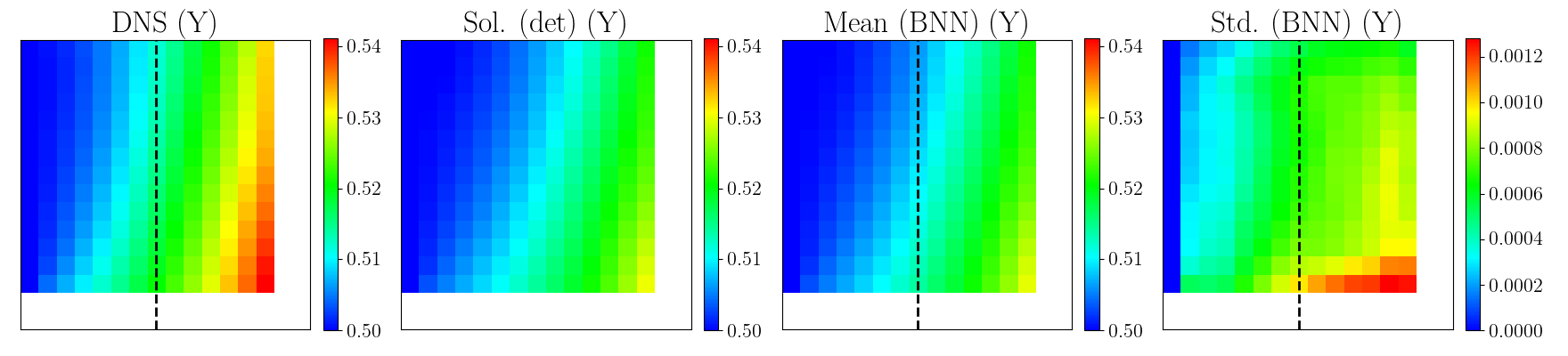}} 
  {\includegraphics[height=0.08\linewidth]{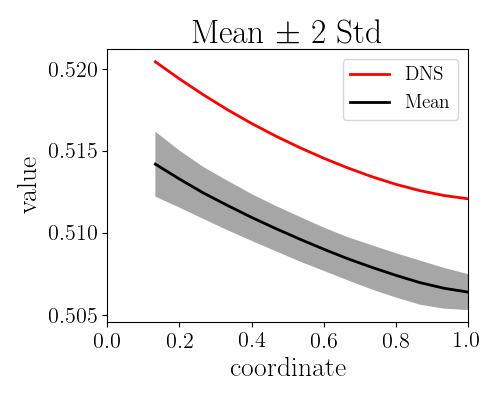}} \\
  {\includegraphics[height=0.08\linewidth]{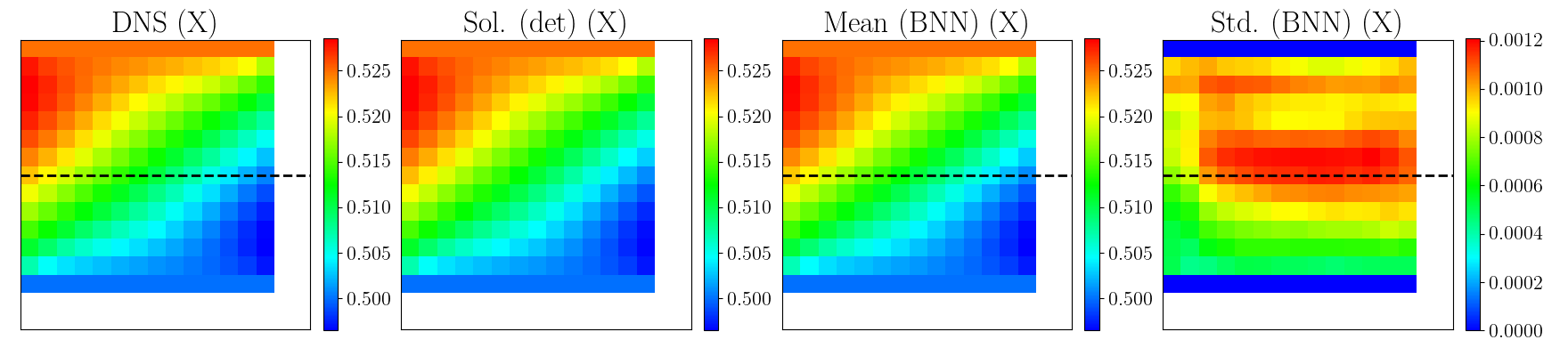}} 
  {\includegraphics[height=0.08\linewidth]{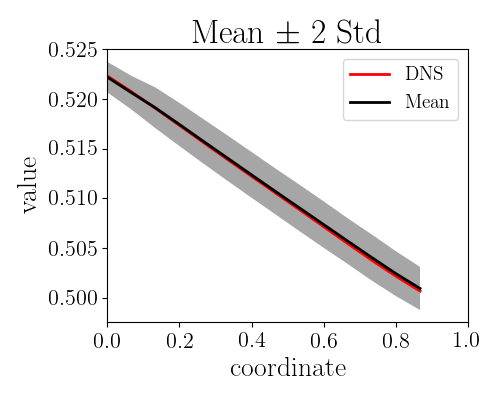}} 
  {\includegraphics[height=0.08\linewidth]{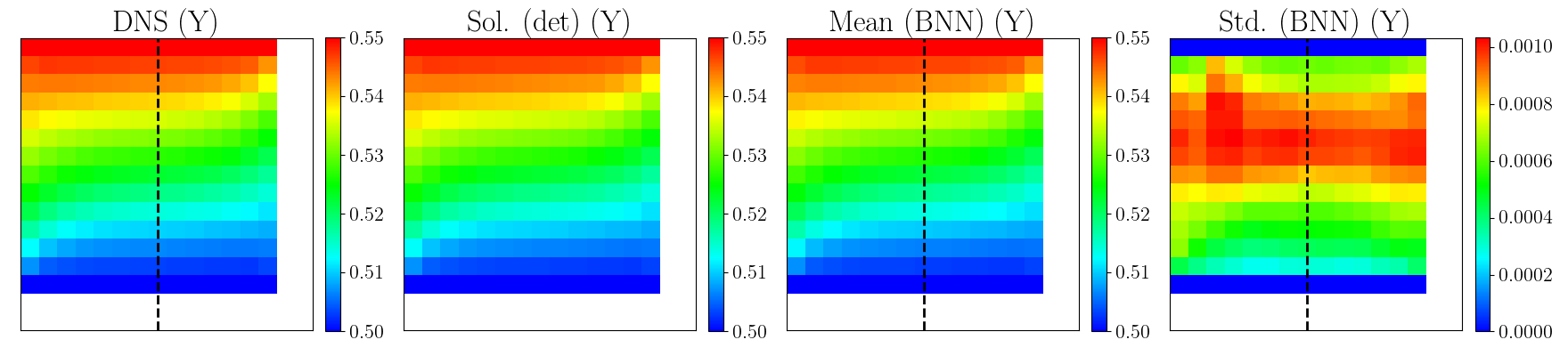}} 
  {\includegraphics[height=0.08\linewidth]{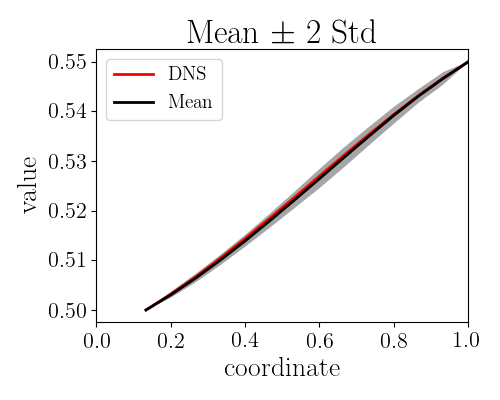}} \\
  {\includegraphics[height=0.08\linewidth]{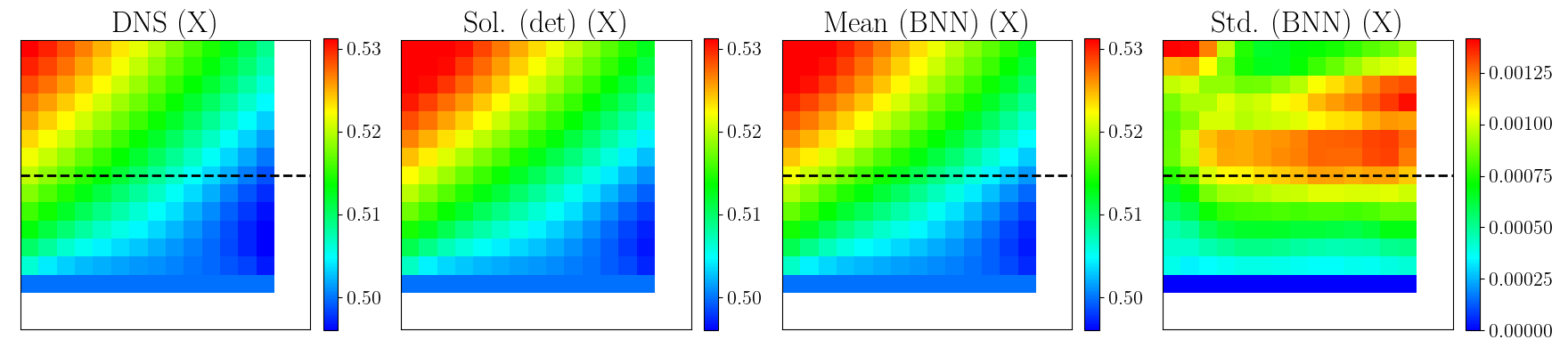}} 
  {\includegraphics[height=0.08\linewidth]{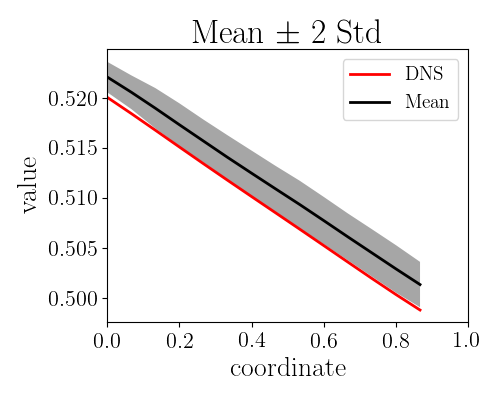}} 
  {\includegraphics[height=0.08\linewidth]{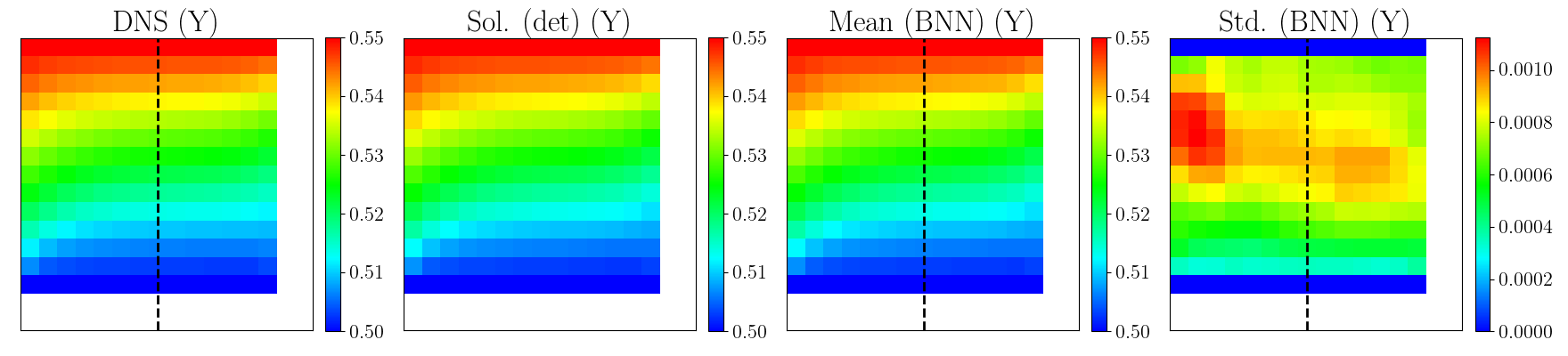}} 
  {\includegraphics[height=0.08\linewidth]{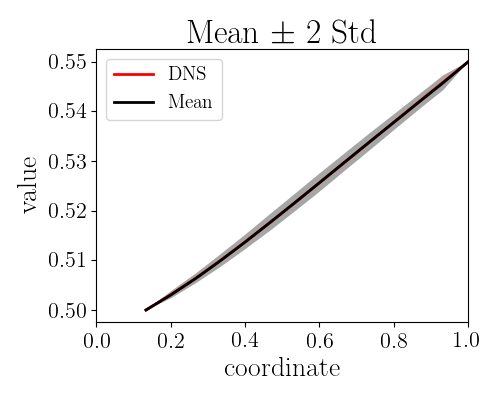}} \\
  {\includegraphics[height=0.08\linewidth]{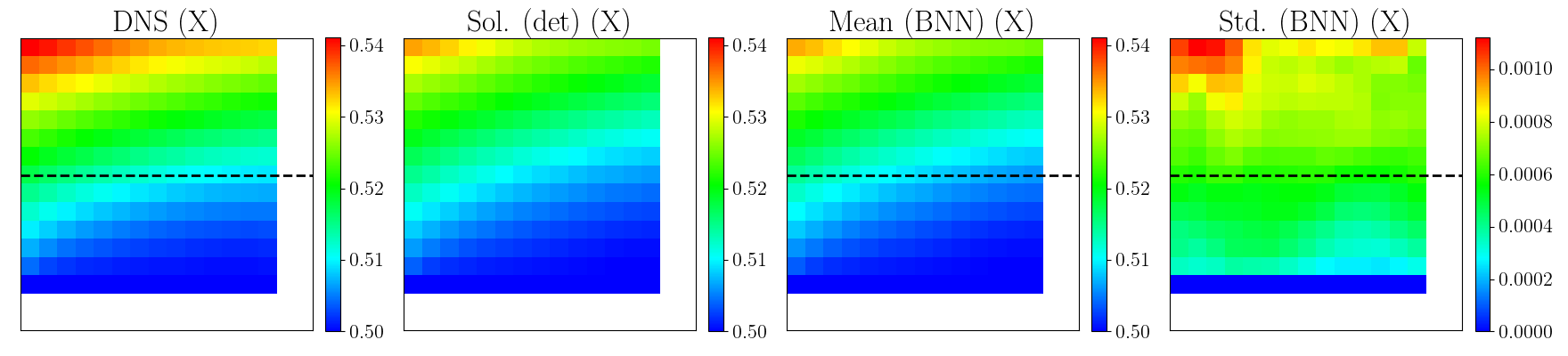}} 
  {\includegraphics[height=0.08\linewidth]{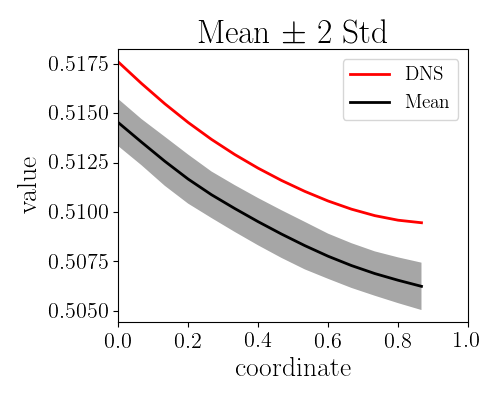}} 
  {\includegraphics[height=0.08\linewidth]{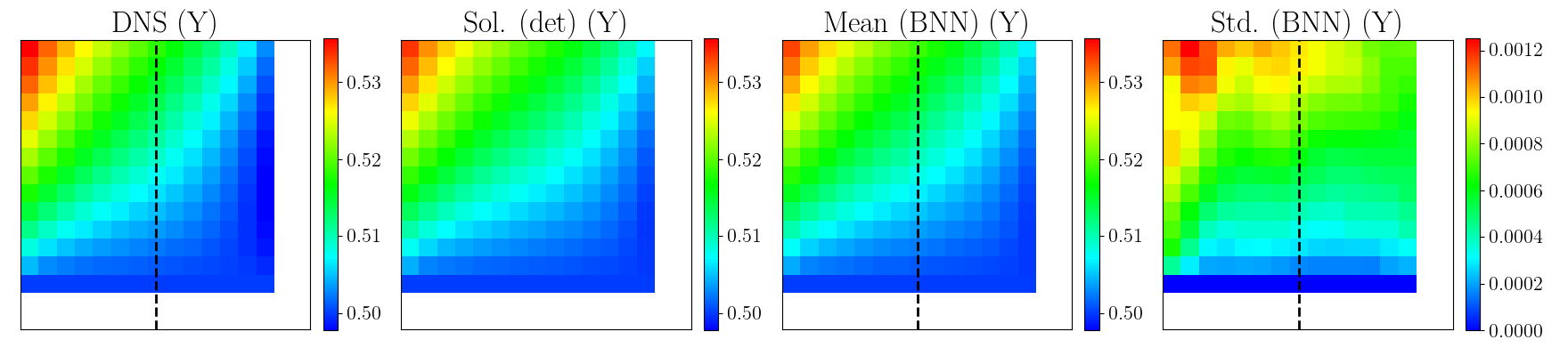}} 
  {\includegraphics[height=0.08\linewidth]{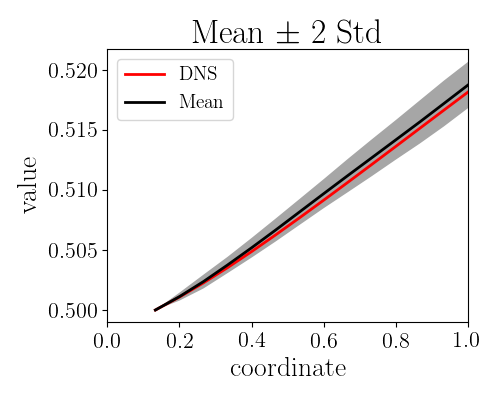}} \\
  {\includegraphics[height=0.08\linewidth]{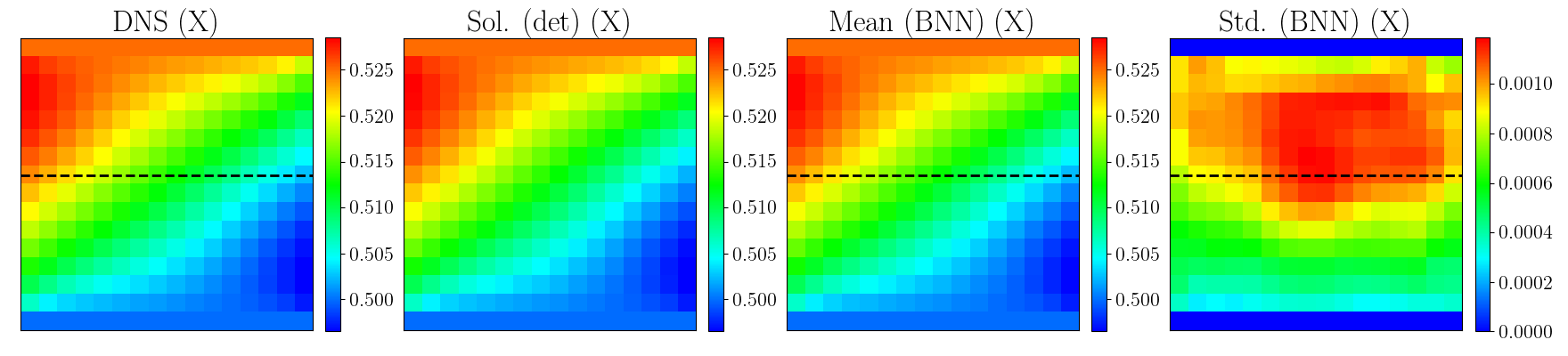}} 
  {\includegraphics[height=0.08\linewidth]{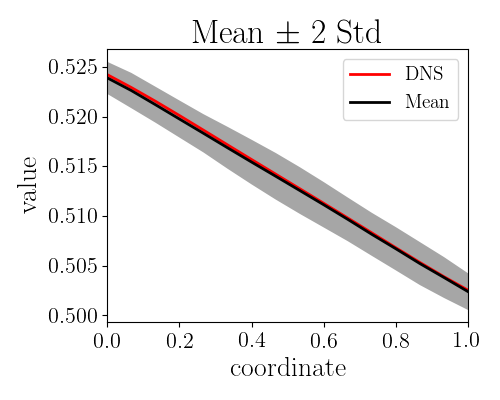}} 
  {\includegraphics[height=0.08\linewidth]{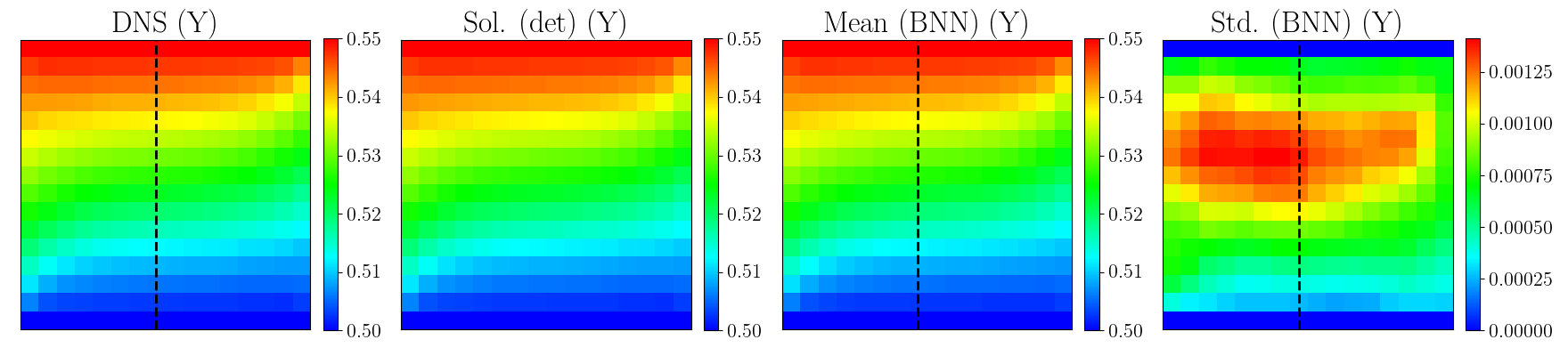}} 
  {\includegraphics[height=0.08\linewidth]{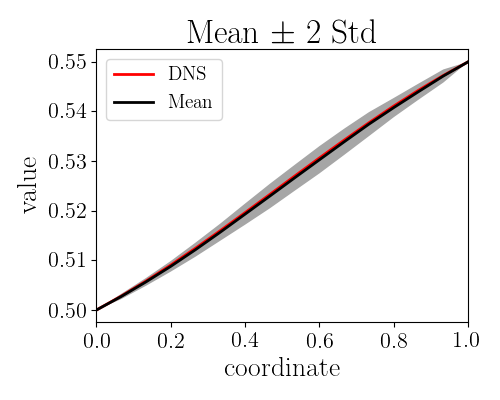}} \\
  {\includegraphics[height=0.08\linewidth]{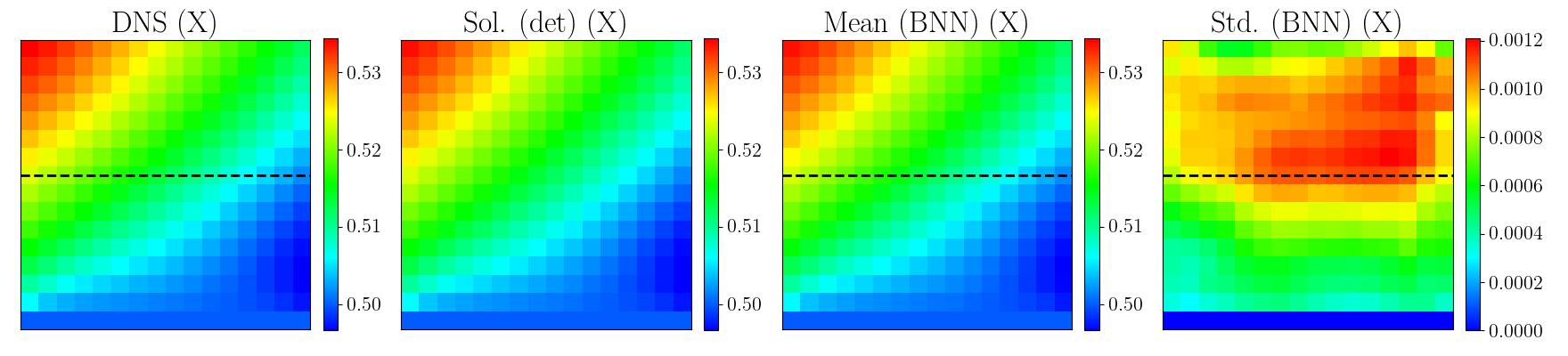}} 
  {\includegraphics[height=0.08\linewidth]{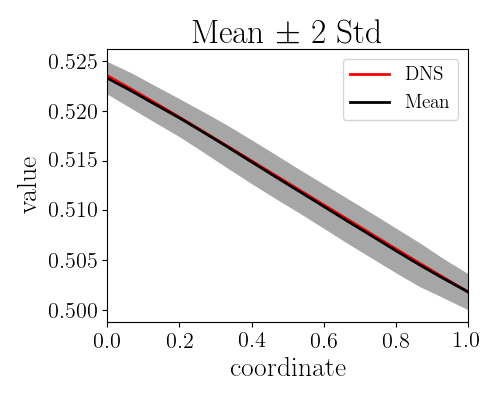}} 
  {\includegraphics[height=0.08\linewidth]{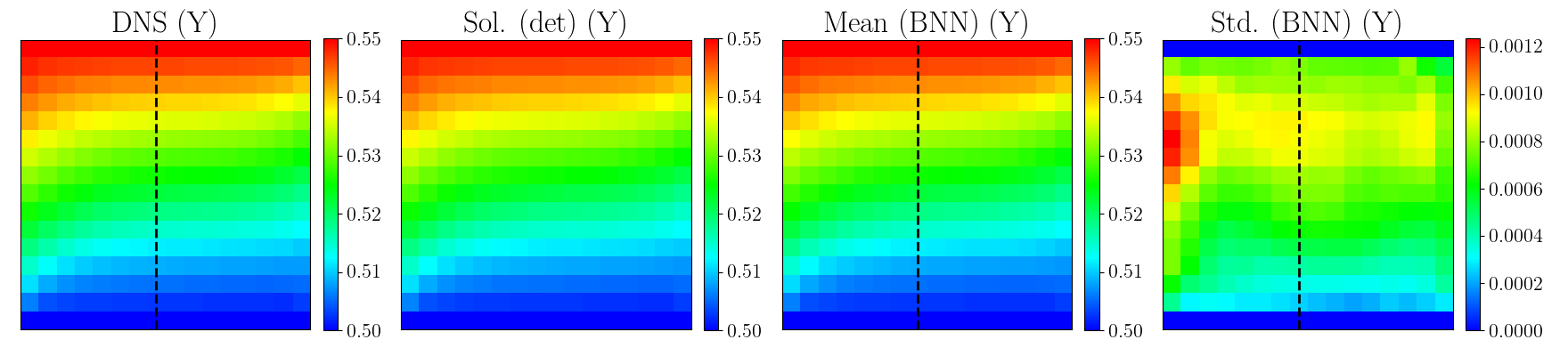}} 
  {\includegraphics[height=0.08\linewidth]{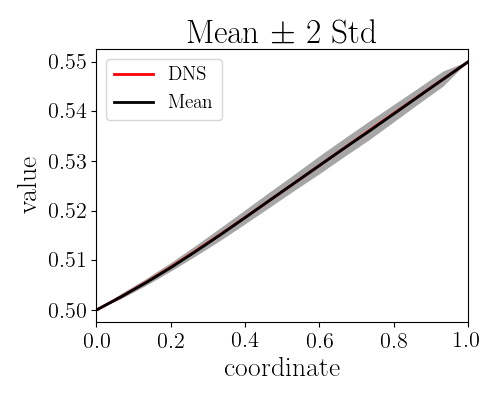}} \\
  {\includegraphics[height=0.08\linewidth]{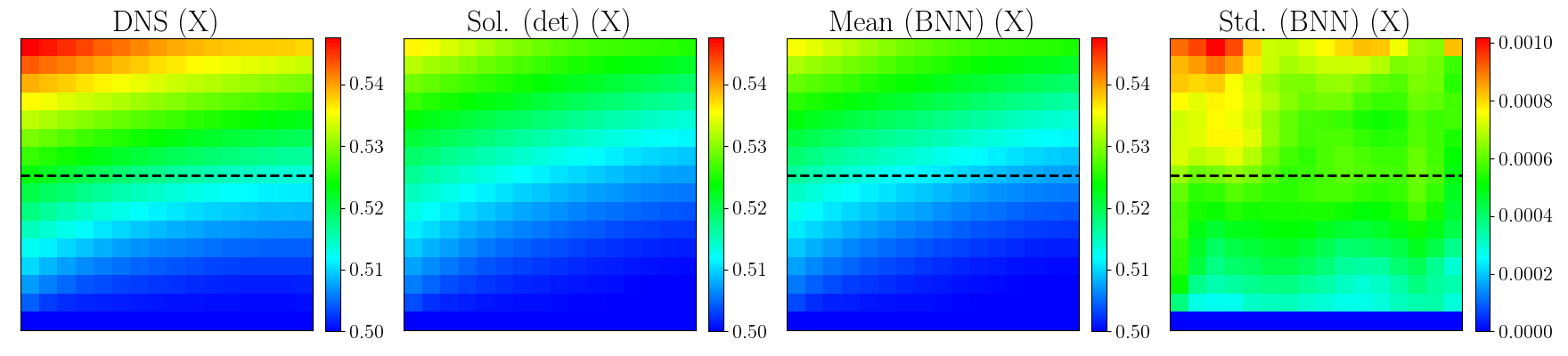}} 
  {\includegraphics[height=0.08\linewidth]{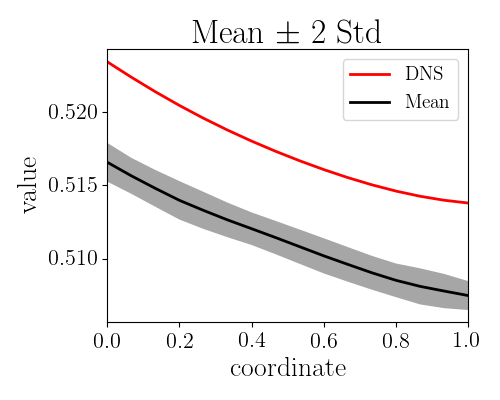}} 
  {\includegraphics[height=0.08\linewidth]{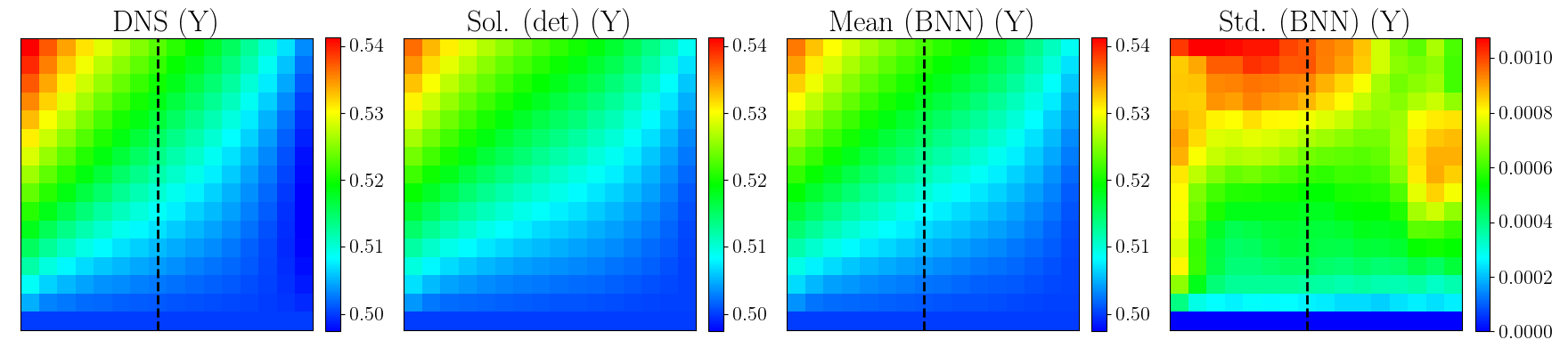}} 
  {\includegraphics[height=0.08\linewidth]{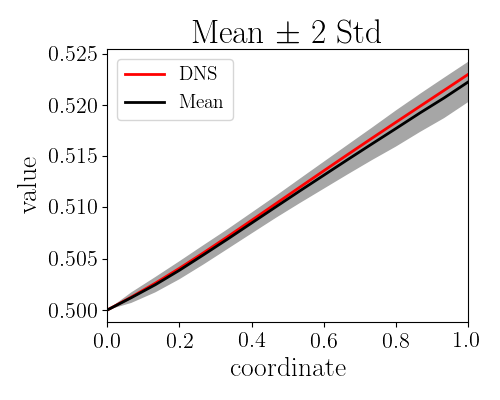}} \\
  {\includegraphics[height=0.08\linewidth]{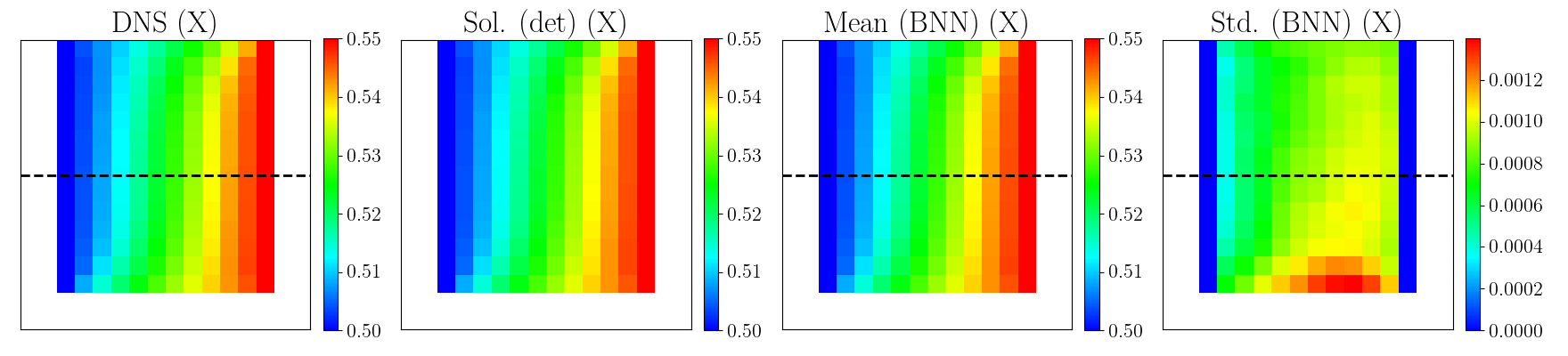}} 
  {\includegraphics[height=0.08\linewidth]{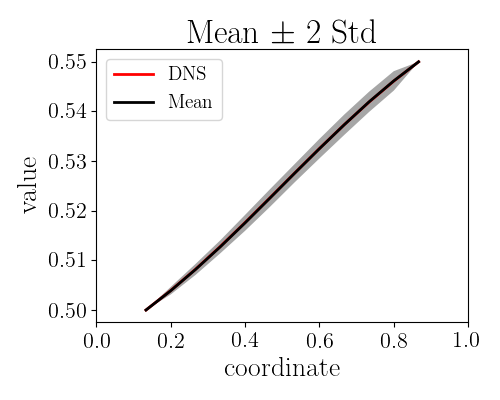}} 
  {\includegraphics[height=0.08\linewidth]{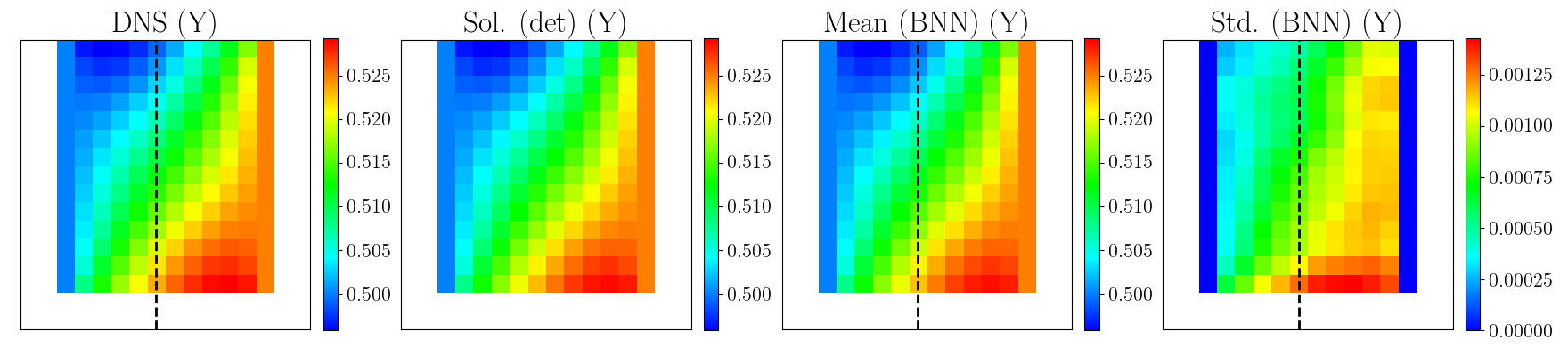}} 
  {\includegraphics[height=0.08\linewidth]{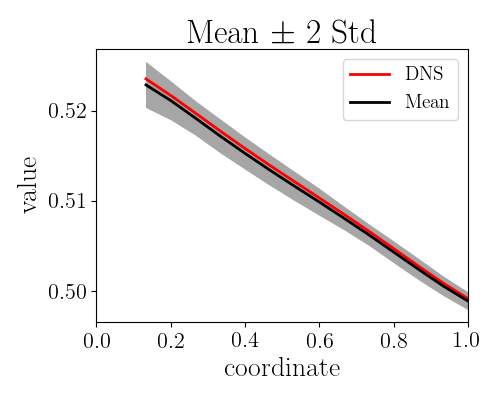}} \\
  {\includegraphics[height=0.08\linewidth]{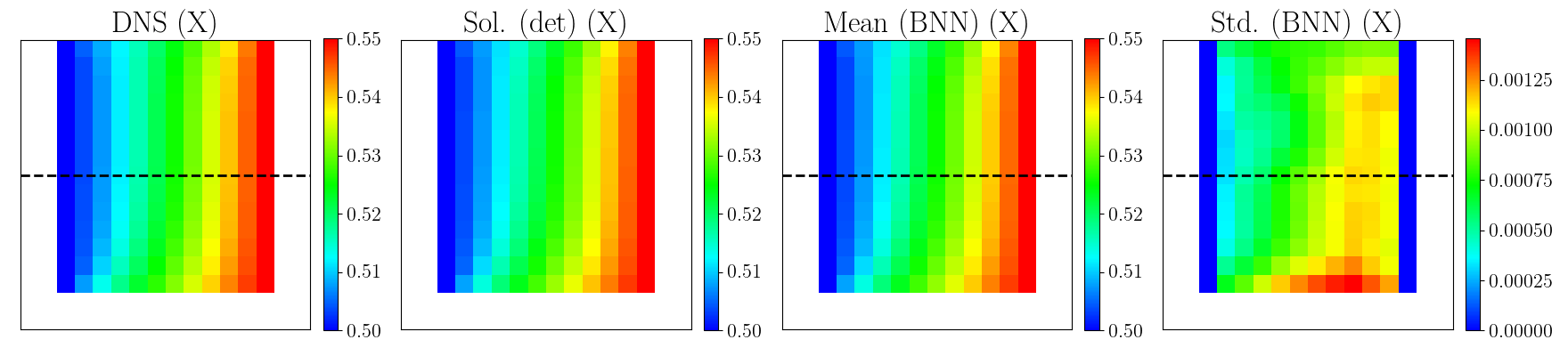}} 
  {\includegraphics[height=0.08\linewidth]{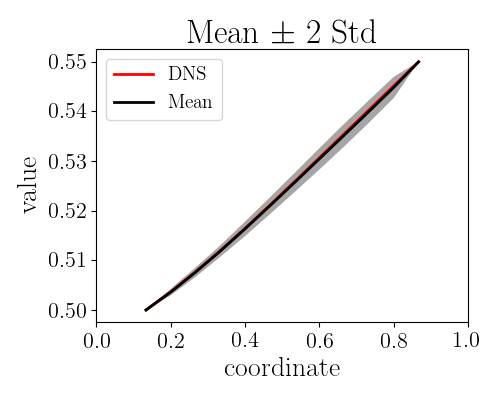}} 
  {\includegraphics[height=0.08\linewidth]{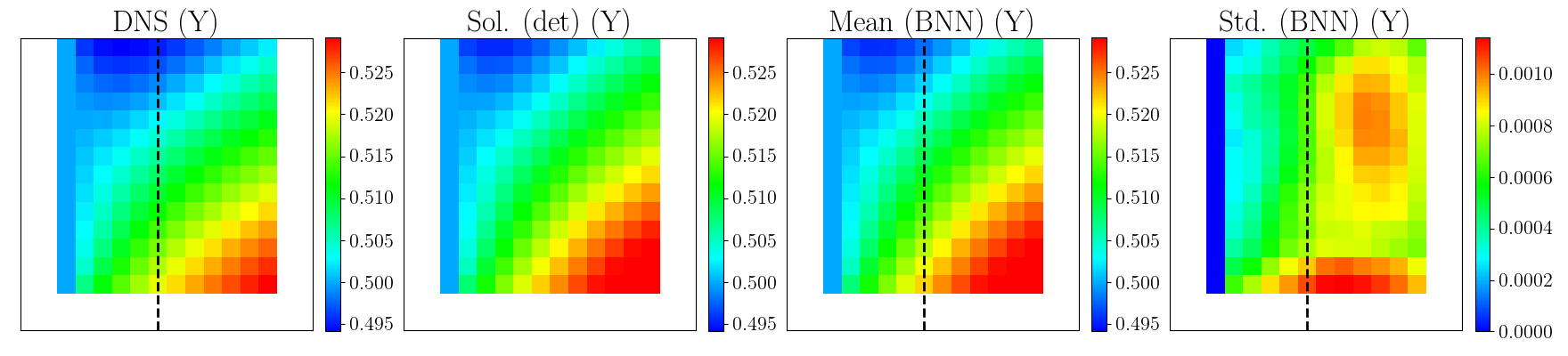}} 
  {\includegraphics[height=0.08\linewidth]{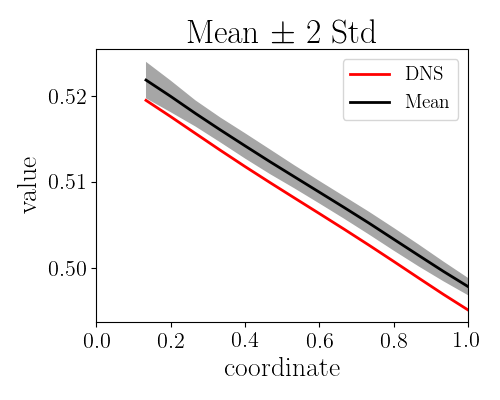}} \\
  {\includegraphics[height=0.08\linewidth]{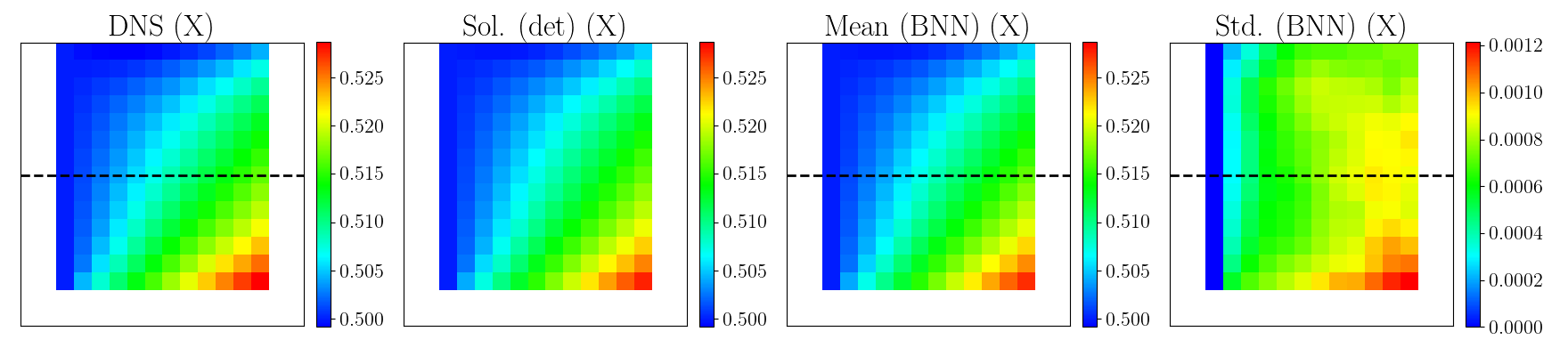}} 
  {\includegraphics[height=0.08\linewidth]{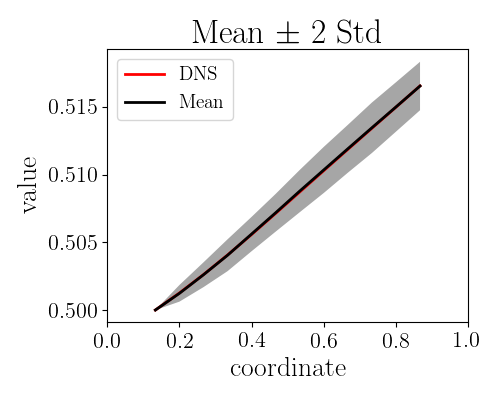}} 
  {\includegraphics[height=0.08\linewidth]{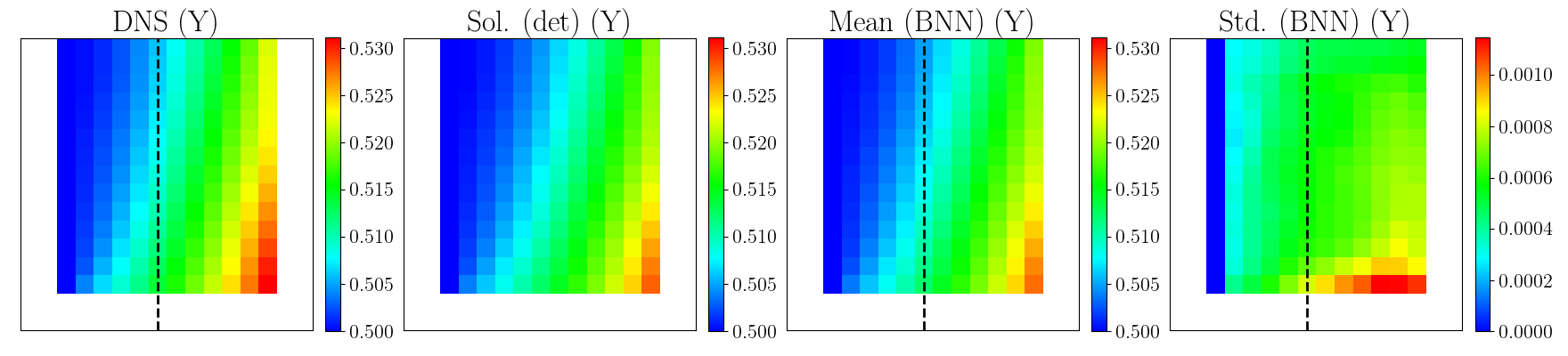}} 
  {\includegraphics[height=0.08\linewidth]{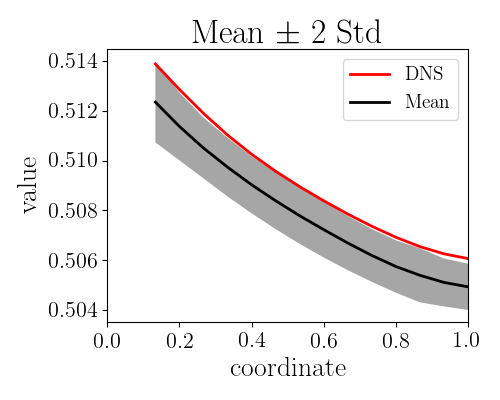}} \\
  {\includegraphics[height=0.08\linewidth]{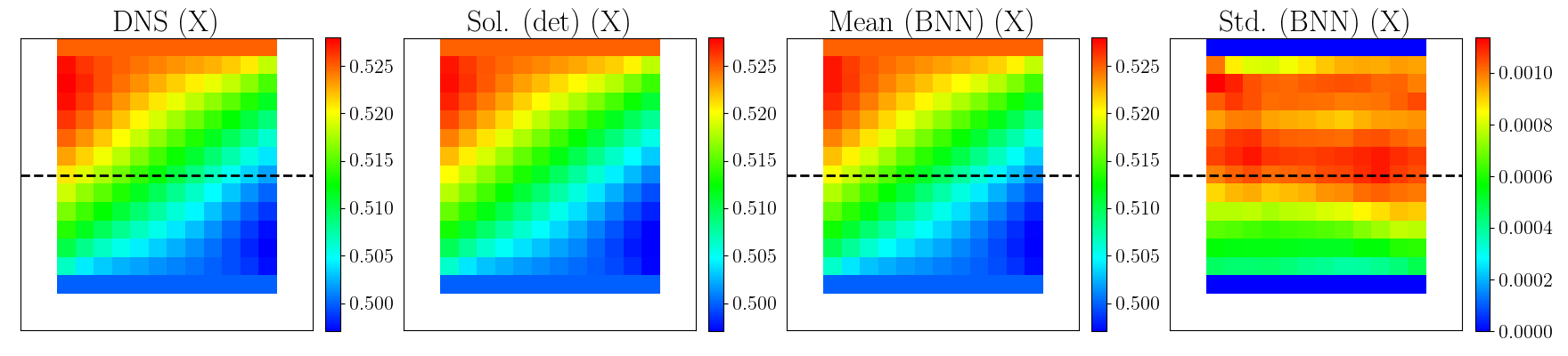}} 
  {\includegraphics[height=0.08\linewidth]{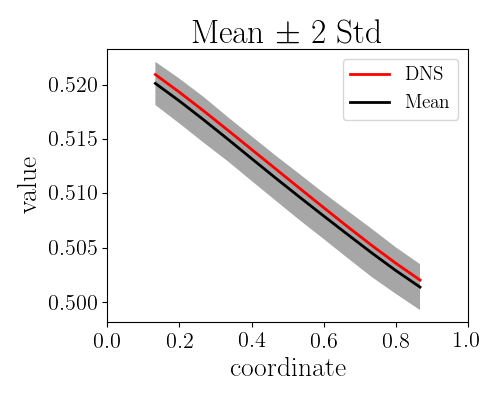}} 
  {\includegraphics[height=0.08\linewidth]{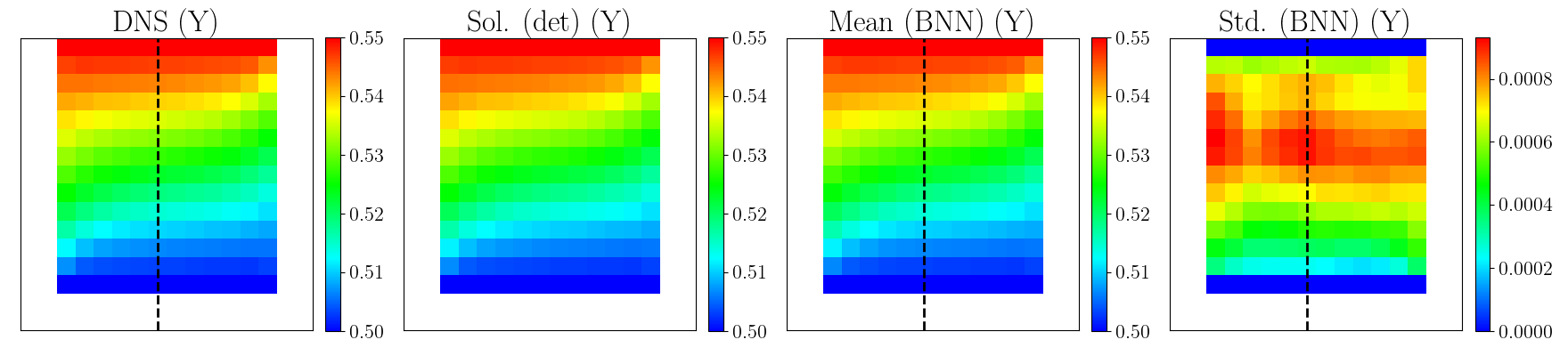}} 
  {\includegraphics[height=0.08\linewidth]{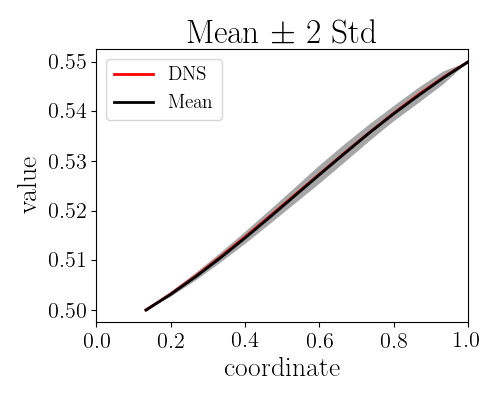}} \\
  {\includegraphics[height=0.08\linewidth]{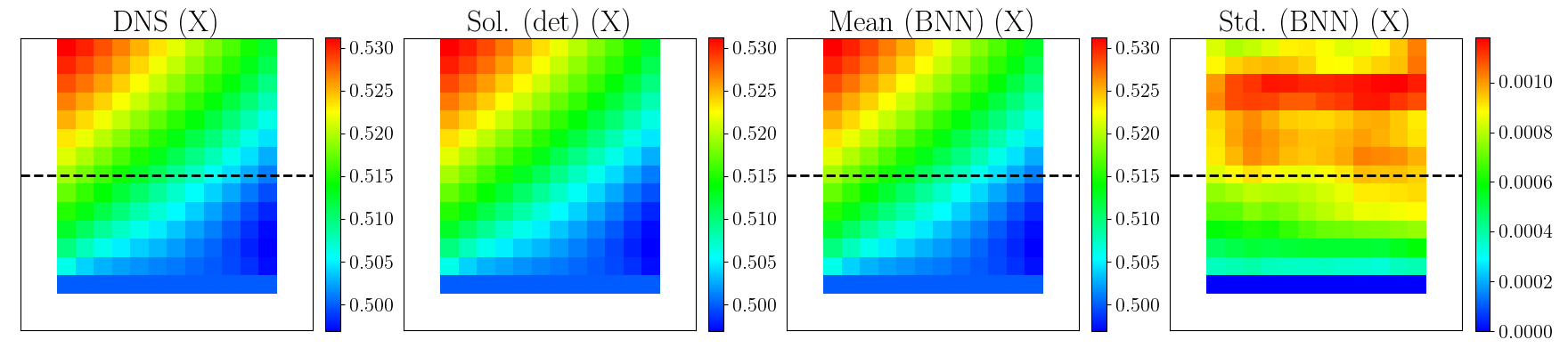}} 
  {\includegraphics[height=0.08\linewidth]{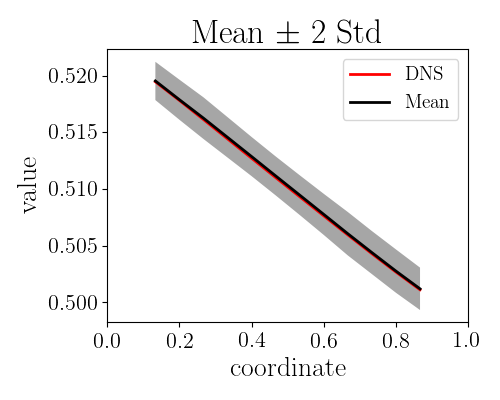}} 
  {\includegraphics[height=0.08\linewidth]{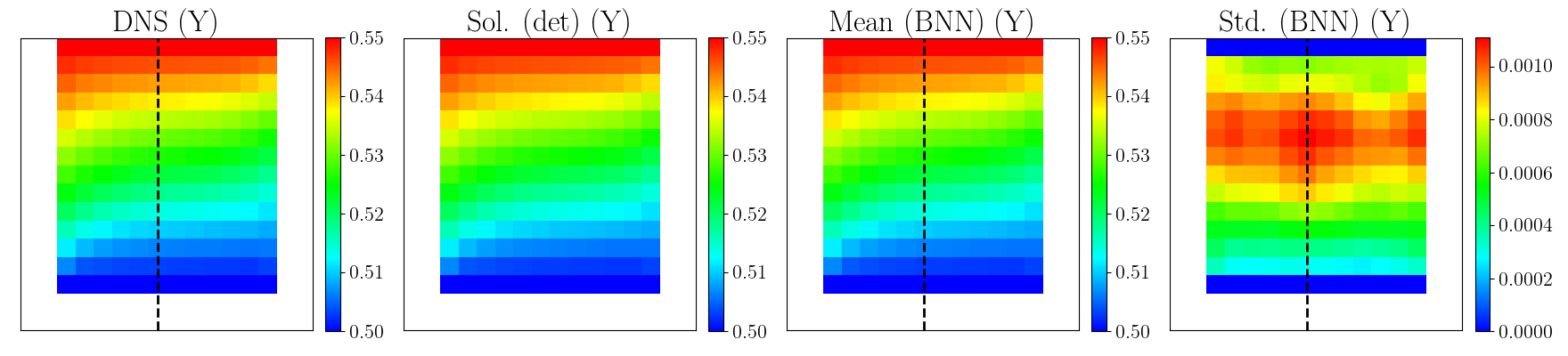}} 
  {\includegraphics[height=0.08\linewidth]{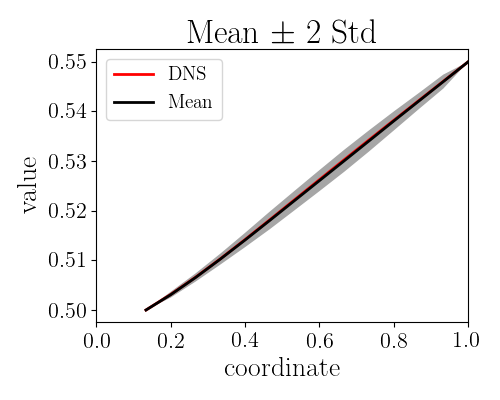}} \\
  \caption{Additional NN results for linear BVPs on rectangle domains.}
  \label{fig:linear-30bvp-results-additional-1}
\end{figure}

\begin{figure}[p!]
  \centering
  {\includegraphics[height=0.08\linewidth]{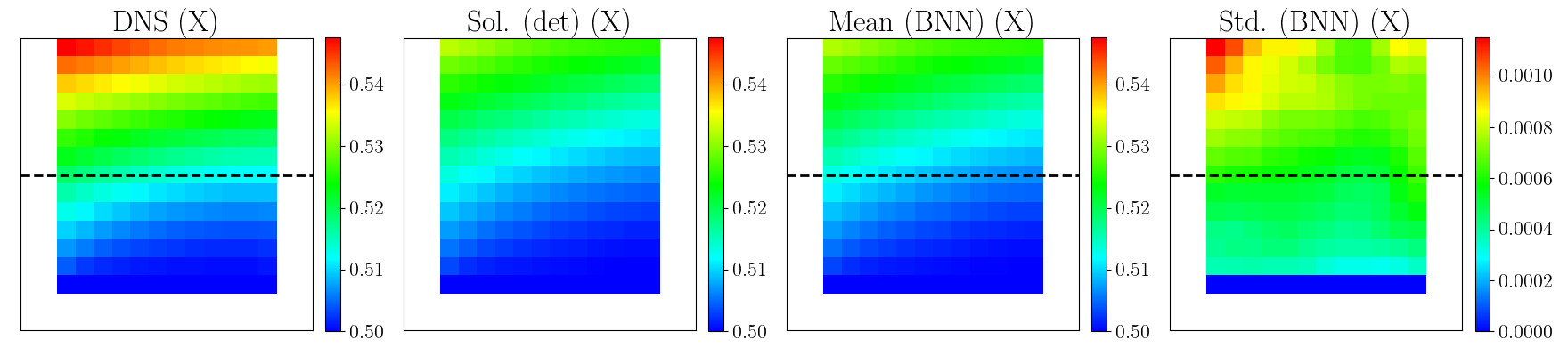}} 
  {\includegraphics[height=0.08\linewidth]{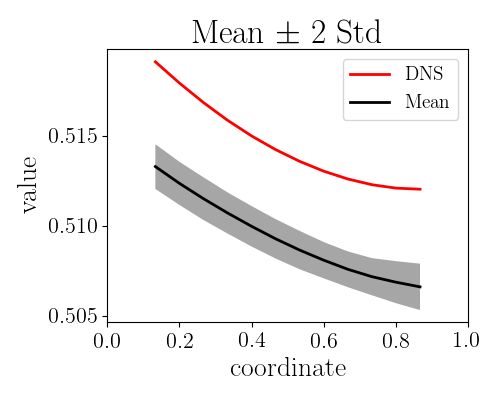}} 
  {\includegraphics[height=0.08\linewidth]{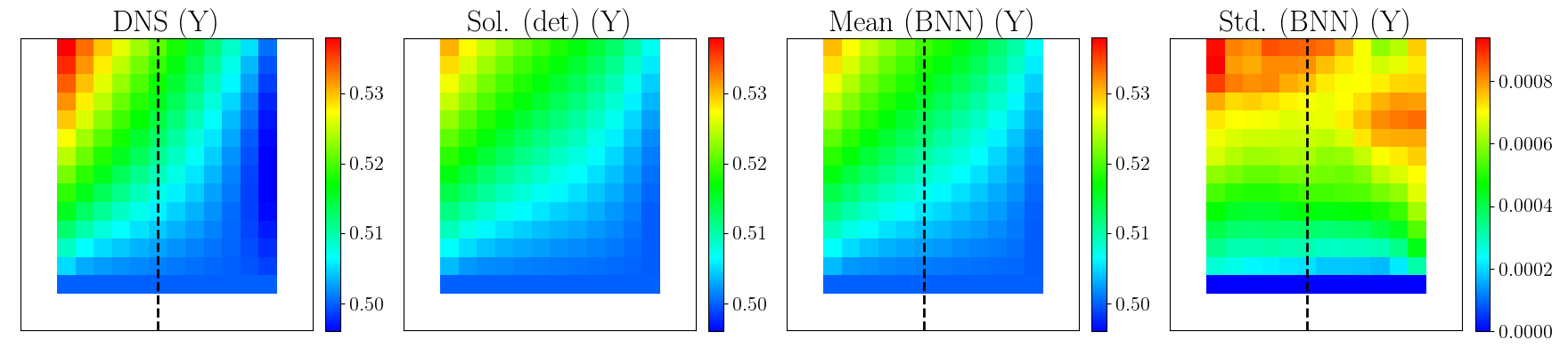}} 
  {\includegraphics[height=0.08\linewidth]{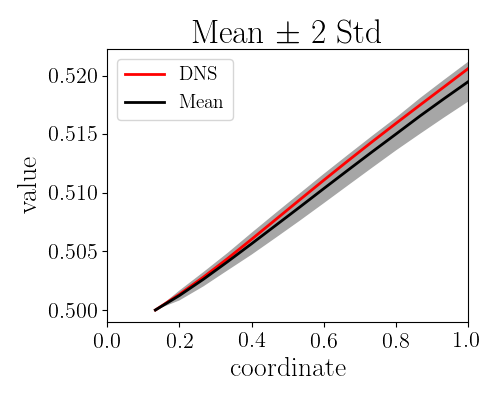}} \\
  {\includegraphics[height=0.08\linewidth]{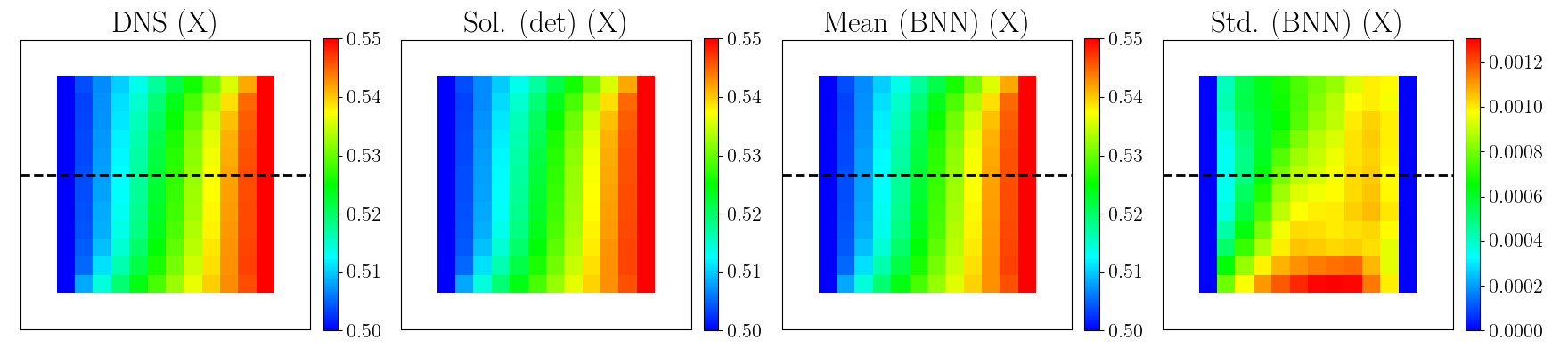}} 
  {\includegraphics[height=0.08\linewidth]{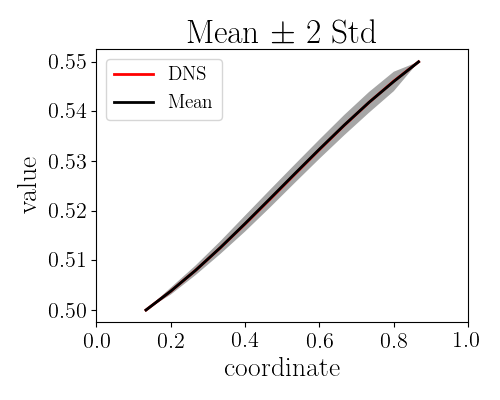}} 
  {\includegraphics[height=0.08\linewidth]{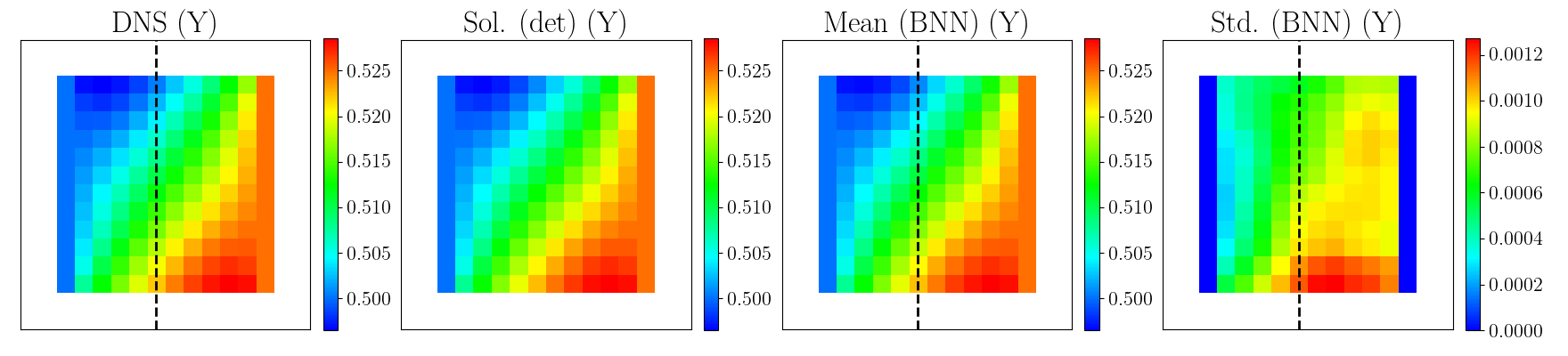}} 
  {\includegraphics[height=0.08\linewidth]{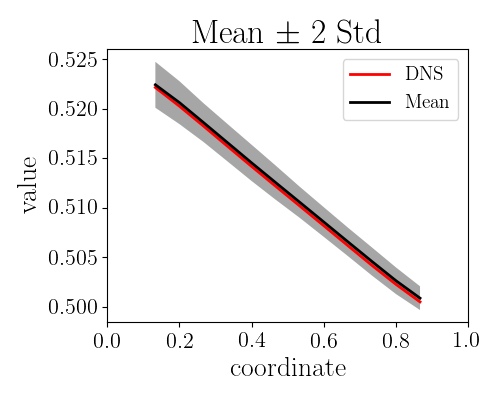}} \\
  {\includegraphics[height=0.08\linewidth]{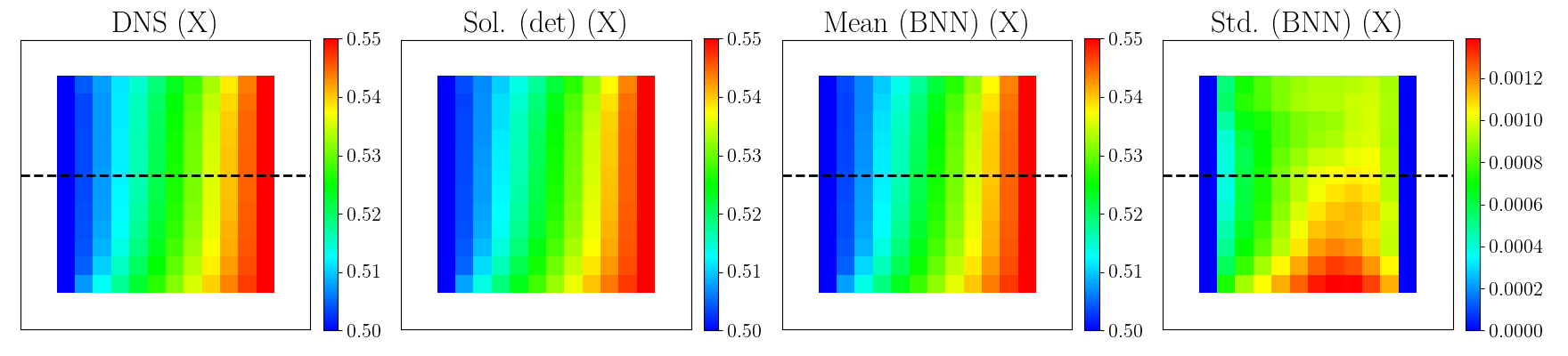}} 
  {\includegraphics[height=0.08\linewidth]{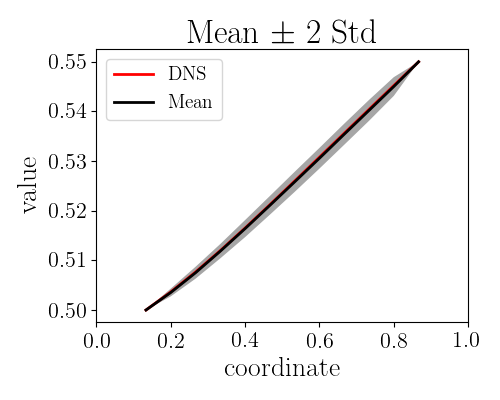}} 
  {\includegraphics[height=0.08\linewidth]{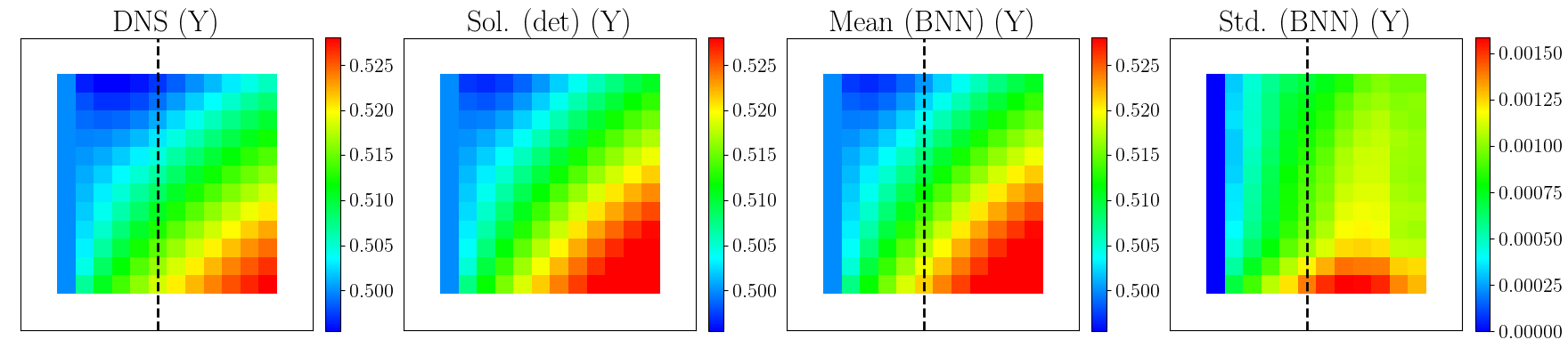}} 
  {\includegraphics[height=0.08\linewidth]{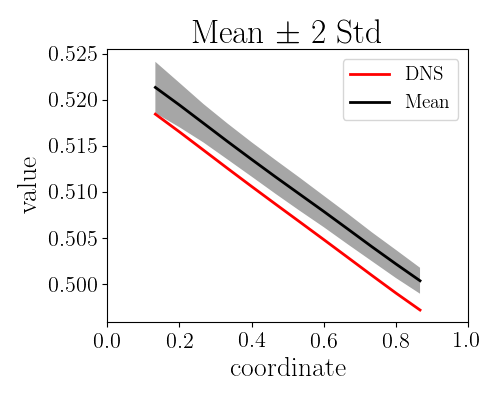}} \\
  {\includegraphics[height=0.08\linewidth]{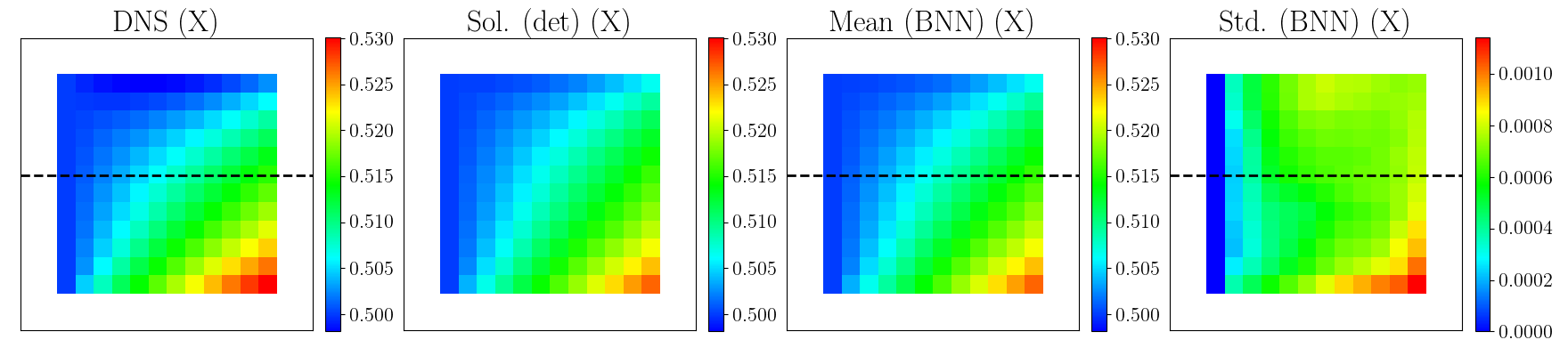}} 
  {\includegraphics[height=0.08\linewidth]{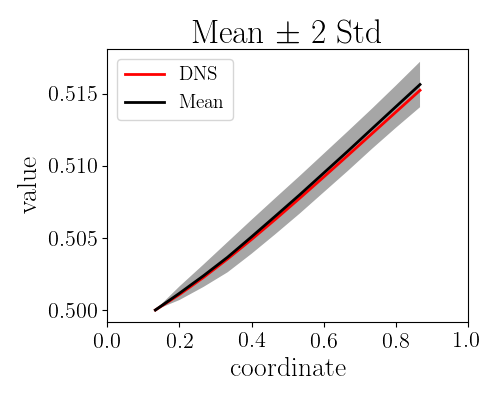}}
  {\includegraphics[height=0.08\linewidth]{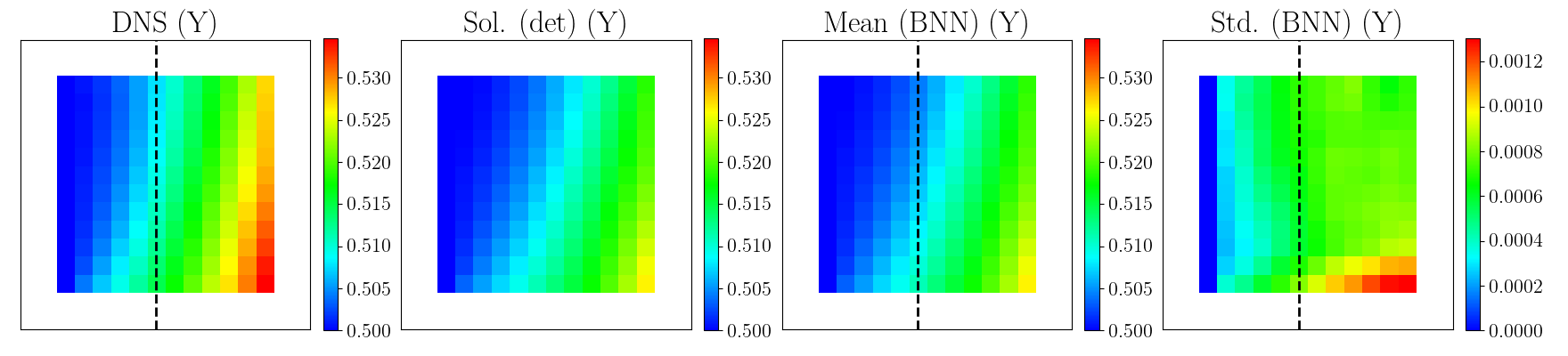}} 
  {\includegraphics[height=0.08\linewidth]{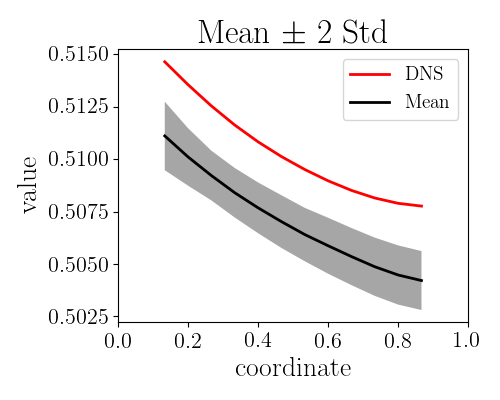}} \\
  {\includegraphics[height=0.08\linewidth]{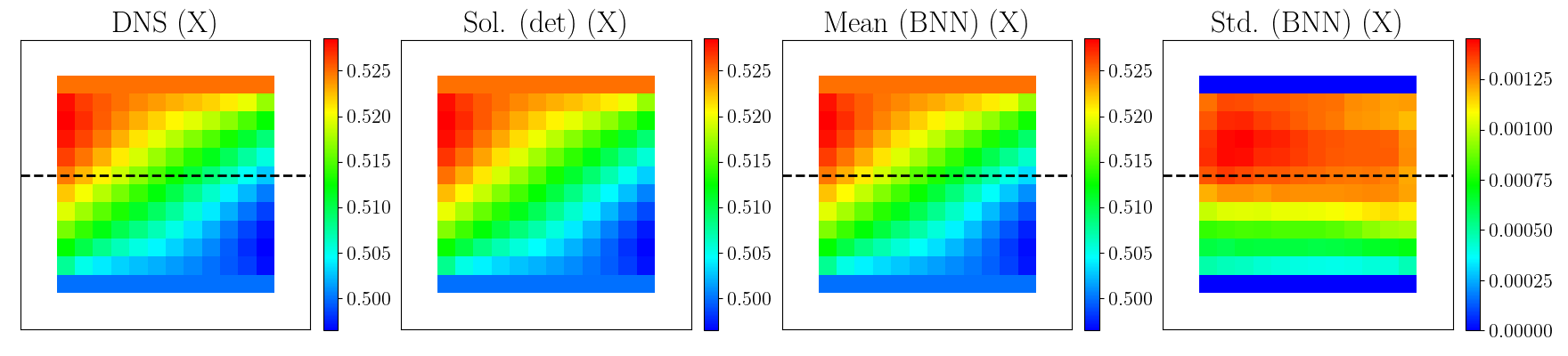}} 
  {\includegraphics[height=0.08\linewidth]{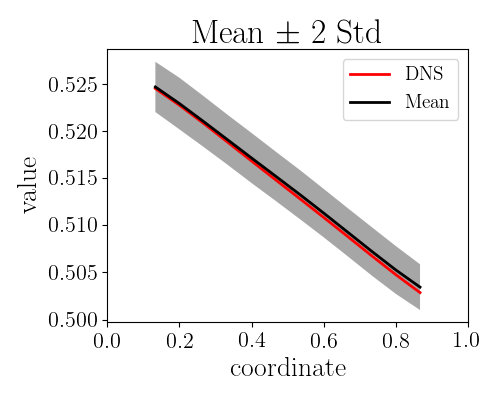}} 
  {\includegraphics[height=0.08\linewidth]{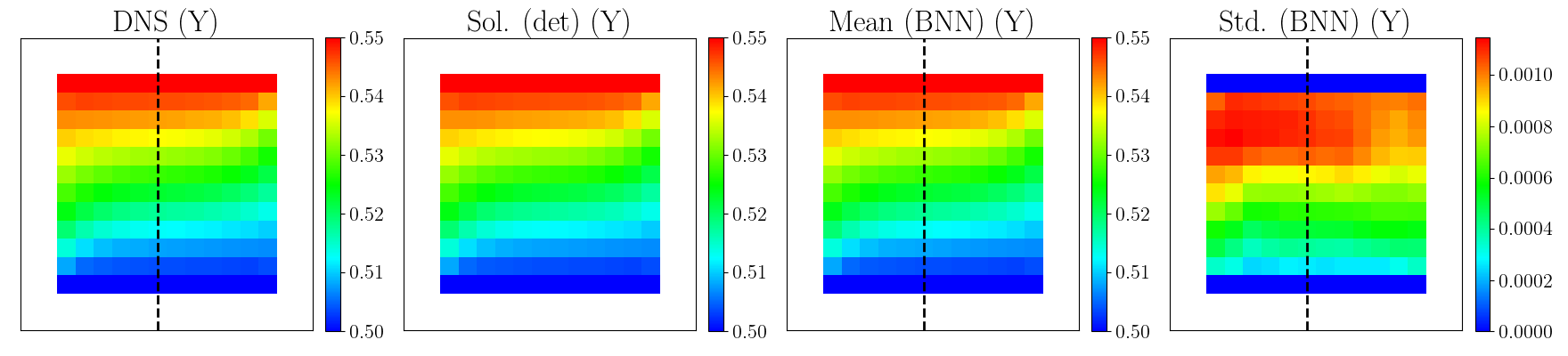}} 
  {\includegraphics[height=0.08\linewidth]{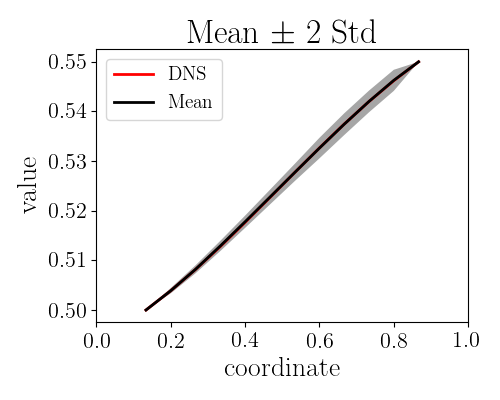}} \\
  {\includegraphics[height=0.08\linewidth]{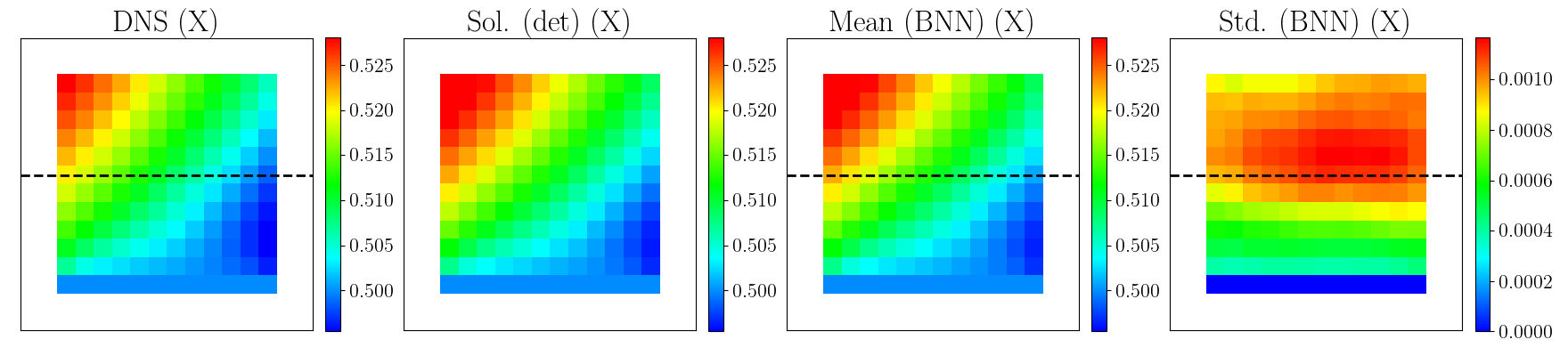}} 
  {\includegraphics[height=0.08\linewidth]{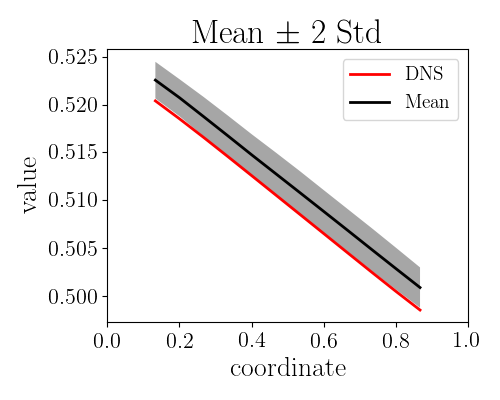}} 
  {\includegraphics[height=0.08\linewidth]{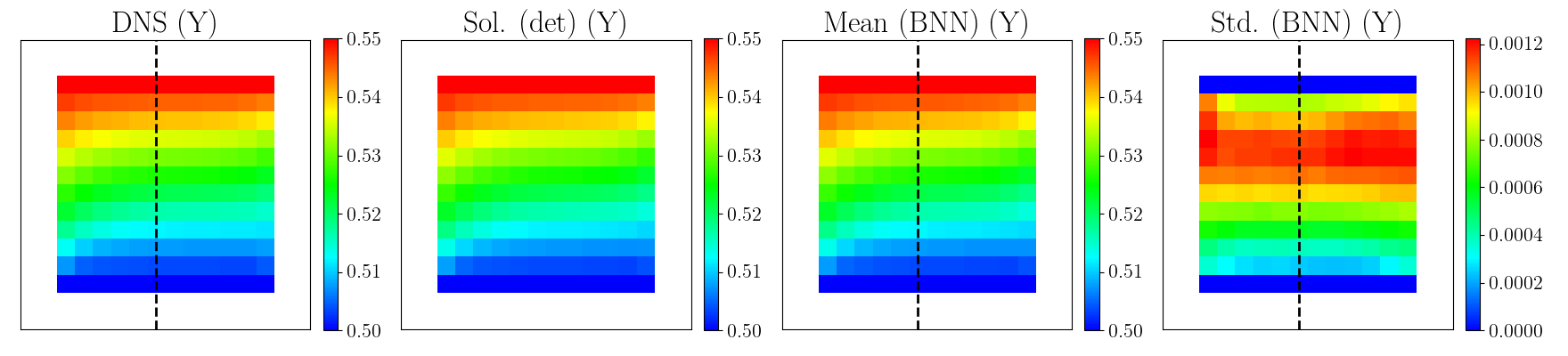}} 
  {\includegraphics[height=0.08\linewidth]{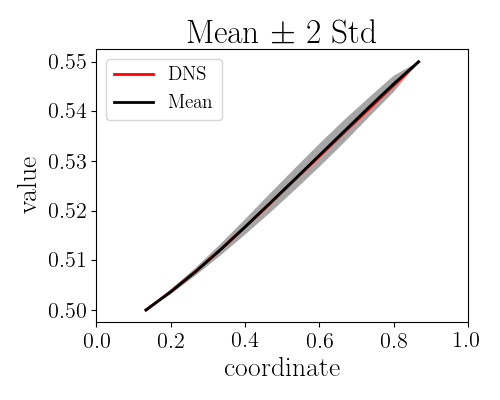}} \\
  {\includegraphics[height=0.08\linewidth]{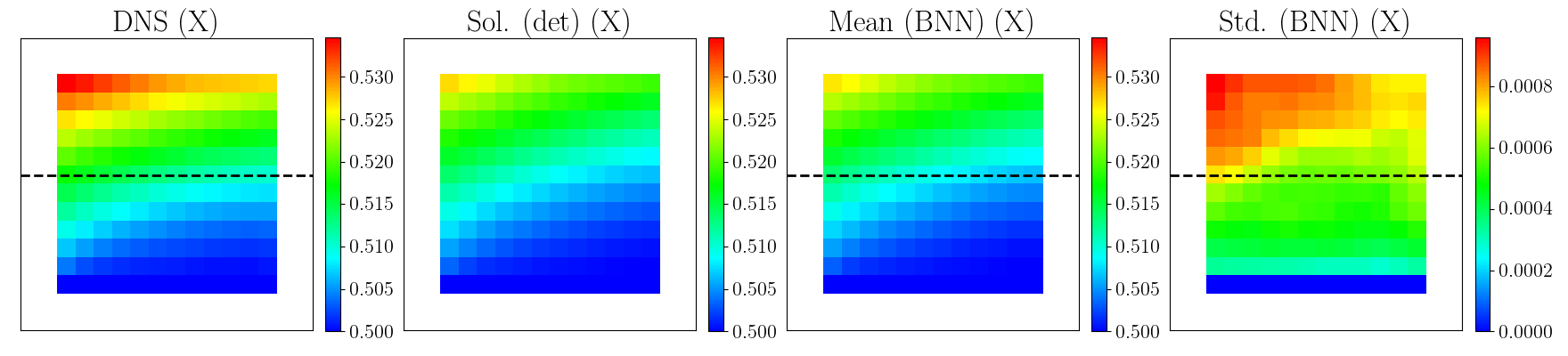}} 
  {\includegraphics[height=0.08\linewidth]{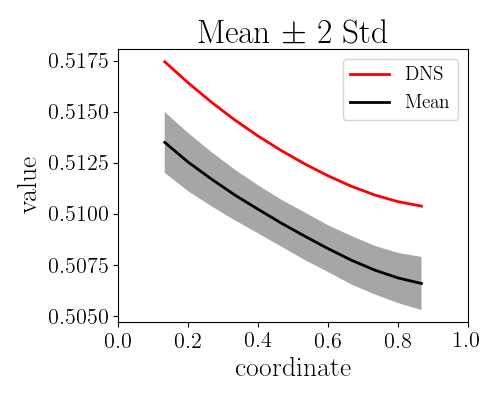}} 
  {\includegraphics[height=0.08\linewidth]{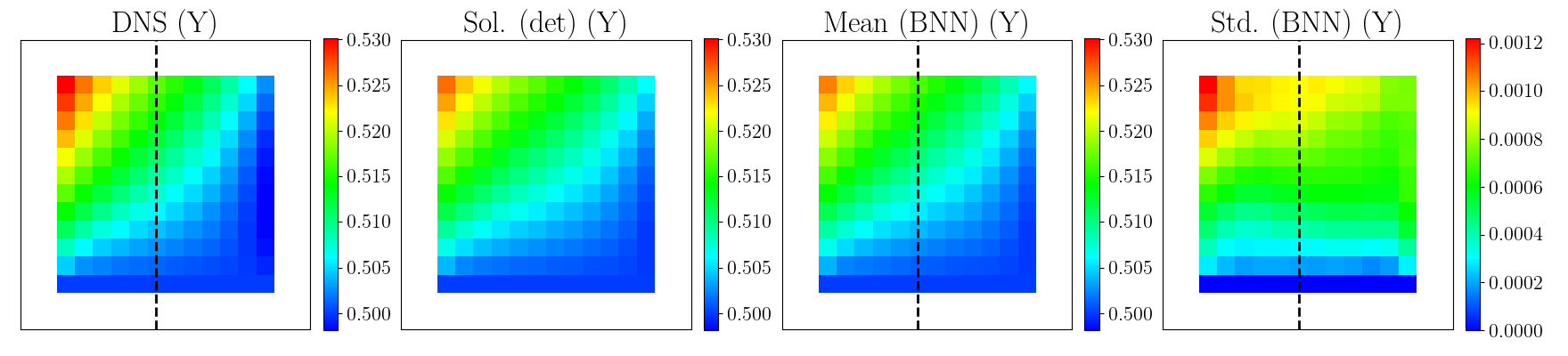}} 
  {\includegraphics[height=0.08\linewidth]{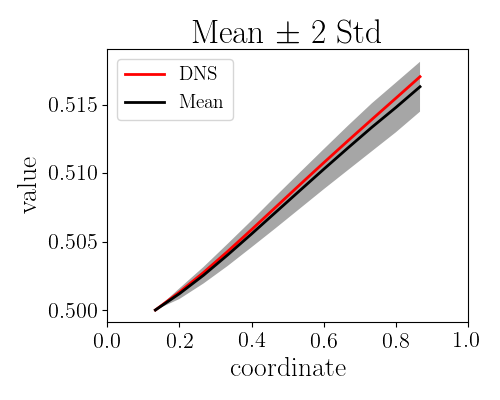}} \\
  {\includegraphics[height=0.08\linewidth]{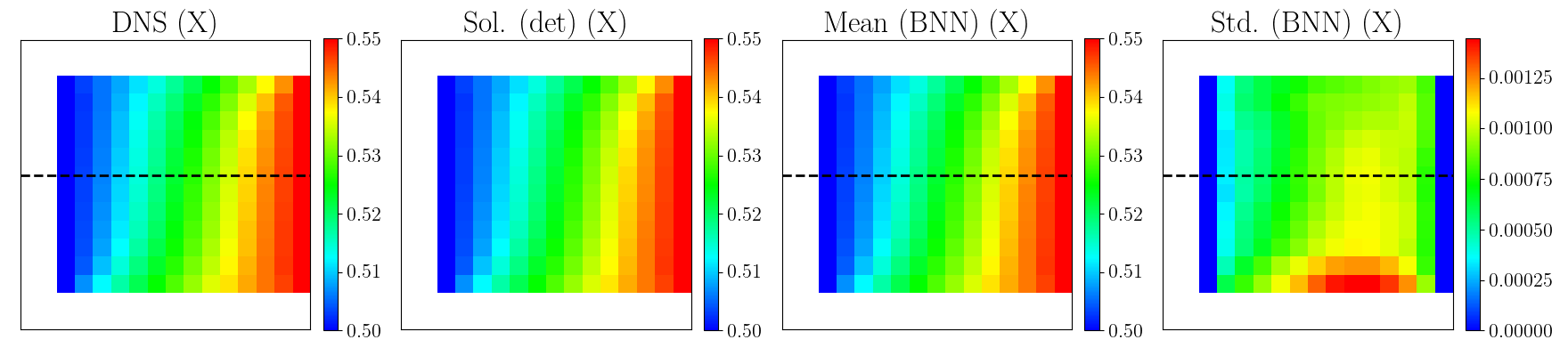}} 
  {\includegraphics[height=0.08\linewidth]{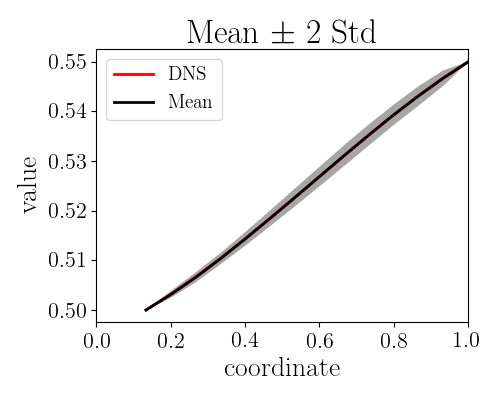}} 
  {\includegraphics[height=0.08\linewidth]{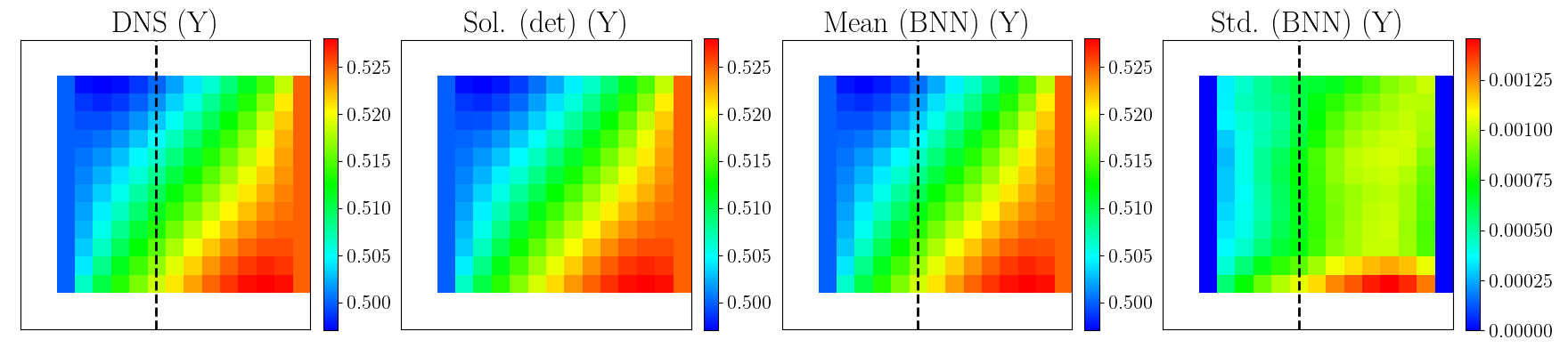}} 
  {\includegraphics[height=0.08\linewidth]{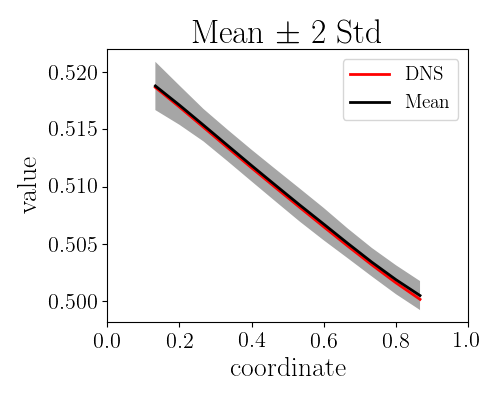}} \\
  {\includegraphics[height=0.08\linewidth]{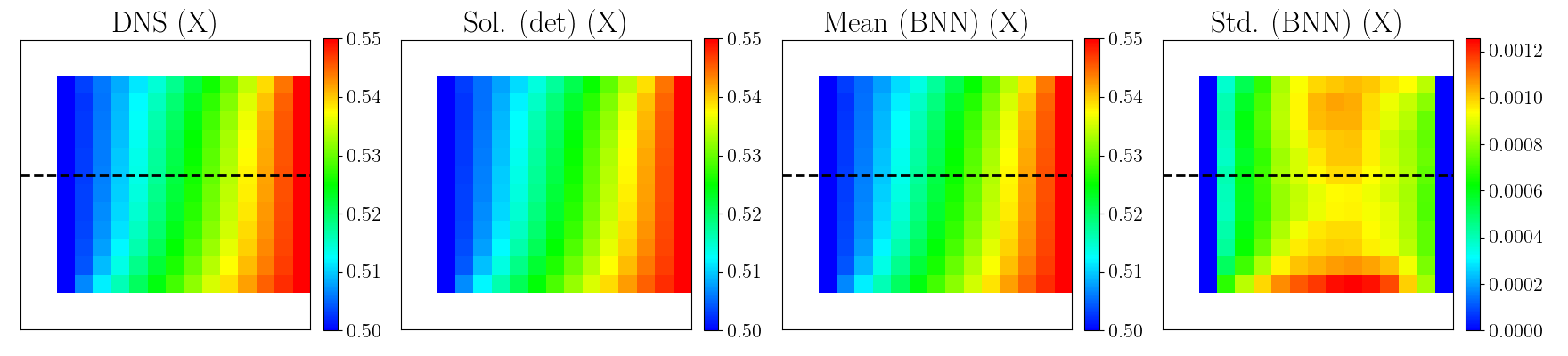}} 
  {\includegraphics[height=0.08\linewidth]{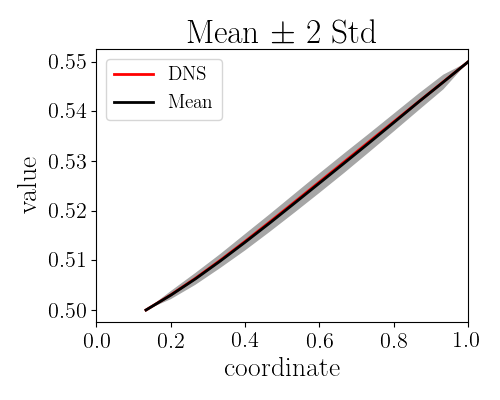}} 
  {\includegraphics[height=0.08\linewidth]{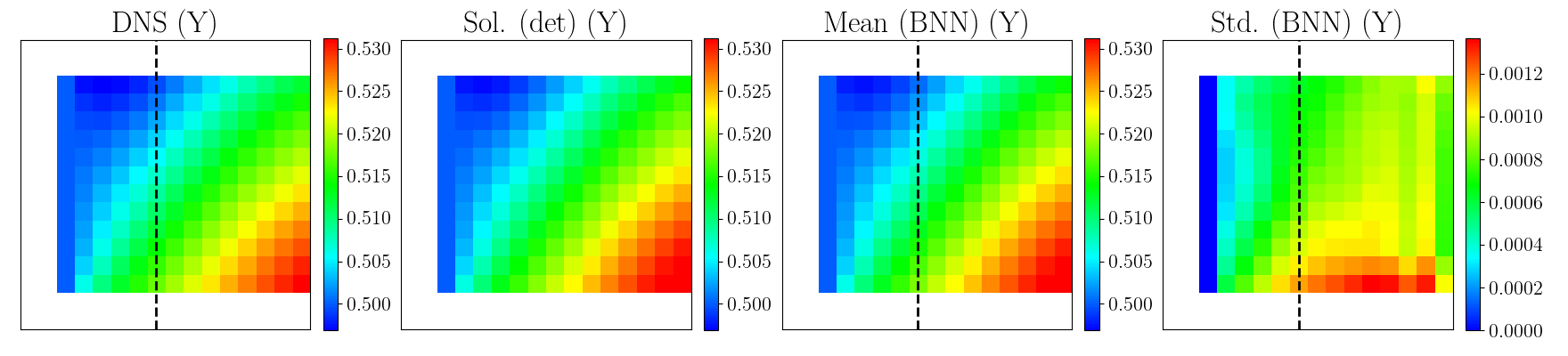}} 
  {\includegraphics[height=0.08\linewidth]{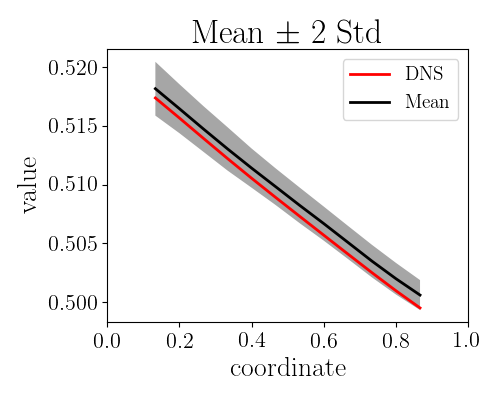}} \\
  {\includegraphics[height=0.08\linewidth]{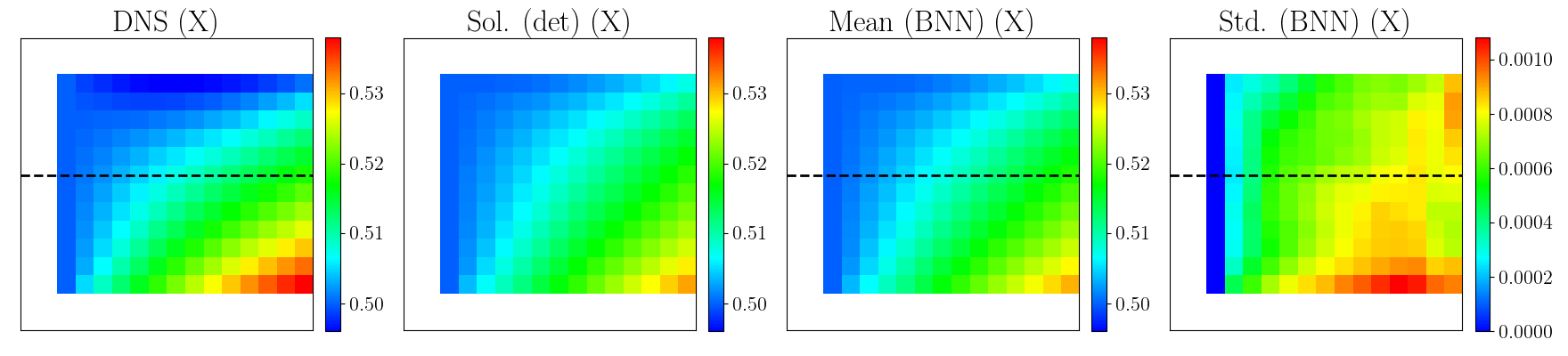}} 
  {\includegraphics[height=0.08\linewidth]{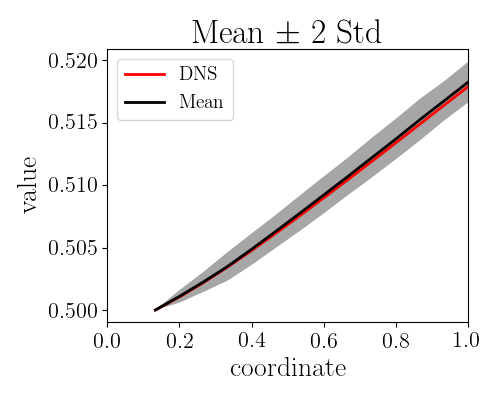}} 
  {\includegraphics[height=0.08\linewidth]{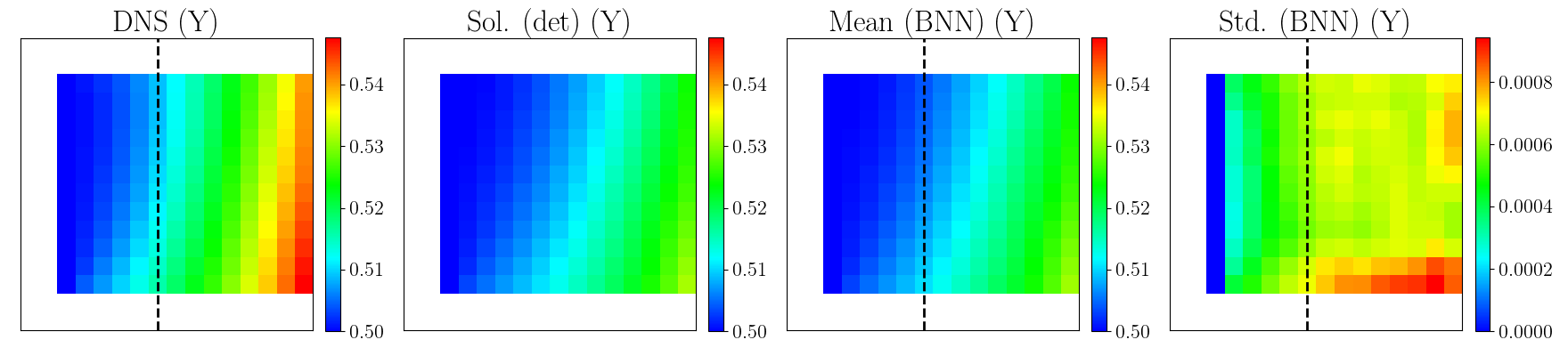}} 
  {\includegraphics[height=0.08\linewidth]{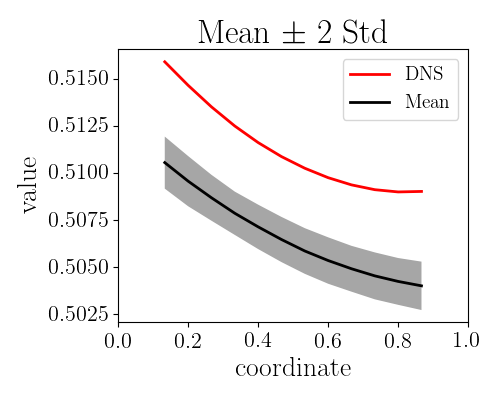}} \\
  {\includegraphics[height=0.08\linewidth]{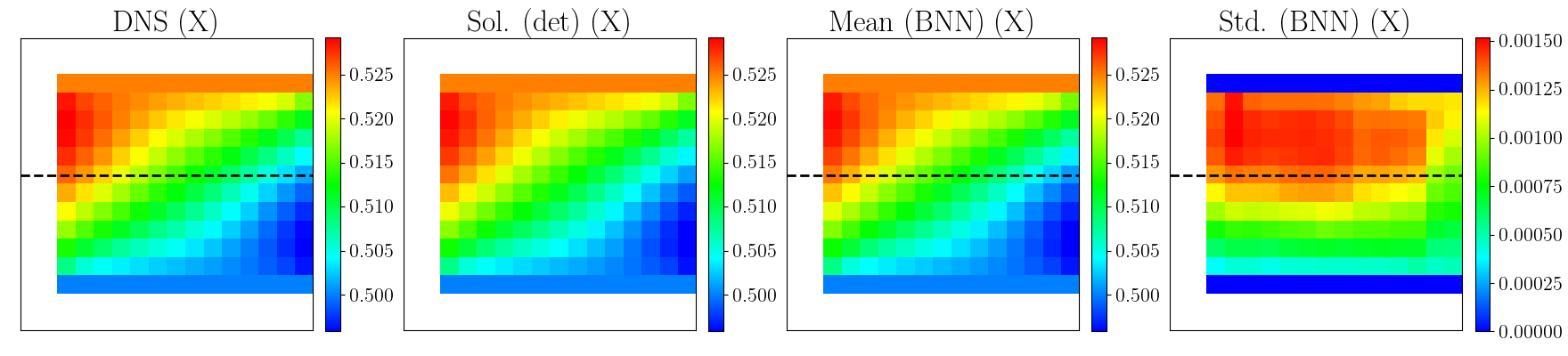}} 
  {\includegraphics[height=0.08\linewidth]{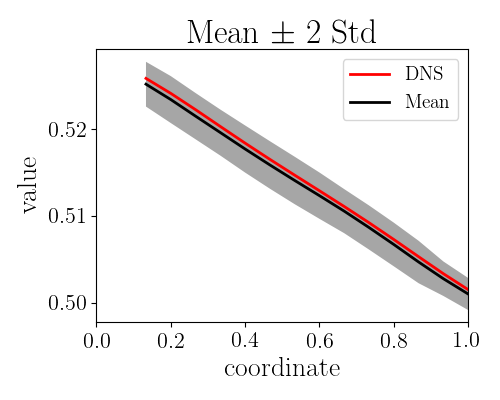}} 
  {\includegraphics[height=0.08\linewidth]{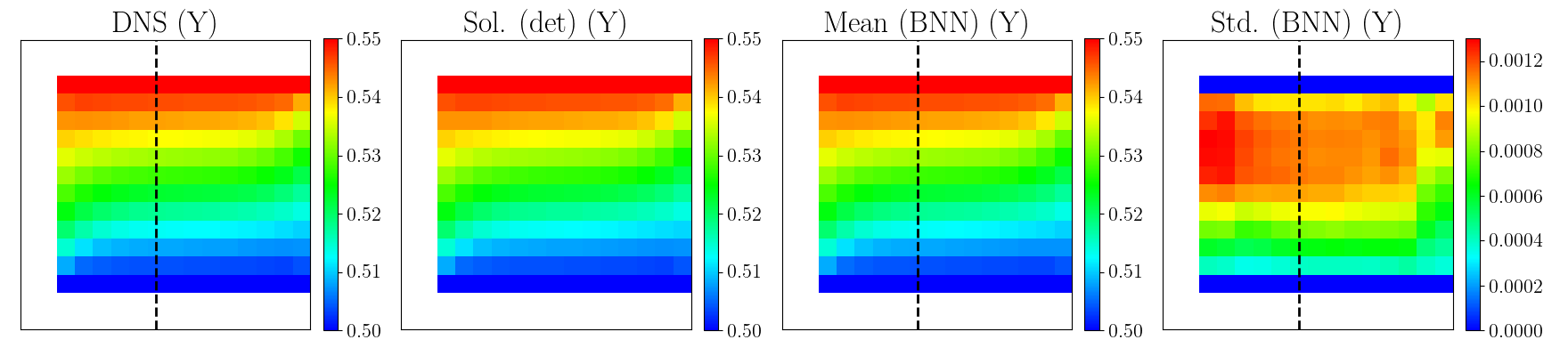}} 
  {\includegraphics[height=0.08\linewidth]{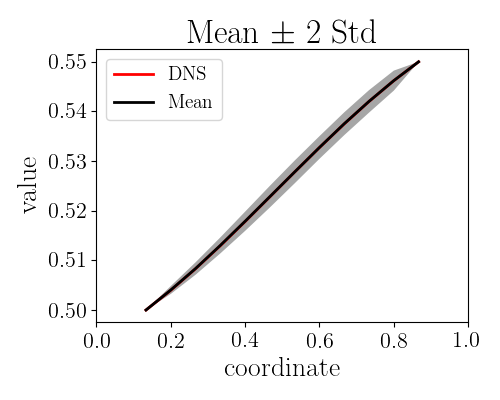}} \\
  {\includegraphics[height=0.08\linewidth]{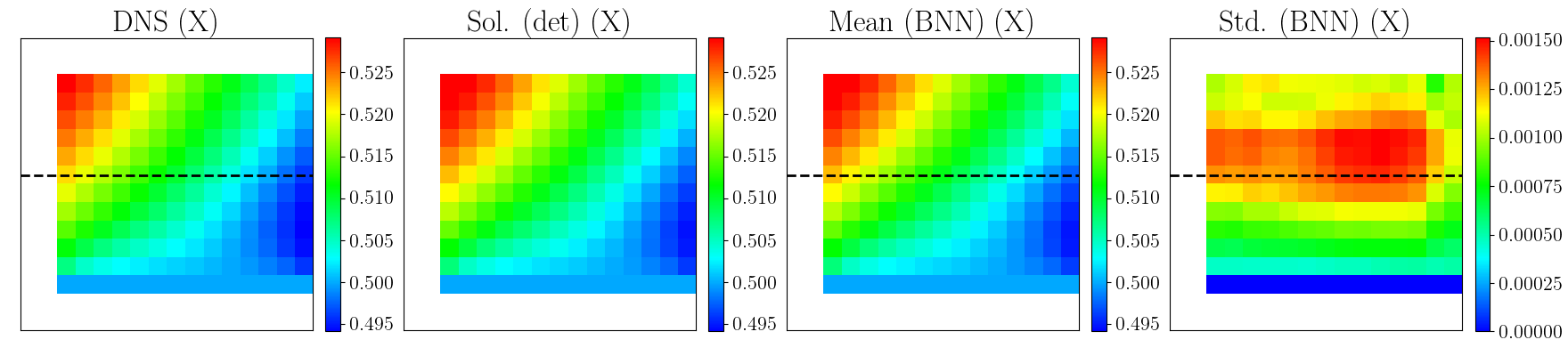}} 
  {\includegraphics[height=0.08\linewidth]{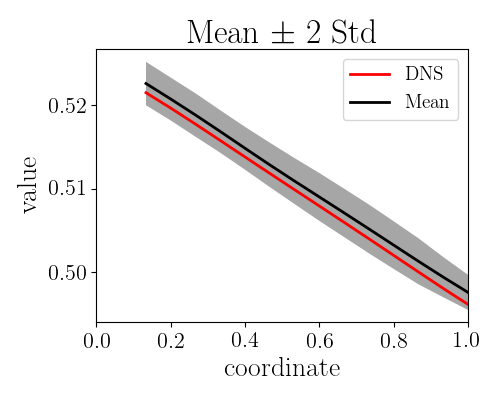}} 
  {\includegraphics[height=0.08\linewidth]{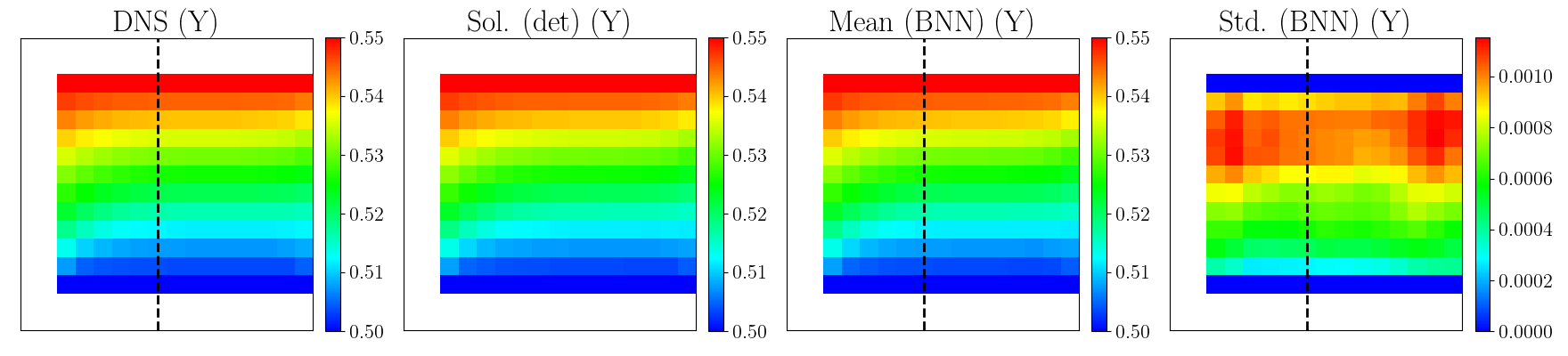}} 
  {\includegraphics[height=0.08\linewidth]{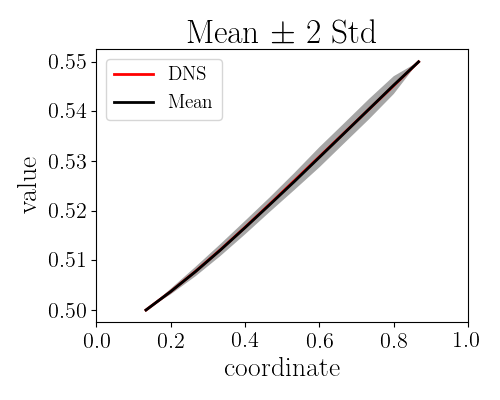}} \\
  {\includegraphics[height=0.08\linewidth]{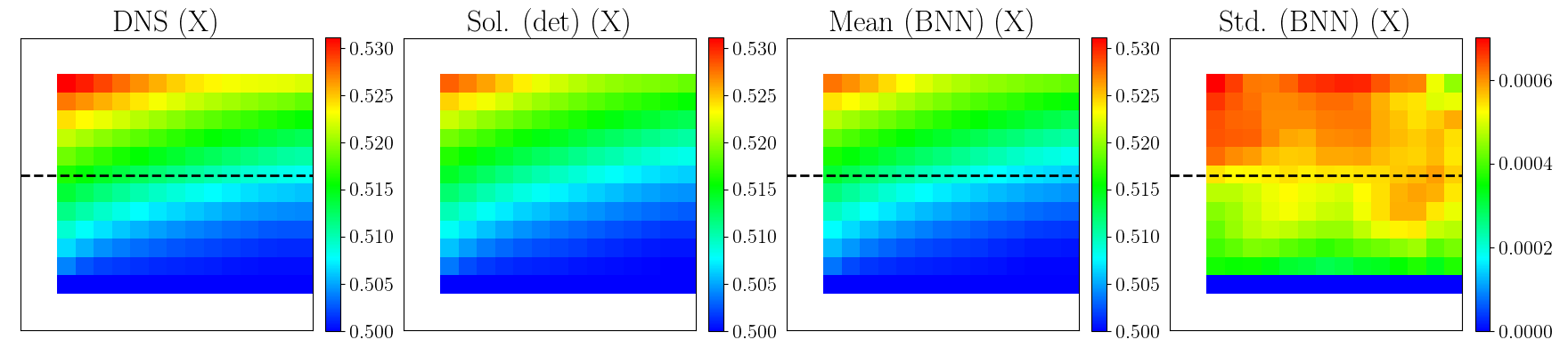}} 
  {\includegraphics[height=0.08\linewidth]{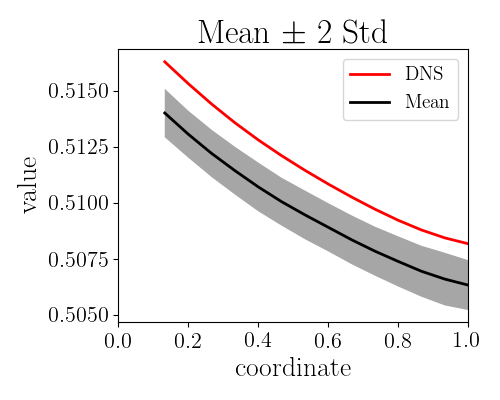}} 
  {\includegraphics[height=0.08\linewidth]{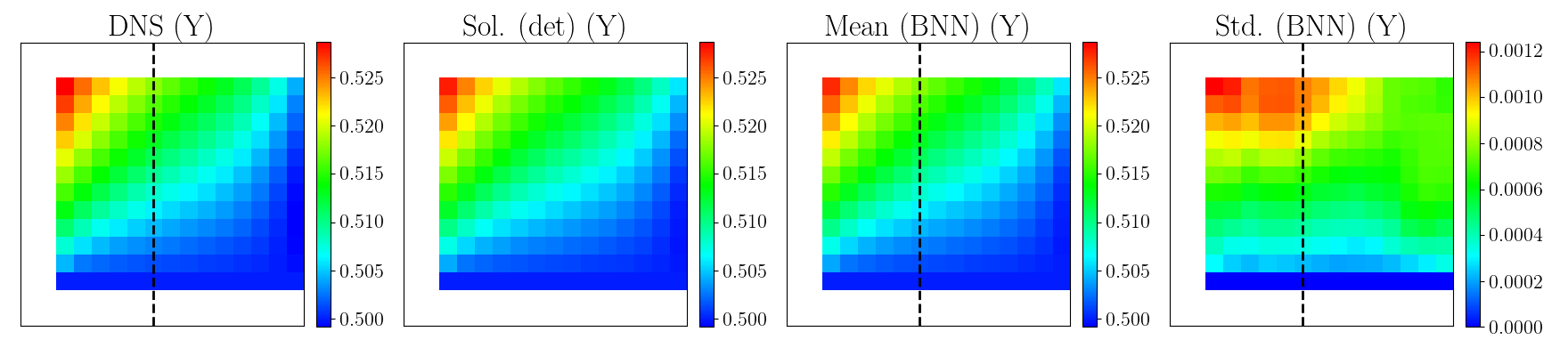}} 
  {\includegraphics[height=0.08\linewidth]{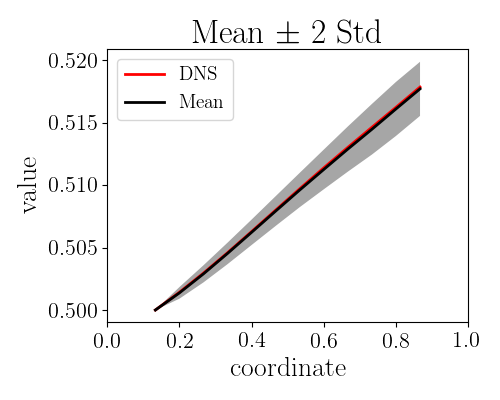}} \\
  \caption{Additional NN results for linear BVPs on rectangle domains (continue).}
  \label{fig:linear-30bvp-results-additional-2}
\end{figure}

Additional results from the 30 BVPs with rectangle domains are summarized in Fig. \ref{fig:linear-30bvp-results-additional-1} and \ref{fig:linear-30bvp-results-additional-2}.

%6,7,8

\subsubsection{L-shape domains} \label{appendix:linear-lshape}

\begin{figure}[h!]
  \centering
  {\includegraphics[height=0.08\linewidth]{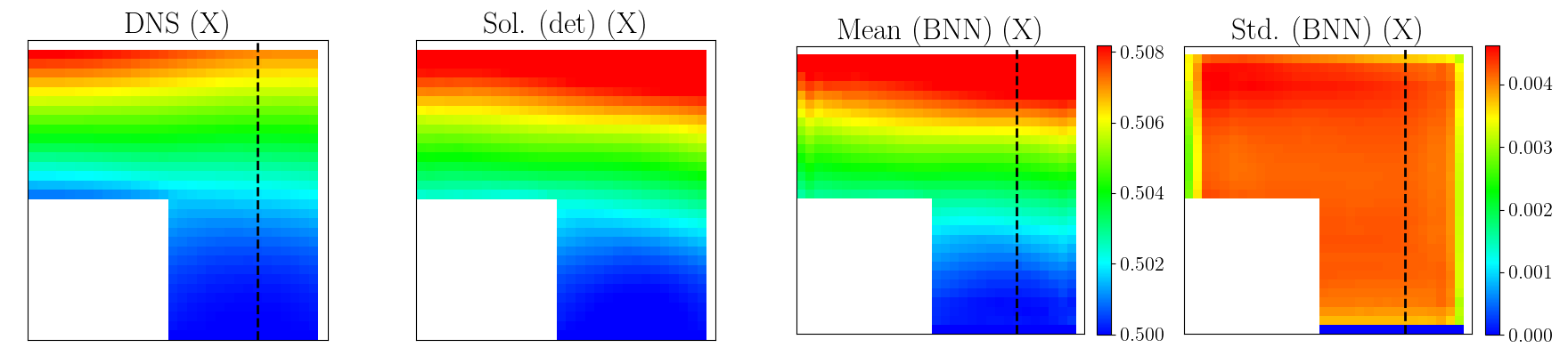}}
  {\includegraphics[height=0.08\linewidth]{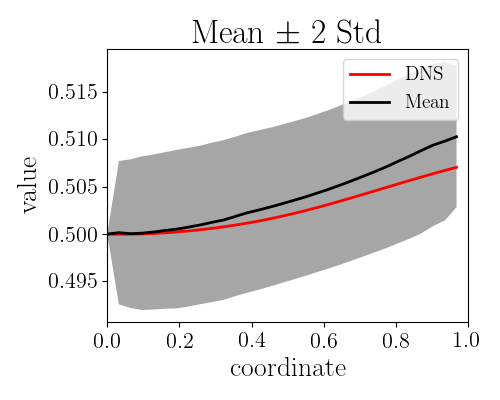}} 
  {\includegraphics[height=0.08\linewidth]{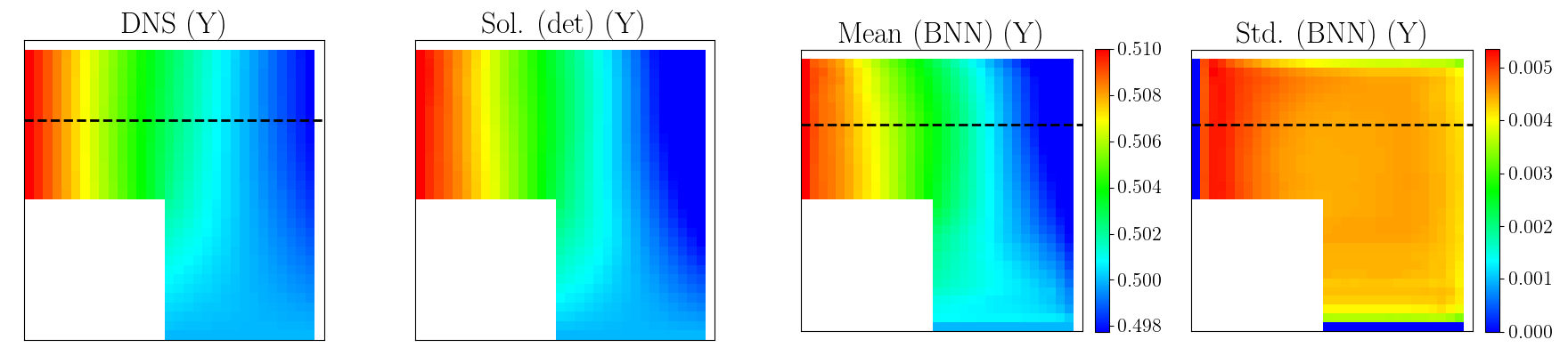}}
  {\includegraphics[height=0.08\linewidth]{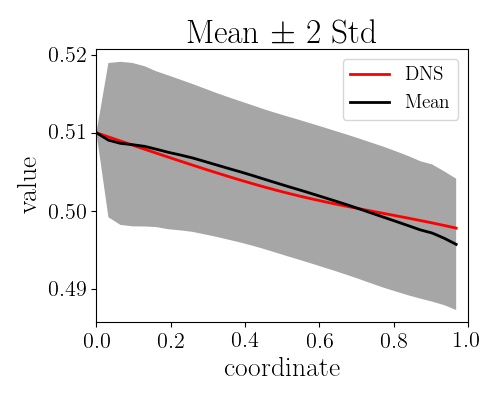}} \\
  %{\includegraphics[height=0.08\linewidth]{linear/linear-results-inter-dns-1-dof-0.png}}
  %{\includegraphics[height=0.08\linewidth]{linear/linear-results-inter-dns-1-uq-x.png}} 
  %{\includegraphics[height=0.08\linewidth]{linear/linear-results-inter-dns-1-dof-1.png}}
  %{\includegraphics[height=0.08\linewidth]{linear/linear-results-inter-dns-1-uq-y.png}} \\
  {\includegraphics[height=0.08\linewidth]{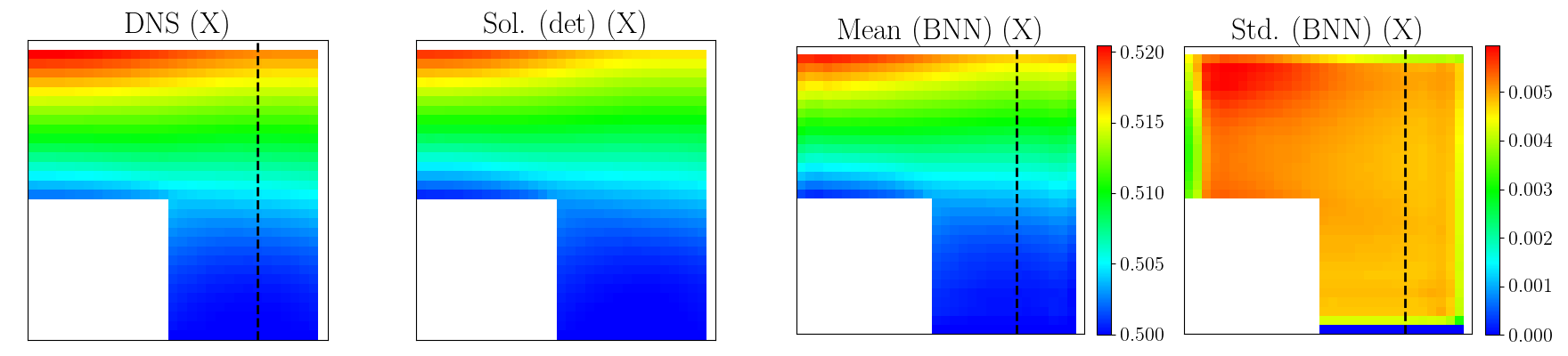}}
  {\includegraphics[height=0.08\linewidth]{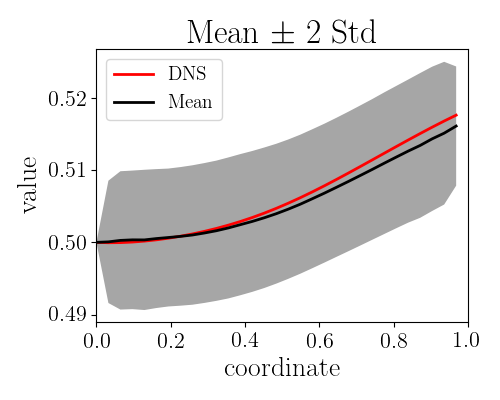}} 
  {\includegraphics[height=0.08\linewidth]{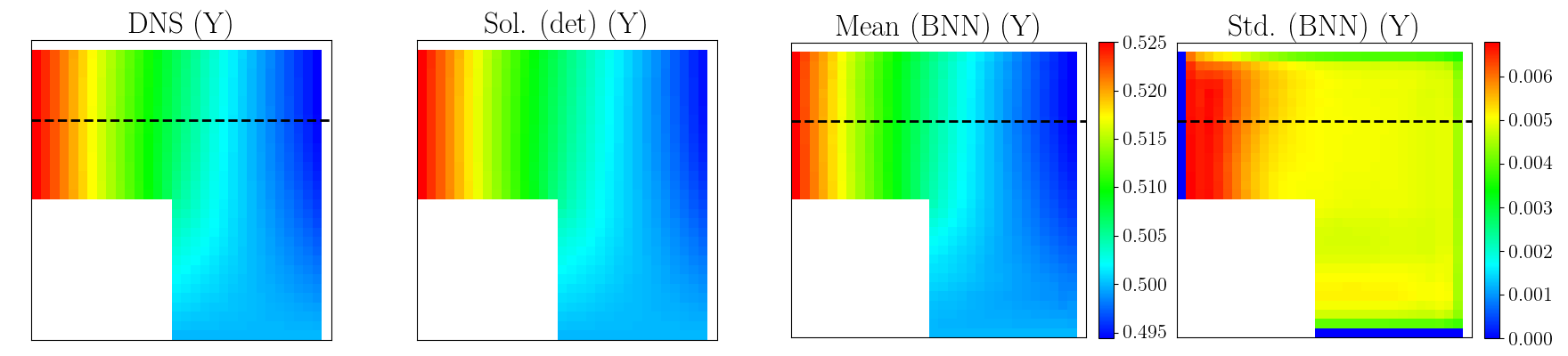}}
  {\includegraphics[height=0.08\linewidth]{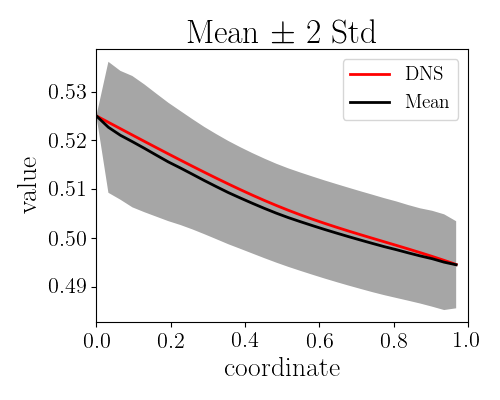}} \\
  {\includegraphics[height=0.08\linewidth]{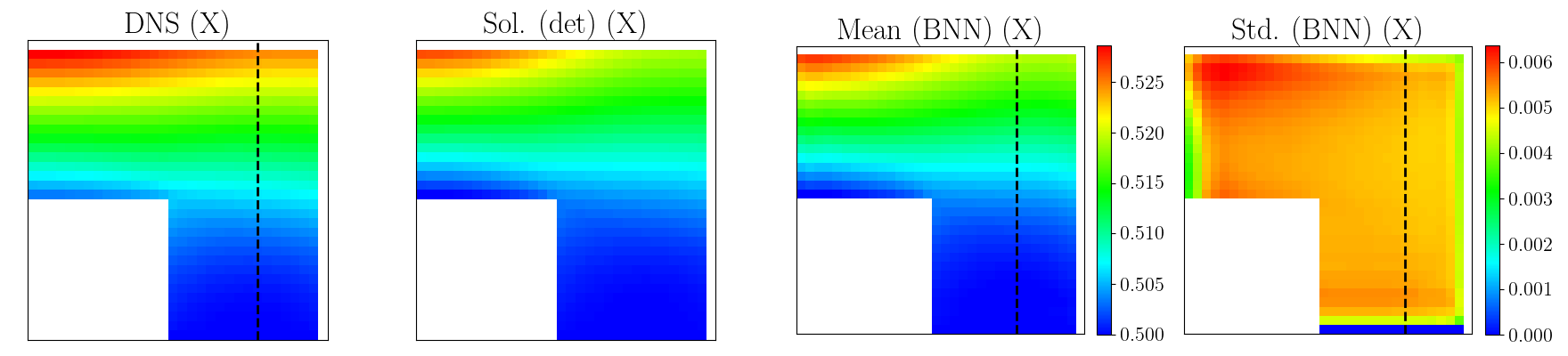}}
  {\includegraphics[height=0.08\linewidth]{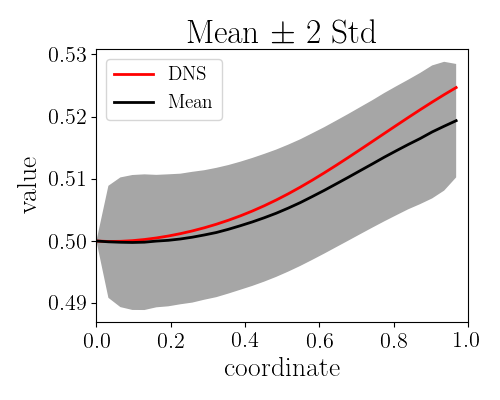}} 
  {\includegraphics[height=0.08\linewidth]{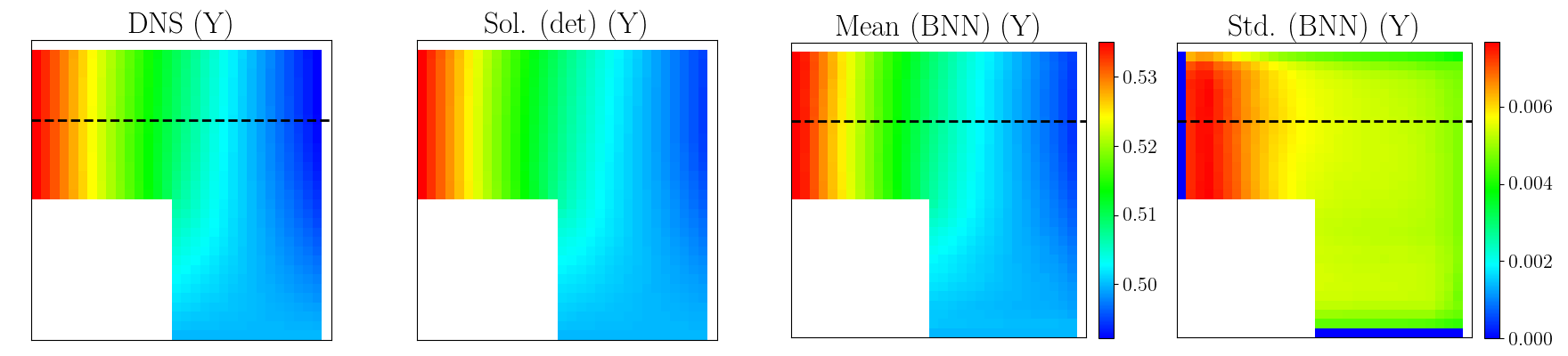}}
  {\includegraphics[height=0.08\linewidth]{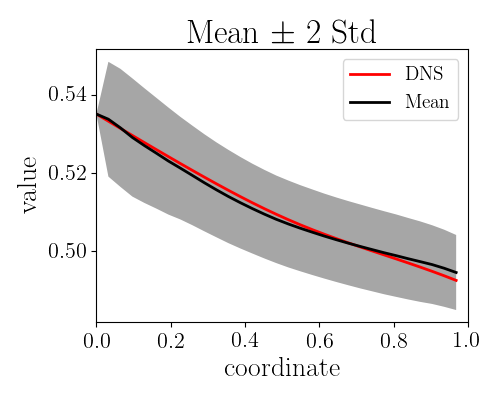}} \\
  {\includegraphics[height=0.08\linewidth]{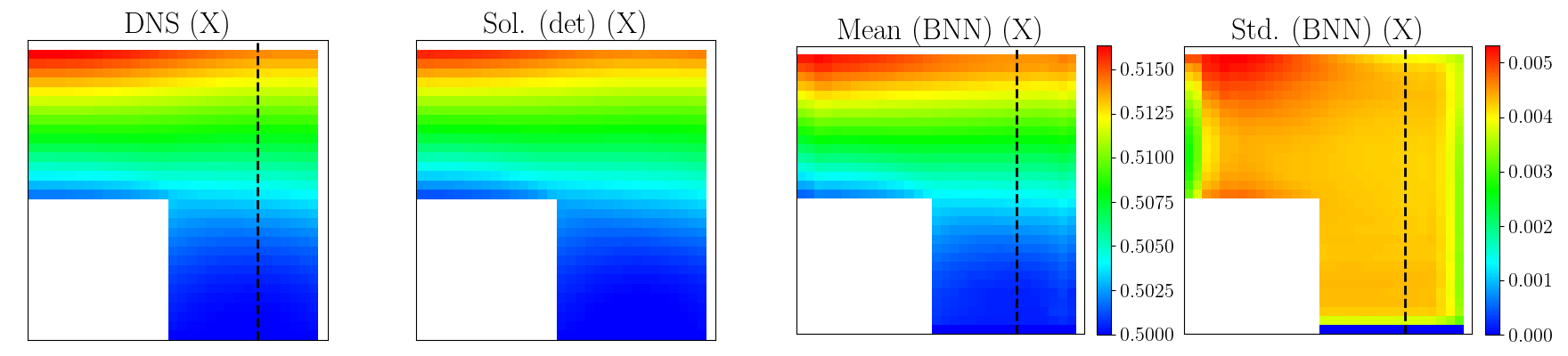}}
  {\includegraphics[height=0.08\linewidth]{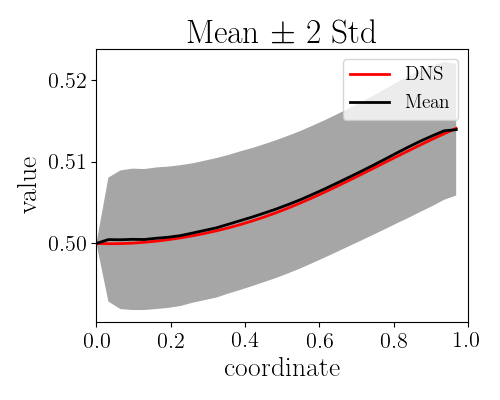}} 
  {\includegraphics[height=0.08\linewidth]{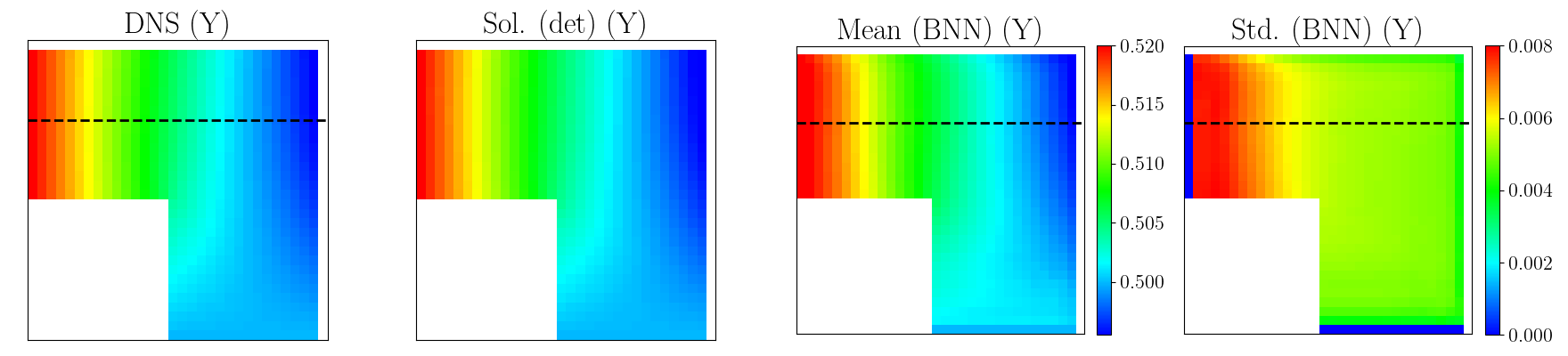}}
  {\includegraphics[height=0.08\linewidth]{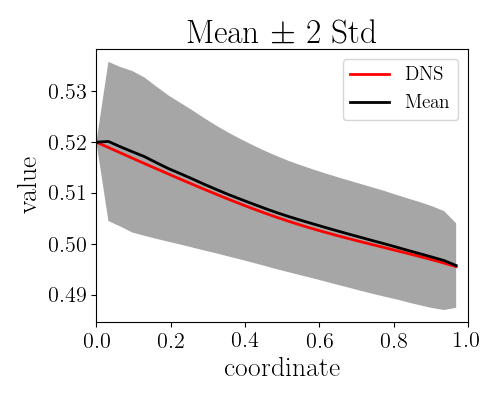}} \\
  \caption{Additional NN results for linear BVPs on L-shape domains.}
  \label{fig:linear-lshape-results-additional}
\end{figure}

NN structure information for the L-shape domain simulation are summarized in Table \ref{tab:linear-lshape-NNs} and \ref{tab:linear-lshape-NNs-others}.
Additional interpolated prediction results from the L-shape domain simulation are summarized in Fig. \ref{fig:linear-lshape-results-additional}.

\begin{table}
  \centering
  \begin{tabular}{l | l | l | l}
    \hline
    Deterministic         & Probabilistic         & Size         & Layer arguments \\ \hline
    Input                 & Input                 & -            & - \\
    LayerFillRandomNumber & LayerFillRandomNumber & -            & - \\
    Conv2D                & Convolution2DFlipout  & filters = 8  & kernel (5,5), padding: same, ReLU \\
    MaxPooling2D          & MaxPooling2D          & -            & kernel (2,2), padding: same\\
    Conv2D                & Convolution2DFlipout  & filters = 8 & kernel (5,5), padding: same, ReLU \\
    MaxPooling2D          & MaxPooling2D          & -            & kernel (2,2), padding: same\\
    Conv2D                & Convolution2DFlipout  & filters = 16 & kernel (5,5), padding: same, ReLU \\
    MaxPooling2D          & MaxPooling2D          & -            & kernel (2,2), padding: same\\
    Flatten               & Flatten               & -            & - \\
    Dense                 & DenseFlipout          & units = 32   & ReLU \\
    Dense                 & DenseFlipout          & units = 128   & ReLU \\
    Reshape               & Reshape               & -            & $[4,4,8]$ \\
    Conv2D                & Convolution2DFlipout  & filters = 16 & kernel (5,5), padding: same, ReLU \\
    UpSampling2D          & UpSampling2D          & -            & size (2,2) \\
    Conv2D                & Convolution2DFlipout  & filters = 16 & kernel (5,5), padding: same, ReLU \\
    UpSampling2D          & UpSampling2D          & -            & size (2,2) \\
    Conv2D                & Convolution2DFlipout  & filters = 16 & kernel (5,5), padding: same, ReLU \\
    UpSampling2D          & UpSampling2D          & -            & size (2,2) \\
    Conv2D                & Convolution2DFlipout  & filters = 16 & kernel (5,5), padding: same, ReLU \\
    Conv2D                & Convolution2DFlipout  & filters = 2  & kernel (5,5), padding: same, ReLU \\
    \hline
  \end{tabular}
  \caption{Details of both deterministic and probabilistic NNs for solving linear elasticity L-shape BVPs.}
  \label{tab:linear-lshape-NNs}
\end{table}

\begin{table}
  \centering
  \begin{tabular}{l | l | l }
    \hline
    Description                     & Deterministic                 & Probabilistic         \\ \hline
    Total parameters                & 41,346                        & 82,435                 \\
    Size of $\calD$                 & 5 $\times$ Aug: $2^{10}$      & 5 $\times$ Aug: $2^{9}$      \\
    Epochs                          & 10,000                        & 100                 \\
    Zero initialization epochs      & 100                           & -                     \\
    Optimizer                       & Adam                          & Nadam                 \\
    Learning Rate                   & 2.5e-4                        & 1e-8                  \\
    Batch Size                      & 256                           & 64                    \\
    $\Sigma_1$                      & -                             & 1e-8                  \\
    Initial value of $\Sigma_2$     & -                             & 1e-8                  \\
    \hline
  \end{tabular}
  \caption{Training related parameters for solving linear elasticity L-shape BVPs. Aug: data augmentation. }
  \label{tab:linear-lshape-NNs-others}
\end{table}

\subsection{Non-linear elasticity}\label{appendix:nonlinear-results-extra}
\subsubsection{Multiple rectangle domains with different BCs} \label{appendix:nonlinear-30bvps}
\begin{figure}[p!]
  \centering
  {\includegraphics[height=0.08\linewidth]{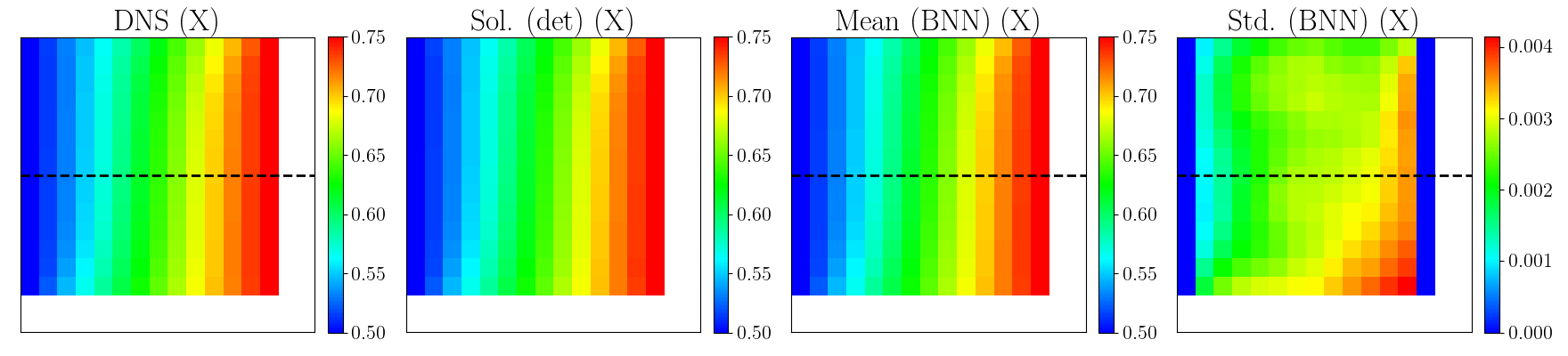}} 
  {\includegraphics[height=0.08\linewidth]{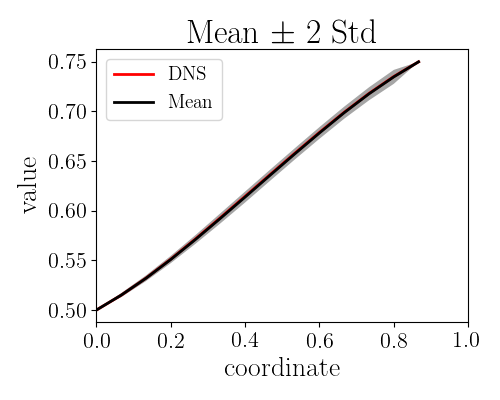}}
  {\includegraphics[height=0.08\linewidth]{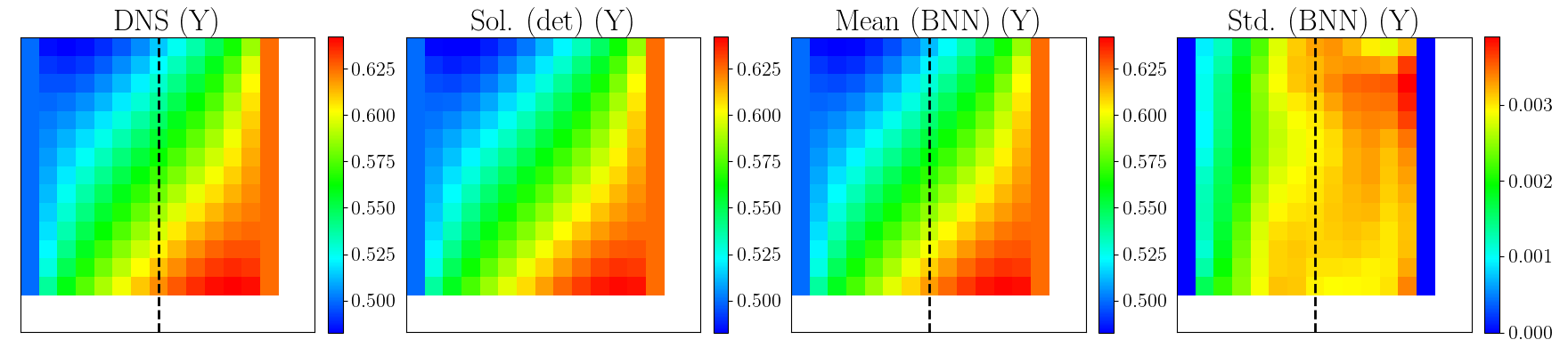}} 
  {\includegraphics[height=0.08\linewidth]{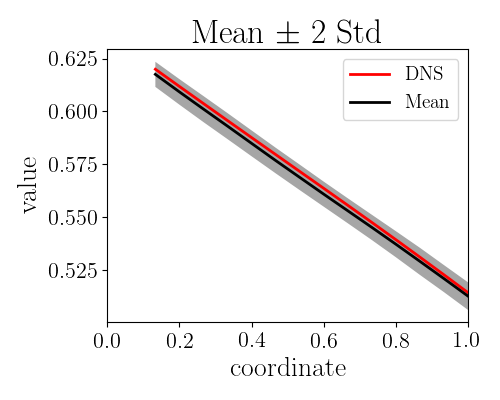}} \\
  {\includegraphics[height=0.08\linewidth]{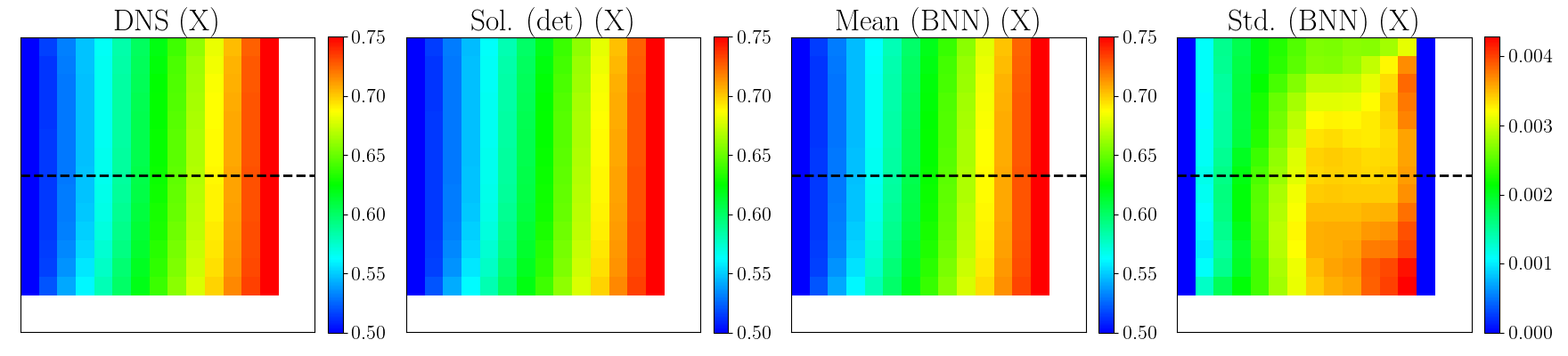}} 
  {\includegraphics[height=0.08\linewidth]{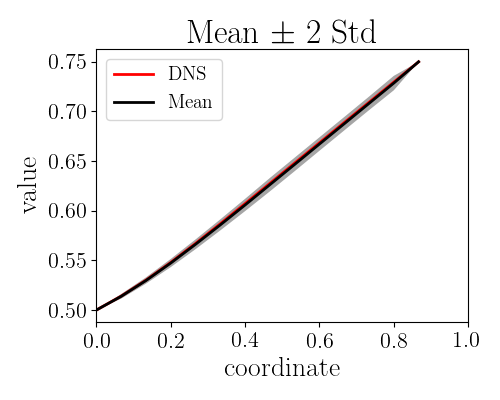}} 
  {\includegraphics[height=0.08\linewidth]{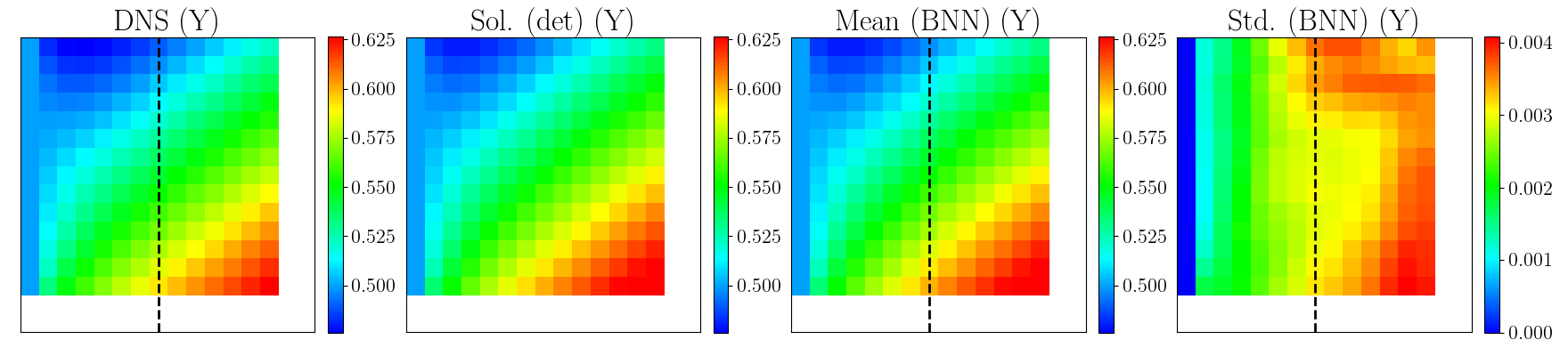}} 
  {\includegraphics[height=0.08\linewidth]{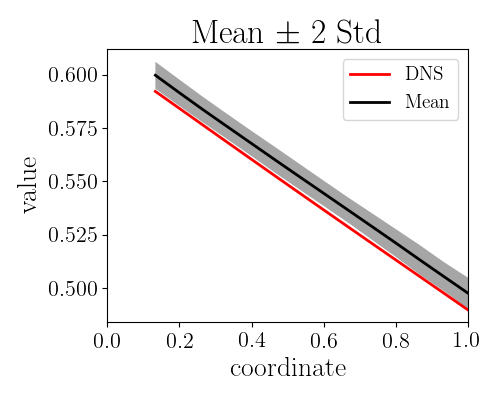}} \\
  {\includegraphics[height=0.08\linewidth]{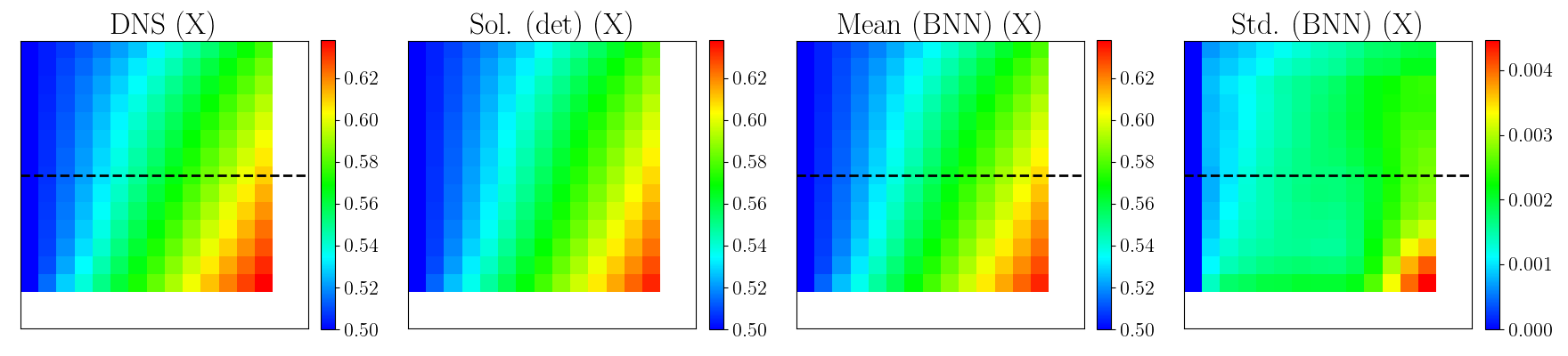}} 
  {\includegraphics[height=0.08\linewidth]{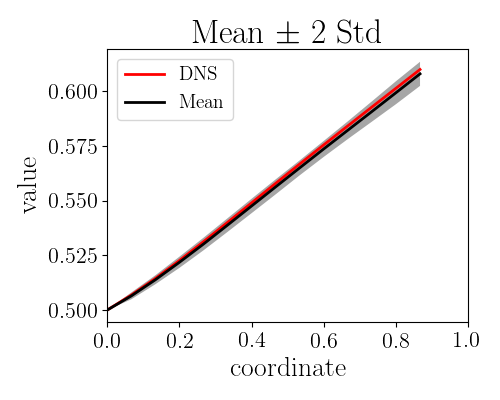}} 
  {\includegraphics[height=0.08\linewidth]{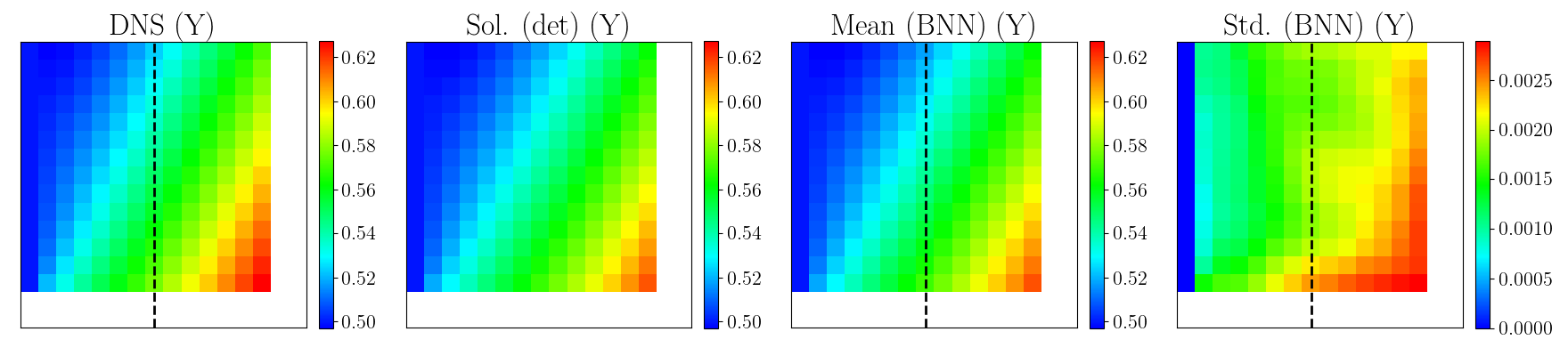}} 
  {\includegraphics[height=0.08\linewidth]{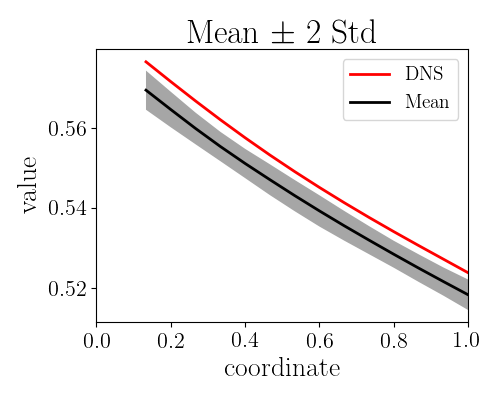}} \\
  {\includegraphics[height=0.08\linewidth]{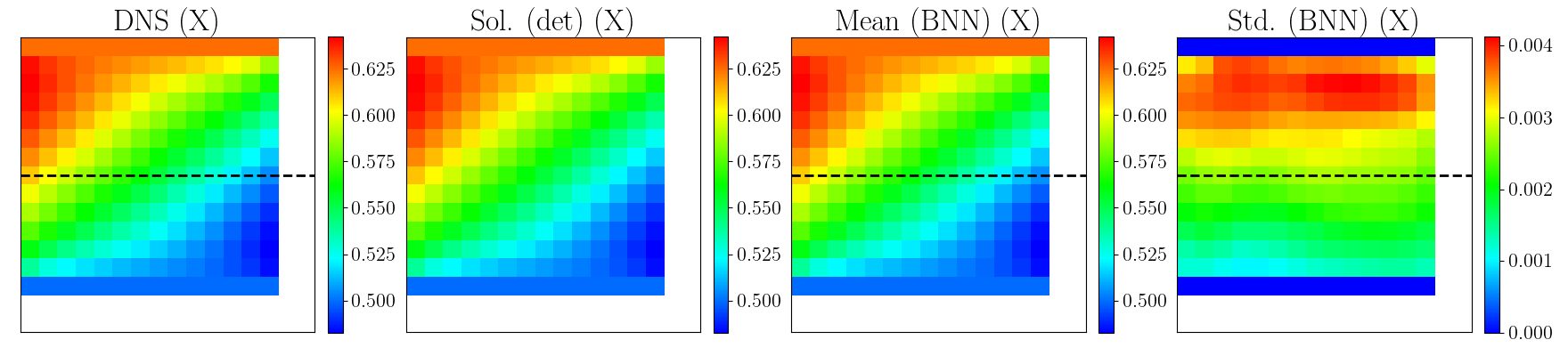}} 
  {\includegraphics[height=0.08\linewidth]{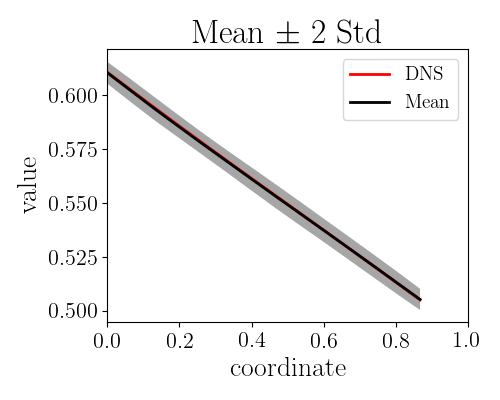}} 
  {\includegraphics[height=0.08\linewidth]{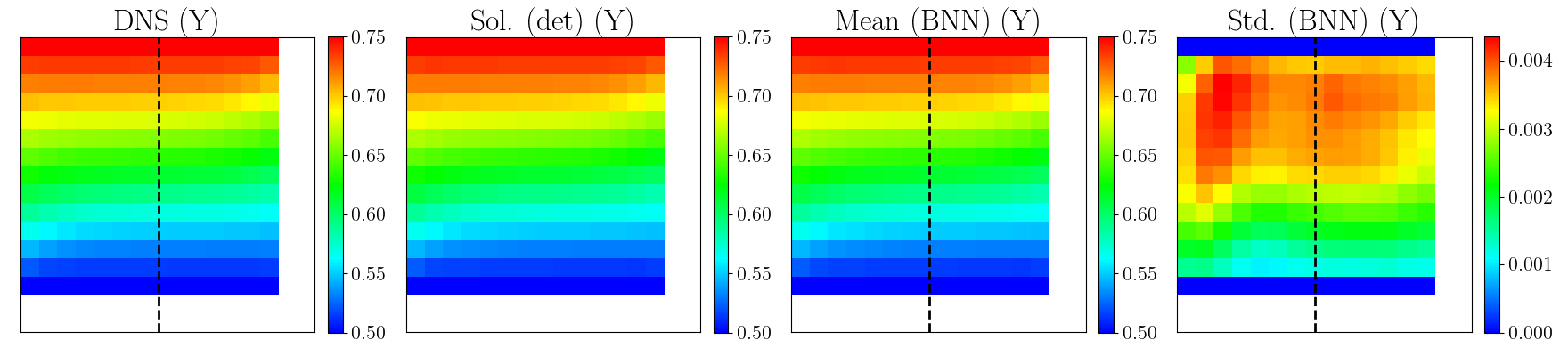}} 
  {\includegraphics[height=0.08\linewidth]{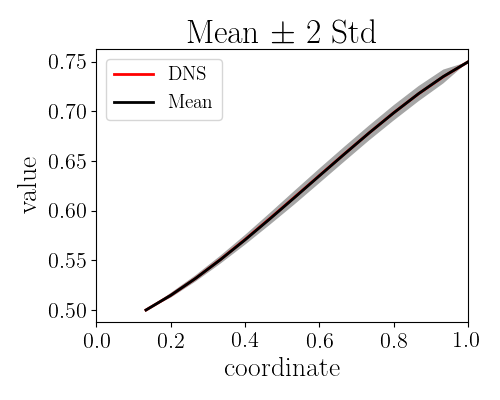}} \\
  {\includegraphics[height=0.08\linewidth]{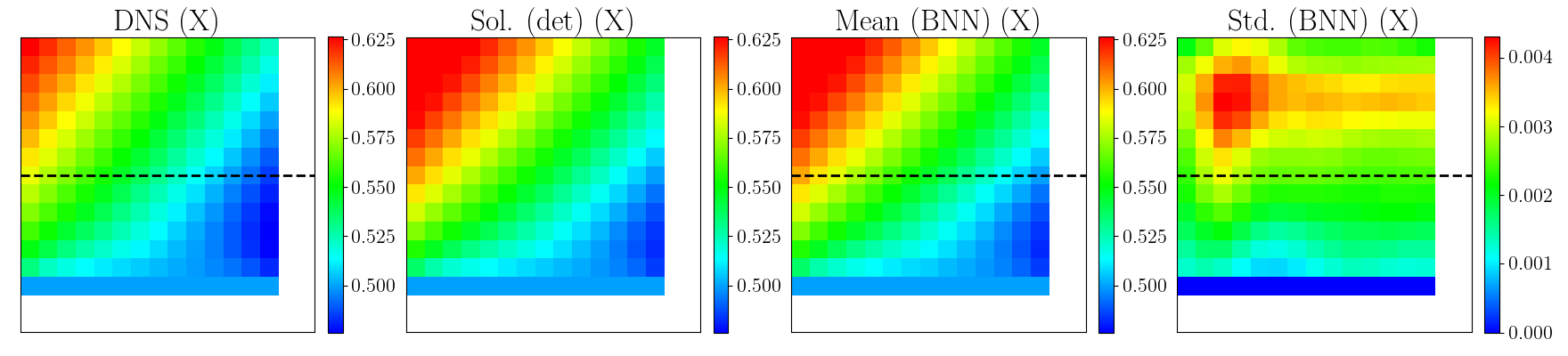}} 
  {\includegraphics[height=0.08\linewidth]{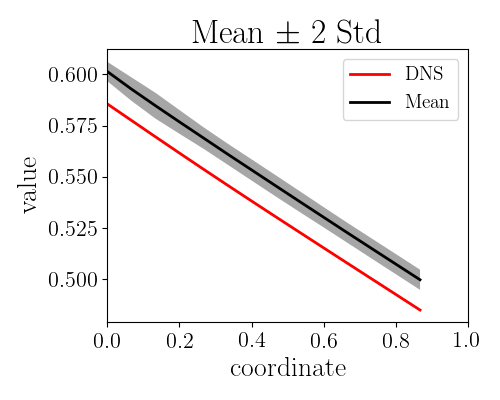}} 
  {\includegraphics[height=0.08\linewidth]{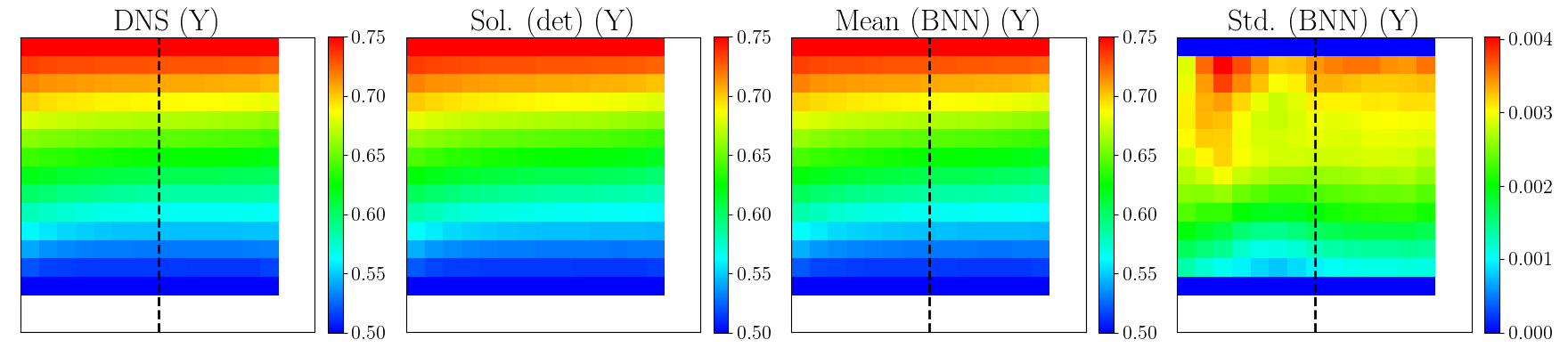}} 
  {\includegraphics[height=0.08\linewidth]{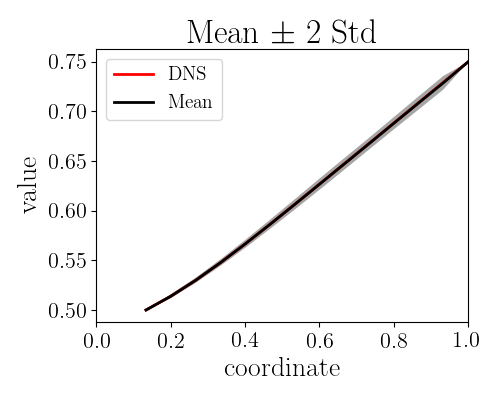}} \\
  {\includegraphics[height=0.08\linewidth]{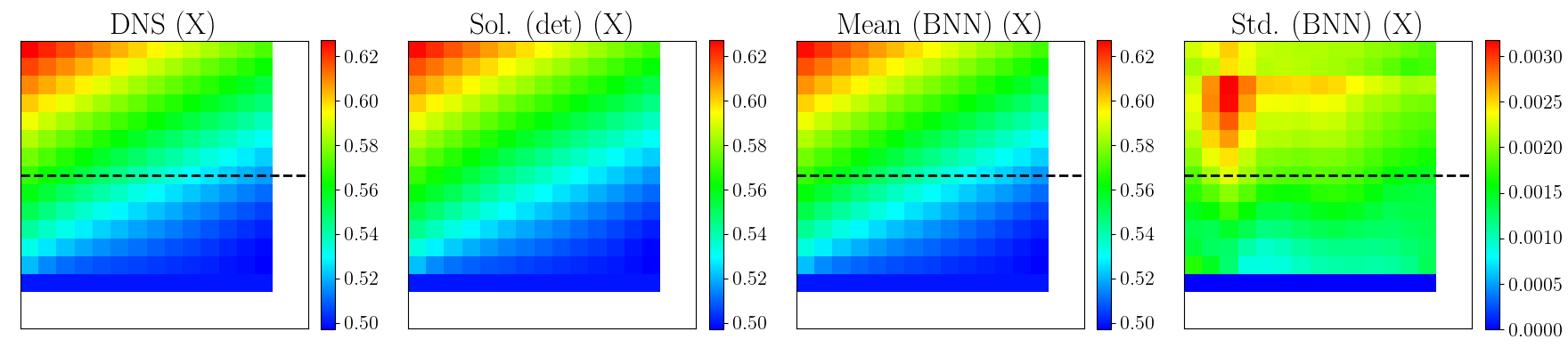}} 
  {\includegraphics[height=0.08\linewidth]{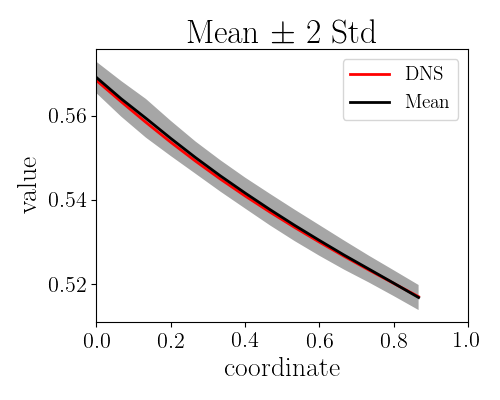}} 
  {\includegraphics[height=0.08\linewidth]{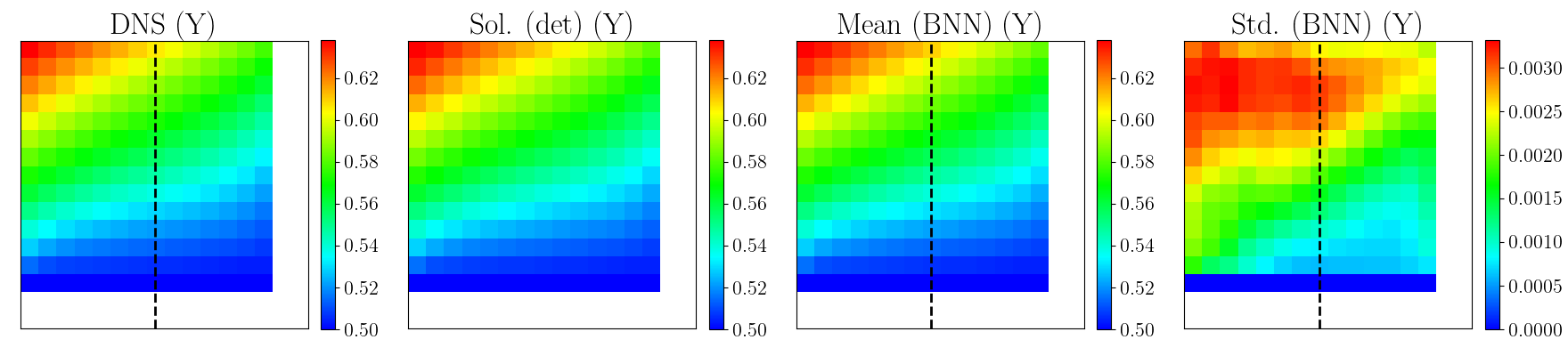}} 
  {\includegraphics[height=0.08\linewidth]{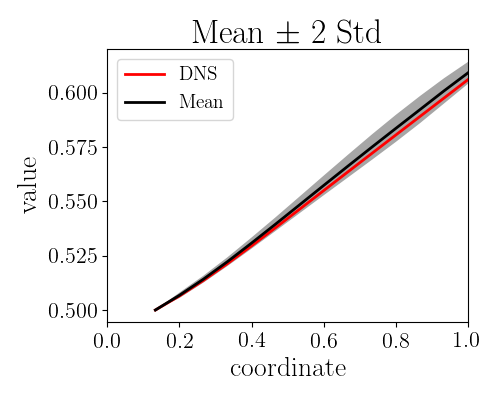}} \\
  {\includegraphics[height=0.08\linewidth]{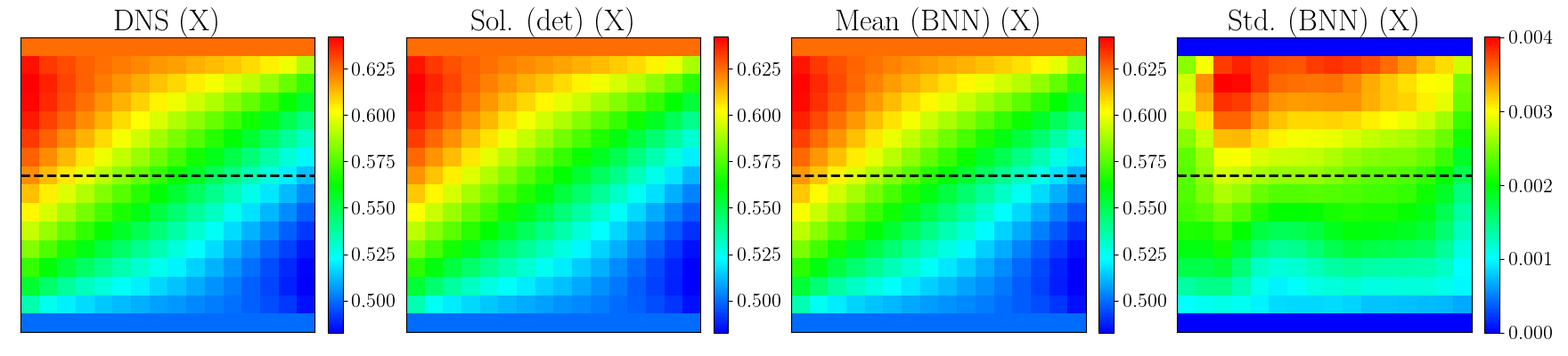}} 
  {\includegraphics[height=0.08\linewidth]{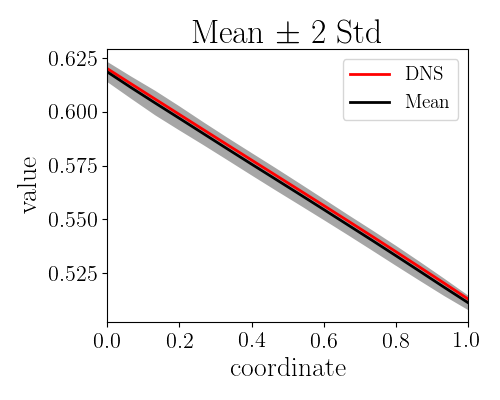}} 
  {\includegraphics[height=0.08\linewidth]{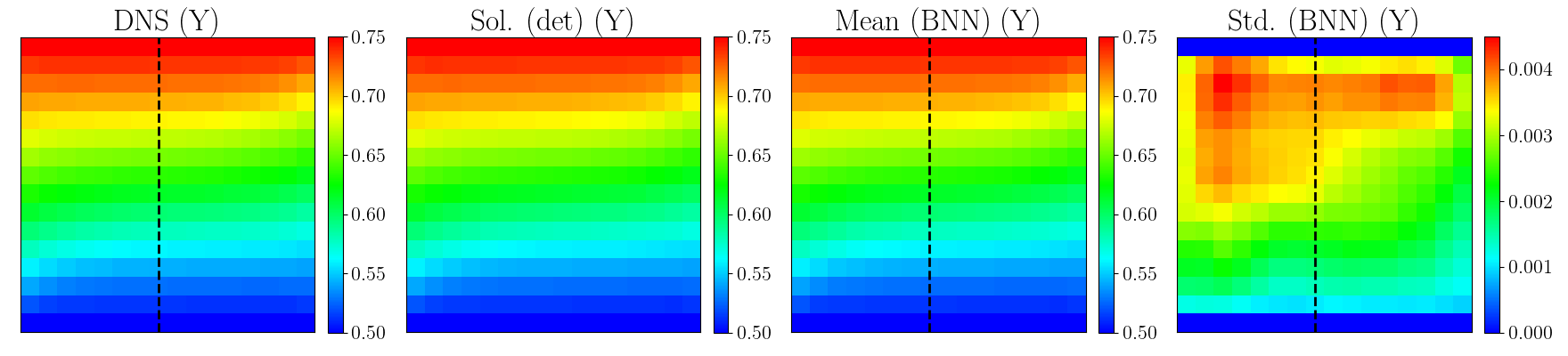}} 
  {\includegraphics[height=0.08\linewidth]{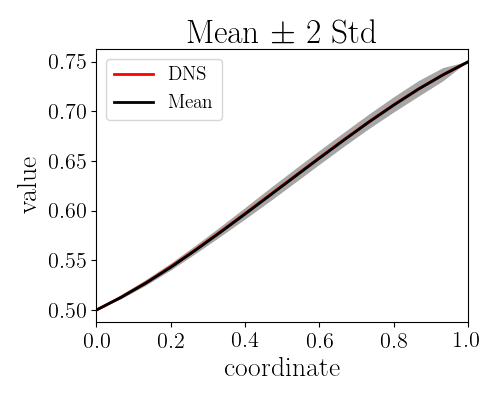}} \\
  {\includegraphics[height=0.08\linewidth]{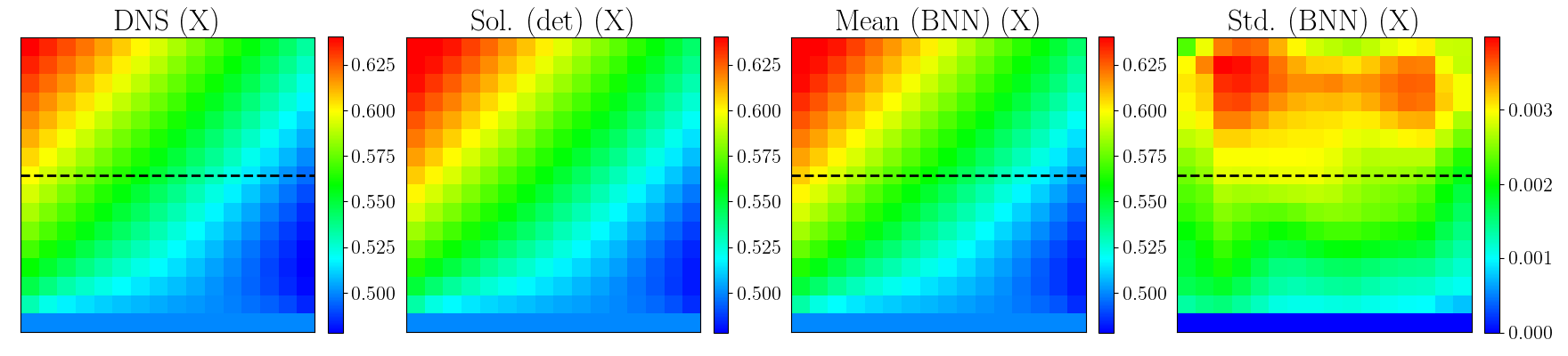}} 
  {\includegraphics[height=0.08\linewidth]{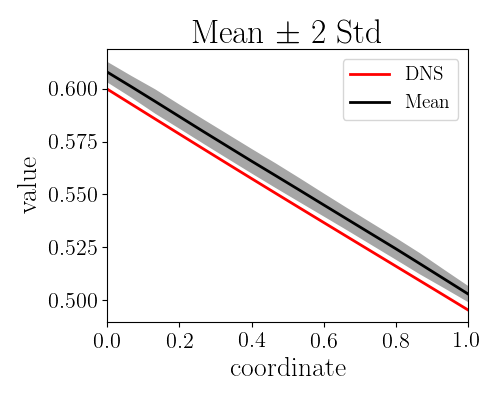}} 
  {\includegraphics[height=0.08\linewidth]{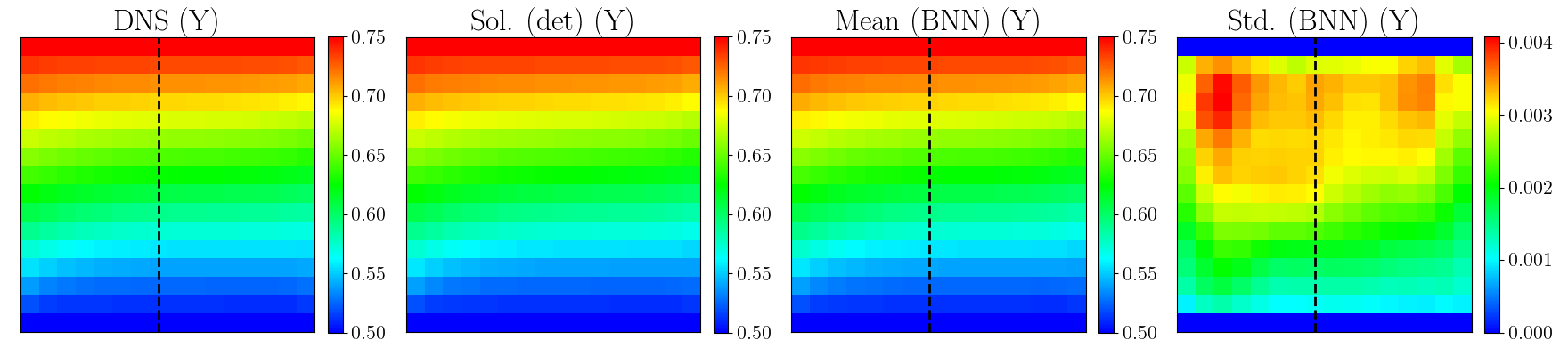}} 
  {\includegraphics[height=0.08\linewidth]{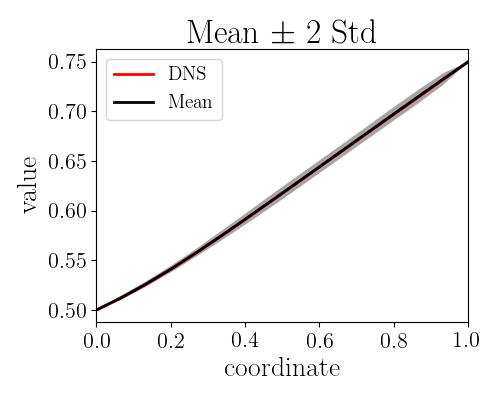}} \\
  {\includegraphics[height=0.08\linewidth]{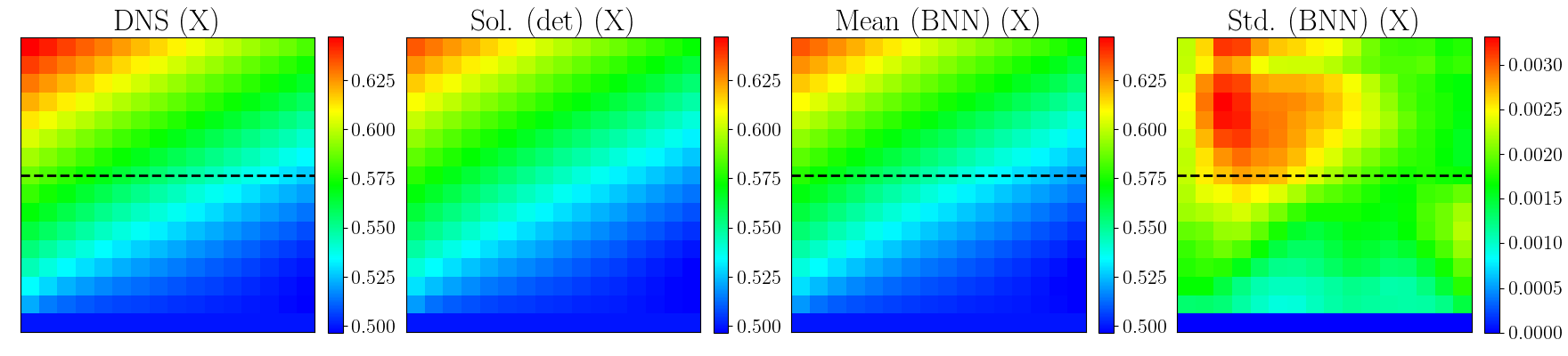}} 
  {\includegraphics[height=0.08\linewidth]{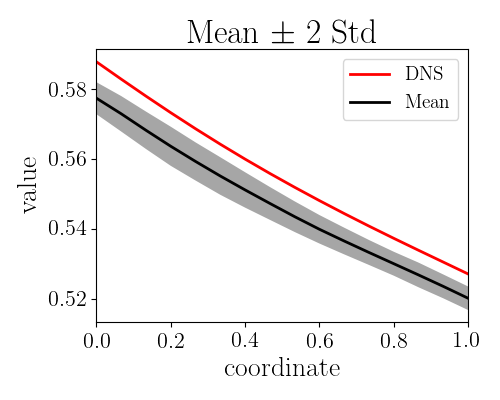}} 
  {\includegraphics[height=0.08\linewidth]{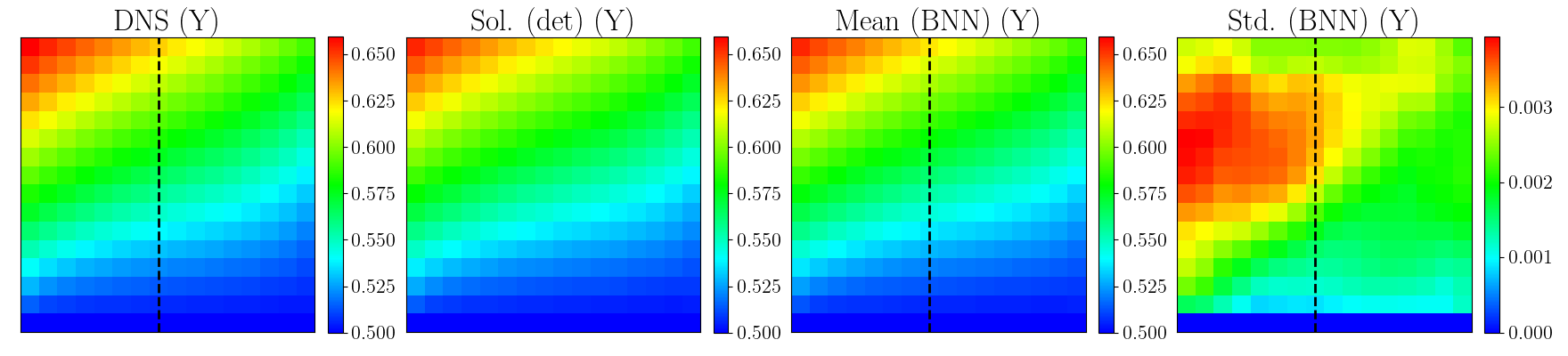}} 
  {\includegraphics[height=0.08\linewidth]{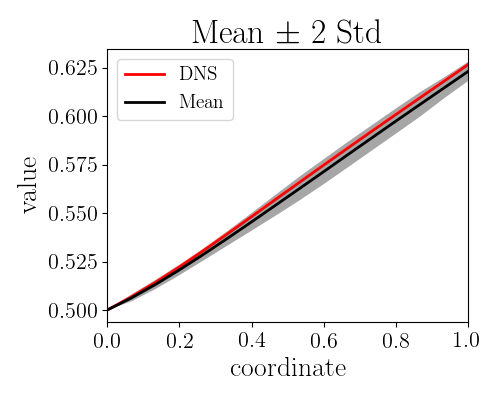}} \\
  {\includegraphics[height=0.08\linewidth]{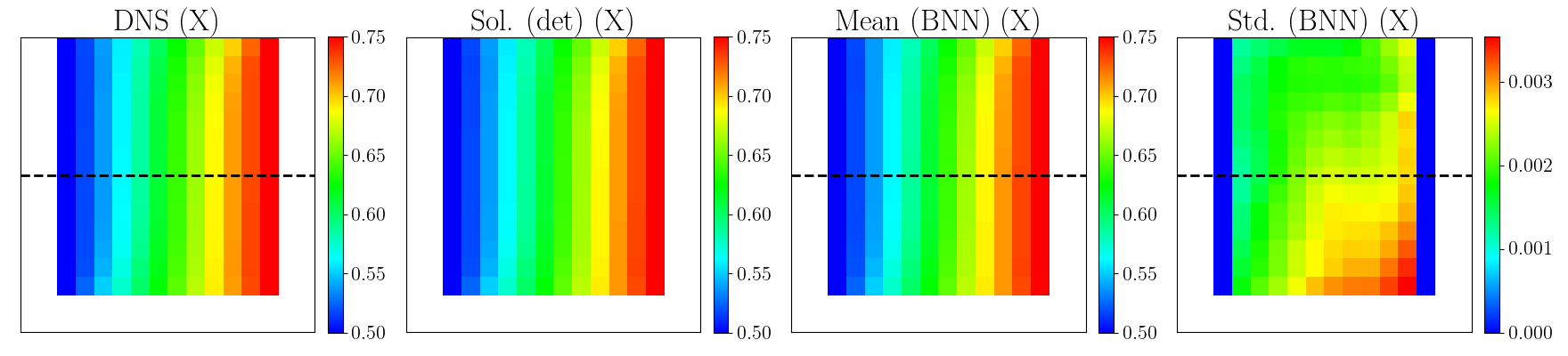}} 
  {\includegraphics[height=0.08\linewidth]{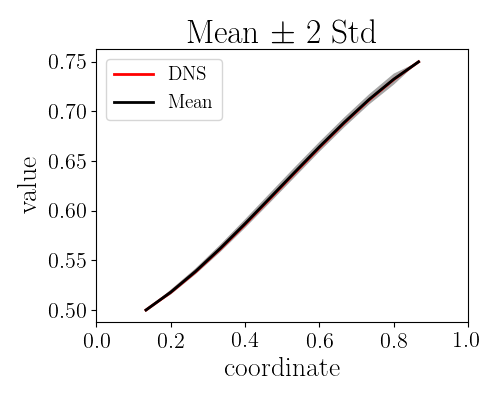}} 
  {\includegraphics[height=0.08\linewidth]{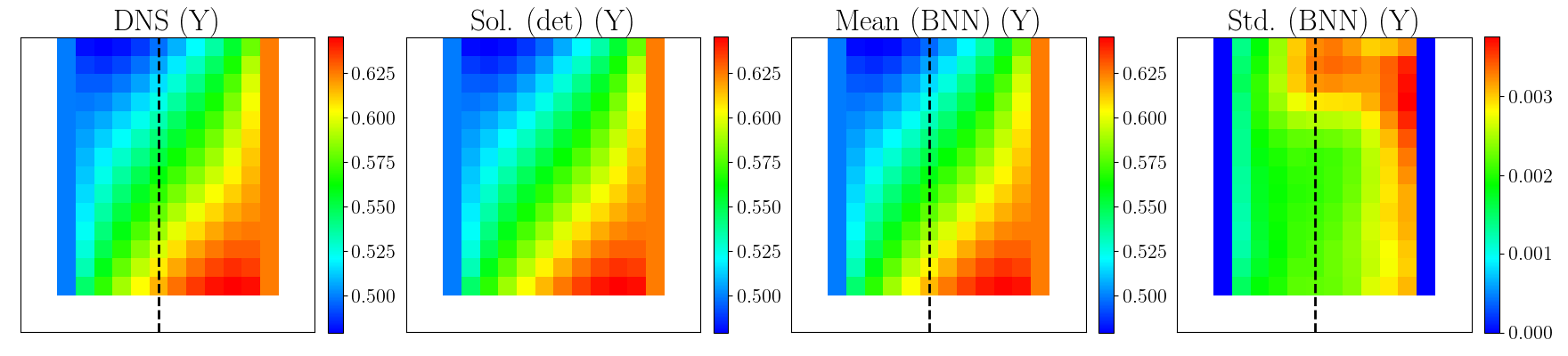}} 
  {\includegraphics[height=0.08\linewidth]{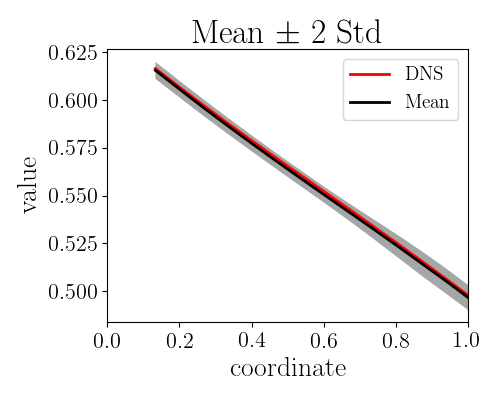}} \\
  {\includegraphics[height=0.08\linewidth]{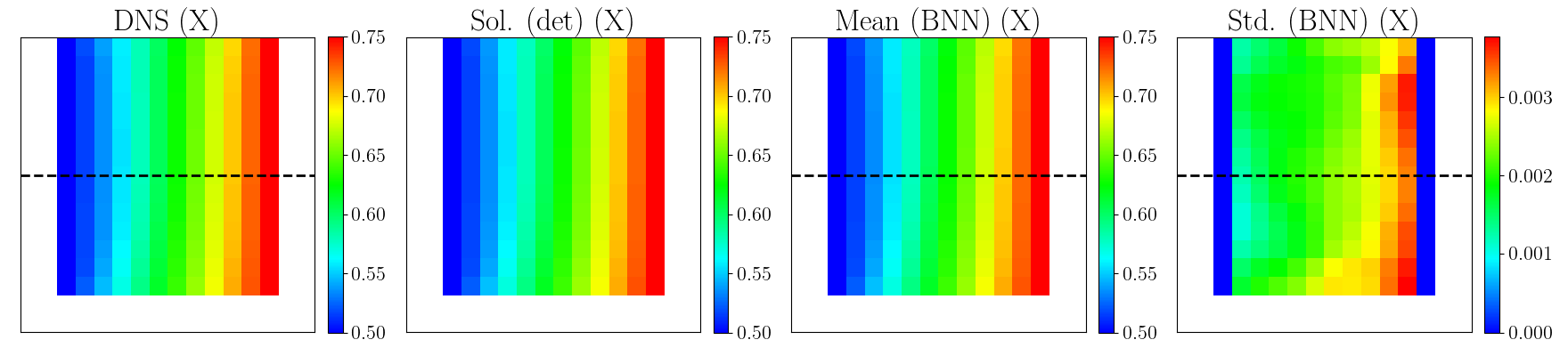}} 
  {\includegraphics[height=0.08\linewidth]{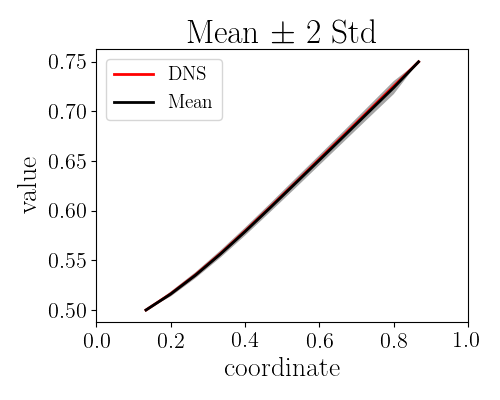}} 
  {\includegraphics[height=0.08\linewidth]{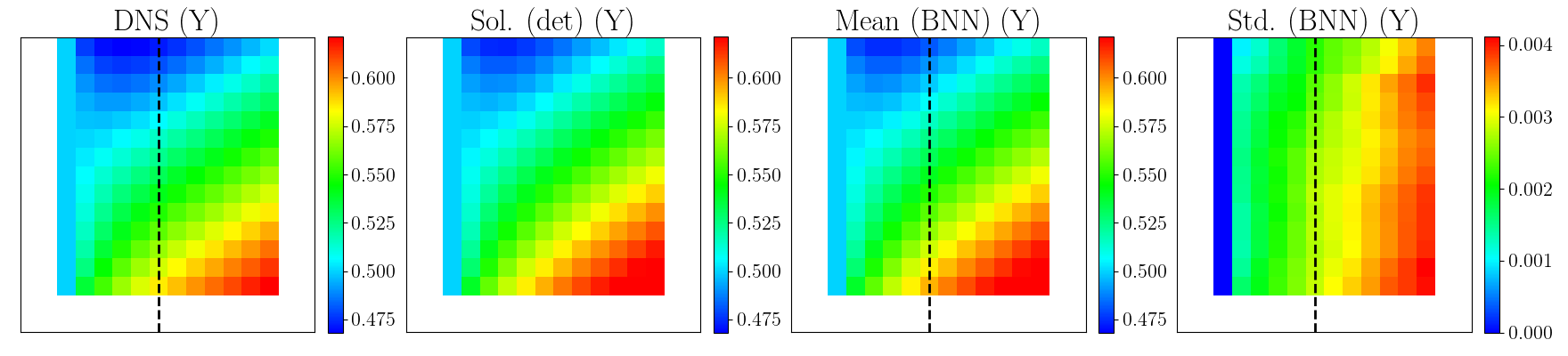}} 
  {\includegraphics[height=0.08\linewidth]{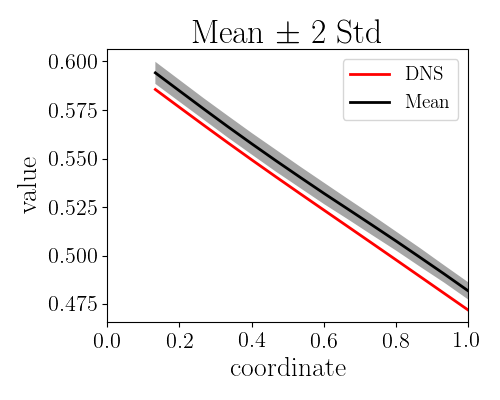}} \\
  {\includegraphics[height=0.08\linewidth]{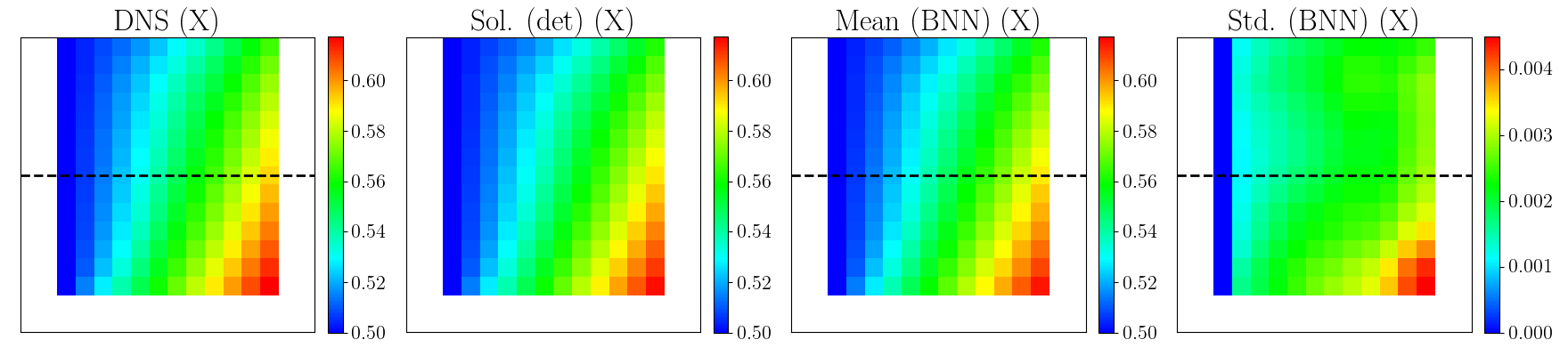}} 
  {\includegraphics[height=0.08\linewidth]{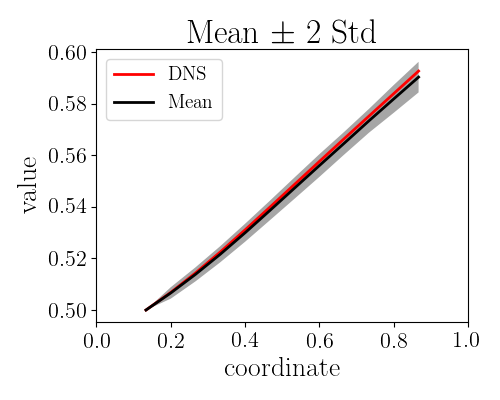}} 
  {\includegraphics[height=0.08\linewidth]{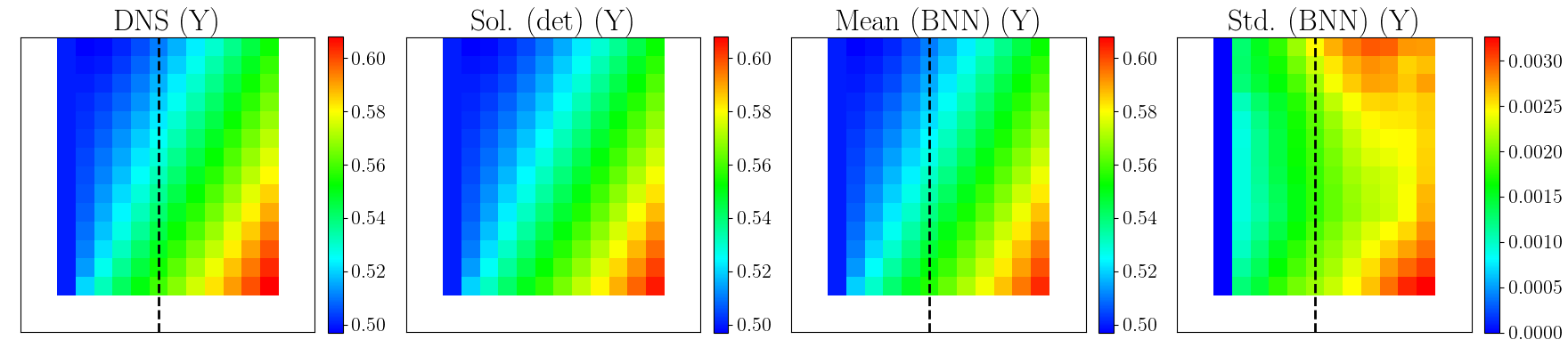}} 
  {\includegraphics[height=0.08\linewidth]{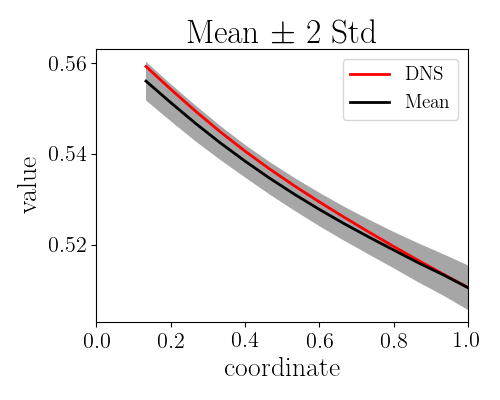}} \\
  {\includegraphics[height=0.08\linewidth]{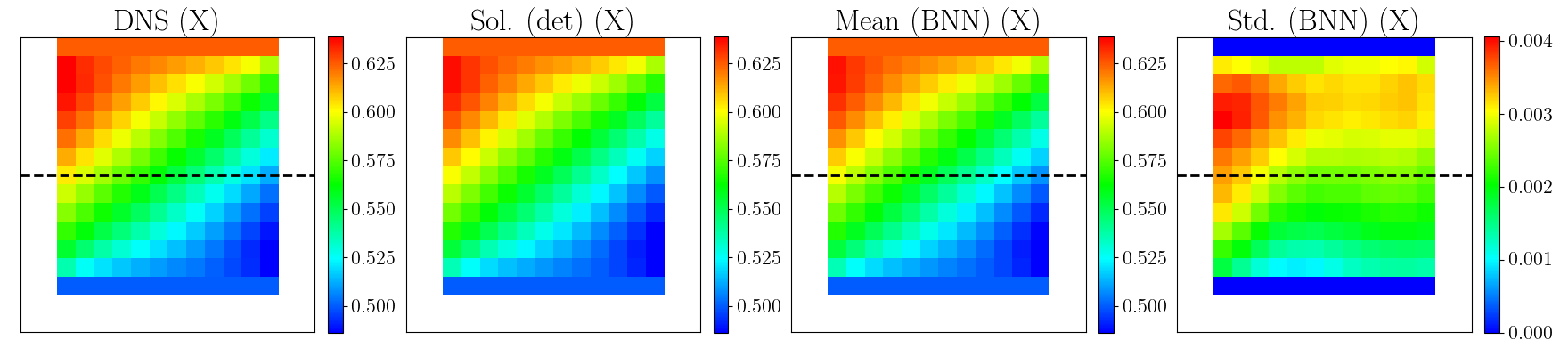}} 
  {\includegraphics[height=0.08\linewidth]{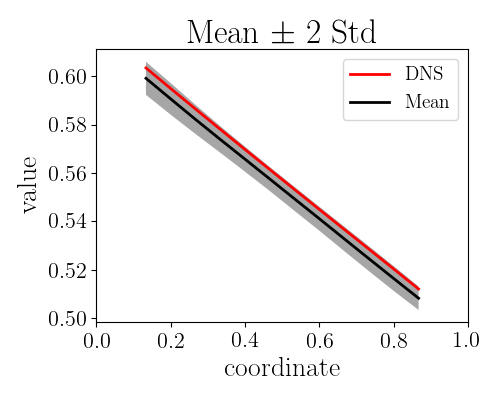}} 
  {\includegraphics[height=0.08\linewidth]{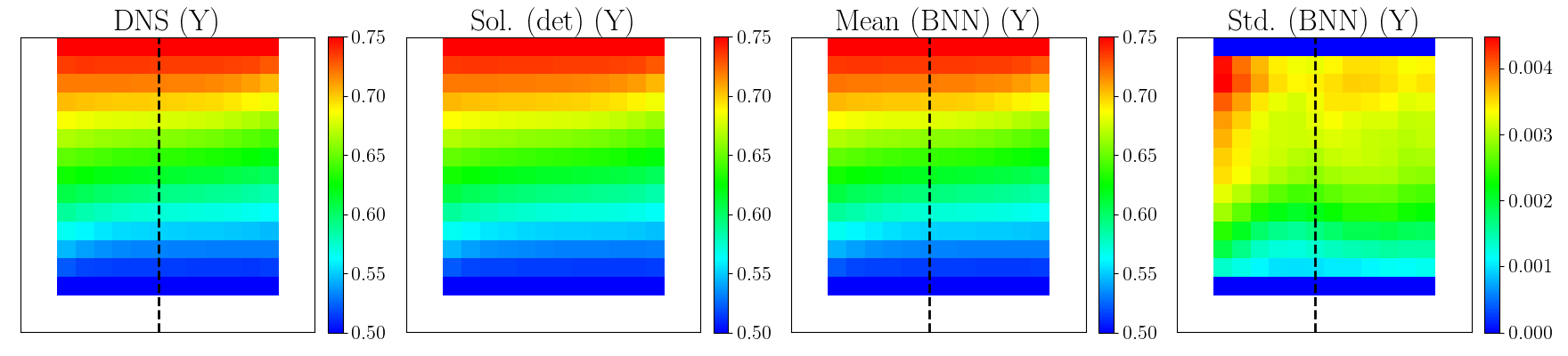}} 
  {\includegraphics[height=0.08\linewidth]{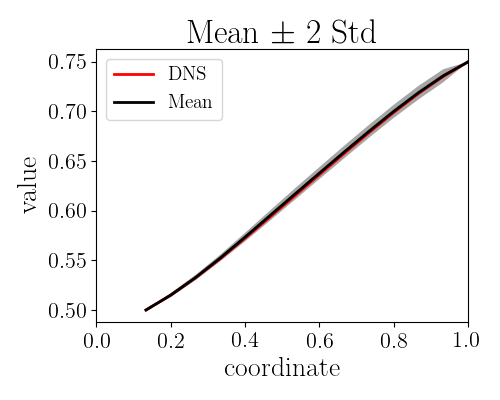}} \\
  {\includegraphics[height=0.08\linewidth]{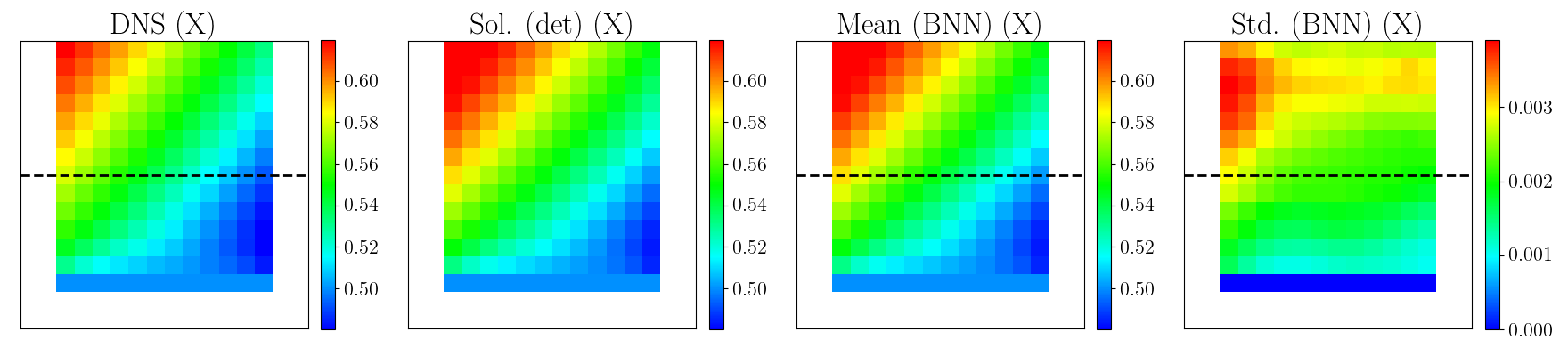}} 
  {\includegraphics[height=0.08\linewidth]{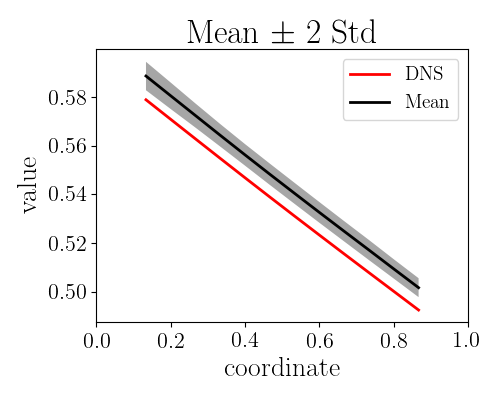}} 
  {\includegraphics[height=0.08\linewidth]{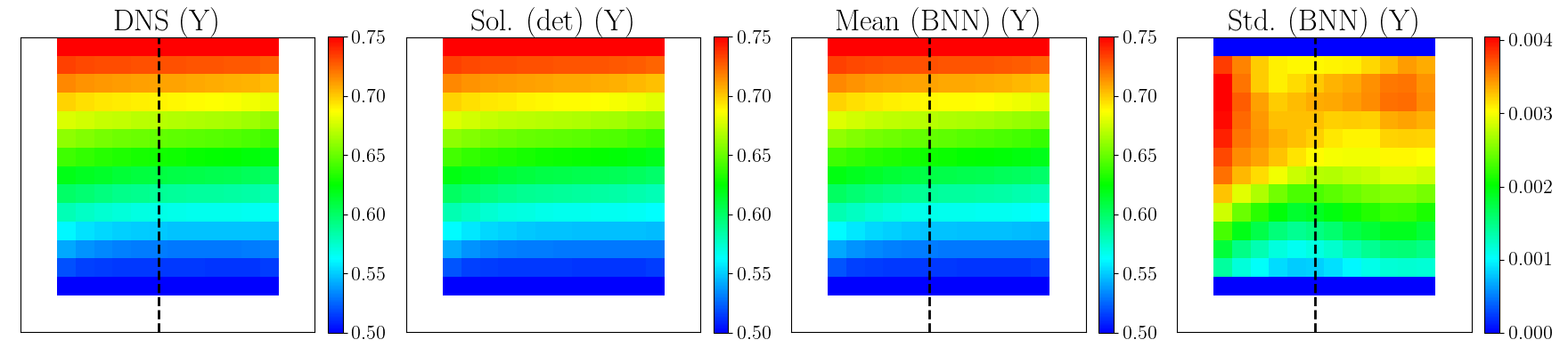}} 
  {\includegraphics[height=0.08\linewidth]{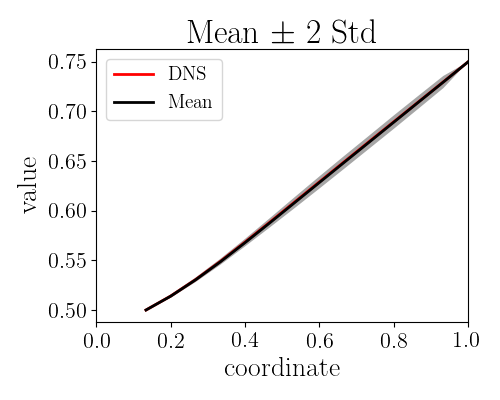}} \\
  \caption{Additional NN results for linear BVPs on rectangle domains.}
  \label{fig:nonlinear-30bvp-results-additional-1}
\end{figure}

\begin{figure}[p!]
  \centering
  {\includegraphics[height=0.08\linewidth]{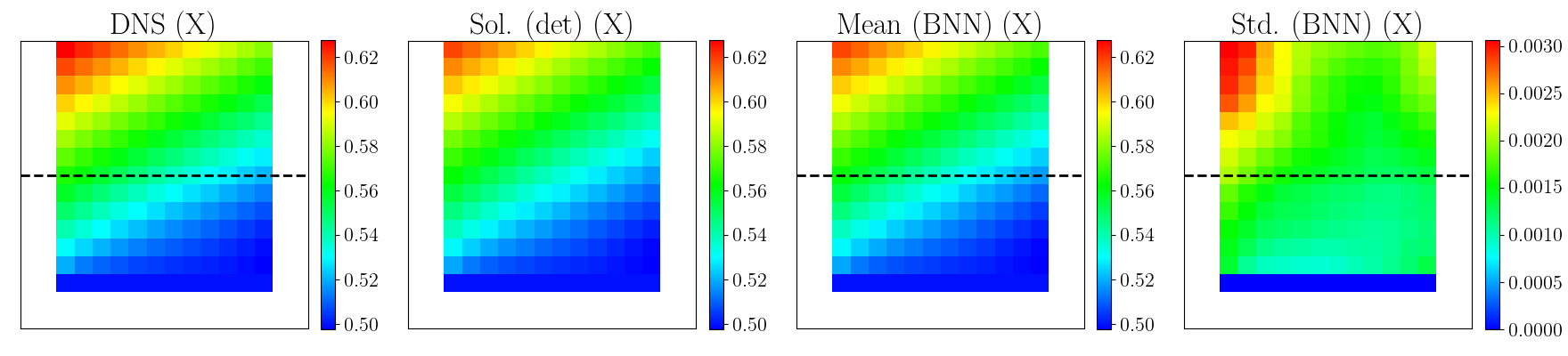}} 
  {\includegraphics[height=0.08\linewidth]{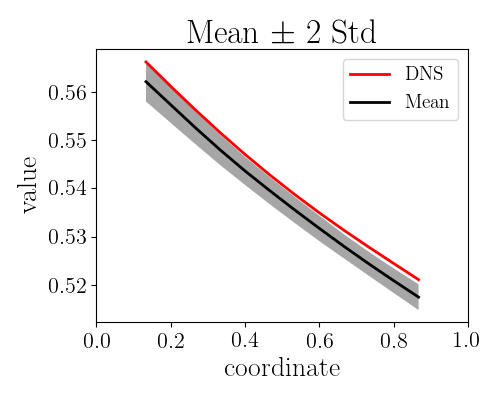}} 
  {\includegraphics[height=0.08\linewidth]{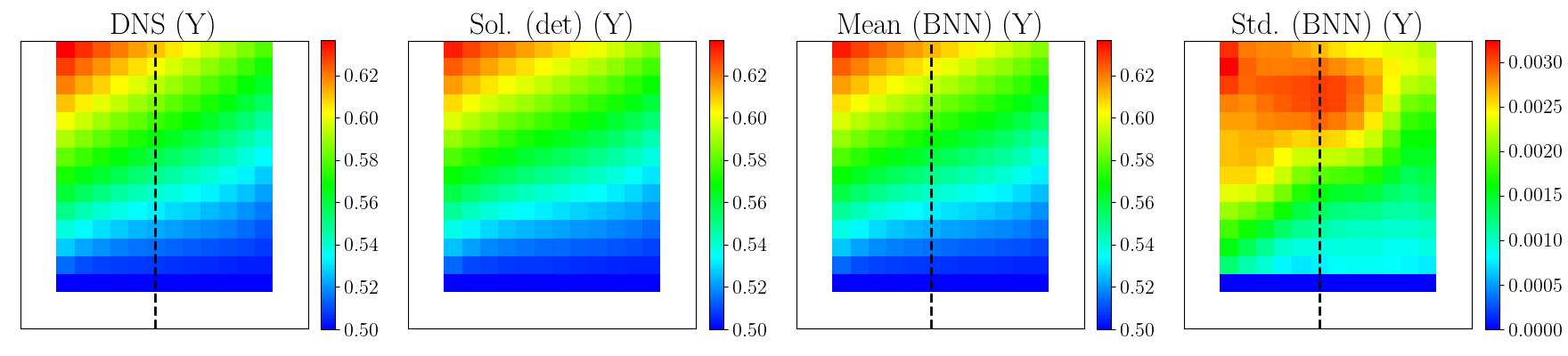}} 
  {\includegraphics[height=0.08\linewidth]{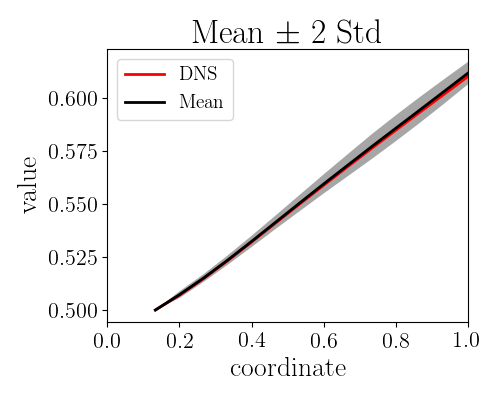}} \\
  {\includegraphics[height=0.08\linewidth]{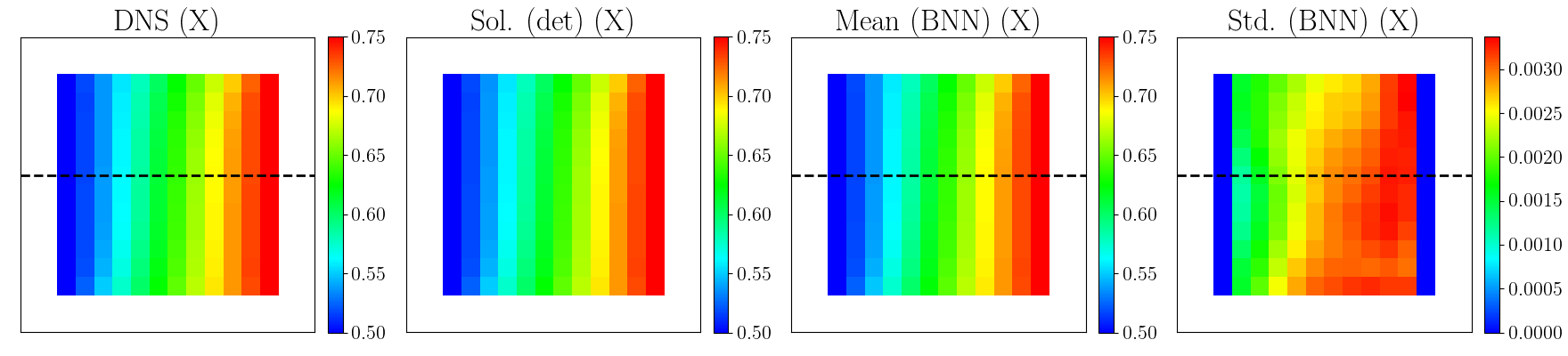}} 
  {\includegraphics[height=0.08\linewidth]{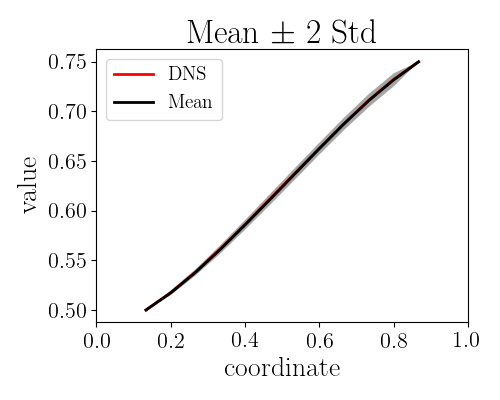}} 
  {\includegraphics[height=0.08\linewidth]{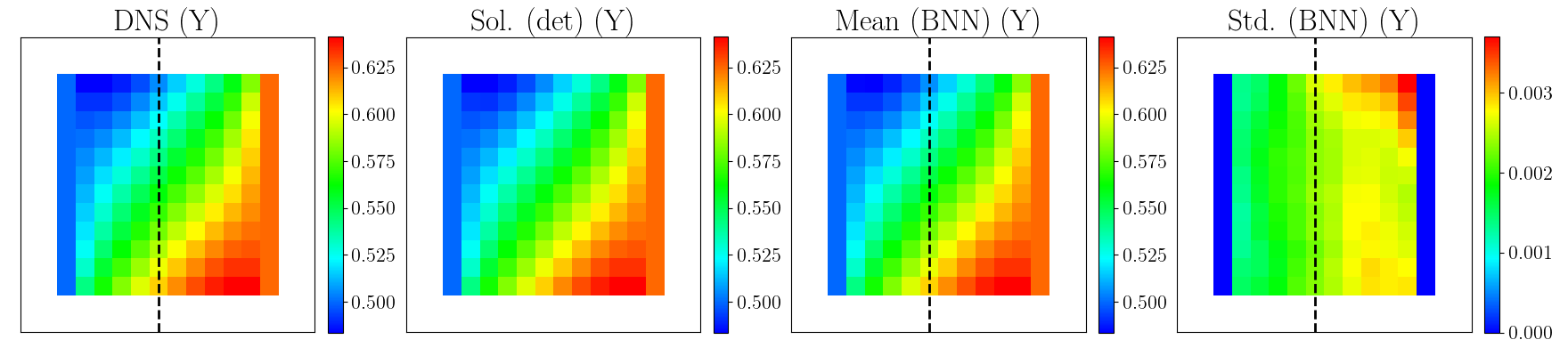}} 
  {\includegraphics[height=0.08\linewidth]{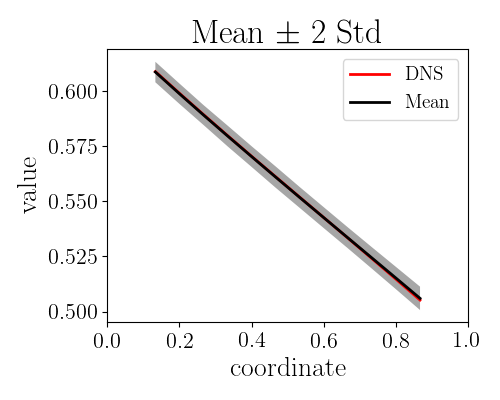}} \\
  {\includegraphics[height=0.08\linewidth]{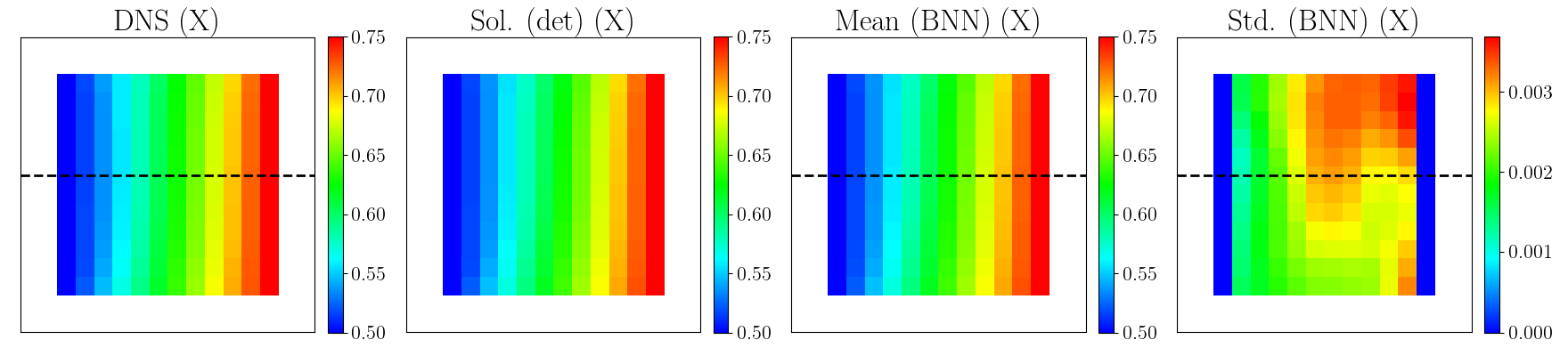}} 
  {\includegraphics[height=0.08\linewidth]{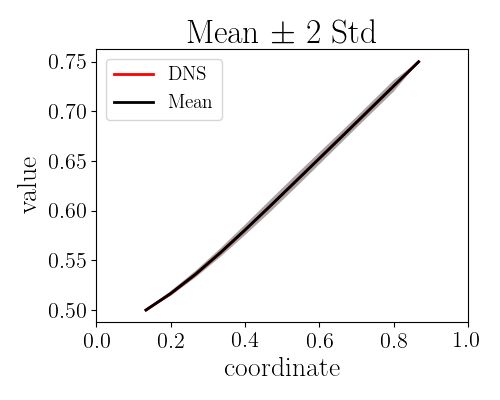}} 
  {\includegraphics[height=0.08\linewidth]{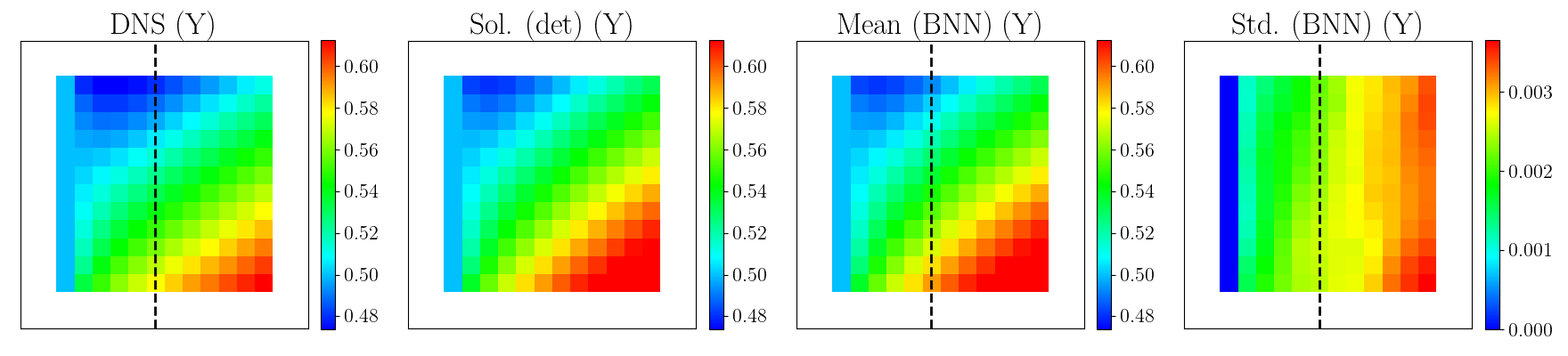}} 
  {\includegraphics[height=0.08\linewidth]{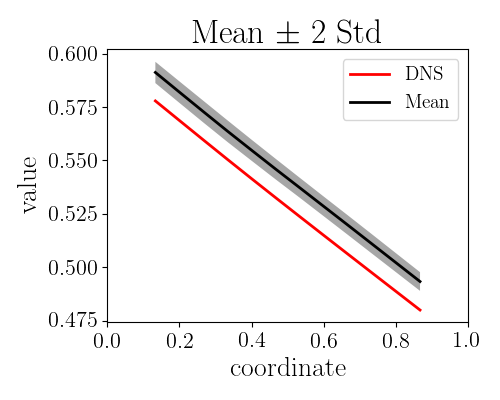}} \\
  {\includegraphics[height=0.08\linewidth]{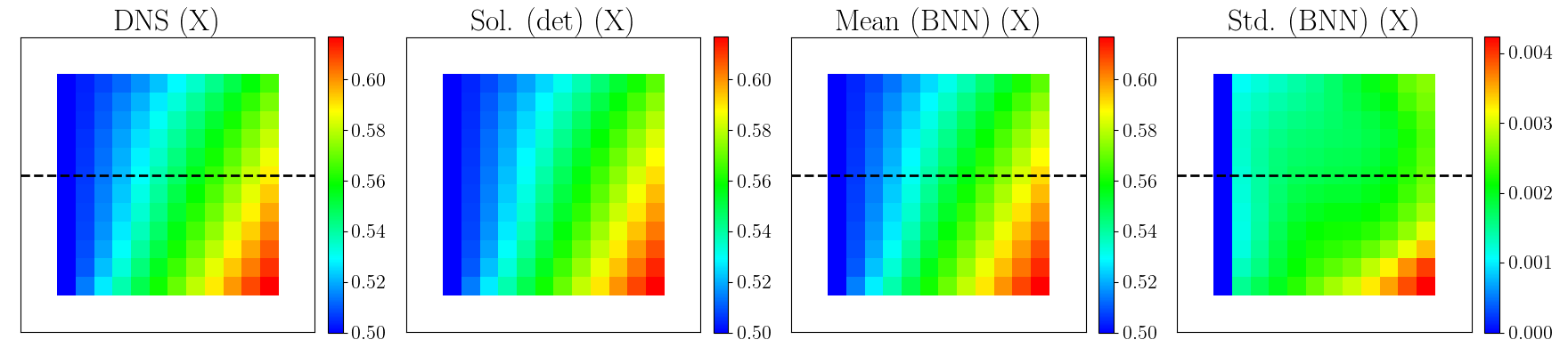}} 
  {\includegraphics[height=0.08\linewidth]{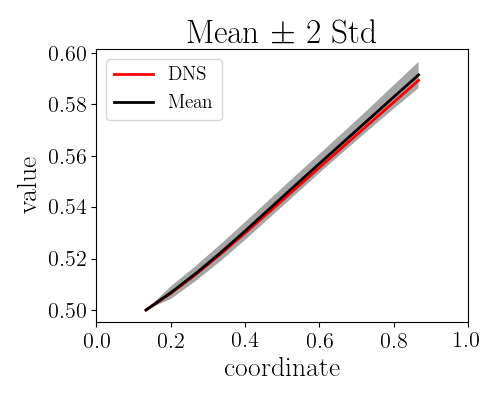}}
  {\includegraphics[height=0.08\linewidth]{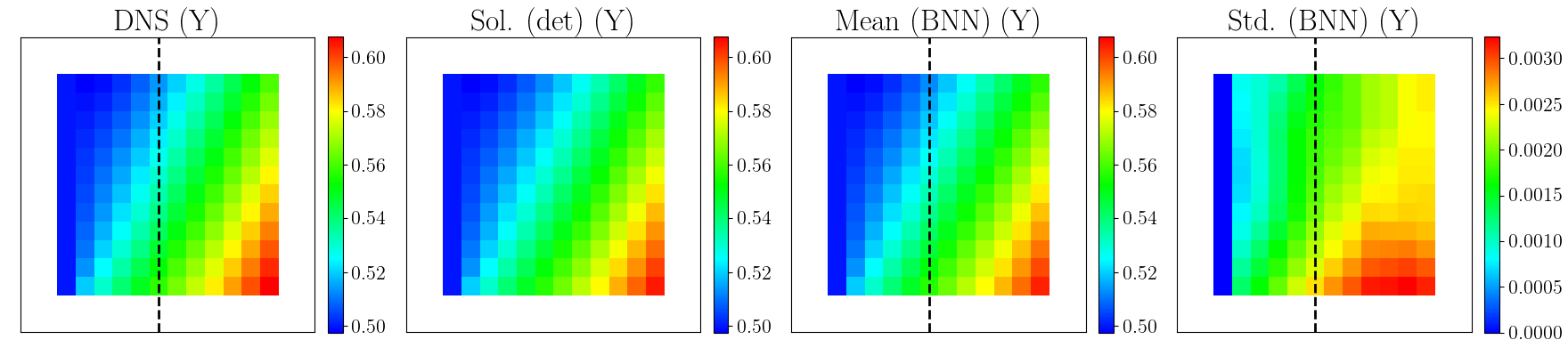}} 
  {\includegraphics[height=0.08\linewidth]{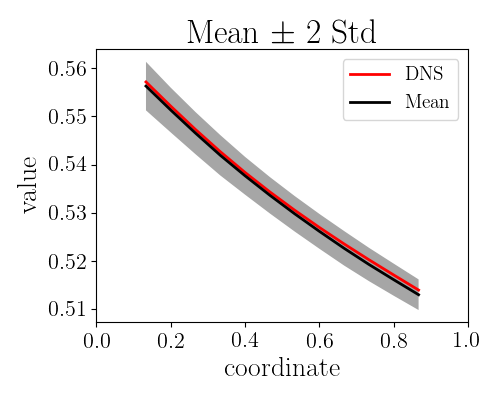}} \\
  {\includegraphics[height=0.08\linewidth]{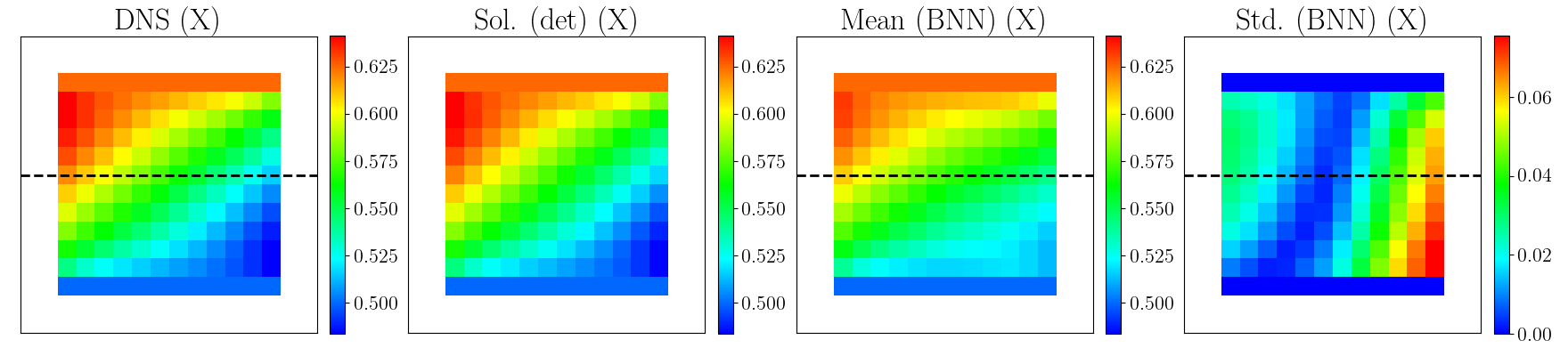}} 
  {\includegraphics[height=0.08\linewidth]{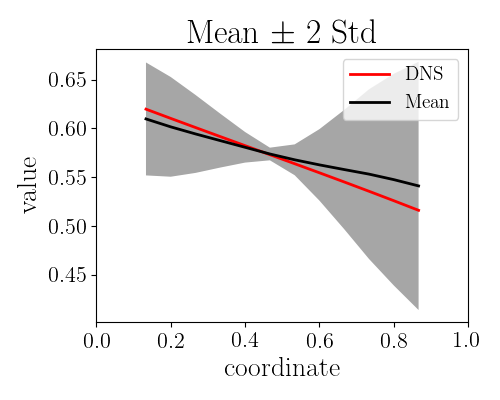}} 
  {\includegraphics[height=0.08\linewidth]{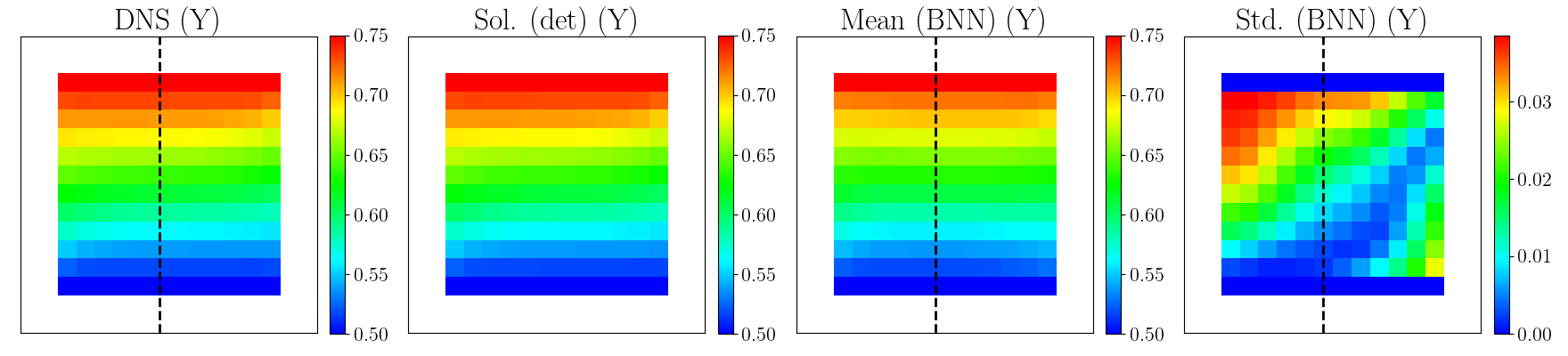}} 
  {\includegraphics[height=0.08\linewidth]{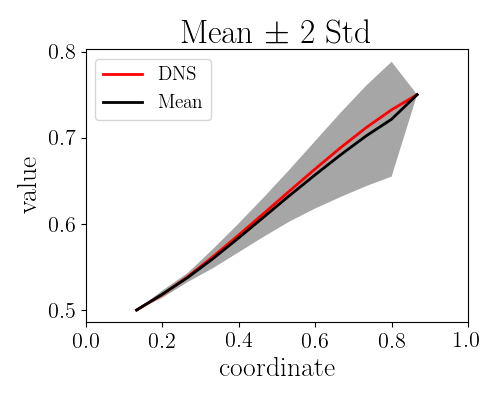}} \\
  {\includegraphics[height=0.08\linewidth]{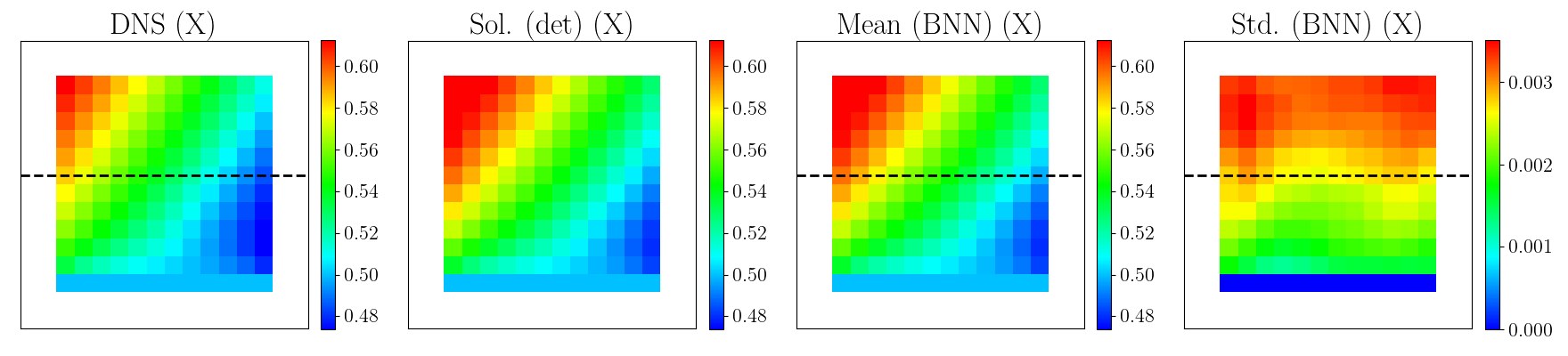}} 
  {\includegraphics[height=0.08\linewidth]{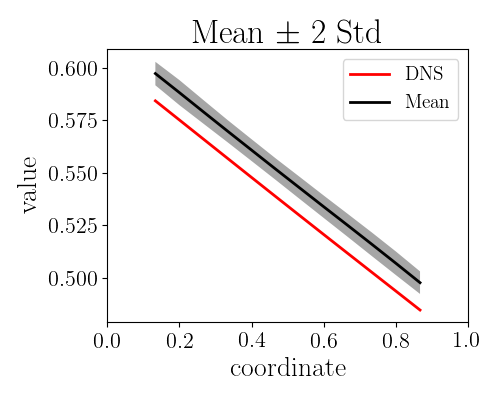}} 
  {\includegraphics[height=0.08\linewidth]{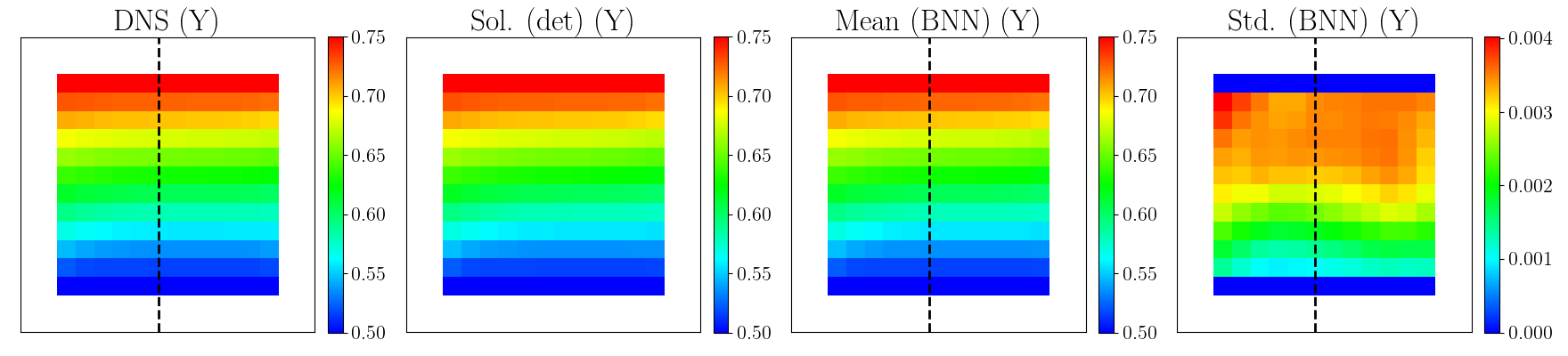}} 
  {\includegraphics[height=0.08\linewidth]{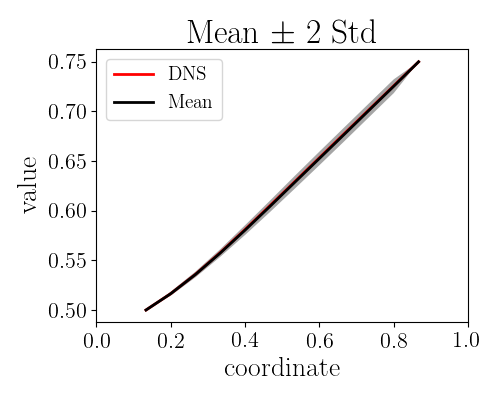}} \\
  {\includegraphics[height=0.08\linewidth]{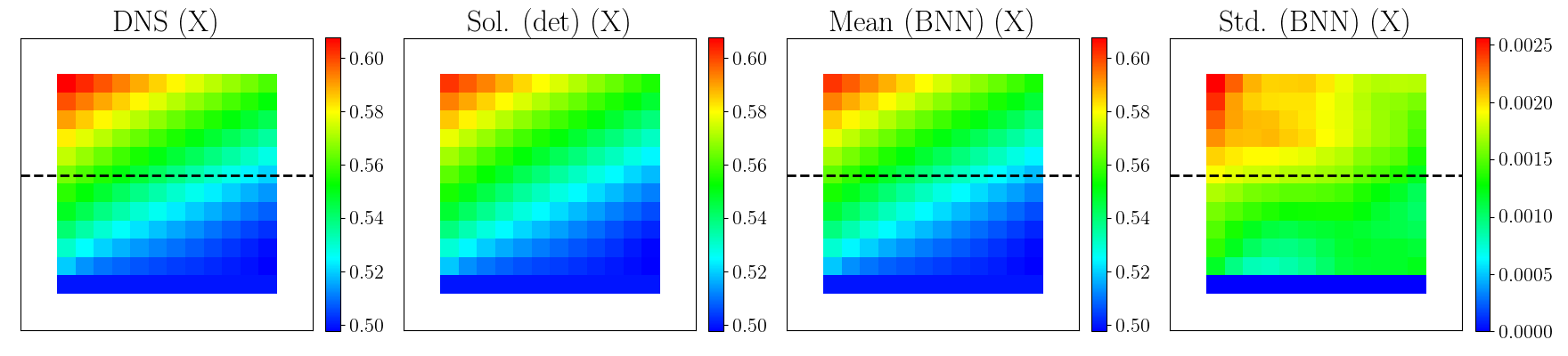}} 
  {\includegraphics[height=0.08\linewidth]{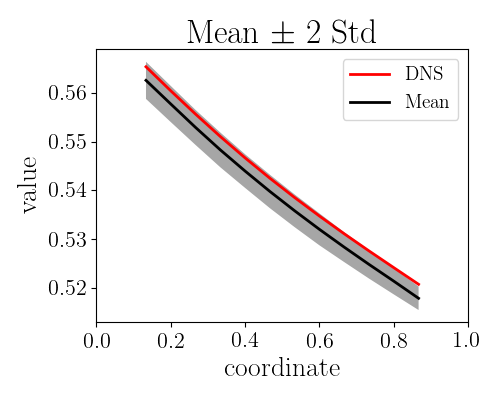}} 
  {\includegraphics[height=0.08\linewidth]{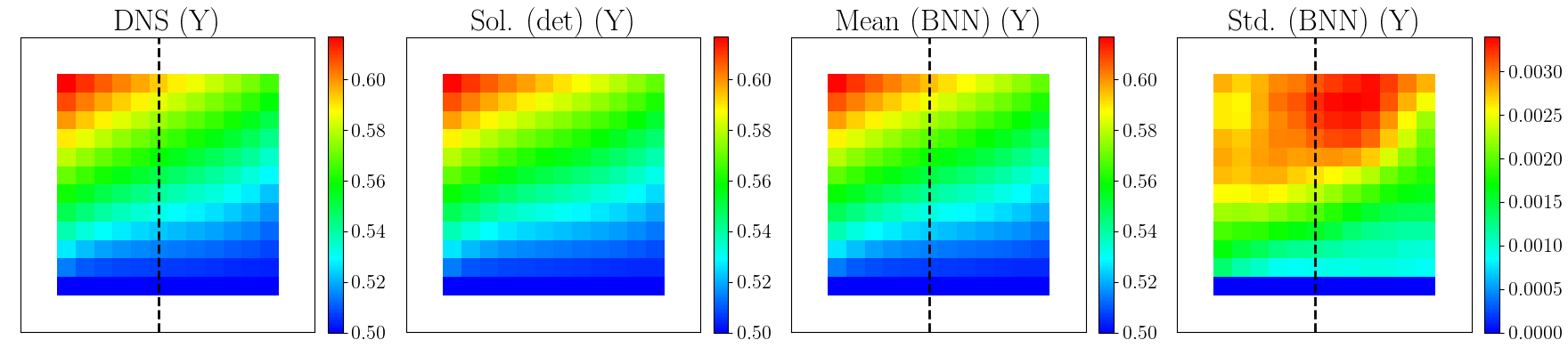}} 
  {\includegraphics[height=0.08\linewidth]{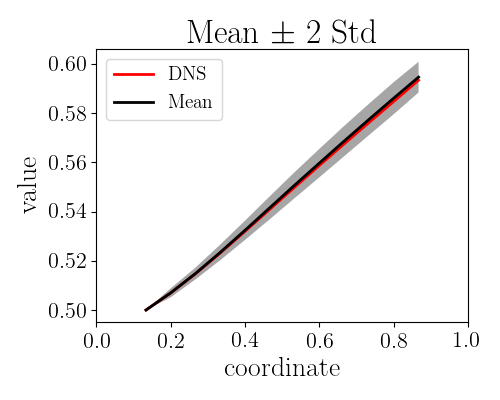}} \\
  {\includegraphics[height=0.08\linewidth]{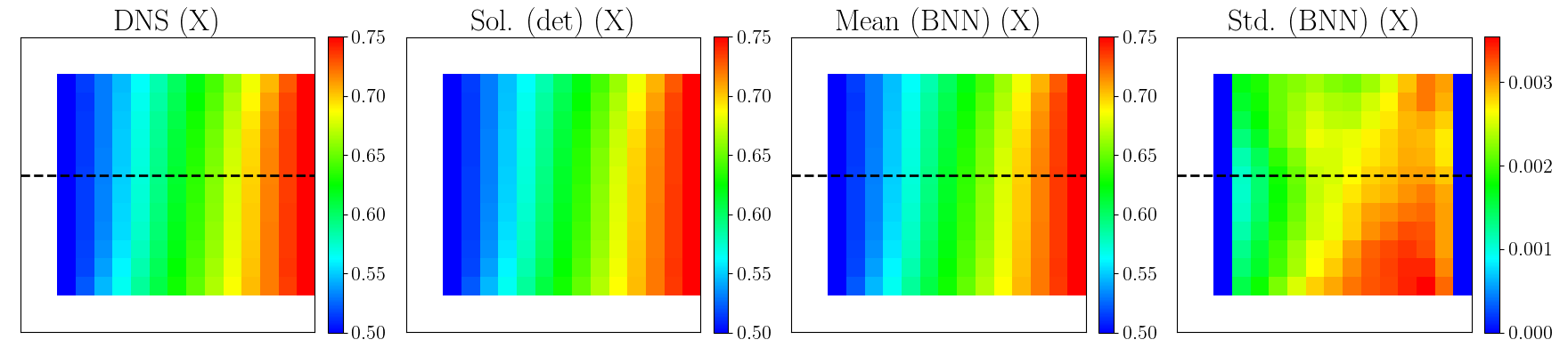}} 
  {\includegraphics[height=0.08\linewidth]{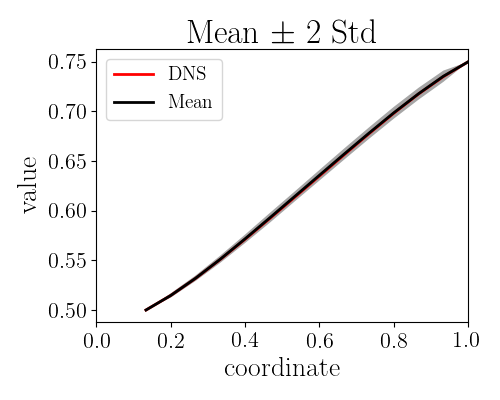}} 
  {\includegraphics[height=0.08\linewidth]{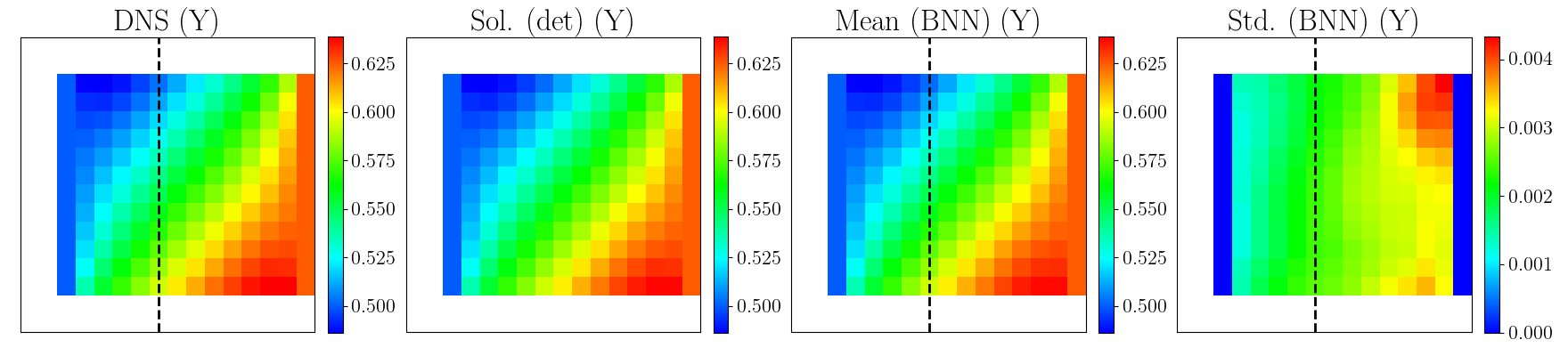}} 
  {\includegraphics[height=0.08\linewidth]{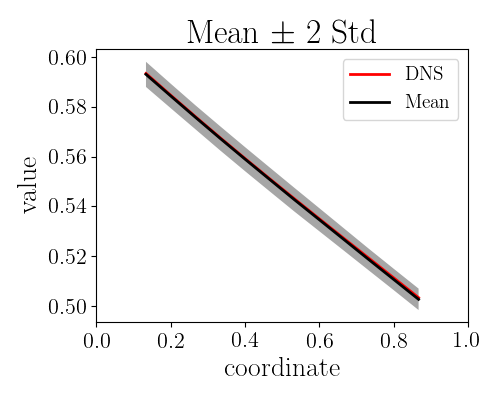}} \\
  {\includegraphics[height=0.08\linewidth]{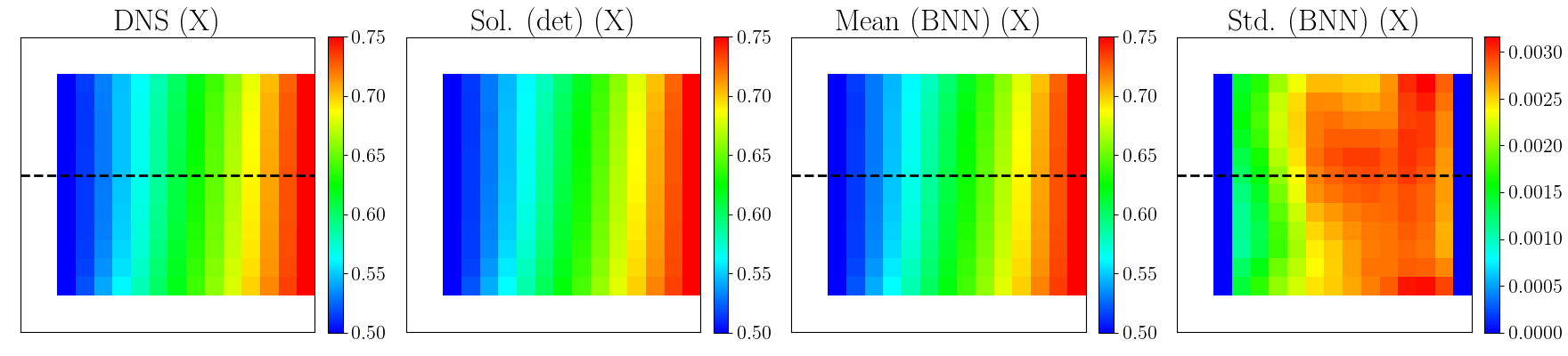}} 
  {\includegraphics[height=0.08\linewidth]{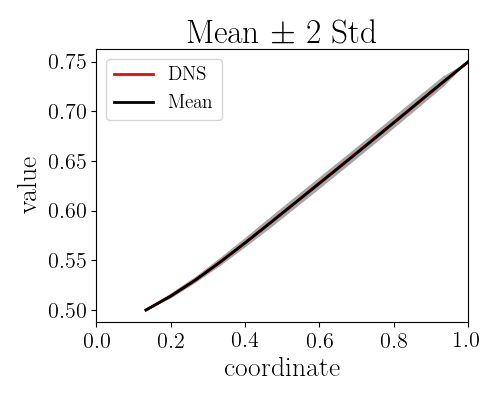}} 
  {\includegraphics[height=0.08\linewidth]{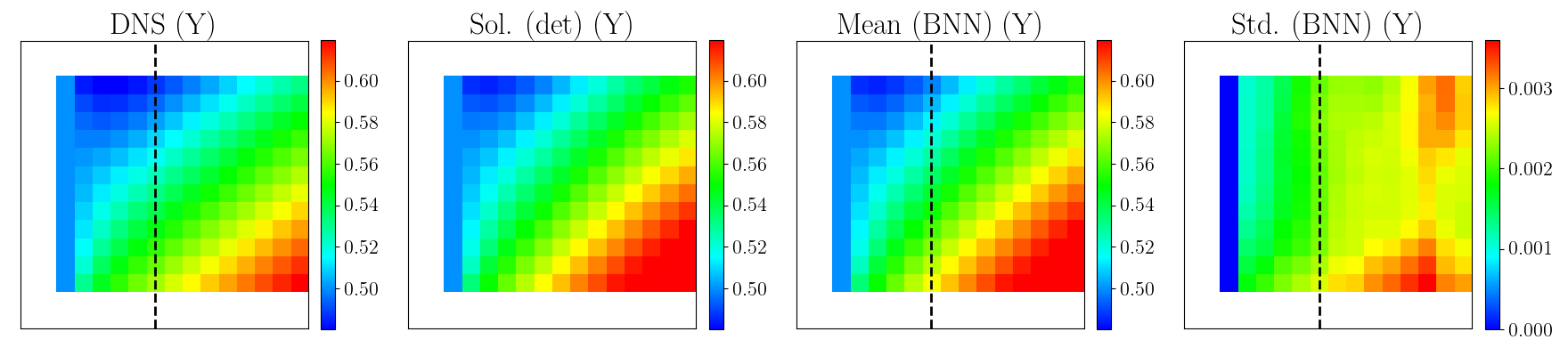}} 
  {\includegraphics[height=0.08\linewidth]{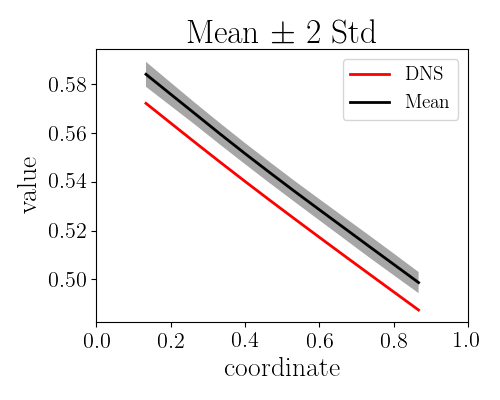}} \\
  {\includegraphics[height=0.08\linewidth]{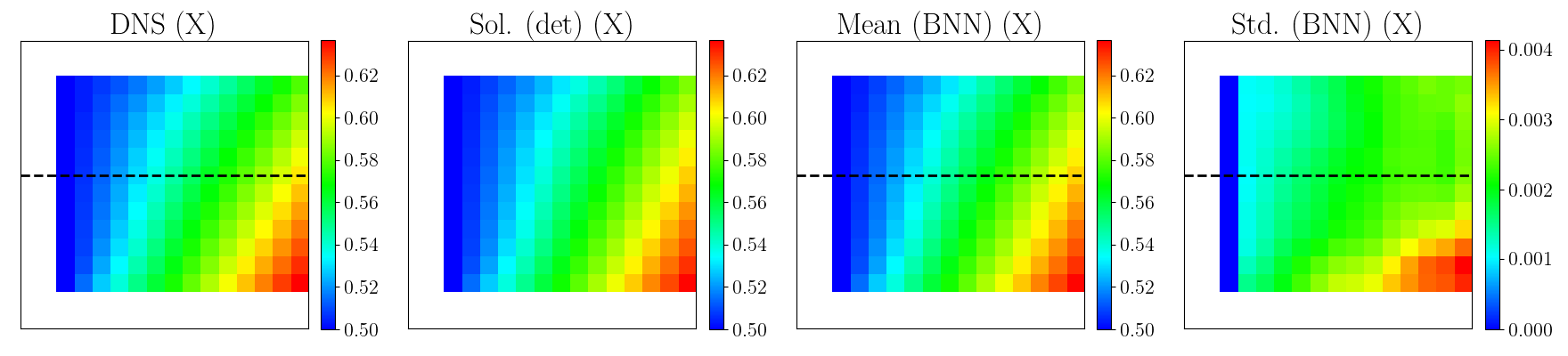}} 
  {\includegraphics[height=0.08\linewidth]{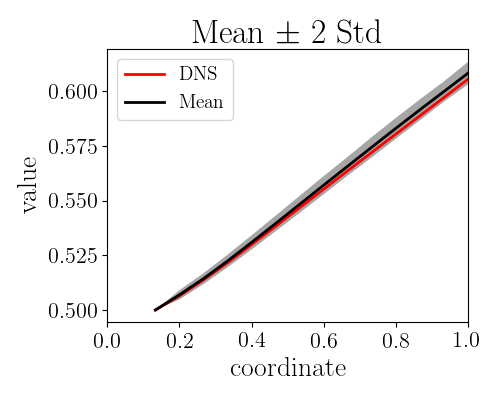}} 
  {\includegraphics[height=0.08\linewidth]{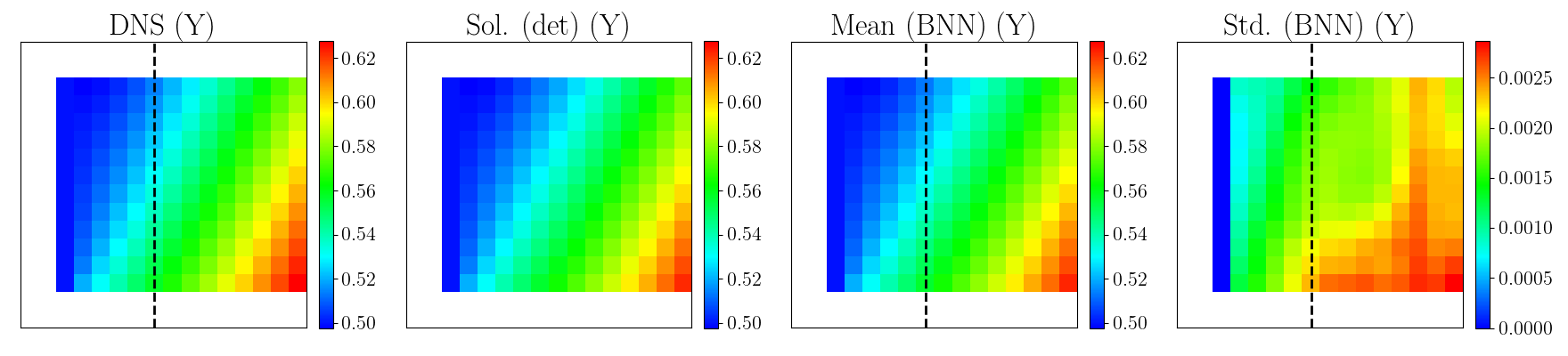}} 
  {\includegraphics[height=0.08\linewidth]{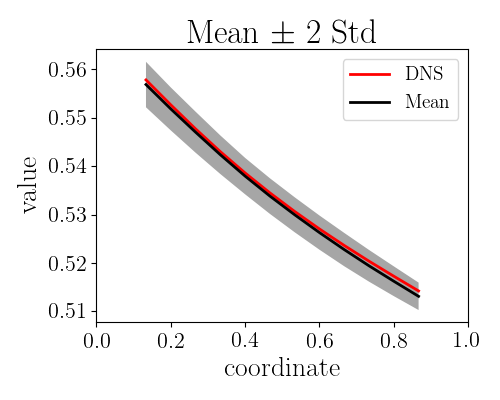}} \\
  {\includegraphics[height=0.08\linewidth]{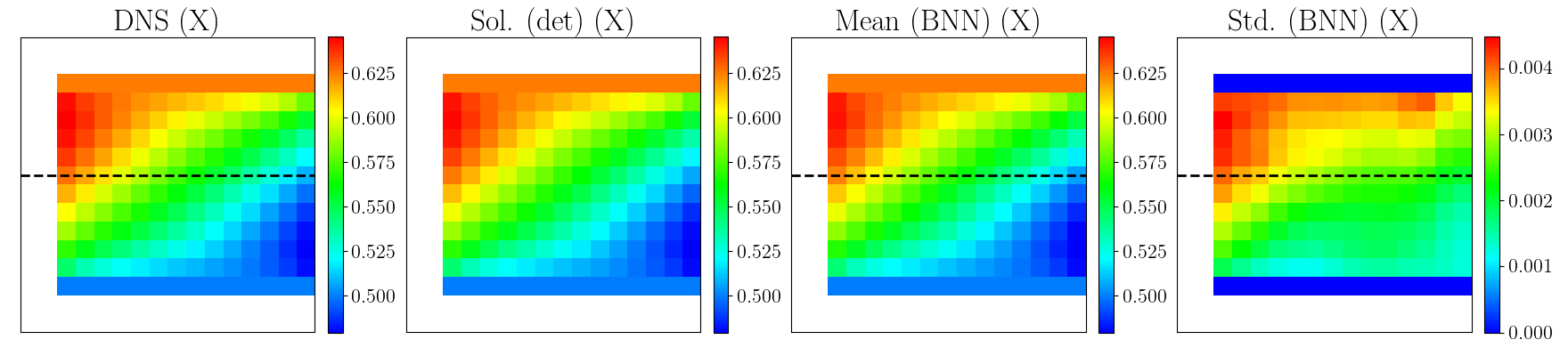}} 
  {\includegraphics[height=0.08\linewidth]{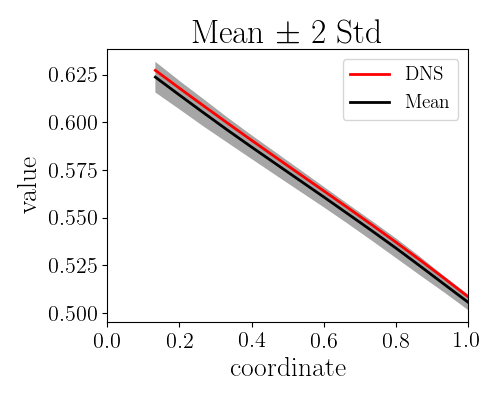}} 
  {\includegraphics[height=0.08\linewidth]{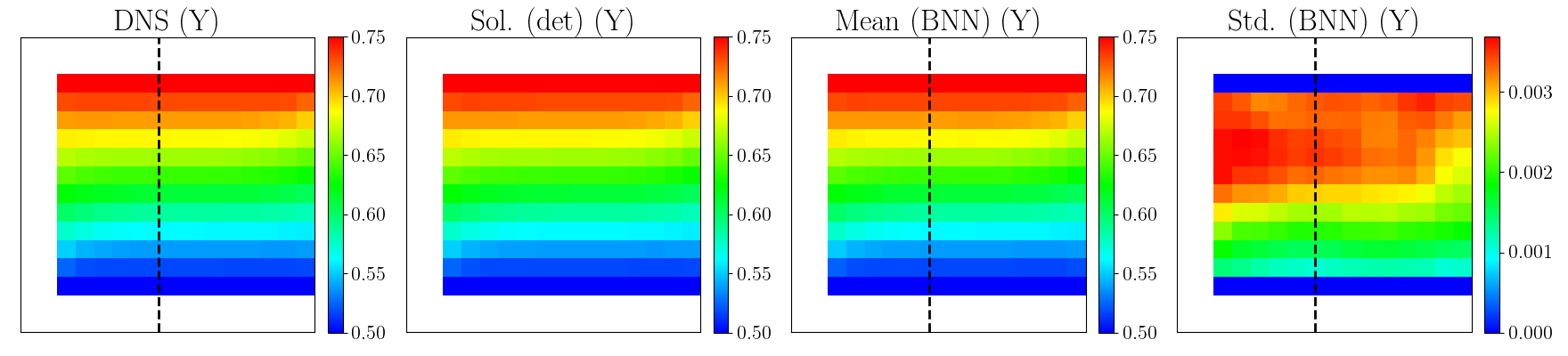}} 
  {\includegraphics[height=0.08\linewidth]{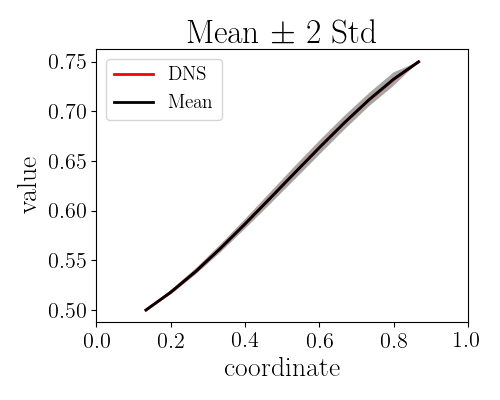}} \\
  {\includegraphics[height=0.08\linewidth]{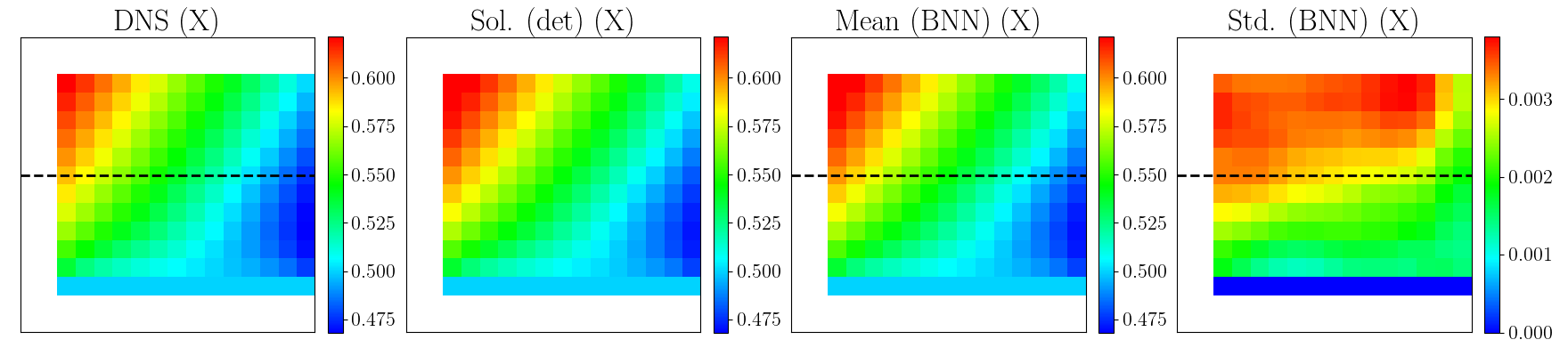}} 
  {\includegraphics[height=0.08\linewidth]{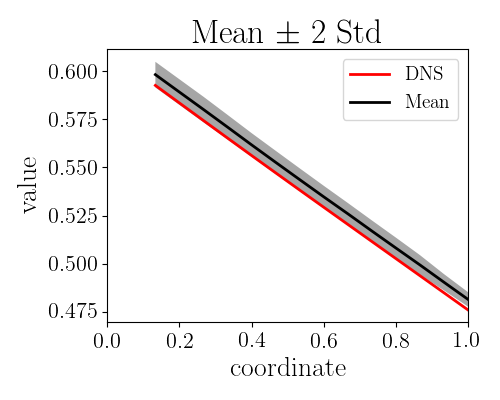}} 
  {\includegraphics[height=0.08\linewidth]{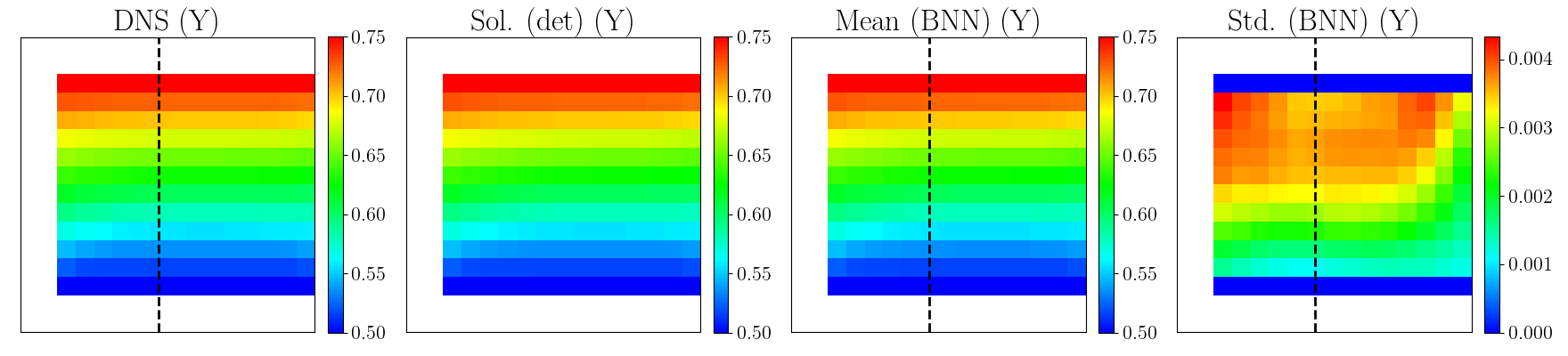}} 
  {\includegraphics[height=0.08\linewidth]{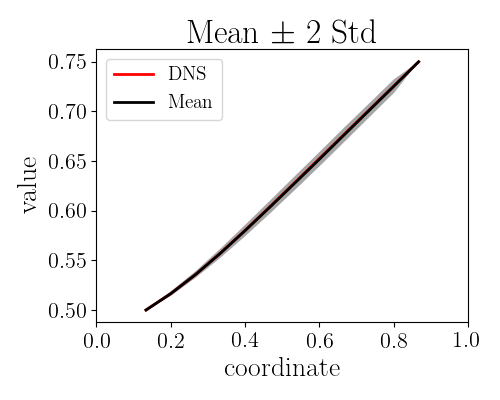}} \\
  {\includegraphics[height=0.08\linewidth]{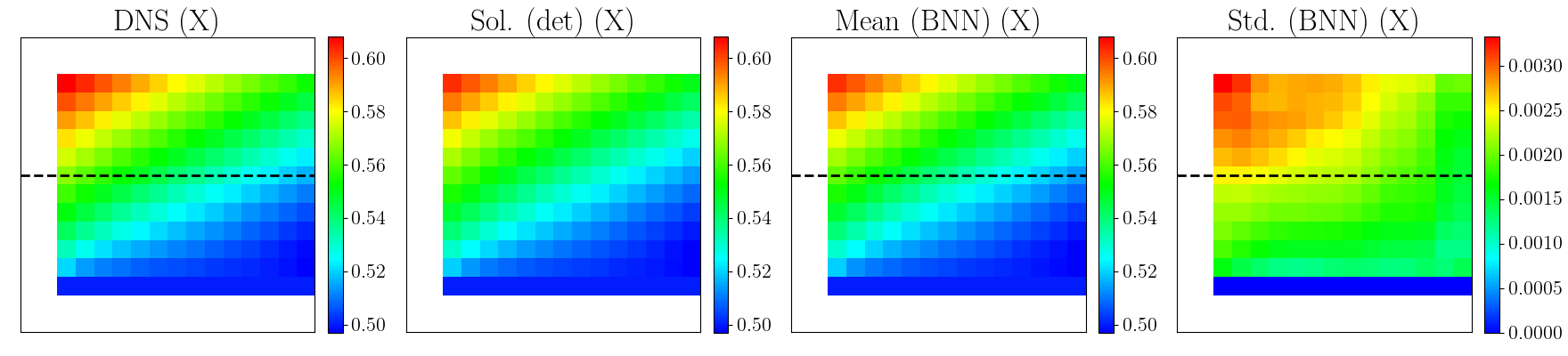}} 
  {\includegraphics[height=0.08\linewidth]{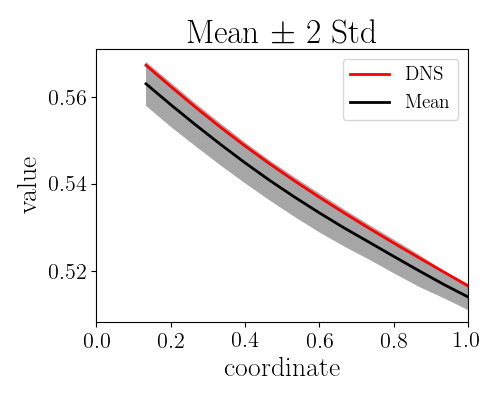}} 
  {\includegraphics[height=0.08\linewidth]{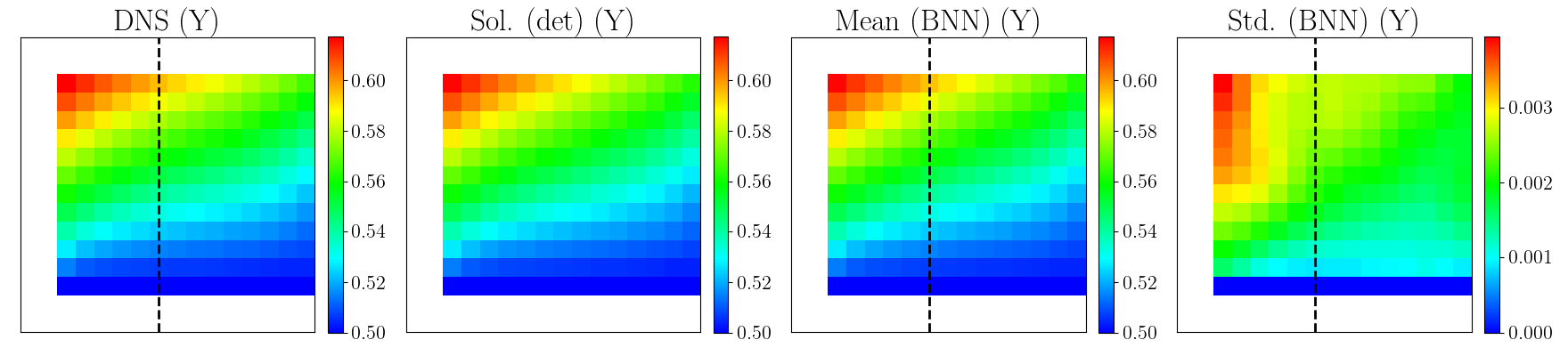}} 
  {\includegraphics[height=0.08\linewidth]{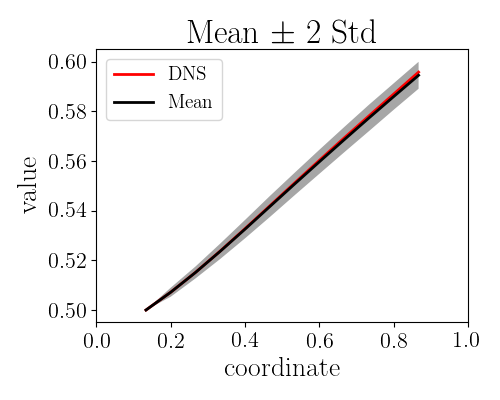}} \\
  \caption{Additional NN results for linear BVPs on rectangle domains (continue).}
  \label{fig:nonlinear-30bvp-results-additional-2}
\end{figure}

Additional results from the 30 BVPs with rectangle domains are summarized in Fig. \ref{fig:nonlinear-30bvp-results-additional-1} and \ref{fig:nonlinear-30bvp-results-additional-2}.

%6,7,8

\subsubsection{Rectangle domain with solution interpolation and extrapolation} \label{appendix:nonlinear-inter-extra-additional}

\begin{figure}[p!]
  \centering
  {\includegraphics[height=0.08\linewidth]{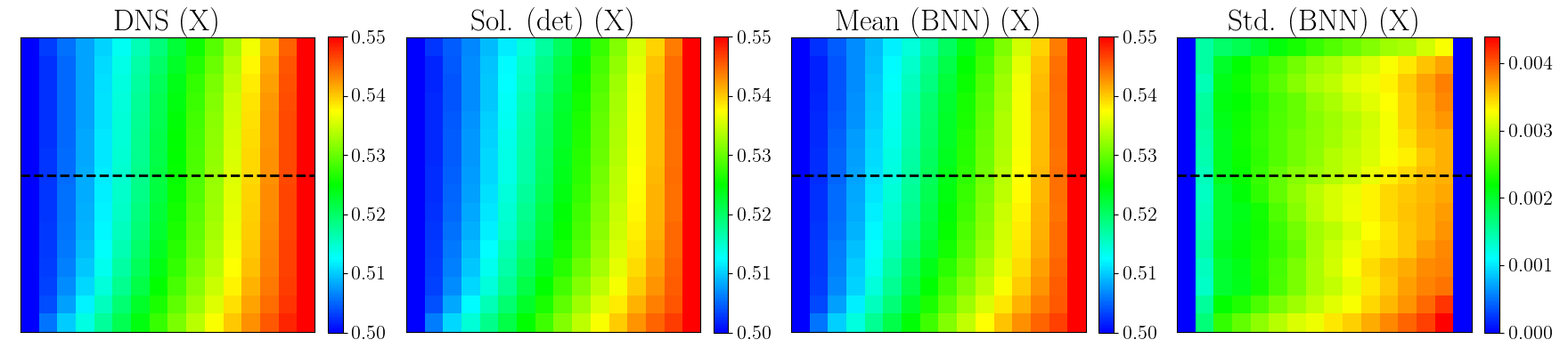}}
  {\includegraphics[height=0.08\linewidth]{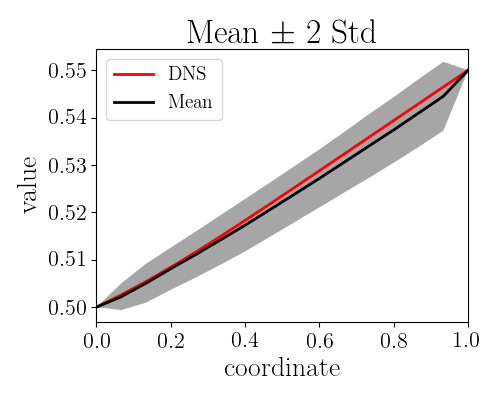}} 
  {\includegraphics[height=0.08\linewidth]{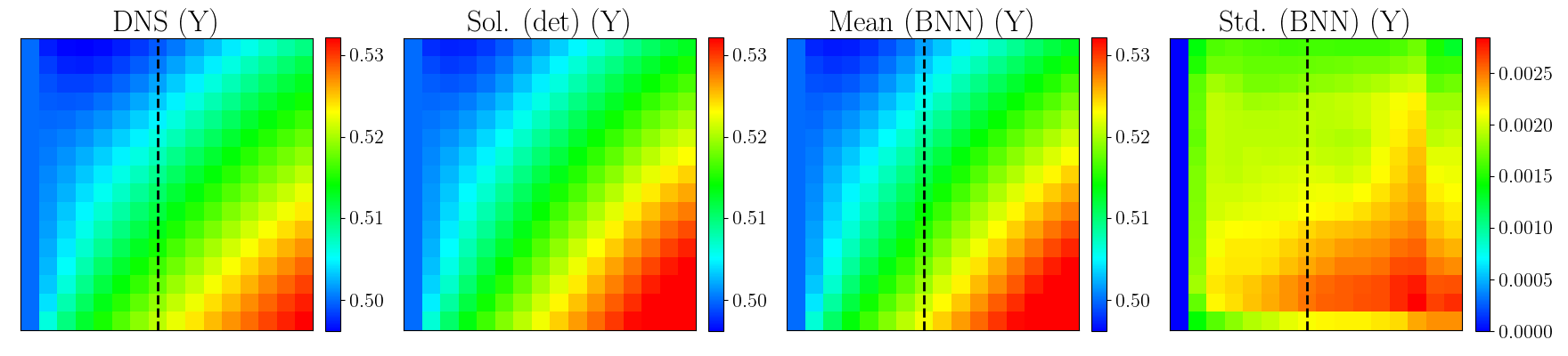}}
  {\includegraphics[height=0.08\linewidth]{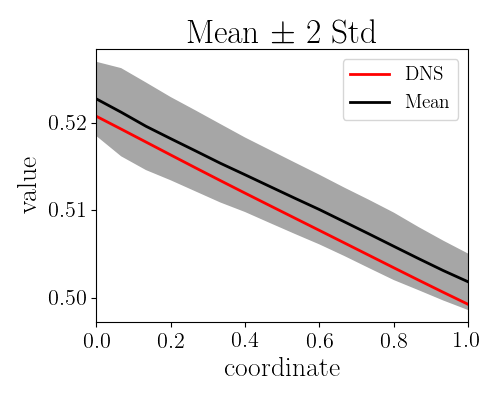}} \\
  {\includegraphics[height=0.08\linewidth]{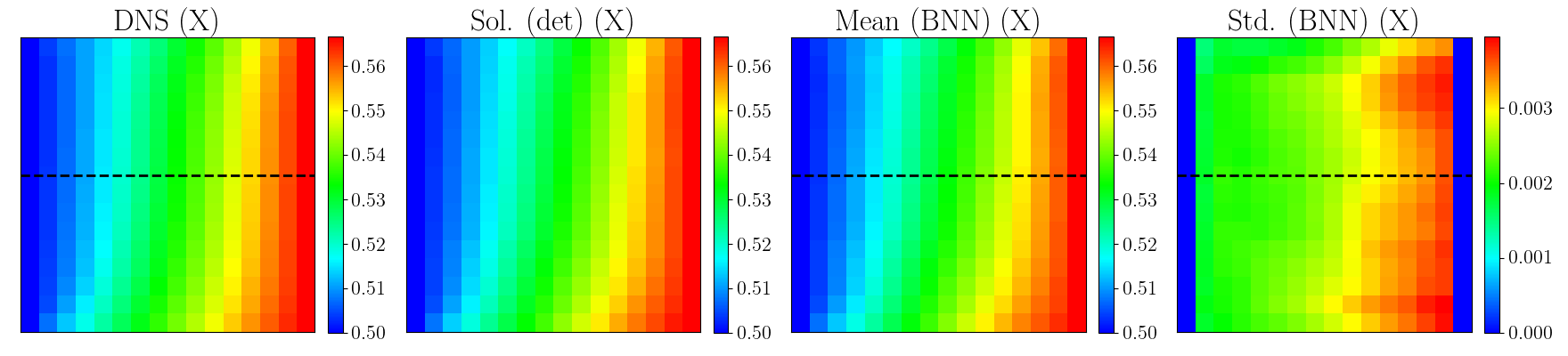}}
  {\includegraphics[height=0.08\linewidth]{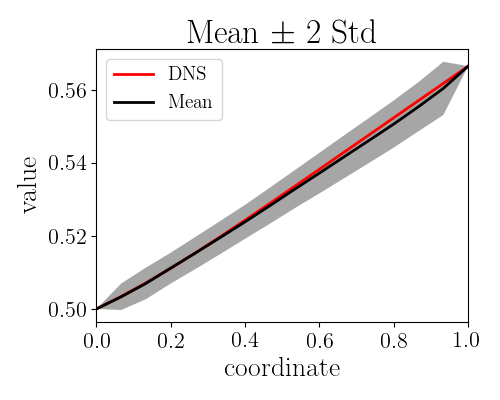}} 
  {\includegraphics[height=0.08\linewidth]{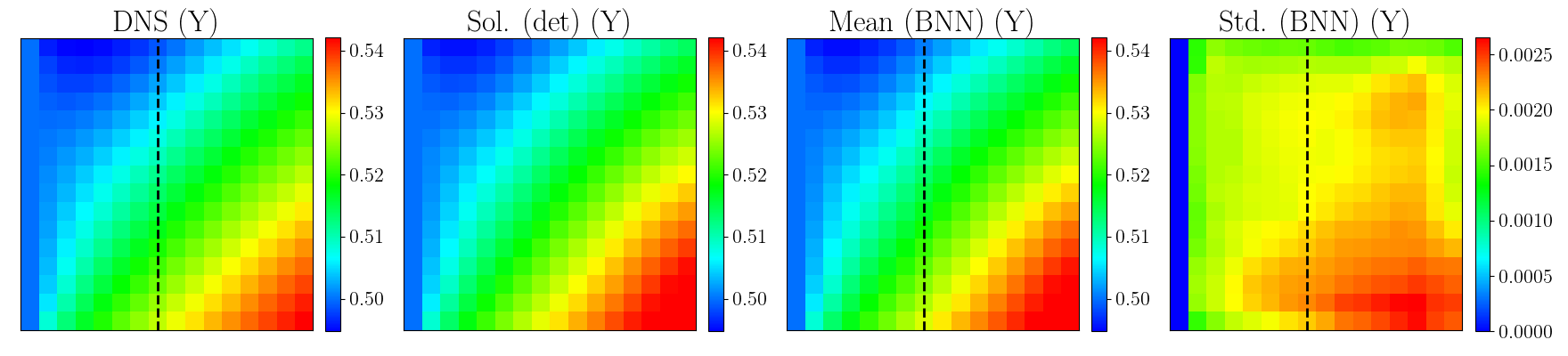}}
  {\includegraphics[height=0.08\linewidth]{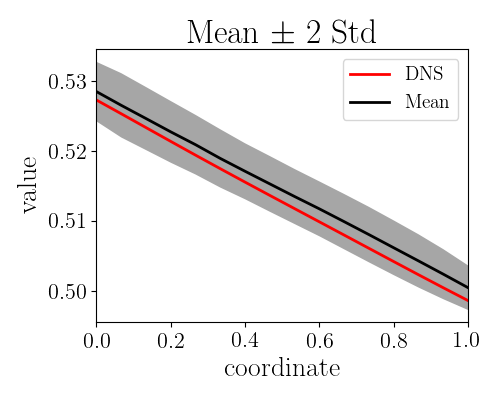}} \\
  {\includegraphics[height=0.08\linewidth]{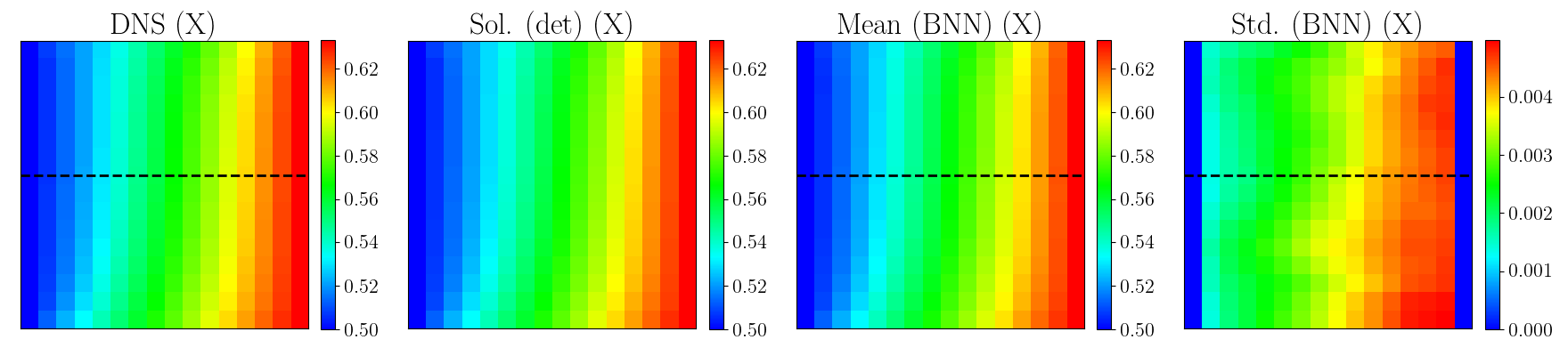}}
  {\includegraphics[height=0.08\linewidth]{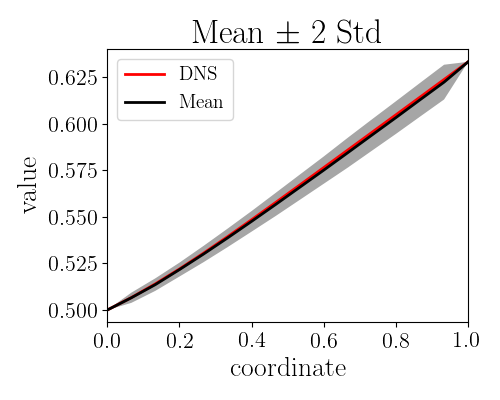}} 
  {\includegraphics[height=0.08\linewidth]{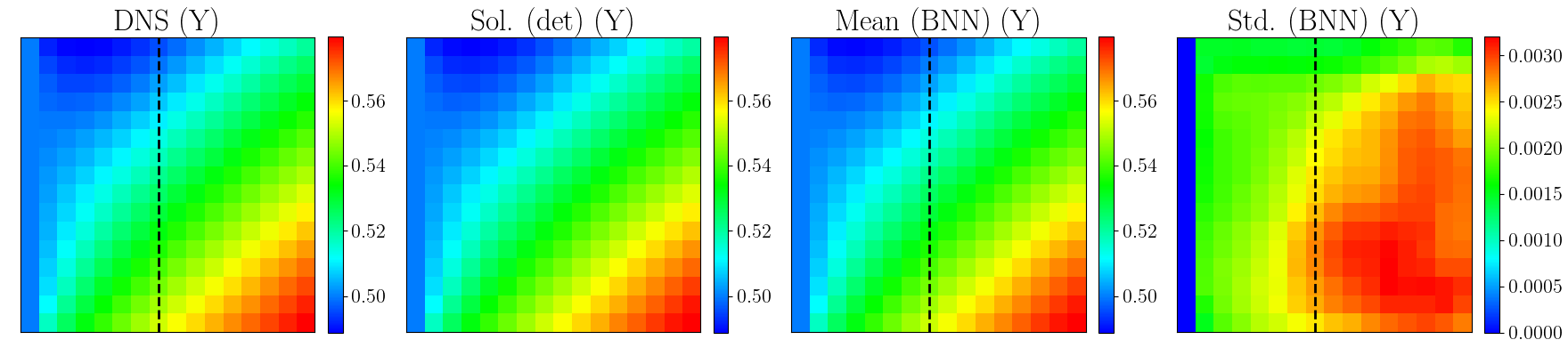}}
  {\includegraphics[height=0.08\linewidth]{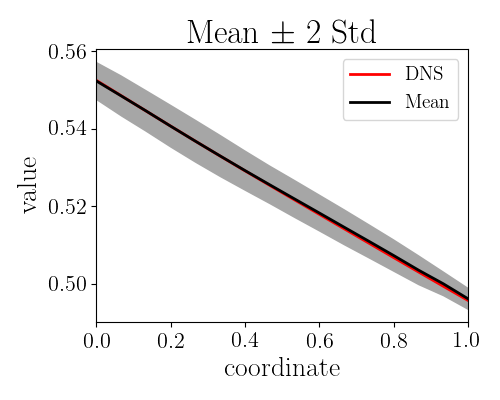}} \\
  {\includegraphics[height=0.08\linewidth]{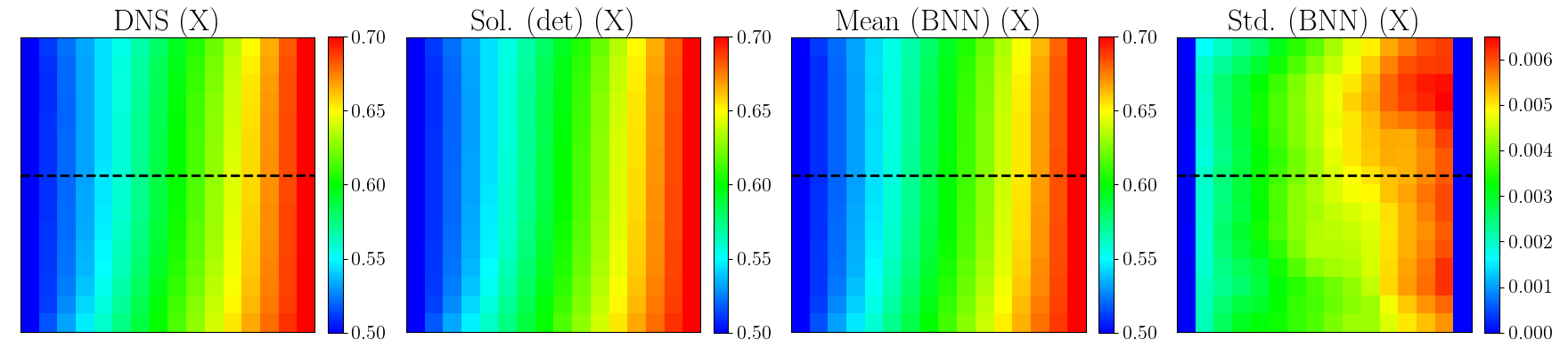}}
  {\includegraphics[height=0.08\linewidth]{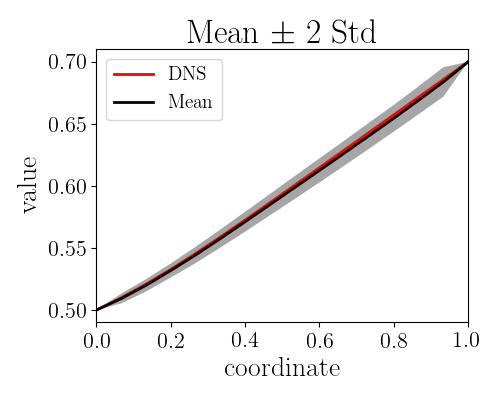}} 
  {\includegraphics[height=0.08\linewidth]{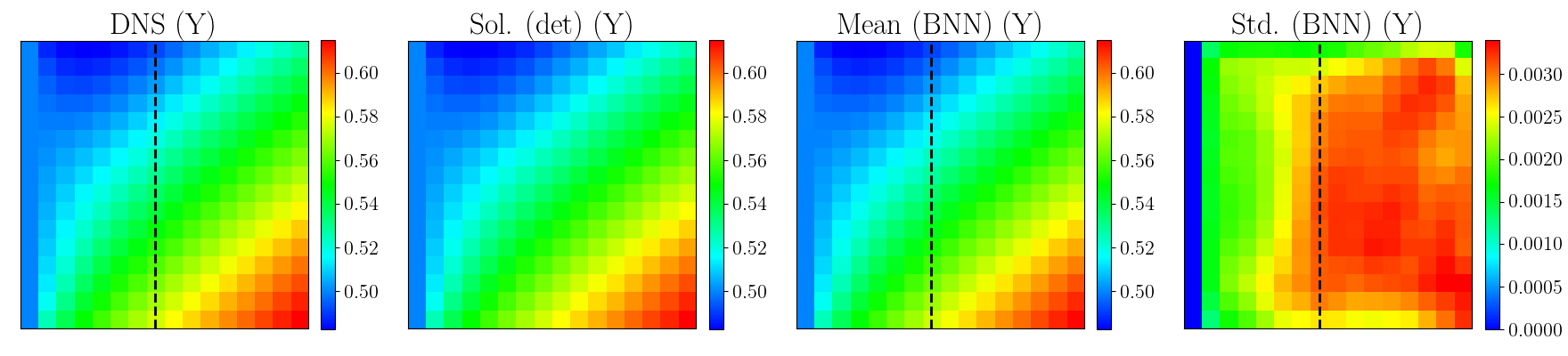}}
  {\includegraphics[height=0.08\linewidth]{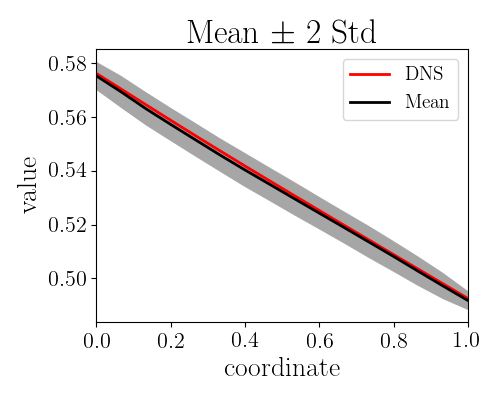}} \\
  {\includegraphics[height=0.08\linewidth]{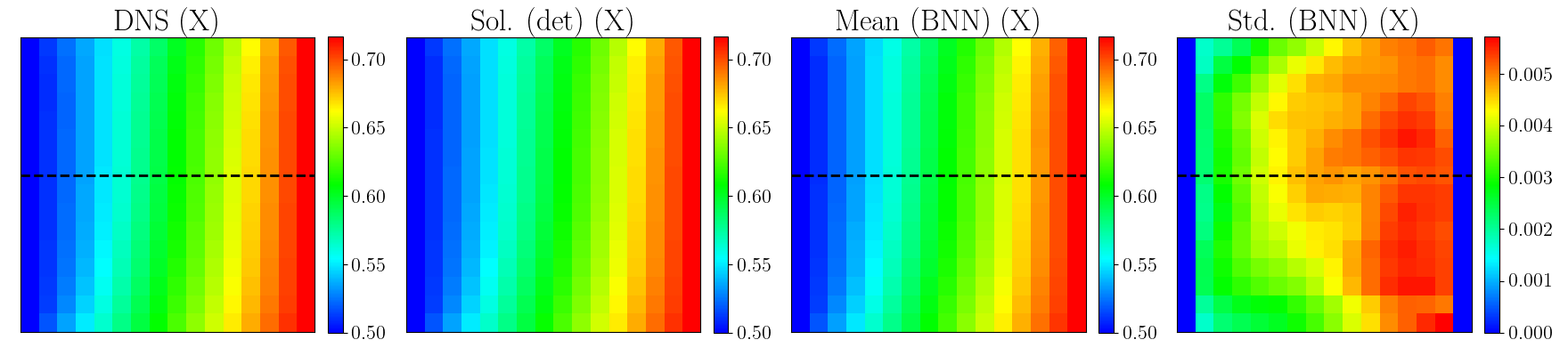}}
  {\includegraphics[height=0.08\linewidth]{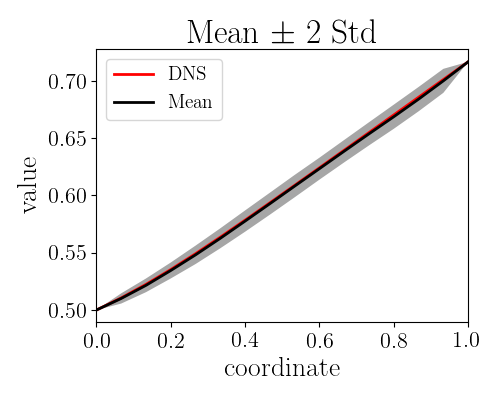}} 
  {\includegraphics[height=0.08\linewidth]{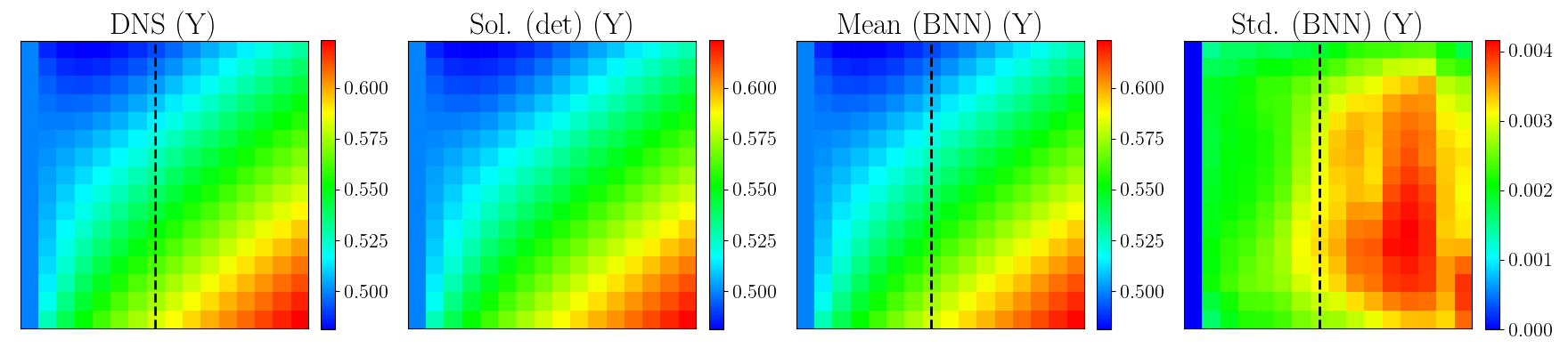}}
  {\includegraphics[height=0.08\linewidth]{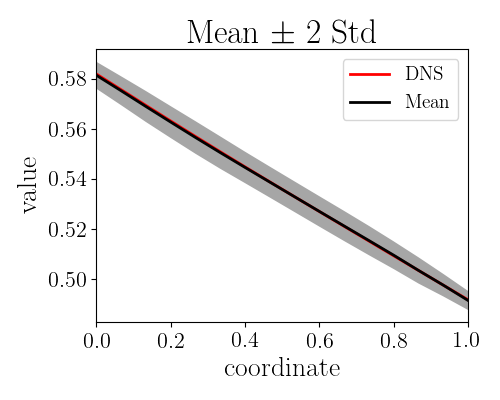}} \\
  {\includegraphics[height=0.08\linewidth]{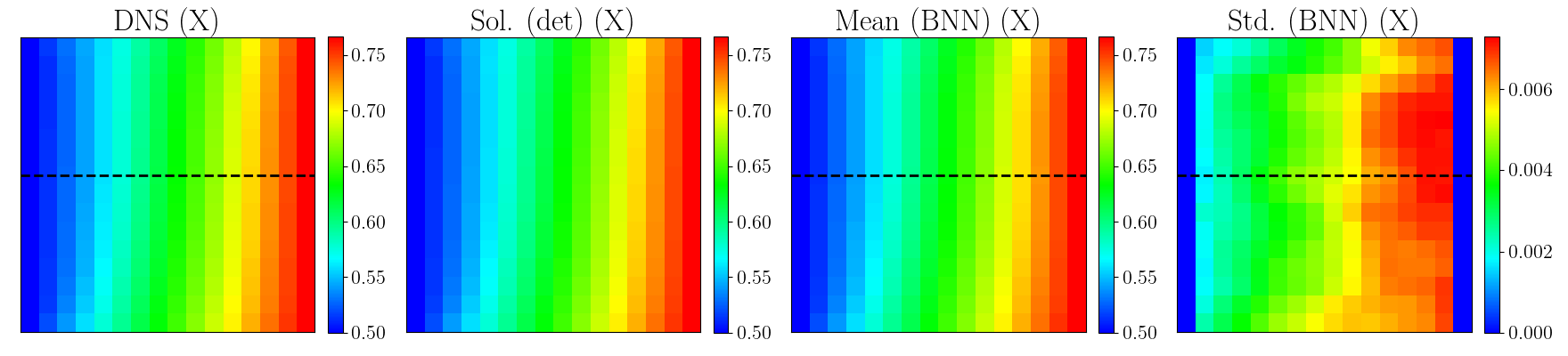}}
  {\includegraphics[height=0.08\linewidth]{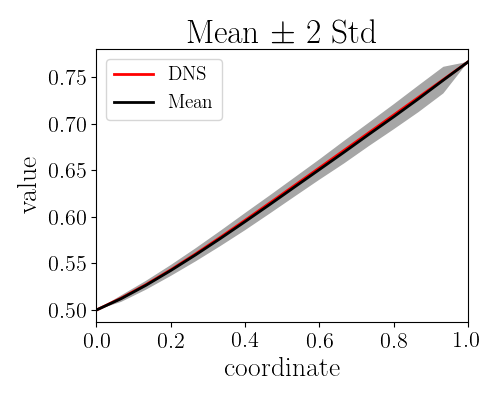}} 
  {\includegraphics[height=0.08\linewidth]{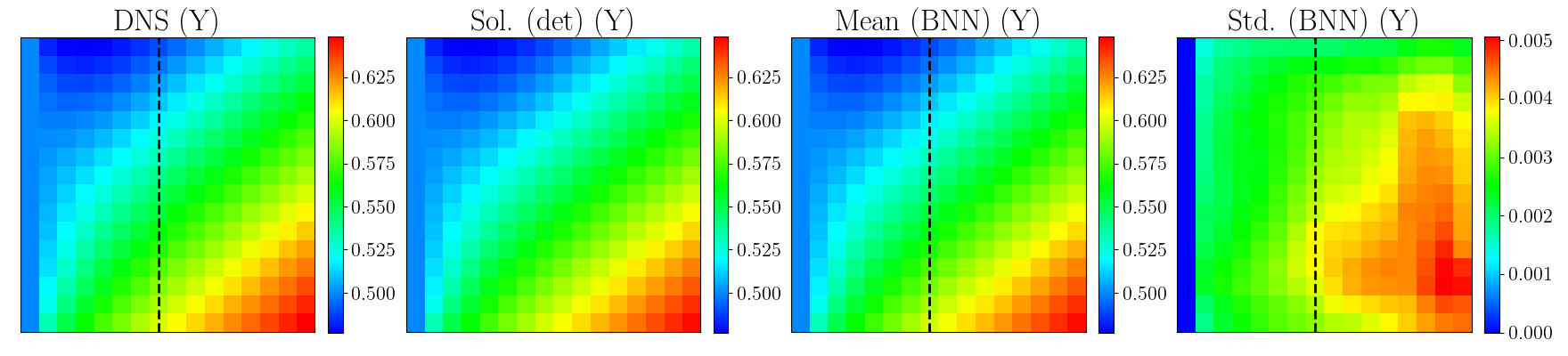}}
  {\includegraphics[height=0.08\linewidth]{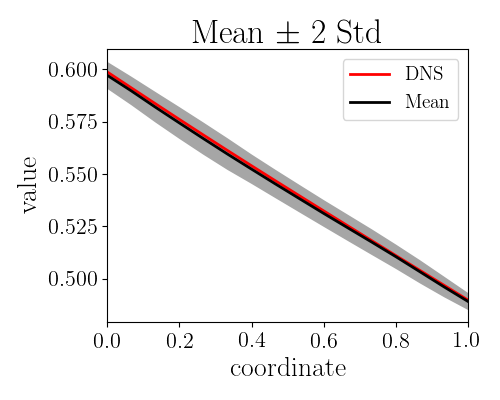}} \\
  {\includegraphics[height=0.08\linewidth]{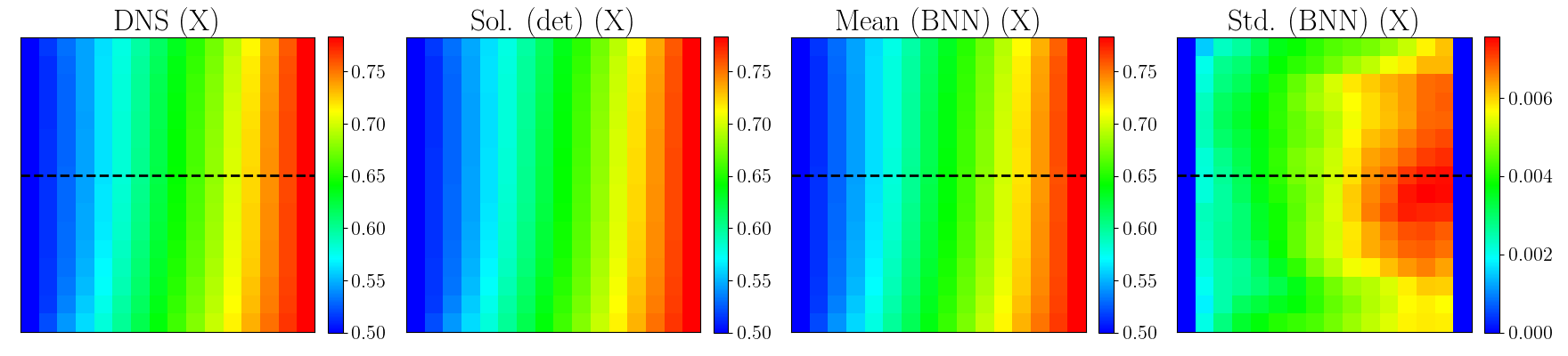}}
  {\includegraphics[height=0.08\linewidth]{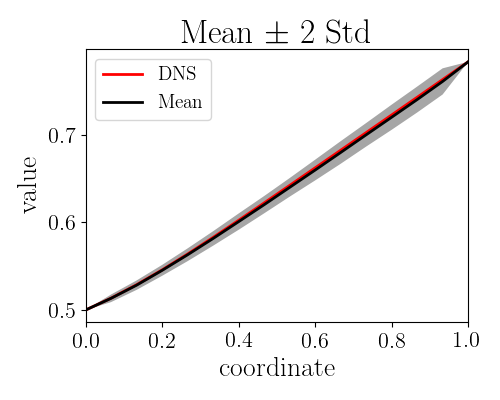}} 
  {\includegraphics[height=0.08\linewidth]{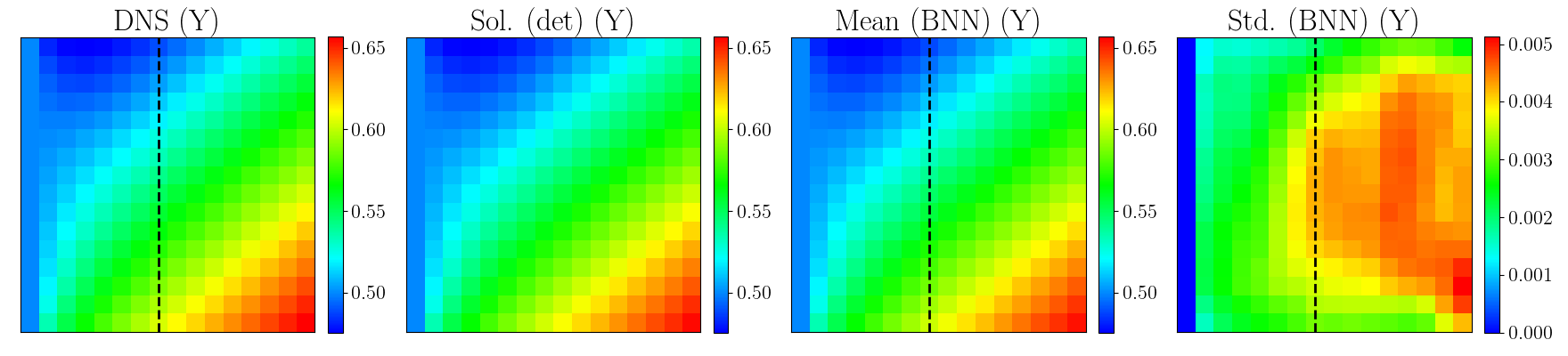}}
  {\includegraphics[height=0.08\linewidth]{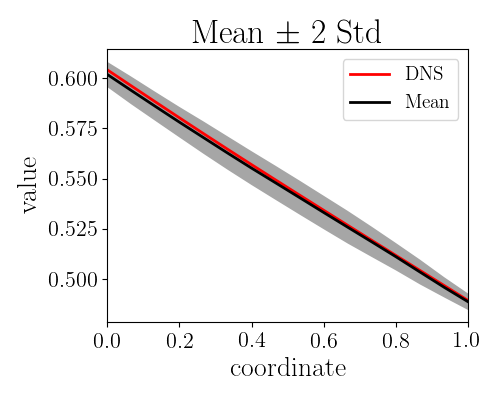}} \\
  {\includegraphics[height=0.08\linewidth]{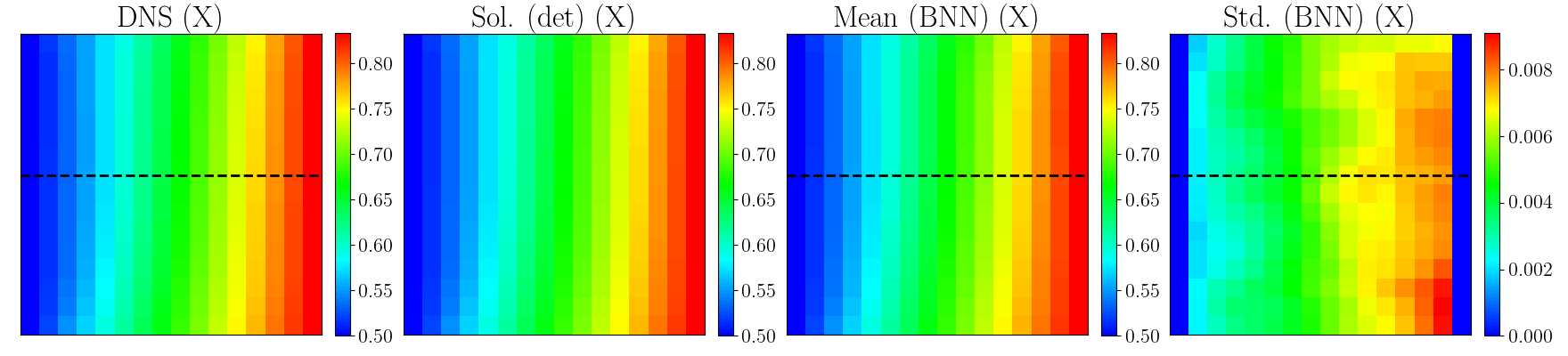}}
  {\includegraphics[height=0.08\linewidth]{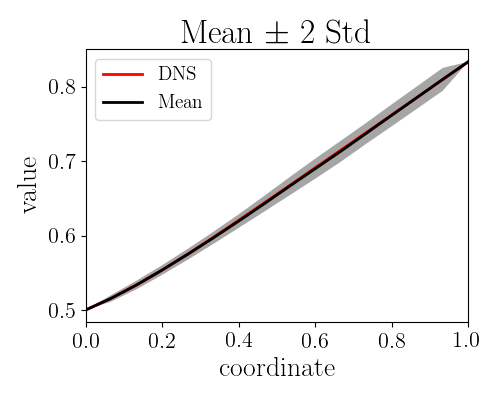}} 
  {\includegraphics[height=0.08\linewidth]{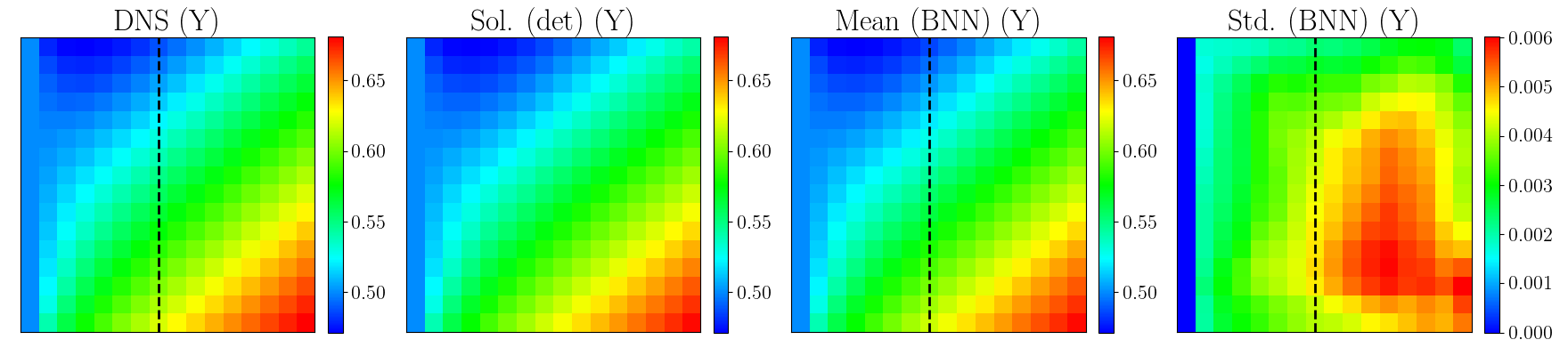}}
  {\includegraphics[height=0.08\linewidth]{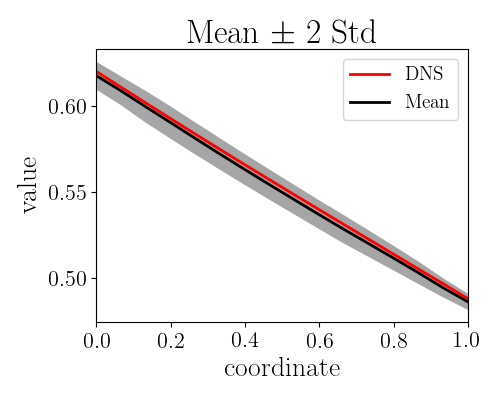}} \\
  {\includegraphics[height=0.08\linewidth]{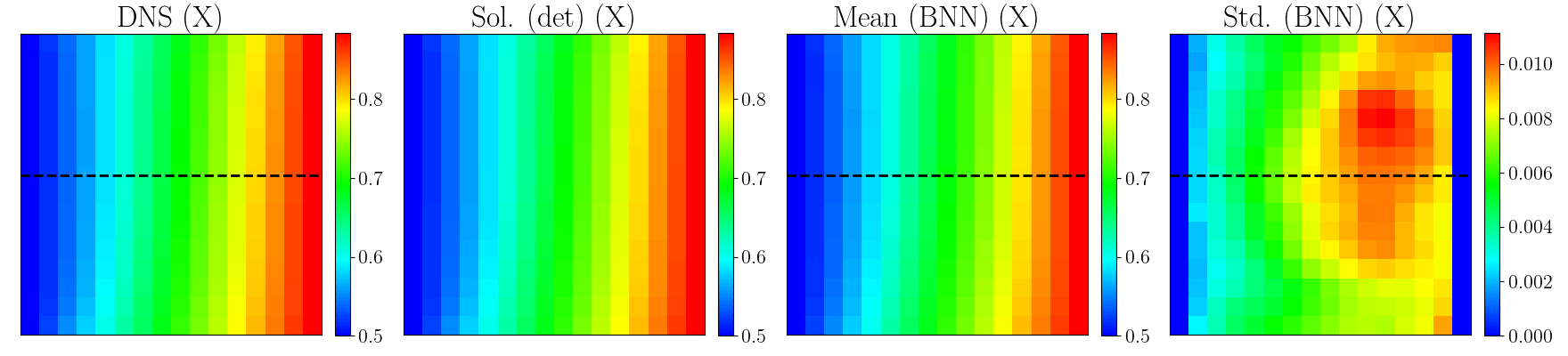}}
  {\includegraphics[height=0.08\linewidth]{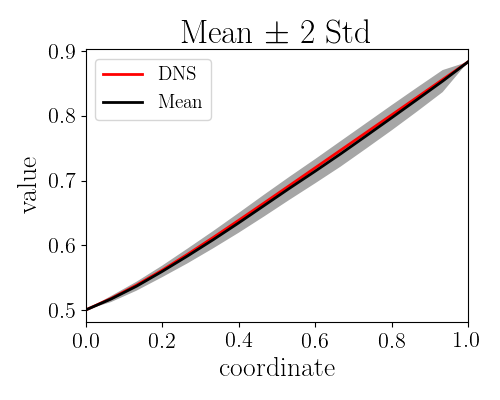}} 
  {\includegraphics[height=0.08\linewidth]{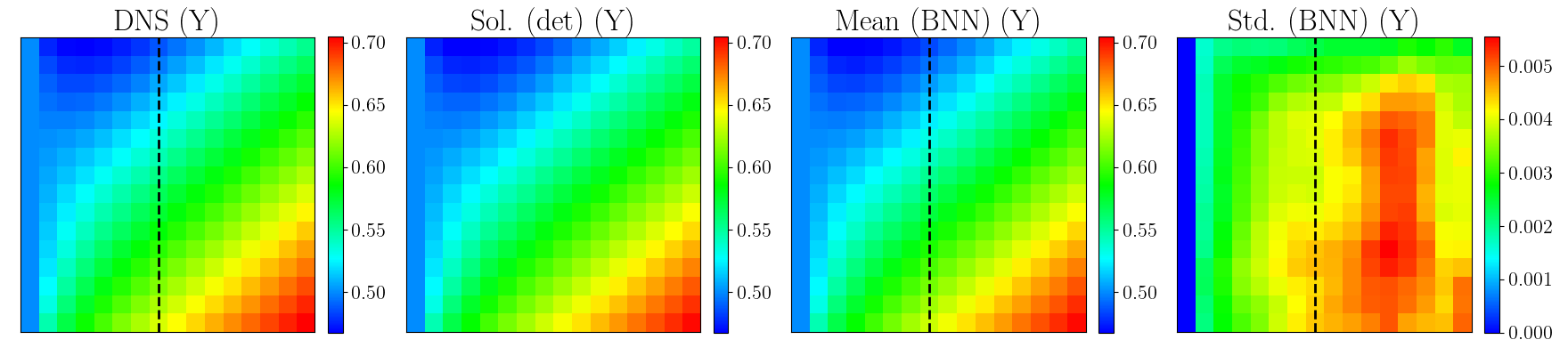}}
  {\includegraphics[height=0.08\linewidth]{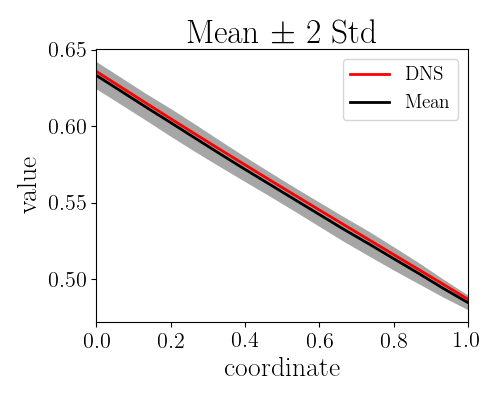}} \\
  \caption{Additional interpolating NN prediction results for case (i).}
  \label{fig:nonlinear-inter-additional}
\end{figure}

\begin{figure}[p!]
  \centering
  {\includegraphics[height=0.08\linewidth]{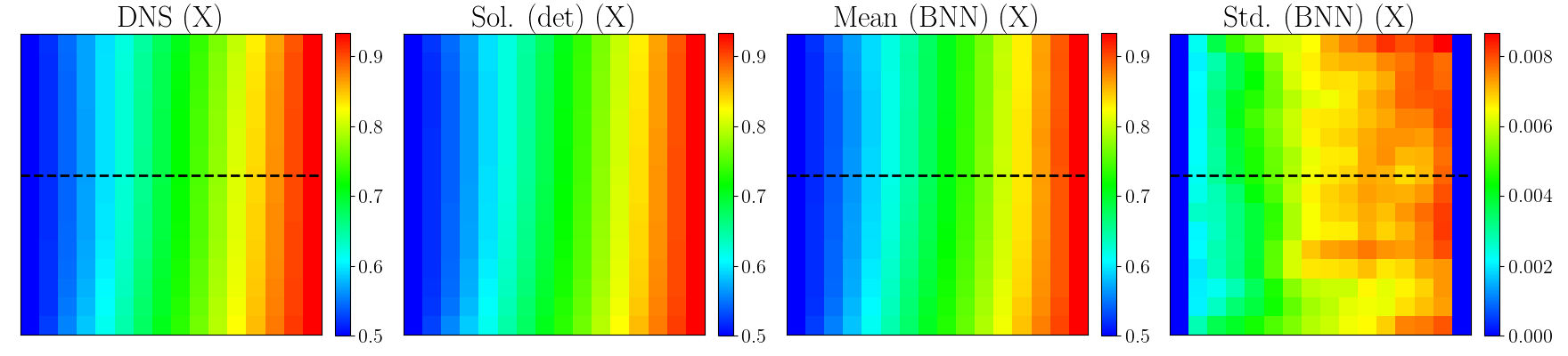}}
  {\includegraphics[height=0.08\linewidth]{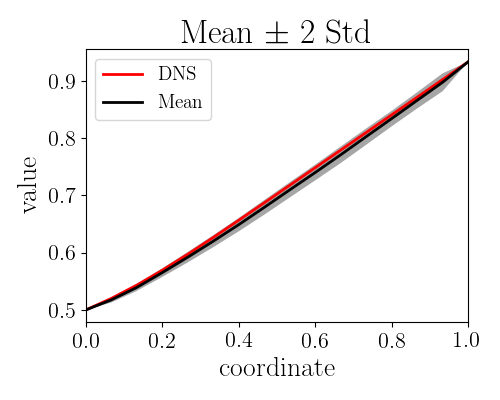}} 
  {\includegraphics[height=0.08\linewidth]{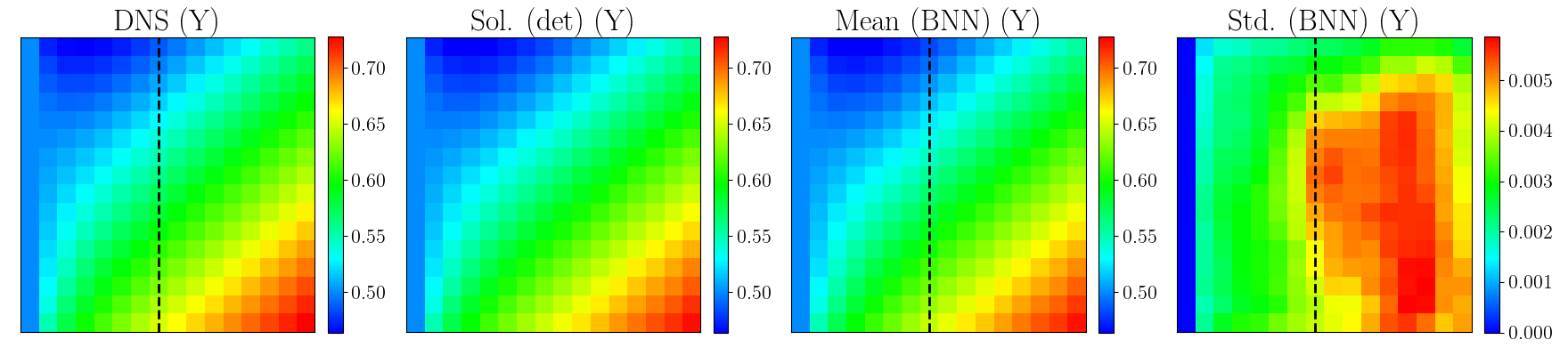}}
  {\includegraphics[height=0.08\linewidth]{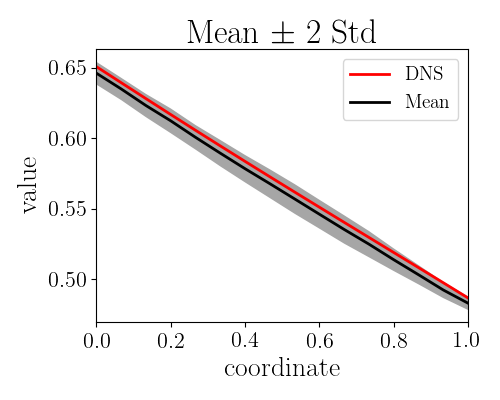}} \\
  {\includegraphics[height=0.08\linewidth]{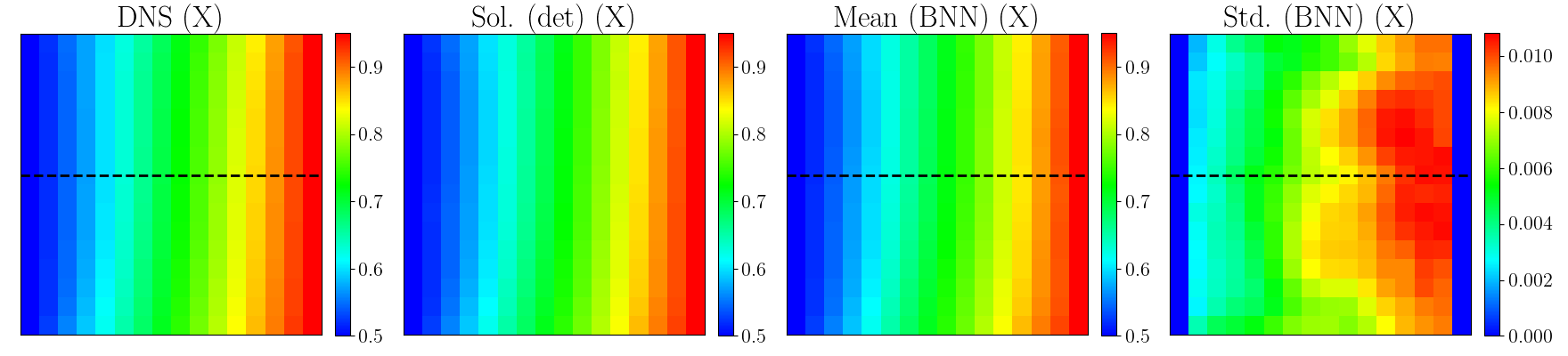}}
  {\includegraphics[height=0.08\linewidth]{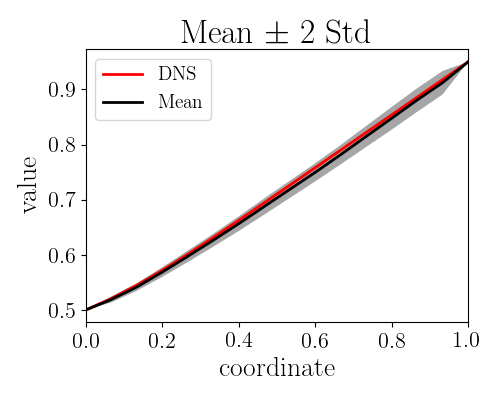}} 
  {\includegraphics[height=0.08\linewidth]{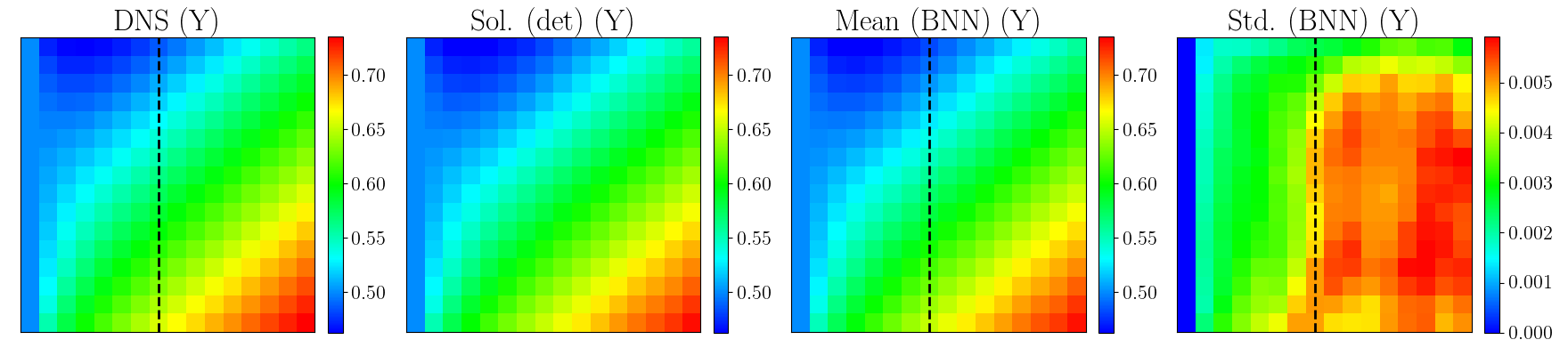}}
  {\includegraphics[height=0.08\linewidth]{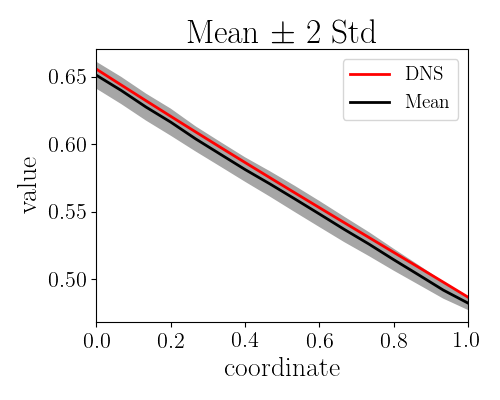}} \\
  {\includegraphics[height=0.08\linewidth]{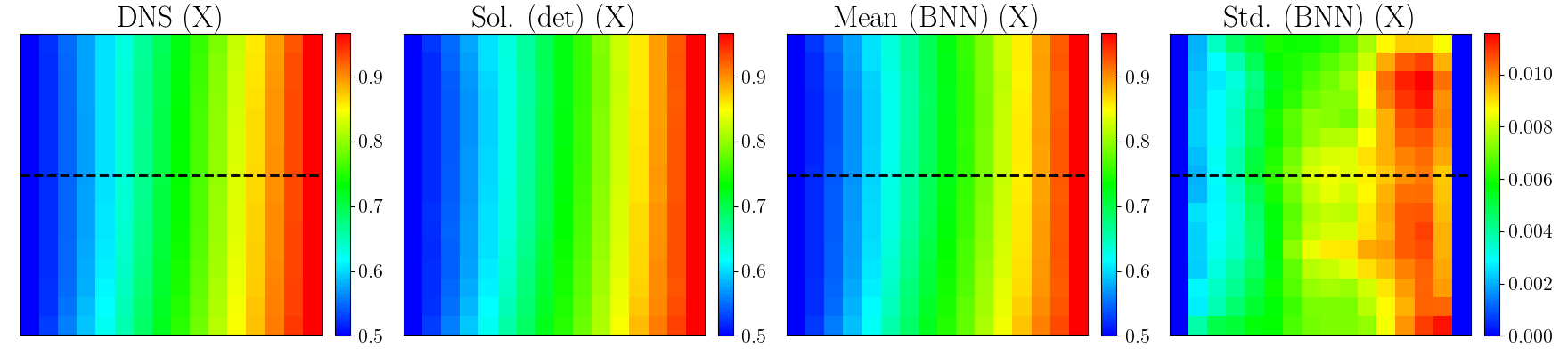}}
  {\includegraphics[height=0.08\linewidth]{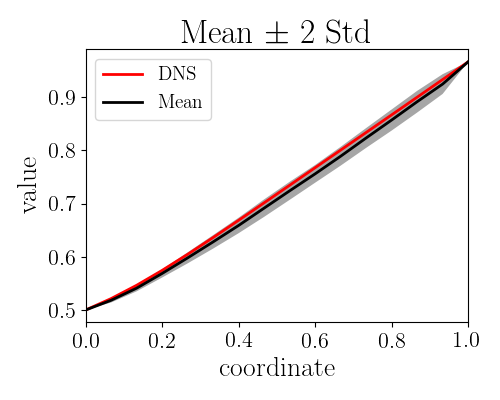}} 
  {\includegraphics[height=0.08\linewidth]{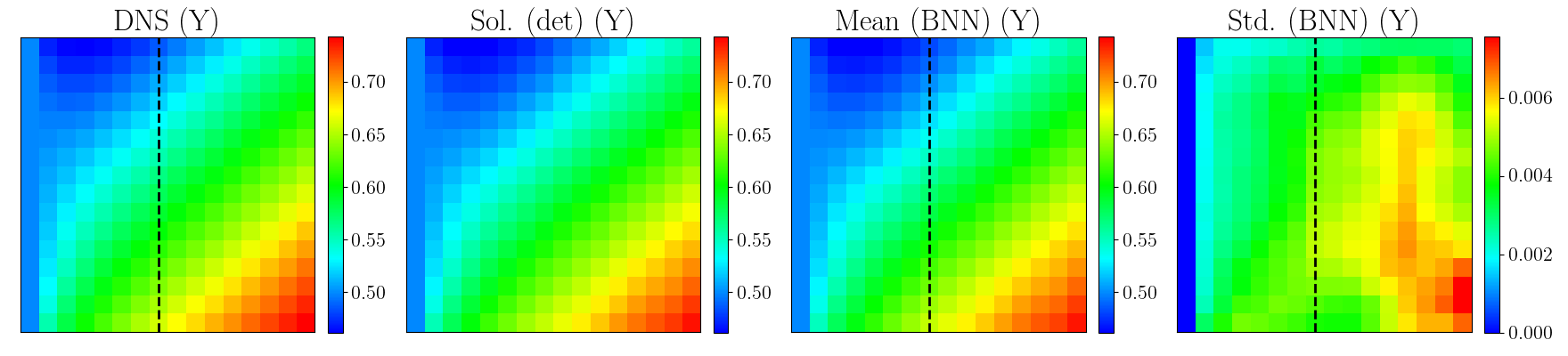}}
  {\includegraphics[height=0.08\linewidth]{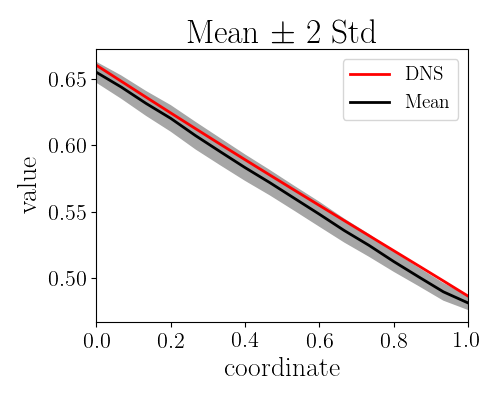}} \\
  {\includegraphics[height=0.08\linewidth]{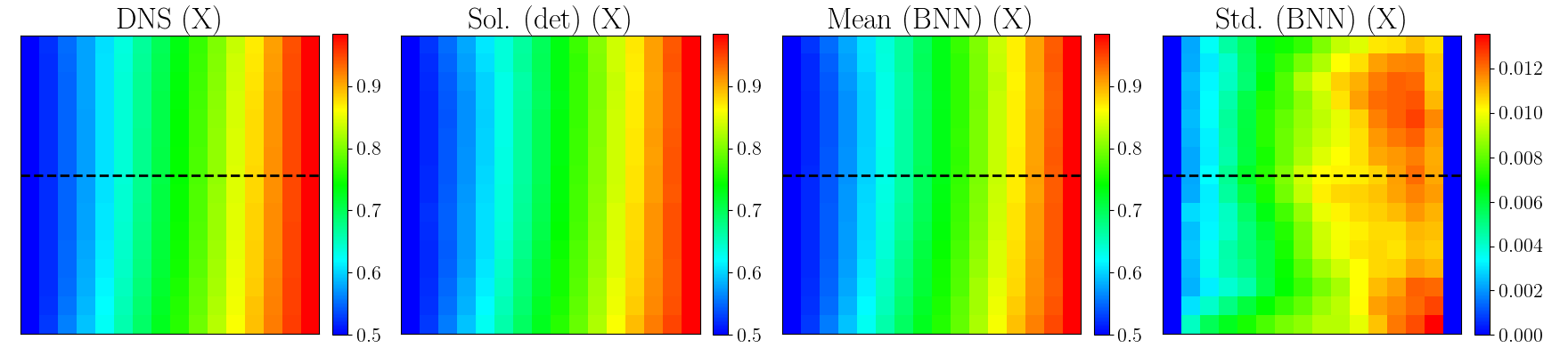}}
  {\includegraphics[height=0.08\linewidth]{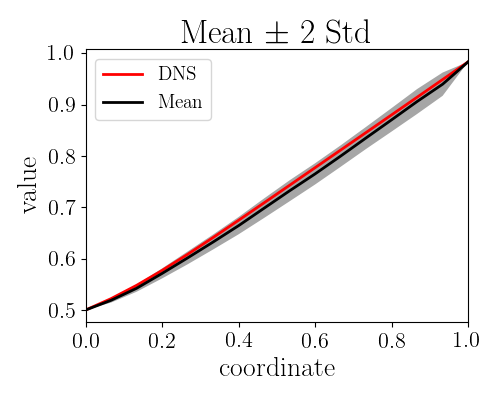}} 
  {\includegraphics[height=0.08\linewidth]{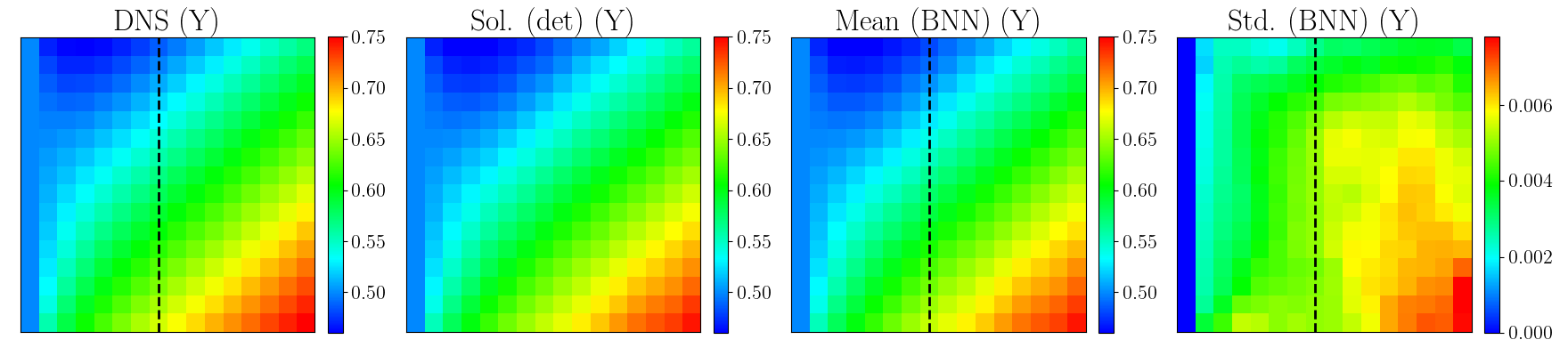}}
  {\includegraphics[height=0.08\linewidth]{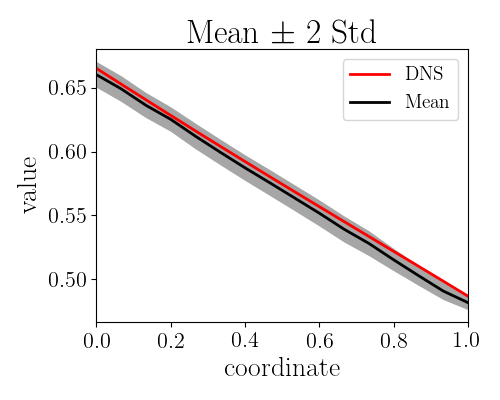}} \\
  \caption{Additional extrapolating NN prediction results for case (i).}
  \label{fig:nonlinear-extra-additional}
\end{figure}

Additional interpolated and extrapolated NN prediction results for case (i) are given in Fig. \ref{fig:nonlinear-inter-additional} and \ref{fig:nonlinear-extra-additional}, respectively.

\bibliographystyle{unsrt} % numbered
\bibliography{lib.bib}
\end{document}